%% file: draft-anec-ope.tex
\documentclass[11pt, oneside]{article}   	

\usepackage{jheppub}

\usepackage{amsmath}
\usepackage{amssymb}
\usepackage{amsfonts}
\usepackage{amsthm}
\usepackage{amsbsy}
\usepackage{array}

\usepackage[OT2,OT1]{fontenc}
\newcommand\cyr
{
\renewcommand\rmdefault{wncyr}
\renewcommand\sfdefault{wncyss}
\renewcommand\encodingdefault{OT2}
\normalfont
\selectfont
}
\DeclareTextFontCommand{\textcyr}{\cyr}
\def\cprime{\char"7E }
\def\cdprime{\char"7F }
\def\eoborotnoye{\char'013}

\usepackage{mathtools}
\usepackage[usenames,dvipsnames]{xcolor}

\usepackage{subcaption}

\usepackage{dsdshorthand}

\usepackage{graphicx}

\usepackage[vcentermath]{youngtab}
\newcommand{\myng}[1]{\,{\tiny\yng #1}\,}

\newcommand\wL{\mathbf{L}}

\renewcommand\vol{\mathop{\mathrm{vol}}}

\usepackage{tikzit}
\input{harmonic_analysis.tikzstyles}
\input{harmonic_analysis.tikzdefs}
\input{contour_integrals.tikzstyles}
\input{styles.tikzstyles}

\newcommand{\tsym}{\cT}
\newcommand{\la}{\langle}
\newcommand{\ra}{\rangle}
\newcommand{\pa}{\partial}
\renewcommand\bm{\mathbf{m}}

\newcommand\ruone{{  \!\!\!\!\!\!\!\!\!\!\!\textrm{\normalsize \cyr \eoborotnoye}}}
\newcommand\rutwo{{  \!\!\!\!\!\!\!\!\!\!\!\textrm{\normalsize \cyr \cprime}}}
\newcommand\ruthree{{\!\!\!\!\!\!\!\!\!\!\!\!\textrm{\normalsize \cyr \cdprime}}}

\newcommand*\link[1]{\hspace*{0em plus 1fill}\makebox{#1}}

\definecolor{energycolor}{RGB}{230,50,10}

\tikzset{
  energy/.style={->,
  energycolor,
  decoration={
      snake,
      amplitude=1pt,
      segment length=6pt,
      post length=1pt
    },
  decorate
  }
}

\makeatletter
\def\@fpheader{\ }
\makeatother

\title{The light-ray OPE and conformal colliders}
\author{Murat Kolo\u{g}lu$^\ruone$, Petr Kravchuk$^\rutwo$, David Simmons-Duffin$^\ruone$, and Alexander Zhiboedov$^\ruthree$}
\affiliation{${}^\ruone$Walter Burke Institute for Theoretical Physics, Caltech, Pasadena, California 91125, USA \\
${}^\rutwo$School of Natural Sciences, Institute for Advanced Study, Princeton, New Jersey 08540, USA \\
${}^\ruthree$CERN, Theoretical Physics Department, 1211 Geneva 23, Switzerland
}

\date{}
\abstract{We derive a nonperturbative, convergent operator product expansion (OPE) for null-integrated operators on the same null plane in a CFT. The objects appearing in the expansion are light-ray operators, whose matrix elements can be computed by the generalized Lorentzian inversion formula. For example, a product of average null energy (ANEC) operators has an expansion in the light-ray operators that appear in the stress-tensor OPE. An important application is to collider event shapes. The light-ray OPE gives a nonperturbative expansion for event shapes in special functions that we call celestial blocks. As an example, we apply the celestial block expansion to energy-energy correlators in $\cN=4$ Super Yang-Mills theory. Using known OPE data, we find perfect agreement with previous results both at weak and strong coupling, and make new predictions at weak coupling through $4$ loops (NNNLO).}

\preprint{CALT-TH 2019-013 \\
\link{CERN-TH-2019-055}}

\begin{document}

\maketitle
\pagenumbering{roman}
\setcounter{page}{2}
\newpage
\pagenumbering{arabic}
\setcounter{page}{1}

\section{Introduction}

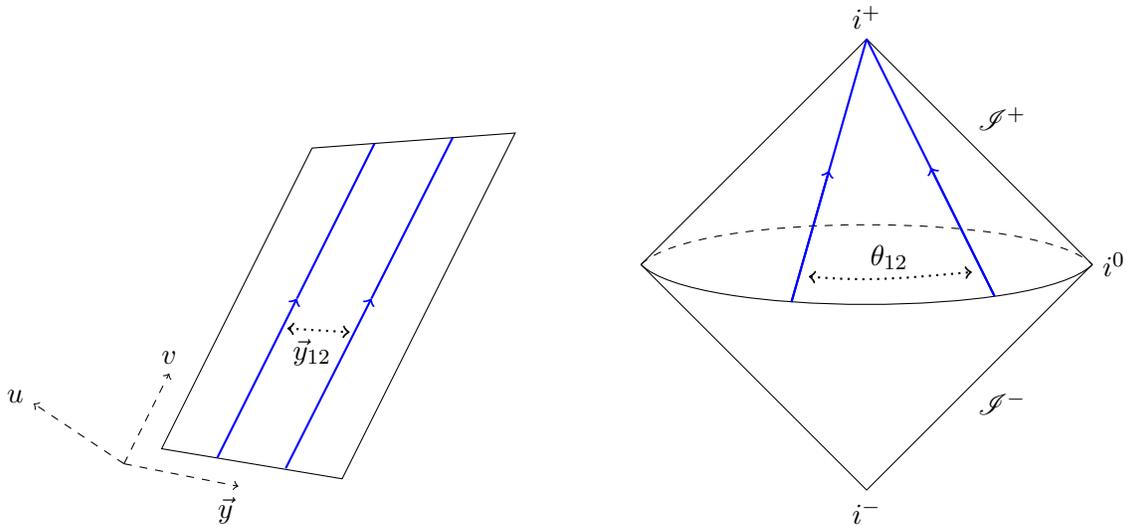
\begin{figure}[t!]
\centering
\begin{subfigure}[t]{0.45\textwidth}
\begin{tikzpicture}
	\draw [] (-2.3,-2) -- (-0.3,2) -- (2.4,2.2) -- (0.1,-2.4) -- cycle;
	\draw [->,thick,blue] (-1.56,-2.12) -- (-0.5,0);
	\draw [thick,blue] (-0.5,0) -- (0.53,2.06);
	\draw [->,thick,blue] (-0.65,-2.26) -- (0.49,0);
	\draw [thick,blue] (0.49,0) -- (1.57,2.14);
	\draw [->,dashed] (-2.8,-2.2) -- (-2.2,-1.0);
	\draw [->,dashed] (-2.8,-2.2) -- (-4.0,-1.4);
	\draw [->,dashed] (-2.8,-2.2) -- (-1.28,-2.5);
	\node [above] at (-2.2,-1.0) {$v$};
	\node [left] at (-4.0,-1.3) {$u$};
	\node [above] at (-1.45,-3.13) {$\vec y$};
	\node [below] at (-.3,-0.4) {$\vec y_{12}$};
	\draw [<->,dotted,thick] (-0.63,-0.4) -- (0.2,-0.45);
\end{tikzpicture}
\end{subfigure}
\hfill
\begin{subfigure}[t]{0.45\textwidth}
\begin{tikzpicture}
	\draw [] (-3,0) -- (0,3) -- (3,0) -- (0,-3) -- cycle;
	\draw [] (-3,0) to[out=-45,in=-135,distance=1.0cm] (3,0);
	\draw [dashed] (-3,0) to[out=45,in=135,distance=1.0cm] (3,0);
	\draw [thick,blue] (-1,-0.5) -- (0,3);
	\draw [thick,blue] (1.7,-0.42) -- (0,3);
	\draw [->,thick, blue] (-1,-0.5) -- (-0.5,1.25);
	\draw [->,thick, blue] (1.7,-0.42) -- (0.85,1.29);
	\draw [<->,dotted,thick] (-0.77,-0.16) to[out=-5,in=-175] (1.4,-0.1);
	\node [above] at (0.3,-0.2) {$\theta_{12}$};
	\node [above] at (1.8,1.65) {$\mathscr{I}^+$};
	\node [below] at (1.8,-1.55) {$\mathscr{I}^-$};
	\node [above] at (0,3) {$i^+$};
	\node [below] at (0,-3) {$i^-$};
	\node [right] at (3,0) {$i^0$};
\end{tikzpicture}
\end{subfigure}
\caption{The local operators $\cO_1$ and $\cO_2$ are integrated along parallel null lines (blue) on the same null plane.
On the left, we show a conformal frame where the null plane is $u=0$, and the operators are at different transverse positions $\vec y_1,\vec y_2\in \R^{d-2}$. 
On the right, we show a conformal frame where the null plane is future null infinity $\mathscr{I}^+$ and the null-integrated operators are separated by an angle $\th_{12}$ on the celestial sphere. We give the relationship between $\th_{12}$ and $\vec y_{12}$ in (\ref{eq:zetacrossratio}). Note that the entire circle at spatial infinity is really a single point $i^0$. Thus, the operators become coincident at the beginnings and ends of their integration contours.
}
\label{fig:nullplanepicture}
\end{figure}

In this work, we study a product of null-integrated operators on the same null plane in a conformal field theory (CFT) in $d>2$ dimensions (figure~\ref{fig:nullplanepicture}):
\be
\label{eq:productofops}
\int_{-\oo}^\oo dv_1\, \cO_{1;v\cdots v}(u=0,v_1,\vec y_1) \int_{-\oo}^\oo dv_2\, \cO_{2;v\cdots v}(u=0,v_2,\vec y_2).
\ee
Here, we use lightcone coordinates
\be
ds^2 = -du\, dv + \vec y^2,\qquad \vec y \in \R^{d-2}.
\ee
The operators are located at different transverse positions $\vec y_1,\vec y_2 \in \R^{d-2}$, and their spin indices are aligned with the direction of integration (the $v$ direction). As an example, when $\cO_1$ and $\cO_2$ are stress-tensors, (\ref{eq:productofops}) is a product of average null energy (ANEC) operators. In \cite{Kologlu:2019bco}, we established sufficient conditions for the existence of the product (\ref{eq:productofops}).

Such null-integrated operators arise naturally in ``event shape" observables in collider physics \cite{Basham:1977iq,Basham:1978zq,Basham:1978bw,Hofman:2008ar,Belitsky:2013xxa}. They also appear in shape variations of information-theoretic quantities in quantum field theory \cite{Faulkner:2016mzt,Casini:2017roe,Ceyhan:2018zfg}, as generators of asymptotic symmetry groups \cite{Cordova:2018ygx}, and in studies of positivity and causality \cite{Hofman:2009ug,Hartman:2016lgu,Chowdhury:2017vel,Cordova:2017dhq,Cordova:2017zej,Delacretaz:2018cfk,Meltzer:2018tnm,Kologlu:2019bco,Belin:2019mnx}. We review event shapes and null-integrated operators in section~\ref{sec:kinematics}.

Each null-integrated operator is pointlike in the transverse plane $\R^{d-2}$, so it is natural to ask whether there exists an operator product expansion (OPE) describing the behavior of the product (\ref{eq:productofops}) at small $|\vec y_{12}|$: 
\be
\label{eq:transverseope}
\int_{-\oo}^\oo dv_1\, \cO_{1;v\cdots v}(u=0,v_1,\vec y_1) \int_{-\oo}^\oo dv_2\, \cO_{2;v\cdots v}(u=0,v_2,\vec y_2) &\stackrel{?}{=} \sum_i |\vec y_{12}|^{\de_i - (\De_1-1) - (\De_2-1)} \mathbb{O}_i.
\ee
Here, the objects $\mathbb{O}_i$ have dimensions $\de_i$ and the powers of $|\vec y_{12}|$ are fixed by dimensional analysis.

The OPE for {\it local} operators is a powerful tool in CFT. It allows one to compute correlation functions and to formulate the bootstrap equations \cite{Ferrara:1973yt,Polyakov:1974gs}. A similar OPE for null-integrated operators (\ref{eq:productofops}) could have myriad applications.
Thus, we would like to ask whether (\ref{eq:transverseope}) exists, whether it is convergent or asymptotic, and what the objects $\mathbb{O}_i$ are.

Hofman and Maldacena analyzed this question in $\cN=4$ SYM and found the leading terms in the small-$|\vec y_{12}|$ expansion where $\cO_1,\cO_2$ are stress tensors and currents \cite{Hofman:2008ar}. At weak-coupling, the leading objects are certain integrated Wilson-line operators. At strong coupling, the leading objects can be described using string theory in AdS: they are certain stringy shockwave backgrounds. What is the analog of these results in a general nonperturbative CFT? Can we extend the leading terms to a systematic convergent expansion?

There is a simple and beautiful argument for the existence of an OPE of {\it local} operators in a nonperturbative CFT (see e.g.\ \cite{polchinski_1998}): Consider a pair of local operators $\cO_1,\cO_2$ in Euclidean signature. We surround the operators with a sphere $S^{d-1}$ (assuming all other operator insertions are outside the sphere) and perform the path integral inside the sphere. This produces a state $|\Psi\>$ on the sphere. In a scale-invariant theory, $|\Psi\>$ can be expanded in dilatation eigenstates
\be
\label{eq:opeargument}
\cO_1\cO_2|0\> = |\Psi\> = \sum_i |\cO_i\>.
\ee
By the state-operator correspondence, these eigenstates are equivalent to insertions of local operators at the origin $|\cO_i\> = \cO_i(0)|0\>$. Thus (\ref{eq:opeargument}) is the desired OPE.

Unfortunately, this argument does not work for the product (\ref{eq:productofops}). There is no obvious way to surround the null-integrated operators with an $S^{d-1}$ such that other operators are outside the sphere. The structure of (\ref{eq:transverseope}) suggests that perhaps we should surround the null-integrated operators with an $S^{d-3}$ in the transverse space $\R^{d-2}$. However there is no obvious Hilbert space of states associated with such an $S^{d-3}$.\footnote{An older argument for the existence of the OPE exists due to Mack \cite{Mack:1976pa}, relying on very different methods. Mack shows that a product of operators acting on the vacuum $\cO_1\cO_2|\Omega\>$ can be expanded in a sum of single operators acting on the vacuum $\sum_i \cO_i|\Omega\>$. However, this result is insufficient for our purposes. One reason is that acting with (\ref{eq:productofops}) on the vacuum immediately gives zero (as we will review shortly). Instead, we would like to act on nontrivial states, and then the theorem of \cite{Mack:1976pa} does not apply.}

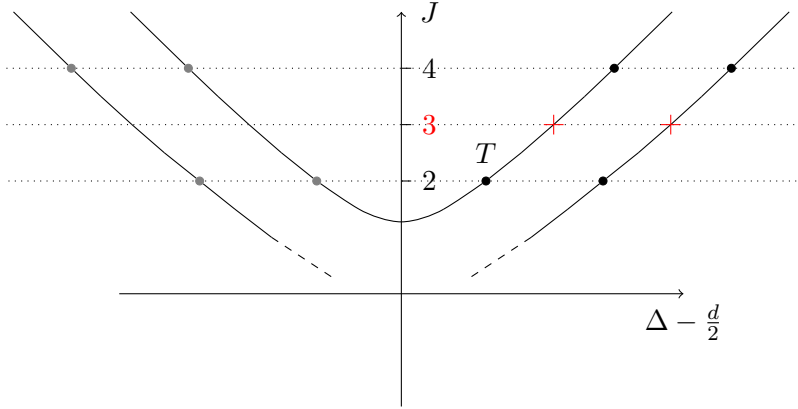
\begin{figure}[t]
	\centering
	{\tikzset{tikzfig/.append style = {scale=1.5}}
	\input{C_F_plot_3_anomalous.tikz}
	}
	\caption{Chew-Frautschi plot of neutral even-spin operators. Local operators are denoted by black dots, gray dots denote shadow operators. Solid lines represent Regge trajectories. The low transverse spin terms in the OPE $\int dv T_{vv}\times\int dv T_{vv}$ are spin-3 light-ray operators on even-spin Regge trajectories, shown here by red crosses.}
	\label{fig:CFplot}
\end{figure}

Nevertheless, using different technology, we will show that a convergent OPE (\ref{eq:transverseope}) for null-integrated operators does exist in a general nonperturbative CFT. The objects appearing on the right-hand side are the light-ray operators $\mathbb{O}_{i,J}^\pm$ defined in \cite{Kravchuk:2018htv}.  Each $\mathbb{O}_{i,J}^\pm$ is obtained by smearing a pair of local operators in a special way in the neighborhood of a light-ray. We review this construction in section~\ref{sec:lightrayreview}. The matrix elements of $\mathbb{O}_{i,J}^\pm$ can be computed via a generalization of Caron-Huot's Lorentzian inversion formula \cite{Caron-Huot:2017vep,Kravchuk:2018htv}. The spectrum of operators $\mathbb{O}^\pm_{i,J}$ is related to the spectrum of local operators by analytic continuation in spin $J$; $i$ labels different Regge trajectories.  In this work, we focus on contributions to the OPE with low spin in the transverse $d-2$-dimensional space (defined in more detail below).\footnote{The generic transverse spin contributions will be derived in \cite{Volume3}.} These contributions are given by operators $\mathbb{O}_{i,J}^\pm$ with spin $J=J_1+J_2-1$. For example, if $\cO_1=\cO_2=T$, then $J=3$ and the low-transverse spin terms are given by the operators $\mathbb{O}_{i,3}^+$, see figure~\ref{fig:CFplot}. Note that thanks to the selection rules, these low-transverse spin contributions are sufficient to compute matrix elements of~\eqref{eq:productofops} between appropriate low-spin states.\footnote{For example, in the case of $\cO_1=\cO_2=T$, interpreted as energy detectors in a conformal collider~\cite{Hofman:2008ar}, the low-transverse contributions described in this paper are sufficient to compute the matrix elements of~\eqref{eq:productofops} between momentum eigenstates with total spin less than or equal to $4$. This includes, in particular, polarization-averaged expectation values, in which case the total spin of initial and final states is zero.}

Our strategy to establish the OPE (\ref{eq:transverseope}) is as follows. First, in section~\ref{sec:harmoniccelestial} we decompose the left-hand side of (\ref{eq:transverseope}) into conformal irreps by smearing the transverse coordinates $\vec y_1,\vec y_2$, using harmonic analysis for the transverse conformal group $\SO(d-1,1)$. In section~\ref{sec:focusonirrep}, we focus on a single irrep and compute its matrix elements. Such matrix elements can be written in terms of an integral of a double commutator. After some manipulation, we express this integral as a linear combination of the generalized Lorentzian inversion formula of \cite{Kravchuk:2018htv}, i.e.\ as a sum of matrix elements of $\mathbb{O}_{i,J}^\pm$'s. Thus, the original product of operators is a sum of $\mathbb{O}_{i,J}^\pm$'s.

As an example, consider the case where $\cO_1=\phi_1$ and $\cO_2=\phi_2$ are scalars, so that $J_1+J_2-1=-1$.\footnote{According to the analysis of \cite{Kologlu:2019bco}, a product of null-integrated scalars is only well-defined in theories with Regge intercept $J_0<-1$. Here, we assume this is the case.} Following the procedure above, we find the OPE\footnote{A more precise expression involves an integral over $\De$ instead of a sum over Regge trajectories. The $\De$ contour can be deformed to pick up singularities in the $\De$ plane. When these singularities are isolated poles, we arrive at the sum of Regge trajectories (\ref{eq:scalar light transform ope intro}). We discuss these points in section~\ref{sec:contour}.}
\be
\label{eq:scalar light transform ope intro}
\int_{-\oo}^\oo dv_1\, \f_{1}(0,v_1,\vec y_1) \int_{-\oo}^\oo dv_2\, \f_{2}(0,v_2,\vec y_2) 
&= \pi i  
\sum_{i}
\cC_{\De_i-1}(\vec y_{12},\ptl_{\vec y_2}) \p{\mathbb{O}^+_{i,-1}(\vec y_2)+\mathbb{O}^-_{i,-1}(\vec y_2)} \nn\\
&\quad + \textrm{nonzero transverse spin}.
\ee
Here, $\cC_{\de}(\vec y,\ptl_{\vec y})$ is the same differential operator that appears in an OPE of local primary scalars in $d-2$ dimensions. It has an expansion
\be
\cC_{\de}(\vec y,\ptl_{\vec y}) &= |\vec y|^{\de - (\De_1-1) - (\De_2-1)}\p{1+ \frac{\De_1-\De_2+\de}{2\de} \vec y\.\ptl_{\vec y} + \dots},
\ee
where the coefficients are fixed by $(d{-}2)$-dimensional conformal invariance.
The objects $\mathbb{O}^\pm_{i,-1}$ are light-ray operators associated to the $i$-th Regge trajectory, evaluated at spin $J=-1$. The superscript $\pm$ is called the ``signature" and it indicates the transformation properties of the light-ray operator under a combination of $\mathsf{CRT}$ and Hermitian conjugation. The ``nonzero transverse spin" terms consist of derivatives of higher-spin light-ray operators $\mathbb{O}_{i,J}$ with $J>-1$, and will be explained in \cite{Volume3}.

In section~\ref{sec:generalizationandcelestialmap}, we generalize (\ref{eq:scalar light transform ope intro}) to arbitrary Lorentz representations for $\cO_1,\cO_2$. The light-ray operators on the right-hand side have spin $J=J_1+J_2-1$, where $J_1,J_2$ are the spins of $\cO_1,\cO_2$.\footnote{For the definition of $J$ in general representations, see appendix~\ref{sec:orthogonalreps}.} For example, when $\cO_1,\cO_2$ are the stress tensor, we have a sum of spin-3 light-ray operators 
\be
\label{eq:stresstensorexample}
\int dv_1 T_{vv}(0,v_1,\vec y_1) \int dv_2 T_{vv}(0,v_2,\vec y_2) &= \pi i\sum_{s=\pm}\sum_{\l,a}\sum_i \cD^{(a),s}_{\De_i-1,\l}(\vec y_{12},\ptl_{\vec y_2}) \mathbb{O}^s_{i,J=3,\l,(a)}(\vec y_2) \nn\\
&\quad + \textrm{higher transverse spin}.
\ee
Here, $\l$ is an $\SO(d-2)$ representation encoding spin in the transverse plane, $s=\pm$ is a signature, $(a)$ labels conformally-invariant three-point structures, and $\cD^{(a),s}_{\de,\l}$ is a differential operator that generalizes $\cC_\de$.

In equation (\ref{eq:stresstensorexample}), the representation $\l$ is restricted to traceless-symmetric transverse spins that can appear in the conventional OPE of local operators $T_{\mu\nu} \x T_{\rho\sigma}$. Specifically, these are the representations $\l\in\{\bullet,\myng{(1)},\myng{(2)},\myng{(3)},\myng{(4)}\}$. The ``higher transverse spin" terms refer to operators with transverse representation $\l$ not included in this set. They will be explained in \cite{Volume3}. Higher transverse spin terms do not contribute to any of the example observables we study in section~\ref{sec:nequalsfour}.

In section~\ref{sec:contactterms} we find that the light-ray OPE also carries information about contact terms in the $\vec y_1\to \vec y_2$ limit. These contact terms are important in at least two aspects. First they are a part of the physical information present in event shape observables. Second, they arise in commutators of null-integrated operators~\cite{Cordova:2018ygx}, leading to an interesting algebra.


\subsection{Commutators and superconvergence}

Our analysis does not assume or require that null-integrated operators commute. Indeed, we can write an expression for a commutator of null-integrated operators using the OPE. For example, the commutator of ANEC operators is given by the odd-signature terms in (\ref{eq:stresstensorexample}),
\be
\label{eq:stresscommutator}
\left[\int dv_1 T_{vv}(0,v_1,\vec y_1), \int dv_2 T_{vv}(0,v_2,\vec y_2)\right] &= \pi i \sum_{\l,a}\sum_i \cD^{(a),-}_{\De_i-1,\l}(\vec y_{12},\ptl_{\vec y_2}) \mathbb{O}^-_{i,J=3,\l,(a)}(\vec y_2) \nn\\
&\quad + \textrm{higher transverse spin}.
\ee
In \cite{Kologlu:2019bco}, we showed that a commutator of ANEC operators vanishes if $J_0 < 3$, where $J_0$ is the Regge intercept of the theory, and furthermore $J_0\leq 1$ in unitary CFTs. It is interesting to understand how vanishing occurs on the right-hand side of (\ref{eq:stresscommutator}). Note that the operators on the right-hand side are light-ray operators with spin 3 and odd signature. We show in section~\ref{sec:commutativity} that if $J_0<3$, such operators must be null integrals of local spin-3 operators.\footnote{This justifies an assumption made in \cite{Cordova:2018ygx}.}  However, local spin-3 operators are not allowed in the $T\x T$ OPE by conservation conditions and Ward identities \cite{Kravchuk:2016qvl}.\footnote{Similar arguments apply to the higher-transverse spin terms, which are derivatives of light-ray operators with odd-integer $J>3$ \cite{Volume3}.} Thus, the commutator vanishes.

As we explain in section~\ref{sec:commutativity}, this argument generalizes to establish vanishing of a commutator of null-integrated operators whenever $J_1+J_2>J_0+1$. It turns out that even if local operators with signature $(-1)^{J_1+J_2-1}$ and spin $J_1+J_2-1$ do appear in the local $\cO_1\x\cO_2$ OPE, they do not survive in the light-ray OPE. This provides another (somewhat circuitous) way to derive the commutativity conditions of \cite{Kologlu:2019bco}. An exception can occur at vanishing transverse separation $\vec y_{12}=0$. In that case, the commutator may contain contact terms, which can be computed by our light-ray OPE formula. As an example, in section~\ref{sec:chargecommutator}, we describe how to compute contact terms in a commutator of null-integrated nonabelian currents (assuming $J_0 < 1$), reproducing results of \cite{Cordova:2018ygx}.

Vanishing of the commutator of ANEC operators means that the odd-signature terms in (\ref{eq:stresstensorexample}) disappear, and the OPE of ANEC operators can be simplified to a sum of even-signature light-ray operators.

Despite the fact that local spin-3 operators are not allowed in the $T\x T$ OPE, we can try to compute their OPE data with the Lorentzian inversion formula. This is equivalent to evaluating matrix elements of the right-hand side of (\ref{eq:stresscommutator}). The result must be zero. However, if we plug the OPE in a different channel (the ``$t$-channel") into the inversion formula, we obtain sums that are not zero term by term. The conditions that these sums vanish are precisely the ``superconvergence" sum rules of \cite{Kologlu:2019bco}. As we explain in section~\ref{sec:superconvergenceinnuspace}, in this language it is simple to argue that (suitably-smeared) superconvergence sum rules have a convergent expansion in $t$-channel blocks.

\subsection{Celestial blocks and event shapes}

An important application of the light-ray OPE is to event shapes \cite{Basham:1977iq,Basham:1978zq,Basham:1978bw,Hofman:2008ar,Belitsky:2013xxa}. For example, to compute a two-point event shape, we place a pair of null-integrated operators (``detectors") along future null infinity (right half of figure~\ref{fig:nullplanepicture}) and evaluate a matrix element in a momentum eigenstate $|\cO(p)\>$. By applying the OPE (\ref{eq:scalar light transform ope intro}), we obtain a sum of matrix elements of individual light-ray operators $\mathbb{O}_{i,J}^\pm$ in momentum eigenstates $|\cO(p)\>$,
\be
\label{eq:precelestialblock}
\cC_{\De_i-1}(\vec y_{12},\ptl_{\vec y_2})\<\cO(p)|\mathbb{O}_{i,J}^\pm(\vec y_2) |\cO(p)\>.
\ee
The quantity (\ref{eq:precelestialblock}) is fixed by conformal symmetry up to a constant. It plays a role for event shapes analogous to the role that conformal blocks play in the usual OPE expansion of local 4-point functions. It is proportional to a function of a single cross-ratio
\be
\label{eq:zetacrossratio}
\z &= \frac{1-\cos\theta_{12}}{2} = \frac{\vec y_{12}^{\,2}}{(1+\vec y_1^{\,2})(1+\vec y_2^{\,2})} \in [0,1],
\ee
where $\th_{12}$ is the angle between detectors on the celestial sphere. We have also written $\z$ in terms of the transverse positions $\vec y_{1},\vec y_2$ in the conventions of \cite{Hofman:2008ar}. In an event shape, $\z\to 0$ is the collinear limit, while $\z\to 1$ corresponds to back-to-back detectors. We call (\ref{eq:precelestialblock}) a ``celestial block." 

In section~\ref{sec:celestialblocks}, we compute celestial blocks by solving an appropriate conformal Casimir equation. For example, when $\cO$ is a scalar, the result is\footnote{Celestial blocks are an analytic continuation of the boundary conformal blocks studied in \cite{McAvity:1995zd,Liendo:2012hy}.}
\be
\label{eq:celestialblockfunction}
f_{\De}^{\Delta_1, \Delta_2}(\z) &= \z^{\frac{\De-\Delta_1 -\Delta_2+1}{2}}{}_2F_1\p{\frac{\De-1+\Delta_1 - \Delta_2}{2},\frac{\De-1-\Delta_1 + \Delta_2}{2},\De+1-\frac{d}{2},\z}.
\ee
Note that $f_{\De}^{\De_1,\De_2}$ becomes a pure power $\z^{\frac{\De-\Delta_1 -\Delta_2+1}{2}}$ in the collinear limit $\z\to 0$.

The light-ray OPE thus yields an expansion for two-point event shapes in celestial blocks. For example, using (\ref{eq:scalar light transform ope intro}) and superconformal symmetry \cite{Belitsky:2014zha, Korchemsky:2015ssa}, an energy-energy correlator (EEC) in $\cN=4$ SYM can be written as
\be
\<\cE(\vec n_1) \cE(\vec n_2)\>_{\psi(p)} &= \frac{(p^0)^2}{8\pi^2} \cF_{\cE}(\zeta) ,\nn \\
\cF_{\cE}(\z) &=\sum_i p_{\Delta_i}  \frac{4 \pi^4 \Gamma(\Delta_i -2)}{ \Gamma(\frac{\Delta_i -1 }{2})^3 \Gamma(\frac{3-\Delta_i }{ 2})} f_{\Delta_i}^{4,4} (\z) + \frac{1}{4}(2\de(\z)-\de'(\z)), \label{eq:finCouplIntro}
\ee
where $\De_i$ runs over dimensions of Regge trajectories at spin $J=-1$, and $p_{\De_i}$ are squared OPE coefficients of operators in the $\mathbf{105}$ representation of $\SO(6)$ in the $\cO_{\mathbf{20'}} \times \cO_{\mathbf{20'}}$ OPE, analytically continued to spin $J=-1$. 
The state $\psi(p)$ carries momentum $p=(p^0,0,0,0)$ and is created by acting with an $\cO_\mathbf{20'}$ operator on the vacuum. The angle between energy detectors is  $\cos\th=\vec n_1 \. \vec n_2$, and $\zeta$ is defined by (\ref{eq:zetacrossratio}). The coupling-independent contact terms $\frac{1}{4}(2\de(\z)-\de'(\z))$ are related to the contribution of protected operators to the EEC.

Thus, (\ref{eq:finCouplIntro}) expresses the EEC in $\cN=4$ SYM in terms of OPE data.
This formula holds nonperturbatively in both the size of the gauge group $N_c$ and the 't Hooft coupling $\lambda$. In section~\ref{sec:nequalsfour}, we check it against previous results at weak and strong coupling and find perfect agreement. Using known results for leading-twist OPE data in $\cN=4$ SYM, we use (\ref{eq:finCouplIntro}) to make new predictions for the small-angle limit of $\cN=4$ energy-energy correlators through 4 loops (NNNLO).

We conclude in section~\ref{sec:discussion} with discussion and future directions. In appendix~\ref{app:notation} we summarize our notation, in appendix~\ref{sec:orthogonalreps} we review general representations of orthogonal groups, and in appendix~\ref{app:moreonanalyticcont} we clarify some points about analytic continuation in spin. Appendices~\ref{sec:triple light transform},~\ref{app:swapping} and~\ref{app:z1contacts} contain details of the calculations described in the main text.

\paragraph{Note added:} During the last stages of this work we learned about \cite{BacktoBackGrisha} and \cite{LanceFuture} which have some overlap with our analysis. Let us briefly describe the results of \cite{BacktoBackGrisha} and \cite{LanceFuture} in relation to our work. 

In \cite{BacktoBackGrisha} the EEC in $\cN=4$ SYM was analyzed using the Mellin space approach of \cite{Belitsky:2013xxa}. We analyze $\cN=4$ SYM  in section \ref{sec:nequalsfour}. It was shown in \cite{BacktoBackGrisha} how the back-to-back $\z \to 1$ limit of the EEC is captured by the double light-cone limit of the correlation function studied in \cite{Alday:2010zy}. It led to the derivation of (\ref{eq:btb}) and identification of the coefficient function $H(a)$ with a certain spin-independent part of the three-point functions of large spin twist-2 operators. We do not analyze $\z \to 1$ limit of the EEC in a generic CFT in the present paper. Similarly, a leading small angle asymptotic of the EEC in $\cN=4$ SYM, the small $\z$ limit of (\ref{eq:fincoupltw2}), was rederived in \cite{BacktoBackGrisha}.\footnote{We reported (\ref{eq:fincoupltw2}) to G.~Korchemsky in September 2018.} Based on  (\ref{eq:fincoupltw2}), the four-loop small angle asymptotic was worked out in \cite{BacktoBackGrisha}, we do it in section \ref{sec:finitecoupling}. This represents the leading small-angle asymptotic of our complete, non-perturbative OPE formula (\ref{eq:finCouplIntro}). 

In \cite{LanceFuture} a factorization formula describing the small $\z \to 0$ limit for the EEC was derived in a generic massless QFT, conformal or asymptotically free, in terms of the time-like data of the theory. The authors \cite{LanceFuture} applied their results to QCD, $\cN=1$, and $\cN=4$ SYM, in particular they analyzed the effects of a non-zero $\beta$-function which goes beyond our considerations in the present paper.  In the conformal case of $\cN=4$ SYM which is relevant to our analysis, the leading small-angle asymptotic was derived in \cite{LanceFuture} through three loops.

In addition, both \cite{BacktoBackGrisha,LanceFuture} emphasized the importance of contact terms in the EEC (we compute these using the OPE in section \ref{sec:contacttermsEE}), the way to compute them from the small angle and back-to-back limits, see appendix \ref{app:z1contacts}, and their importance to the Ward identities (\ref{eq:energyWI},\ref{eq:momentumWI}). In particular,  \cite{BacktoBackGrisha, LanceFuture} checked that the $\cN=4$ SYM NLO result \cite{Belitsky:2013ofa} satisfies Ward identities, we do this in section \ref{sec:twoloopscontact}. In \cite{BacktoBackGrisha} it was also checked that the NNLO result \cite{Henn:2019gkr} satisfies Ward identities, which we do in section \ref{sec:threeloopsN4}.

\section{Kinematics of light-ray operators and event shapes}
\label{sec:kinematics}

\subsection{Null integrals and symmetries}
\label{sec:boostselection}

Let us begin by examining the symmetries of a product of light-ray operators (\ref{eq:productofops}).  This analysis will already give a hint why the objects $\mathbb{O}_{i,J}^\pm$ appear in the OPE.

Firstly, consider a boost
\be
(u,v,\vec y) &\ \to\  (\l^{-1} u, \l v, \vec y),\quad \l\in \R^+.
\ee
Each null-integrated operator $\int dv_i\, \cO_{i;v\cdots v}$ has boost eigenvalue $1-J_i$, where $1$ comes from the measure $dv_i$ and $-J_i$ comes from the lowered $v$-indices. Thus, the product (\ref{eq:productofops}) has boost eigenvalue $(1-J_1)+(1-J_2) = 1-(J_1+J_2-1)$. In other words, it transforms like the null integral of an operator with spin $J_1+J_2-1$ \cite{Hofman:2008ar}.

Another important symmetry is $\mathsf{CRT}$, which is an anti-unitary symmetry taking
\be
(u,v,\vec y) &\ \to\  (-u,-v,\vec y).
\ee
Combining $\mathsf{CRT}$ with Hermitian conjugation, we obtain a linear map on the space of operators. It is easy to check that
\be
\p{(\mathsf{CRT}) \int_{-\oo}^\oo dv\, \cO_{i;v\cdots v}(0,v,\vec y) (\mathsf{CRT})^{-1}}^\dag &= (-1)^{J_i}\int_{-\oo}^\oo dv\, \cO_{i;v\cdots v}(0,v,\vec y).
\ee
We call the eigenvalue with respect to the combination of $\mathsf{CRT}$ and Hermitian conjugation the ``signature" of the operator. Applying $\mathsf{CRT}$ and Hermitian conjugation to (\ref{eq:productofops}), we find
\be
\label{eq:commutator}
&\left[\int_{-\oo}^\oo dv_1\, \cO_{1;v\cdots v}(0,v_1,\vec y_1), \int_{-\oo}^\oo dv_2\, \cO_{2;v\cdots v}(0,v_2,\vec y_2)\right] &&\textrm{has signature $(-1)^{J_1+J_2-1}$} \\
&\left\{\int_{-\oo}^\oo dv_1\, \cO_{1;v\cdots v}(0,v_1,\vec y_1), \int_{-\oo}^\oo dv_2\, \cO_{2;v\cdots v}(0,v_2,\vec y_2)\right\} &&\textrm{has signature $(-1)^{J_1+J_2}$},
\label{eq:anticommutator}
\ee
where $[\ ,\ ]$ and $\{\ ,\ \}$ denote a commutator and anticommutator, respectively. The extra minus sign in the commutator appears because Hermitian conjugation reverses operator ordering.

It often happens (under circumstances described in \cite{Kologlu:2019bco} and discussed in section~\ref{sec:commutativity}) that the commutator (\ref{eq:commutator}) vanishes. For example, a commutator of ANEC operators on the same null plane vanishes.
For simplicity, suppose that the commutator vanishes. In this case, the product (\ref{eq:productofops}) is the same as the anticommutator (\ref{eq:anticommutator}).
Thus, (\ref{eq:productofops}) transforms like the null-integral of an operator with spin $J_1+J_2-1$ and signature $(-1)^{J_1+J_2}$. An integrated local operator can never have these quantum numbers. This shows that the OPE (\ref{eq:transverseope}) cannot be computed by simply performing the usual OPE between $\cO_1$ and $\cO_2$ inside the integral.

\subsection{Review: embedding formalism and the Lorentzian cylinder}
\label{eq:reviewlorentziancyl}

It is instructive to re-derive the selection rule $J=J_1+J_2-1$ in a different way, using conformal transformation properties of null-integrated operators. These properties are easiest to understand in the embedding formalism \cite{Dirac:1936fq,Mack:1969rr,Boulware:1970ty,Ferrara:1973eg,Ferrara:1973yt,Cornalba:2009ax,Weinberg:2010fx,Costa:2011mg}.

In the embedding formalism, Minkowski space is realized as a subset of the projective null cone in $\R^{d,2}$. Let us choose coordinates $X=(X^+,X^-,X^\mu)=(X^+,X^-,X^0,\cdots, X^{d-1})$ on $\R^{d,2}$, with metric
\be
\label{eq:embeddingmetric}
X\.X &= -X^+ X^- -(X^0)^2 + (X^1)^2 +\cdots + (X^{d-1})^2.
\ee
The projective null cone is the locus $X\.X=0$, modulo positive rescalings $X\sim \l X$ $(\l\in \R_{+})$. This space is topologically $S^1 \x S^{d-1}$. Lorentzian CFTs live on the universal cover of the projective null cone $\tl{\cM}_d$, which is topologically $\R \x S^{d-1}$ --- sometimes called the Lorentzian cylinder. The conformal group $\tl{\SO}(d,2)$ is the universal cover of $\SO(d,2)$.

Minkowski space corresponds to the locus $X=(X^+,X^-,X^\mu)=(1,x^2,x^\mu)\in \R^{d,2}$, where $x\in \R^{d-1,1}$. Spatial infinity $i^0$ is obtained by taking $x\to \oo$ in a spacelike direction and rescaling $X$ so it remains finite, yielding $X_{i^0}=(0,1,0)$. Timelike infinity $i^\pm$ corresponds to $X_{i^\pm}=(0,-1,0)$. (Note that future and past infinity $i^\pm$ correspond to the same embedding-space vector, but they are distinguished on the universal cover of the projective null cone.) Finally, null infinity corresponds to the points $X_\mathrm{\mathscr{I}^\pm}(\s,z)=(0,-2\s,z)$, $z=(\pm 1,\vec n)$, where $\vec n \in S^{d-2}$ is a point on the celestial sphere and $\s$ is retarded time.

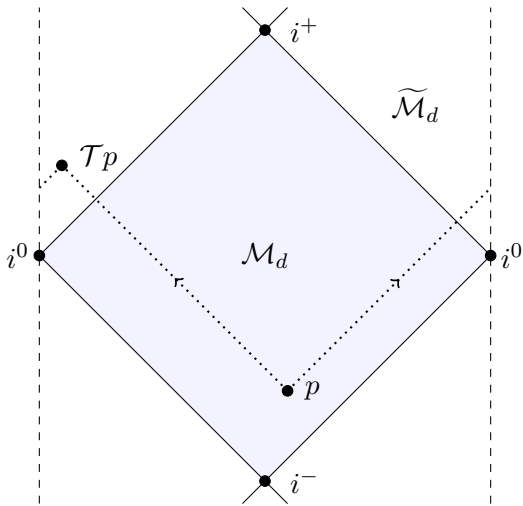
\begin{figure}[t!]
	\centering
	\begin{tikzpicture}

	\draw[fill=yellow, opacity = 0.05,blue] (3,0) -- (0,3) -- (-3,0) -- (0,-3) -- cycle;
	
	\draw[] (-3,0) -- (0,3) -- (3,0) -- (0,-3) -- cycle;
	\draw[] (0,3) -- (0.3,3.3);
	\draw[] (0,3) -- (-0.3,3.3);
	\draw[] (0,-3) -- (0.3,-3.3);
	\draw[] (0,-3) -- (-0.3,-3.3);
		
	\draw[dashed] (-3,-3.3) -- (-3,3.3);				
	\draw[dashed] (3,-3.3) -- (3,3.3);	
	
	\draw[fill=black] (-3,0) circle (0.07);
	\draw[fill=black] (3,0) circle (0.07);
	\draw[fill=black] (0,3) circle (0.07);
	\draw[fill=black] (0,-3) circle (0.07);
	\draw[fill=black] (0.3,-1.8) circle (0.07);
	\draw[fill=black] (-2.7,1.2) circle (0.07);
	\node[right] at (0.4,-1.8) {$p$};
	\node[right] at (-2.6,1.3) {$\cT p$};
\begin{scope}[dotted,thick,decoration={
    markings,
    mark=at position 0.5 with {\arrow{>}}}
    ]
	\draw[postaction={decorate}] (0.3,-1.8) -- (-2.7,1.2);
\end{scope}
\begin{scope}[dotted,thick,decoration={
    markings,
    mark=at position 0.55 with {\arrow{>}}}
    ]
	\draw[postaction={decorate}] (0.3,-1.8) -- (3,0.9);
\end{scope}
	\draw[dotted,thick] (-3,0.9) -- (-2.7,1.2);
		
	\node[left] at (-3,0) {$i^0$};
	\node[right] at (3,0) {$i^0$};
	\node[right] at (0.2,3) {$i^+$};
	\node[right] at (0.2,-3) {$i^-$};
	
	\node at (0,0) {$\cM_d$};
	
	\node at (2,2) {$\tl\cM_d$};

	\end{tikzpicture}
	\caption{Minkowski patch $\cM_d$ (blue, shaded) inside the Lorentzian cylinder $\tl\cM_d$ in the case of 2 dimensions. Spacelike infinity of $\cM_d$ is marked by $i^0$ and future/past infinity are marked by $i^\pm$. The dashed lines should be identified. The point $\cT p$ is obtained from $p$ by shooting light-rays in all possible future directions (dotted lines) and finding the first point where they converge.}
	\label{fig:lorentziancylinder}
\end{figure}

The Lorentzian cylinder $\tl \cM_d$ is tiled by Minkowski ``patches" (figure~\ref{fig:lorentziancylinder}). To every point $p\in \tl \cM_d$, there is an associated point $\cT p$ obtained by shooting light rays in all future directions from $p$ and finding the point where they converge in the next patch. In embedding coordinates, $\cT$ takes $X\to -X$. For example, $\cT$ takes spatial infinity $i^0$ to future infinity $i^+$. We sometimes write $p^+\equiv \cT p$ and $p^-\equiv \cT^{-1} p$.

To describe operators with spin, it is helpful to introduce index-free notation. Given a traceless symmetric tensor $\cO^{\mu_1\cdots\mu_J}(x)$, we can contract its indices with a  future-pointing null polarization vector $z_\mu$ to form
\be
\label{eq:indexfree}
\cO(x,z) &\equiv \cO^{\mu_1\cdots\mu_J}(x) z_{\mu_1}\cdots z_{\mu_J}.
\ee
When $\cO^{\mu_1\cdots\mu_J}(x)$ is an integer-spin local operator, $\cO(x,z)$ is a homogeneous polynomial of degree $J$.

In the embedding formalism, the operator $\cO(x,z)$ gets lifted to a homogeneous function $\cO(X,Z)$ of coordinates $X,Z\in \R^{d,2}$, subject to the relations $X^2 = X\.Z=Z^2 = 0$  \cite{Costa:2011mg}. It is defined by
\be
\label{eq:tofromflatdict}
\cO(X,Z) &= (X^+)^{-\De}\cO\p{x=\frac{X}{X^+},z=Z-\frac{Z^+}{X^+}X},
\ee
where $\De$ is the dimension of $\cO$.
The advantage of $\cO(X,Z)$ is that conformal transformations act linearly on the coordinates $X,Z$.
Note that $\cO(X,Z)$ has gauge invariance
\be
\label{eq:gaugeredunancyembedding}
\cO(X,Z) &= \cO(X,Z+\b X),
\ee
and homogeneity
\be
\label{eq:homogeneityembedding}
\cO(\l X,\a Z) = \l^{-\De} \a^J \cO(X,Z).
\ee
The operator $\cO(x,z)$ on $\R^{d-1,1}$ can be recovered by the dictionary
\be
\label{eq:tofromflatdict}
\cO(x,z) &= \cO\p{X=(1,x^2,x),Z=(0,2x\.z,z)}.
\ee

Index-free notation and the procedure of lifting operators to the embedding space can be generalized to arbitrary representations of the Lorentz group. We describe this construction in appendix~\ref{sec:orthogonalreps}.

\subsection{Review: the light transform}
\label{sec:lighttransformreview}

Null-integrated operators like those in (\ref{eq:productofops}) can be understood in terms of a conformally-invariant integral transform called the ``light-transform" \cite{Kravchuk:2018htv}. In embedding-space language, the light-transform is defined by
\be
\label{eq:definitionlighttransform}
\wL[\cO](X,Z) &\equiv \int_{-\oo}^\oo d\a\, \cO(Z-\a X, -X).
\ee
This transform is invariant under $\tl\SO(d,2)$ because (\ref{eq:definitionlighttransform}) only depends on the embedding-space vectors $X,Z$. It respects the gauge redundancy (\ref{eq:gaugeredunancyembedding}) because a shift $Z\to Z+\b X$ can be compensated by shifting $\a\to \a+\b$ in the integral. 
The initial point of the integration contour in (\ref{eq:definitionlighttransform}) is $X$, since $Z-(-\oo)X$ is projectively equivalent to $X$. Furthermore, if $\cO(X,Z)$ has homogeneity (\ref{eq:homogeneityembedding}), then its light-transform has homogeneity
\be
\label{eq:lighttransformhomogeneity}
\wL[\cO](\l X,\a Z) &= \l^{-(1-J)} \a^{1-\De} \wL[\cO](X,Z).
\ee
 Thus, $\wL[\cO]$ transforms like a primary at $X$ with dimension $1-J$ and spin $1-\De$:
\be
\label{eq:affineweyllight}
\wL:(\De,J) &\to (1-J,1-\De).
\ee
Note that the light-transform naturally gives rise to operators with non-integer spin. 

In Minkowski coordinates, $\wL$ becomes
\be
\label{eq:minkowskilighttransform}
\wL[\cO](x,z) &= \left.\int_{-\oo}^\oo d\a\, \cO(Z-\a X, -X)\right|_{\substack{X=(1,x^2,x) \\ Z=(0,2x\.z,z)}} \nn\\
&= \left.\int_{-\oo}^\oo d\a\, (-\a)^{-\De-J}\cO\p{X-\frac Z \a, Z}\right|_{\substack{X=(1,x^2,x) \\ Z=(0,2x\.z,z)}} \nn\\
&= \int_{-\oo}^\oo d\a\, (-\a)^{-\De-J}\cO\p{x-\frac z \a,z}.
\ee
In the second line above, we used gauge invariance (\ref{eq:gaugeredunancyembedding}) to shift $-X \to -X -(Z-\a X)/\a = -Z/\a$ and then homogeneity (\ref{eq:homogeneityembedding}) to pull out factors of $(-\a)$. In the third line, we used (\ref{eq:tofromflatdict}). The integration contour in (\ref{eq:minkowskilighttransform}) starts at $x$ when $\a=-\oo$ and reaches the boundary of Minkowski space when $\a=0$. The correct prescription there is to continue the contour into the next Poincare patch to the point $\cT x \in \tl{\cM}_d$. The expression (\ref{eq:minkowskilighttransform}) makes it clear that $\wL[\cO]$ converges whenever $\De+J > 1$, as long as there are no other operators at $x$ or $\cT x$.\footnote{More precisely, $\wL[\cO]$ converges as an operator-valued tempered distribution when $\De+J>1$. To define $\wL[\cO](x,z)$ as a distribution, we must show how to smear it against a test function, $\int d^d x f(x) \wL[\cO](x,z)$. We do so by integrating the light-transform by parts $\int d^d x (\cT^{-1} \wL)[f](x,z) \cO(x,z)$. This makes it clear that $\wL[\cO]$ converges whenever $\wL[f]$ converges for any test function $f$. This again leads to the condition $\De+J>1$.}  Note that $\wL[\cO](x,z)$ is not a polynomial in $z$ and thus cannot be written in terms of an underlying tensor with $1-\De$ indices.

For any local operator $\cO$ satisfying $\De+J>1$, the light-transform $\wL[\cO]$ annihilates the vacuum $|\Omega\>$. The reason is that if $\wL[\cO]|\Omega\>$ did not vanish, then it would be a primary state with spin $1-\De \notin \Z_{\geq 0}$, which is not allowed in a unitary CFT \cite{Mack:1975je}. One can also verify that $\wL[\cO]|\Omega\>=0$ by deforming the $\a$ contour in the complex plane inside a Wightman correlation function \cite{Kravchuk:2018htv}.

Let us now return to the boost selection rule $J=J_1+J_2-1$ from section~\ref{sec:boostselection}.
To recover the setup of that section, we can set 
\be
\label{eq:firstsetofcoordinates}
X_0 &= -(0,0,\tfrac 1 2,\tfrac 1 2,\vec 0) \in \mathscr{I}^-,
\nn\\
Z_0 &= (1,\vec y^2,0,0,\vec y),
\ee
where $\vec 0,\vec y\in \R^{d-2}$. Note that these satisfy the conditions $X_0^2=X_0\.Z_0=Z_0^2=0$. The light-transform becomes
\be
\label{eq:lighttransformkinematicsone}
\wL[\cO](X_0,Z_0) &= \int_{-\oo}^\oo d\a\,\cO_{v\cdots v}(u=0,v=\a,\vec y),
\ee
Thus, we should think of $\int dv\, \cO_{v\cdots v}$ as a primary operator associated to the point $X_0$ at past null infinity.

Consider now a product of null-integrated operators
\be
\label{eq:coincidentlightray}
\wL[\cO_1](X_0,Z_0) \wL[\cO_2](X_0,Z_0') &= \int_{-\oo}^\oo dv_1\, \cO_{1;v\cdots v}(0,v_1,\vec y_1) \int_{-\oo}^\oo dv_2\, \cO_{2;v\cdots v}(0,v_2,\vec y_2).
\ee
This is a product of primaries associated to the {\it same} point $X_0$ at past null infinity (with different polarization vectors $Z_0,Z_0'$). Thus, the dimension of the product (assuming it is well-defined) is the sum of the dimensions: $(1-J_1)+(1-J_2)=1-(J_1+J_2-1)$.\footnote{Ordinarily in CFT, we do not consider a product of operators at coincident points. Instead, we place them at separated points and study the singularity as they approach each other, for example
\be
\f_1(x) \f_2(0) &\sim \sum_k x^{\De_k-\De_1-\De_2} \f_k(0).
\ee
The dimensionful factor $x^{\De_k-\De_1-\De_2}$ allows the scaling dimension $\De_k$ to be different from $\De_1+\De_2$. However, if the coincident limit $x\to 0$ is nonsingular, the only operators that survive the limit must obey the selection rule $\De_k=\De_1+\De_2$.} This is the same as the dimension of the light-transform of an operator with spin $J_1+J_2-1$. Hence, we have recovered the selection rule from section~\ref{sec:boostselection}.

The relationship between this argument and the one in section~\ref{sec:boostselection} is that the dilatation generator that measures dimension around the point $X_0$ is the same as the boost generator discussed in section~\ref{sec:boostselection}.

An important observation is that the product (\ref{eq:coincidentlightray}) transforms like a primary operator at the point $X_0$. This is because both factors $\wL[\cO_1](X_0,Z_0)$ and $\wL[\cO_2](X_0,Z_0')$ are killed by the special conformal generators that fix $X_0$. (Alternatively, we can simply observe that (\ref{eq:coincidentlightray}) is a homogeneous function of $X_0$ on the null cone in the embedding space, which again implies that it transforms like a primary.) Thus, when we consider the OPE expansion of (\ref{eq:coincidentlightray}) in the limit $Z_0\to Z_0'$, the only terms appearing will be other primary operators at the point $X_0$.

\subsection{Review: event shapes and the celestial sphere}
\label{sec:eventshapes}

\begin{figure}
\centering
\begin{tikzpicture}
	\node [] at (0,0) {\includegraphics[scale=0.085]{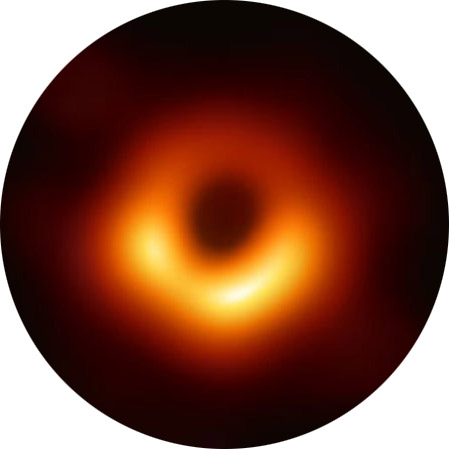}};
	\draw [energy] (-0.1,0.1) -- (-1.3,1.3);
	\draw [energy] (0.1,0.1) -- (1.2,1.2);
	\draw [energy] (-0.025,0.137) -- (-0.25,1.9);
	\draw [energy] (0.06,0.13) -- (0.6,1.2);
	\draw [energy] (-0.06,0.13) -- (-0.5,1);
	\draw [energy] (0.02,0.137) -- (0.2,1.5);
	\draw [] (-3,0) -- (0,3) -- (3,0) -- (0,-3) -- cycle;
	\draw [] (-3,0) to[out=-45,in=-135,distance=1.5cm] (3,0);
	\draw [dashed] (-3,0) to[out=45,in=135,distance=1.5cm] (3,0);
\begin{scope}[thick,blue,decoration={
    markings,
    mark=at position 0.5 with {\arrow{>}}}]
	\draw[postaction={decorate}] (-1.1,-0.72) -- (0,3);
\end{scope}	
	\node [above] at (1.8,1.65) {$\mathscr{I}^+$};
	\node [below] at (1.8,-1.55) {$\mathscr{I}^-$};
	\node [above] at (0,3) {$i^+$};
	\node [below] at (0,-3) {$i^-$};
	\node [right] at (3,0) {$i^0$};
\end{tikzpicture}
\caption{A one-point event shape \cite{Akiyama:2019cqa}. The detector $\cO=\cO_\mathrm{EHT}$ is integrated along a null line (blue) along future null infinity, starting at spatial infinity $i^0$ and ending at future timelike infinity $i^+$. (Note that the circle at spatial infinity is really a single point.) The red wavy lines indicate energy propagating from the interior of Minkowski space out to null infinity, created by the insertion of the source $\f_1(p)$.}
\label{eq:eventshapekinematics}
\end{figure}
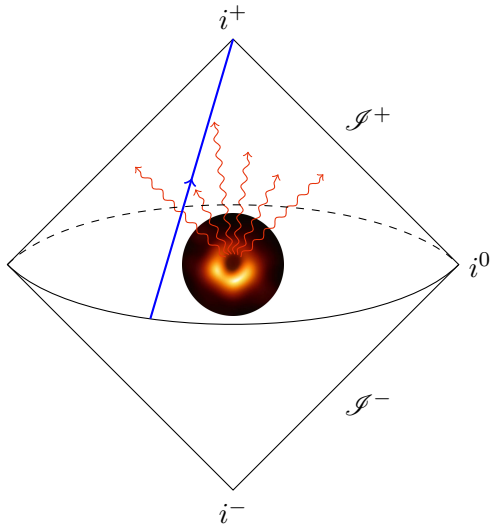

The symmetries of light-ray operators on a null plane are easiest to understand when we take the null plane to be $\mathscr{I}^+$. This corresponds to choosing the embedding-space coordinates
\be
X_{\oo} &= (0,1,0), \nn\\
Z_\oo(z) &= (0,0,z),
\ee
where $z\in \R^{d-1,1}$ is a future-pointing null vector. The integration contour for the light-transform now lies inside $\mathscr{I}^+$, running from $i^0$ to $i^+$ along the $z$ direction (figure~\ref{eq:eventshapekinematics}).

The operator $\wL[\cO](\oo,z) \equiv \wL[\cO](X_\oo, Z_\oo(z))$ transforms like a primary inserted at spatial infinity, which means it is killed by momentum generators
\be
[P^\mu, \wL[\cO](\oo,z)] &= 0.
\ee
Consequently, its matrix elements with other operators are translationally-invariant, for example
\be
\<\Omega|\phi_2(x_2) \wL[\cO](\oo,z)  \phi_1(x_1)|\Omega\> &= f(x_1-x_2).
\ee
(Throughout this work, we use $\f$ to denote scalar operators and $\cO$ to denote operators in general Lorentz representations.)
Thus, it is natural to go to momentum space,
\be
\<\phi_2(q)| \wL[\cO](\oo,z) |\phi_1(p)\> &= (2\pi)^d \de^d(p-q) \tl f(p),
\ee
where
\be
|\phi_i(p)\> &\equiv \int d^d x\, e^{ip\.x} \phi_i(x)|\Omega\>.
\ee
Note that $|\phi_i(p)\>$ vanishes unless $p$ is inside the forward lightcone $p>0$, by positivity of energy.\footnote{It is sometimes hard to keep track of signs in Lorentzian signature, so let us explain this point. Ignoring position-dependence for simplicity, we have $\f(t)=e^{iHt}\f(0)e^{-i H t}$. The minus sign is in the right-hand exponential $e^{-i H t}$ because that operator generates Schrodinger time-evolution. Acting on the vacuum, we obtain $e^{iHt}\f(0)|\Omega\>$, which is a sum of positive-imaginary exponentials $e^{iEt}$. To get a nonzero result under the Fourier transform, we must multiply by $e^{-iEt}$, which is contained in the factor $e^{ip\.x}$.}
We often abuse notation by writing
\be
\<\phi_2(p)| \wL[\cO](\oo,z) |\phi_1(p)\>=\tl f(p),
\ee
where it is understood that we have stripped off the momentum-conserving factor $(2\pi)^d \de^{(d)}(p+q)$. 

More generally, we can consider a product of light-transforms along $\mathscr{I}^+$, inserted between momentum eigenstates
\be
\label{eq:eventshapeexample}
\<\phi_2(p)| \wL[\cO_1](\oo,z_1) \cdots \wL[\cO_n](\oo,z_n) |\phi_1(p)\>.
\ee
Following \cite{Belitsky:2013bja}, we call such matrix elements ``event shapes." This terminology comes from interpreting (\ref{eq:eventshapeexample}) as the expectation value of a product of ``detectors" $\cO_1,\cdots,\cO_n$ in a ``source" state $|\phi_1(p)\>$ and ``sink" state $\<\phi_2(p)|$. The detectors sit at points on the celestial sphere and are integrated over retarded time to capture signals that propagate to null infinity. 

In addition to being translationally-invariant, $\wL[\cO](\oo,z)$
transforms in a simple way under $d$-dimensional Lorentz transformations $\SO(d-1,1)$: they act linearly on the polarization vector $z$. The Lorentz group in $d$-dimensions is the same as the Euclidean conformal group on the $(d{-}2)$-dimensional celestial sphere. Indeed, we can think of $z\in \R^{d-1,1}$ as an embedding-space coordinate for the celestial sphere $S^{d-2}$. Furthermore, $\wL[\cO](\oo,z)$ is homogeneous of degree $1-\De$ in $z$, due to (\ref{eq:lighttransformhomogeneity}). Thus, $\wL[\cO](\oo,z)$ transforms like a primary operator on the celestial sphere with dimension $\de=\De-1$.

In the previous coordinates (\ref{eq:firstsetofcoordinates}), the group $\SO(d-1,1)$ acted by conformal transformations on the transverse direction $\vec y$. The coordinates $\vec y$ are stereographic coordinates on $S^{d-2}$. Thus, we have proven the claim from section~\ref{sec:boostselection} that $\int dv\, \cO_{v\cdots v}$ transforms as a primary in the transverse space.

The event shape (\ref{eq:eventshapeexample}) is similar to a correlator of operators with dimensions $\de_i=\De_i-1$ in a Euclidean $(d{-}2)$-dimensional CFT. However, the presence of a timelike momentum $p$ breaks the Lorentz group further to $\SO(d-1)$. In the language of $(d{-}2)$-dimensional CFT, this is similar to the symmetry-breaking pattern that occurs in the presence of a spherical codimension-$1$ boundary or defect \cite{McAvity:1995zd,Liendo:2012hy}. This fact will play an important role in section~\ref{sec:celestialblocks}. We can choose a center-of-mass frame $p=(p^0,0,\dots,0)$ and write $z_i=(1, \vec n_i)$ with $ \vec n_i\in S^{d-2}$. The dependence on $p^0$ is fixed by dimensional analysis, so we can additionally set $p^0=1$. The event shape then becomes a nontrivial function of dot-products $\vec n_i\.\vec n_j$.

In addition to respecting symmetries, event shapes are useful for studying positivity conditions. For example, consider the average null energy operator $\cE=2\wL[T]$, where $T_{\mu\nu}$ is the stress tensor. $\cE$ is positive-semidefinite \cite{Hofman:2008ar,Faulkner:2016mzt,Hartman:2016lgu}. To test this, we could compute the expectation value of $\cE$ in several different states (primaries and descendents at different points, etc.) and then aggregate the resulting positivity conditions. However, it is sufficient to study event shapes $\<\cO_i(p)|\cE|\cO_j(p)\>$ for the following reason. The Hilbert space of a CFT is densely spanned by states of the form
\be
\sum_i \int d^d x f_i(x) \cO_i(x) |\Omega\>,
\ee
where $\cO_i$ are primary operators and $f_i(x)$ are test functions. Positivity of $\cE$ is thus equivalent to the statement that for any collection of test functions $f_i(x)$,
\be
\sum_{i,j} \int d^d x_1 d^d x_2 f_i^*(x_1) f_j(x_2) K_{ij}(x_1-x_2) \geq 0,
\ee
where
\be
K_{ij}(x_1-x_2)&\equiv \<\Omega|\cO_i(x_1) \cE(\oo,z) \cO_j(x_2)|\Omega\>.
\ee
This is the same as saying that $K_{ij}(x_1-x_2)$ is a positive-semidefinite integral kernel. To determine whether a kernel is positive-semidefinite, we should compute its eigenvalues and check that they are positive. Because $K_{ij}(x_1-x_2)$ is translation-invariant, it can be partially diagonalized by going to Fourier space. Thus, $\cE$ is positive-semidefinite if and only if its one-point event shapes are positive-semidefinite.

\subsubsection{1-point event shapes}

As an example, let us compute a one-point event shape $\<\f_2|\wL[\cO]|\f_1\>$, where $\cO$ has dimension $\De$ and spin $J$, and $\f_1,\f_2$ are scalars. We start from the Wightman function\footnote{We use the same conventions for two- and three-point structures as \cite{Kravchuk:2018htv}. These include some extra factors of $2^J$ that ensure that three-point structures glue together into a conventionally-normalized conformal block. These conventions are convenient when discussing inversion formulas. We also use correlators $\<0|\cdots|0\>$ in the fictitious state $|0\>$ to indicate functions whose form is fixed by conformal invariance (as opposed to correlators in a physical theory). See appendix~\ref{app:notation} for a summary of our notation.}
\be
\label{eq:wightmanstart}
&\<0|\f_2(x_2) \cO(x_3,z_3) \f_1(x_1)|0\> \nn\\&= \frac{(2 V_{3,12})^J}{
(x_{12}^2+i\e x_{21}^0)^{\frac{\De_1+\De_2-\De-J}{2}}(x_{13}^2+i\e x_{31}^0)^{\frac{\De_1-\De_2+\De+J}{2}}(x_{32}^2+i\e x_{23}^0)^{\frac{\De_2-\De_1+\De+J}{2}}},
\ee
where
\be
V_{3,12} &\equiv \frac{z_3\.x_{13} x_{23}^2 - z_3\.x_{23} x_{13}^2}{x_{12}^2}.
\ee
In (\ref{eq:wightmanstart}), we have written the $i\e$ prescription appropriate for the given operator ordering. This is obtained by introducing small imaginary parts $x_i^0 \to x_i^0 - i\e_i$ with $\e_2>\e_3>\e_1$ in the same order as the operators appearing in the Wightman function. We often omit explicit $i\e$'s, restoring them only when necessary during a computation. In these cases, the $i\e$ prescription should be inferred from the ordering of the operators in the correlator.

\begin{figure}[t]
	\centering
	\begin{tikzpicture}
	\draw[dashed] (-3,-3.3) -- (-3,3.3);
	\draw[dashed] (3,-3.3) -- (3,3.3);
	\draw[gray] (-0.5-2.3,1+2.3) -- (-0.5+3.5,1-3.5);
	\draw[gray] (-3,1-3.5) -- (-3+0.8,1-3.5-0.8);
	\draw[gray] (-0.5+2.3,1+2.3) -- (-0.5-2.5,1-2.5);
	\draw[gray] (3,1-2.5) -- (3-1.8,1-2.5-1.8);
	\draw[purple] (1.5-2.3,1+2.3) -- (1.5+1.5,1-1.5);
	\draw[purple] (-3,1-1.5) -- (-3+2.8,1-1.5-2.8);
	\draw[purple] (1.5-4.3,1-4.3) -- (1.5+1.5,1+1.5);
	\draw[purple] (-3,1+1.5) -- (-3+0.8,1+1.5+0.8);	
	\draw[fill=black] (-0.5,1) circle (0.05);
	\node[above] at (-0.5,1.1) {$2$};
	\draw[fill=black] (-0.5+3,1-3) circle (0.05);
	\node[below] at (-0.5+3+0.05,1-3-0.1) {$2^-$};	
	\draw[fill=black] (1.5,1) circle (0.05);
	\node[above] at (1.5,1+0.1) {$1$};	
	\draw[fill=black] (1.5-3,1-3) circle (0.05);
	\node[below] at (1.5-3+0.05,1-0.1-3) {$1^-$};
	\draw[fill=black] (-1.2,-0.8) circle (0.05);
	\node[left] at (-1.2,-0.8) {$3$};	
	\draw[fill=black] (-1.2+3,-0.8+3) circle (0.05);
	\node[right] at (-1.2+3,-0.8+3+0.1) {$3^+$};	
\begin{scope}[blue,thick,decoration={
    markings,
    mark=at position 0.5 with {\arrow{>}}}
    ]
	\draw[postaction={decorate}] (-1.2,-0.8) -- (-1.2+3,-0.8+3);
\end{scope}	
	\end{tikzpicture}
\caption{The causal relationship between points $2>3>1^-$ used in  (\ref{eq:lighttransformsimplekinematics}). The lightcone of $2$ is drawn in gray and the lightcone of $1$ in purple.}
\label{fig:light}	
\end{figure}
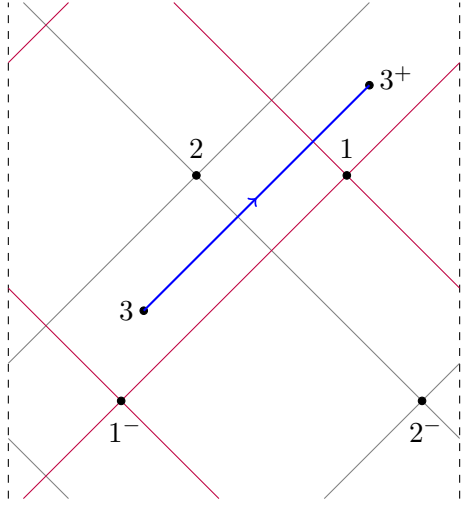

 The light-transform of (\ref{eq:wightmanstart}) is \cite{Kravchuk:2018htv}
\be
\label{eq:lighttransformsimplekinematics}
&\<0|\f_2(x_2)\wL[\cO](x_3,z_3) \f_1(x_1)|0\>\nn\\
&= \frac{ L(\f_1\f_2[\cO])(2V_{3,12})^{1-\De}}{
(x_{12}^2)^{\frac{\De_1+\De_2-(1-J)-(1-\De)}{2}} (x_{13}^2)^{\frac{\De_1-\De_2+(1-J)+(1-\De)}{2}} (-x_{23}^2)^{\frac{\De_2-\De_1+(1-J)+(1-\De)}{2}}}\qquad (2>3>1^-),
\ee
where
\be
\label{eq:lighttransformcoefficient}
L(\f_1\f_2[\cO])&\equiv
-2\pi i \frac{\Gamma(\De+J-1)}{\Gamma(\frac{\De+\De_1-\De_2+J}{2})\Gamma(\frac{\De-\De_1+\De_2+J}{2})}.
\ee
This indeed has the form of a conformally-invariant three-point function with an operator with dimension $1-J$ and spin $1-\De$. The notation $i>j$ means ``$x_i$ is inside the future lightcone of $x_j$."  Below, we will also use the notation $i\approx j$ to indicate that $x_i$ is spacelike from $x_j$.
We have written (\ref{eq:lighttransformsimplekinematics}) in the kinematics $2>3>1^-$ (figure~\ref{fig:light}), where all the quantities in parentheses are positive. This time, we have left the $i\e$ prescription implicit.

We should now take $x_3$ to spatial infinity. Keeping track of $i\e$ prescriptions, we find
\be
\<0|\f_2(x_2) \wL[\cO](\oo,z) \f_1(x_1)|0\> &= L(\f_1\f_2[\cO])\frac{ e^{i\pi \De_2} (-2z\.x_{12}+i\e)^{1-\De}}{(-x_{12}^2 + i\e x_{12}^0)^{\frac{\De_1+\De_2-(1-J)+(1-\De)}{2}}}.
\ee
This is indeed translation-invariant.
It is straightforward to compute the Fourier transform
\be
	&\int d^dx e^{ipx}\frac{(-2x\.z + i\e)^{1-\De}}{(-x^2+i\e x^0)^{\tfrac{\De_1+\De_2-(1-J)+(1-\De)}{2}}}\nn\\
	&=\hat\cF_{\De_1+\De_2-(1-J),1-\De}(-2p\.z)^{1-\De}(-p^2)^{\tfrac{\De_1+\De_2-(1-J)-(1-\De)-d}{2}}\theta(p),
\ee
where
\be
	\hat \cF_{\De,J}\equiv \frac{e^{-i\pi\frac{\De}{2}}2^{d+1-\De}\pi^{\frac{d+2}{2}}}{\Gamma(\frac{\De+J}{2})\Gamma(\frac{\De+2-d-J}{2})}.
\ee
The theta function $\th(p)\equiv\th(-p^2)\th(p^0)$ restricts $p$ to lie in the forward lightcone.
Overall, the one-point event shape is given by
\be
\label{eq:onepointeventshape}
&\int d^d x\, e^{ip\.x}\<0|\f_2(0) \wL[\cO](\oo,z) \f_1(x)|0\>\nn\\
& = 
\frac{2^{d-\De_{1}-\De_{2}-J+3}\pi ^{\frac{d}{2}+2} e^{i \pi  \frac{\De_{2}-\De_{1}-J}{2}} \Gamma (J+\Delta -1) 
(-2p\.z)^{1-\De}(-p^2)^{\tfrac{\De_1+\De_2+\De+J-2-d}{2}}\theta(p)
}{\Gamma (\frac{J+\Delta +\De_{1}-\De_{2}}{2} ) \Gamma (\frac{J+\Delta -\De_{1}+\De_{2}}{2}) \Gamma (\frac{J-\Delta +\De_{1}+\De_{2}}{2}) \Gamma (\frac{J+\Delta +\De_{1}+\De_{2}-d}{2})}.
\ee
Note that this is consistent with dimensional analysis in $p$, homogeneity in $z$, and Lorentz invariance. In~\cite{Kologlu:2019bco} we describe an algorithm for computing more general one-point event shapes.

\subsubsection{2-point event shapes}
\label{sec:twopteventshape}
A two-point event shape is constrained by dimensional analysis,
homogeneity, and Lorentz invariance to take the form
\be\label{eq:scalar2pteventshape}
\<\phi_4(p)|\wL[\cO_1](\oo,z_1) \wL[\cO_2](\oo,z_2)|\phi_3(p)\> &=
\frac{(-p^2)^{\frac{\De_1+\De_2+\De_3+\De_4-4-d}{2}} \th(p) }{(-2z_1\. p)^{\Delta_1 - 1} (-2z_2\. p)^{\Delta_2 - 1}}  \mathcal{G}_{\cO_1 \cO_2}(\zeta) ,
\ee
where $\cG_{\cO_1 \cO_2}(\z)$ is a function of the cross-ratio
\be
\label{eq:twopointcrossratio}
\zeta &\equiv \frac{(-2z_1\.z_2) (-p^2)}{(-2 p\.z_1)(- 2 p\.z_2)} = \frac{1-\vec n_1\.\vec n_2}{2}.
\ee
$\zeta$ takes values between $0$ and $1$.  In the last step of (\ref{eq:twopointcrossratio}) we evaluated $\zeta$ in a center-of-mass frame where $p=(p^0,\vec 0)$ and $z_i=(1,\vec n_i)$. The limit $\zeta\to 0$ corresponds to the detector directions $z_1$ and $z_2$ becoming parallel, which is described by the light-ray-light-ray OPE discussed in section~\ref{sec:schannel}. The limit $\z\to 1$ corresponds to the detectors becoming back-to-back in the frame of $p$.

\section{The light-ray-light-ray OPE}
\label{sec:schannel}

\subsection{Summary of computation}
\label{sec:lightrayopesummary}

In this section, we compute an expansion for
\be
\label{eq:theproductwestudy}
\wL[\cO_1](x,z_1) \wL[\cO_2](x,z_2)
\ee
as $z_1\to z_2$. Here, we summarize the key steps of the computation. Our summary will be schematic. We omit details and illustrate calculations using diagrams (which do not capture some subtleties).

The first step is to decompose (\ref{eq:theproductwestudy}) into irreducible representations of the conformal group. As discussed in section~\ref{sec:lighttransformreview}, (\ref{eq:theproductwestudy}) transforms like a primary at the point $x$ with scaling dimension $(1-J_1)+(1-J_2)$. However, it does not transform irreducibly under the Lorentz group $\SO(d-1,1)$ that fixes $x$. The appropriate set of irreducible representations are principal series representations labeled by $\de\in\frac{d-2}{2}+i\R$. To obtain such a representation, we smear the polarizations $z_1,z_2$ against a kernel $t_\de$\footnote{The actual kernel can also depend on a finite-dimensional representation $\l$ of $\SO(d-2)$, but we suppress that detail here for simplicity.}
\be
\label{eq:definitionofL}
\mathbb{W}_\de(x,z_0) &\propto \int Dz_1 Dz_2 \wL[\cO_1](x,z_1) \wL[\cO_2](x,z_2) t_{\de}(z_1,z_2,z_0) \nn \\
&= \int dx_1 dx_2 Dz_1 Dz_2 L_{\de}(x_1,z_1,x_2,z_2;x,z_0) \cO_1(x_1) \cO_2(x_2),
\ee
where $Dz$ is a measure on the projective null cone defined in (\ref{eq:spheremebeddingmeasure}).
We write $t_{\de}$ explicitly in (\ref{eq:explicitequationfortdelta}).

On the second line of (\ref{eq:definitionofL}), we implicitly defined a kernel $L_{\de}$ that combines the light transforms with smearing in $z_1,z_2$. 
We can represent $L_\de$ pictorially by
\be
\label{eq:lkerneldef}
\input{LKernelDef.tikz} \, .
\ee
The incoming arrows labeled $1$ and $2$ indicate that $L_\de$ acts on the representations of $\cO_1,\cO_2$. The outgoing arrow labeled $\cO^L$ indicates that $L_\de$ produces an object transforming with the quantum numbers of $\wL[\cO]$, i.e.\ $(1-J,1-\De)$ where $\cO$ has dimension and spin $(\De,J)=(\de+1,J_1+J_2-1)$. On the right-hand side, the boxes labeled $\wL$ take the light-transform of $\cO_1$ and $\cO_2$. Then, we split each representation into two lines; the solid blue line denotes the Minkowski position $x_i$ of the representation, and the dashed red line denotes the null polarization $z_i$ --- equivalently, the position on the celestial sphere. The reason for this split is to accommodate for the next two operations, which act only on either Minkowski or celestial coordinates. 
The blue triangle represents making the points $x_i$ coincident. The red three-point kernel represents smearing polarization vectors with $t_\de$.

The next step is to compute matrix elements of $\mathbb{W}_\de$. Because a light-transformed operator kills the vacuum, we have
\be
\<\Omega|\cO_4 \mathbb{W}_\de \cO_3|\Omega\> &= \int dx_1 dx_2 Dz_1 Dz_2 L_\de \< \O| \cO_4 \cO_1 \cO_2 \cO_3 |\O\> \nn \\ 
&= \int dx_1 dx_2 Dz_1 Dz_2 L_\de \< \O| [\cO_4, \cO_1] [\cO_2,\cO_3] |\O\>. \label{eq:W matrix element}
\ee 
The appearance of the double commutator suggests that we could relate the matrix elements of $\mathbb{W}_\de$ to the Lorentzian inversion formula. To see this relation, first note that by conformal invariance we have
\be
\label{eq:whatisa}
\<\Omega|\cO_4 \mathbb{W}_\de \cO_3|\Omega\> &= A_b(\de) \<0|\cO_4 \wL[\cO] \cO_3|0\>^{(b)}\ ,
\ee
where $\<0|\cO_4 \cO \cO_3|0\>^{(b)}$ are conformally-invariant three-point structures for the given representations, and in (\ref{eq:whatisa}) we have their light-transforms. The different structures have a label $b$, and summation over $b$ is implicit. Diagrammatically, we can express (\ref{eq:W matrix element}) and (\ref{eq:whatisa}) as
\be
\input{LcontracteddDiscCartoon.tikz}
&= A_b(\de) \times \input{L034str.tikz}\ ,
\ee
where ``dDisc" indicates the double-commutator.

The function $A_b(\de)$ contains the matrix elements we are interested in. To extract it, we pair with a dual structure (the pairing will be defined in (\ref{eq:lorentzianpairing}))
\be
\label{eq:threeptpairingA}
A_b(\de) &= \p{\<\Omega|\cO_4 \mathbb{W}_\de \cO_3|\Omega\>,(\<0|\cO_4 \wL[\cO] \cO_3|0\>^{(b)})^{-1}}.
\ee The dual structure $(\<0|\cO_4 \wL[\cO] \cO_3|0\>^{(a)})^{-1}$ is the one that satisfies
\be
\left(\ \input{L034str.tikz}\ , \ \input{L034strInv.tikz} \ \right) = \delta_a^b\ .
\ee 
We denote the operation of inverting a structure by an enclosing green circle, $\input{greenCircleOfInversion.tikz}$, suggestively labeled by a green inverse ($^{-1}$).
In pictorial language, (\ref{eq:threeptpairingA}) is
\be
A_b(\de) 
&= \mkern-72mu \input{AdeltaFromPairingCartoon.tikz} \mkern-72mu .
\vspace{-1cm}
\ee 
This is a four-point pairing between the double-commutator and a particular conformal block, as can be seen by cutting along the lines of the operators $1$, $2$, $3$, and $4$:
\be
\label{eq:blockwithl}
A_b(\de) 
&= \left( \ \input{dDisc.tikz} \quad, \quad \input{BlockWithLCartoon.tikz} \ \right) \.
\ee
The generalized Lorentzian inversion formula \cite{Kravchuk:2018htv} also has this form, 
\be
\label{eq:blockforlorentzinv}
& C^+_{ab}(\De,J) + C^-_{ab}(\De,J) \nn \\
&\quad = \left( \ \input{dDisc.tikz} \quad, \quad \input{BlockInLorentzianInvCartoon.tikz} \ \right) \ .
\ee
Here, the cross represents the formation of a conformal block from a pair of three-point structures by summing over descendent operators and dividing by their norms. The norms are computed using a two-point structure, which in this case is $\<\wL[\cO]\wL[\cO]\>^{-1}$, defined in (\ref{eq:twoptdefinition}).

Therefore, we can relate $A_b(\de)$ to $C^\pm_{ab}(\de+1,J_1+J_2-1)$ by relating the two conformal blocks in (\ref{eq:blockwithl}) and (\ref{eq:blockforlorentzinv}),
\be
\label{eq:relateblocks}
\input{BlockWithLCartoon.tikz} = \gamma^a  \input{BlockInLorentzianInvCartoon.tikz} \ .
\ee
Both conformal blocks are obtained by gluing three-point structures. The structure appearing on the right is the same for both blocks, so we only need to relate the structures on the left, 
\be \label{decomposition of S[L]}
\input{SofLKernel.tikz} = \gamma^a   \input{aLofO12strInv.tikz}.
\ee
The inverse of the cross on the right-hand side of (\ref{eq:relateblocks}) is integration against a two-point structure.\footnote{The correct two-point structure is actually $\<\wL[\cO]\wL[\cO]\>^{-1}$, but this detail is not reflected in the diagrams for the sake of simplicity.} Here, the two-point structure is indicated by a dot on the left-hand side of (\ref{decomposition of S[L]}). The operation of integrating against a two-point structure is a Lorentzian shadow transform, which changes the quantum numbers from $(1-J,1-\De)$ (labeled as $\cO^L$ with an outgoing arrow) to $(J+d-1,\De-d+1)$ (labeled as $\cO^L$ with an ingoing arrow).

Thus, we can compute $\gamma^a$ by pairing both sides of (\ref{decomposition of S[L]}) with the structure $\< \cO_4 \wL[\cO]\cO_3\>^{(a)}$,
\be 
\gamma^a 
&= \input{SofLKernelPairedWithStrCartoon.tikz} 
= \left( \input{LKernelPairedWithStrTwoPtCartoon.tikz}\, , \, \input{twoPtFn.tikz} \right)\ .
\ee
Here, we rearranged our diagram into a pairing of two-point structures.
Finally, we must compute the bubble diagram on the right-hand side. After substituting the definition of $L_\de$ (\ref{eq:lkerneldef}), we obtain an expression involving a triple light transform of the three-point structure~$a$,
\be
\label{eq:tripleLofStr}
\input{tripleLofStr.tikz} \, .
\ee
The superscripts $\wL^\pm$ are related to a subtlety not captured in the diagrams. The double-discontinuity produces additional $\theta$-functions in the expression for the block on the right-hand side of (\ref{eq:blockwithl}). On the left-hand side of (\ref{decomposition of S[L]}), these theta functions modify the kernel $L_\de$ so that the light-transforms become ``half light-transforms" $\wL^\pm$, i.e.\ null integrals over semi-infinite lines. These are what appear in (\ref{eq:tripleLofStr}).\footnote{If we took three full light-transforms of a time-ordered three-point structure in an appropriate causal configuration, we would get two pieces, one of which would be the object appearing in~\eqref{eq:tripleLofStr}, and the other would differ by a permutation.}

It turns out that the result of (\ref{eq:tripleLofStr}), and therefore also $\g^a$, is remarkably simple. In section~\ref{sec:generalizationandcelestialmap}, we conjecture a formula for it in the case of an arbitrary three-point structure $\<0|\cO_1\cO\cO_2|0\>^{(a)}$ of operators in arbitrary representations. Putting everything together, we obtain
\be
A_b(\de) &= \g^a (C^+_{ab}(\de+1,J_1+J_2-1)+C^-_{ab}(\de+1,J_1+J_2-1)),
\ee
which can be written
\be
\label{eq:partialwavelightray}
\<\Omega|\cO_4\mathbb{W}_\de\cO_3|\Omega\> &= -\g^a \<\Omega|\cO_4 \p{\mathbb{O}^+_{\de+1,J_1+J_2-1(a)} + \mathbb{O}^-_{\de+1,J_1+J_2-1(a)}} \cO_3|\Omega\>.
\ee
This argument needs to be modified in a subtle way for the ``higher transverse spin" terms in the light-ray OPE. We explain briefly how this modification arises below, and give more detail in \cite{Volume3}. Equation (\ref{eq:partialwavelightray}) expresses matrix elements of the smeared product $\mathbb{W}_\de$ in terms of matrix elements of light-ray operators. The smearing can be undone by suitably integrating over $\de$,
\be
\<\Omega|\cO_4\wL[\cO_1](x,z_1) \wL[\cO_2](x,z_2)\cO_3|\Omega\> &= \int d\de \, \cC_\de(z_1,z_2,\ptl_z) \<\Omega|\cO_4\mathbb{W}_\de(x,z)\cO_3|\Omega\> \nn\\
&\quad+\textrm{higher transverse spin},
\ee
where $\cC_\de$ is a differential operator. Lifting this to an operator equation, we have
\be
&\wL[\cO_1](x,z_1) \wL[\cO_2](x,z_2) \nn\\
&\quad = -\int d\de\, \g^a \cC_\de(z_1,z_2,\ptl_z)\p{\mathbb{O}^+_{\de+1,J_1+J_2-1(a)}(x,z) + \mathbb{O}^-_{\de+1,J_1+J_2-1(a)}(x,z)},\nn\\
&\quad\quad+\textrm{higher transverse spin}.
\ee
Finally, the $\de$-contour can be closed to the right, picking up a sum over light-ray operators, as discussed in section~\ref{sec:contour}.

\subsection{Review: light-ray operators and the Lorentzian inversion formula}
\label{sec:lightrayreview}

Let us now proceed with the detailed computation.
The objects that will ultimately appear in the OPE expansion of $\wL[\cO_1](x,z_1) \wL[\cO_2](x,z_2)$ are light-ray operators \cite{Kravchuk:2018htv}. In this section, we collect some facts about these operators that will be needed below.

For simplicity, consider first the case where $\cO_1=\f_1$ and $\cO_2=\f_2$ are scalars. Light-ray operators are defined by starting with a bi-local object that transforms as a primary under the conformal group $\tl \SO(d,2)$,
\be
\label{eq:bilocalobj}
\mathbb{O}_{\De,J}^\pm(x,z) &= \int d^d x_1 d^d x_2 K^\pm_{\De,J}(x_1,x_2,x,z) \f_1(x_1) \f_2(x_2).
\ee
The object $\mathbb{O}^\pm_{\De,J}$ has dimension $1-J$ and spin $1-\De$, which are the quantum numbers of the light-transform of an operator with dimension $\De$ and spin $J$. The $\pm$ sign is the signature, which is the eigenvalue under a combination of $\mathsf{CRT}$ and Hermitian conjugation, as discussed in section~\ref{sec:boostselection}.

The object $\mathbb{O}_{\De,J}^\pm$ is meromorphic in $\De$ and $J$ and has poles of the form
\be
\mathbb{O}_{\De,J}^\pm(x,z) &\sim \frac{1}{\De-\De_i^\pm(J)} \mathbb{O}_{i,J}^\pm(x,z).
\ee
Its residues $\mathbb{O}_{i,J}^\pm$ are light-ray operators. Light-ray operators are analytic continuations in spin of light-transforms of local operators. When $J$ is a nonnegative integer, we have
\be
\mathbb{O}_{i,J}^{(-1)^J} &= f_{12\cO_{i,J}} \wL[\cO_{i,J}],\qquad J\in \Z_{\geq 0}.
\ee
Here, $\cO_{i,J}$ is a spin-$J$ operator appearing in the $\f_1\x\f_2$ OPE with coefficient $f_{12\cO_{i,J}}$, and $i$ labels different Regge trajectories. Note that the even-signature light-ray operators $\mathbb{O}^+_{i,J}$ are analytic continuations in $J$ of light-transformed even-spin operators, while $\mathbb{O}^-_{i,J}$ are analytic continuations in $J$ of light-transformed odd-spin operators.

Matrix elements of light-ray operators can be computed via a Lorentzian inversion formula. Let $\f_3,\f_4$ be primary scalars for simplicity. A time-ordered correlator involving the object $\mathbb{O}_{\De,J}^\pm$ is given by
\be
\label{eq:timeorderedO}
\<\f_4 \mathbb{O}_{\De,J}^\pm(x,z) \f_3\>_\Omega &= -C^\pm(\De,J) \<0|\f_4 \wL[\cO](x,z)\f_3|0\>.
\ee
We use the shorthand notation that $\f_i$ is at position $x_i$ unless otherwise specified. We also use the notation from \cite{Kravchuk:2018htv} where correlators in the state $|\Omega\>$ are physical, while correlators in the state $|0\>$ are conformally-invariant structures for the given representations. The structure on the right-hand side of (\ref{eq:timeorderedO}) is the light-transform of the standard three-point structure for two scalars and a spin-$J$ operator, analytically continued in $J$,
\be
\label{eq:analyticallycontinuedlighttransformstructure}
\<0|\f_4 \wL[\cO](x_0,z_0)\f_3|0\> 
&=  
\frac{
	L(\f_3\f_4[\cO])\p{2 V_{0,34}}^{1-\De}
	}{(x_{34}^2)^{\frac{\De_3+\De_4-(1-J)-(1-\De)}{2}}(x_{30}^2)^{\frac{\De_3+(1-J)-\De_4+(1-\De)}{2}}(-x_{40}^2)^{\frac{\De_4+(1-J)-\De_3+(1-\De)}{2}}}.
\ee
 The coefficient $L(\f_3\f_4[\cO])$ is given in (\ref{eq:lighttransformcoefficient}).

In (\ref{eq:timeorderedO}), the time-ordering acts on $\f_1,\f_2$  inside $\mathbb{O}^\pm_{\De,J}$. Thus the object $\mathbb{O}^\pm_{\De,J}$ is not really an operator. However, its singularities as a function of $\De$ come only from the region where $\f_4$ acts on the future vacuum and $\f_3$ acts on the past vacuum, so upon taking residues, we obtain a genuine operator
\be
\label{eq:residuescalarcase}
\<\Omega|\f_4\mathbb{O}_{i,J}^{\pm}(x,z)\f_3|\Omega\> &= \Res_{\De=\De^\pm_i(J)} \<\f_4 \mathbb{O}_{\De,J}^\pm(x,z) \f_3\>_\Omega \nn\\
&= -\Res_{\De=\De^\pm_i(J)} C^\pm(\De,J) \<0|\f_4 \wL[\cO](x,z)\f_3|0\>.
\ee

The coefficient function $C^\pm(\De,J)$ is given by Caron-Huot's formula \cite{Caron-Huot:2017vep}
\be
\label{eq:simonsformula}
C^\pm(\De,J) &= \frac{\kappa_{\De+J}}{4}\left[ \int_0^1 \int_0^1 \frac{d z d\bar z}{ z^2 \bar z^2} \left|\frac{\bar z- z}{ z\bar z}\right|^{d-2} \mathrm{dDisc}_t[g(z,\bar z)] G_{J+d-1,\De-d+1}^{\tl\De_i}( z,\bar z)\right.\nn\\
&\left.\qquad\quad\quad \pm \int_{-\oo}^0\int_{-\oo}^0 \frac{d z d\bar z}{ z^2 \bar z^2} \left|\frac{\bar z- z}{ z\bar z}\right|^{d-2} \mathrm{dDisc}_u[g(z,\bar z)] \hat G_{J+d-1,\De-d+1}^{\tl\De_i}( z,\bar z)\right],
\ee
where
\be
\kappa_{\De+J} &= \frac{\G(\frac{\De+J+\De_1-\De_2}{2})\G(\frac{\De+J-\De_1+\De_2}{2})\G(\frac{\De+J+\De_3-\De_4}{2}) \G(\frac{\De+J-\De_3+\De_4}{2})}{2\pi^2 \G(\De+J)\G(\De+J-1)}.
\ee
Here, we have defined a stripped four-point function $g(z,\bar z)$, which is a function of conformal cross-ratios\footnote{We use the letter $z$ both for future-pointing null vectors and for conformal cross-ratios. We hope that this does not cause confusion.}
\begin{align}
\label{eq:fourpointscalars}
\<\f_1 \f_2 \f_3 \f_4\>_\Omega &= T^{\De_i}(x_i) g(z,\bar z) \nn\\
T^{\De_i}(x_i) &\equiv \frac{1}{(x_{12}^2)^{\frac{\De_1+\De_2}{2}}(x_{34}^2)^{\frac{\De_3+\De_4}{2}}}\left(\frac{x_{14}^2}{x_{24}^2}\right)^{\frac{\De_2-\De_1}{2}}\left(\frac{x_{14}^2}{x_{13}^2}\right)^{\frac{\De_3-\De_4}{2}}.
\end{align}
The $t$-channel double-discontinuity $\mathrm{dDisc}_{t}$ is defined by
\be
\label{eq:ddisct}
-2\mathrm{dDisc}_t[g](z,\bar z) &\equiv \frac{\<\Omega|[\f_4,\f_1][\f_2,\f_3]|\Omega\>}{|T^{\De_i}(x_i)|}
= -2\cos(\pi\f)\, g( z,\bar z) + e^{i\pi\phi} g^\circlearrowleft( z,\bar z) + e^{-i\pi\phi}g^\circlearrowright( z,\bar z),\nn\\
\f &= \tfrac{\De_2-\De_1+\De_3-\De_4}{2},
\ee
where $g^{\circlearrowleft}$ or $g^{\circlearrowright}$ indicates we should take $\bar z$ around $1$ in the direction shown, leaving $ z$ held fixed. Similarly,
\be
-2\mathrm{dDisc}_u[g](z,\bar z) &\equiv\frac{\<\Omega|[\f_4,\f_2][\f_1,\f_3]|\Omega\>}{|T^{\De_i}(x_i)|}
= -2\cos\p{\pi\f'}g( z,\bar z) + e^{i\pi\f'}g^\circlearrowright( z,\bar z)+ e^{-i\pi\f'}g^\circlearrowleft( z,\bar z),\nn\\
\f' &= \tfrac{\Delta_2-\Delta_1+\Delta_4-\Delta_3}{2}.
\ee
where now $g^\circlearrowleft$ or $g^\circlearrowright$ indicates we should take $\bar z$ around $-\infty$ in the direction shown, leaving $ z$ held fixed.

Finally, $G^{\tl \De_i}_{\De,J}(z,\bar z)$ denotes a conformal block for external scalars with dimensions $\tl \De_i\equiv d-\De_i$, exchanging an operator with dimension $\De$ and spin $J$. In our conventions, it  behaves as $z^{\frac{\De-J}{2}} \bar z^{\frac{\De+J}{2}}$ for positive cross-ratios satisfying $z\ll \bar z \ll 1$. Similarly, $\hat G_{\De,J}^{\tl \De_i}( z,\bar z)$ is a solution to the Casimir equation that behaves like $(- z)^{\frac{\De-J}{2}}(-\bar z)^{\frac{\De+J}{2}}$ for negative cross-ratios satisfying $| z|\ll|\bar z|\ll 1$. In Caron-Huot's formula (\ref{eq:simonsformula}), $G$ and $\hat G$ appear with dimension and spin swapped according to $(\De,J)\to (J+d-1,\De-d+1)$.

\subsubsection{More general representations}
\label{sec:moregeneralrepresentations}

Before generalizing to non-scalar $\cO_1,\cO_2$, we must establish some notation for conformal representations. A  primary operator $\cO$ is labeled by a dimension $\De$ and a representation $\rho$ of $\SO(d-1,1)$, which we can think of as a list of weights under the Cartan subalgebra of $\SO(d-1,1)$.

When $\cO$ is local, $\rho$ is finite-dimensional. In this case, we define shadow and Hermitian conjugate representations to have weights
\be
\tl \cO &: (d-\De,\rho^R), \nn\\
\cO^\dag &: (\De,(\rho^R)^*),
\ee
where $\rho^R$ denotes the reflection of $\rho$ and $(\rho^R)^*$ is the dual of $\rho^R$.
The conjugate shadow representation $\tl \cO^\dag$ has weights
\be
\label{eq:tildedag}
\tl \cO^\dag &:(d-\De,\rho^*),
\ee
and thus admits a conformally-invariant pairing with $\cO$:
\be
\label{eq:finitedimpairing}
\int d^d x\, \cO(x) \tl \cO^\dag(x),
\ee
where the $\SO(d-1,1)$ indices of $\cO(x)$ and $\tl \cO^\dag(x)$ are implicitly contracted. 

For continuous-spin operators, $\rho$ is no longer finite-dimensional. It has weights $\rho=(J,\l)$, where $J\in \C$ is spin and $\l$ is a finite-dimensional representation of $\SO(d-2)$. We can think of $J$ as the length of the first row of the Young diagram of $\rho$, while $\l$ encodes the remaining rows. Altogether, we specify the multiplet of $\cO$ by a triplet $(\De,J,\l)$.

Operators with non-integer $J$ admit a different kind of conformally-invariant pairing
\be
\label{eq:lorentzianpairingbasic}
\int d^d x D^{d-2} z\, \cO(x,z) \cO^{S\dag}(x,z).
\ee
Here, $\cO^{S\dag}$ has weights 
\be
\label{eq:lorentzianshadow}
\cO^{S\dag} &: (d-\De,2-d-J,\l^*).
\ee
In (\ref{eq:lorentzianpairingbasic}), we implicitly contract the $\SO(d-2)$ indices in the representations $\l$ and $\l^*$. The measure $D^{d-2}z$ is defined by
\be
\label{eq:spheremebeddingmeasure}
D^{d-2} z &\equiv \frac{2 d^d z \de(z^2)\th(z^0)}{\vol \R_+},
\ee
where $\mathbb{R}_+$ acts by rescaling $z$. Note that $D^{d-2} z\, \cO(x,z) \cO^{S\dag}(x,z)$ is homogeneous of degree 0 in $z$, so that the integral is well-defined.  Using the pairings (\ref{eq:finitedimpairing}) for integer-spin operators and (\ref{eq:lorentzianpairingbasic}) for continuous-spin operators, we can construct conformally-invariant pairings between two- and three-point structures, as we will see below.

In the diagrams in section \ref{sec:lightrayopesummary} and below, we use an outgoing arrow labeled $\cO$ to denote a representation $\cO$, and an ingoing arrow labeled $\cO$ to denote the dual representation, either $\tl{\cO}^{\dagger}$ or $\cO^{S\dagger}$ as appropriate to $\cO$. Joining lines represents the conformally-invariant pairing appropriate for the representations.

When $\cO_1,\cO_2$ are not scalars, the OPE $\cO_1\x\cO_2$ can contain operators $\cO$ with weights $(\De,J,\l)$, where $\l$ is nontrivial. In addition, $\cO$ can appear with multiple tensor structures. Physical three-point correlators are linear combinations of the possible structures, labeled by indices $a,b$
\be
\<\cO_1 \cO_2 \cO^\dag\>_\Omega &= f_{12\cO^\dag(a)}\<\cO_1 \cO_2 \cO^\dag\>^{(a)},
\nn\\
\label{eq:structb}
\<\cO_3 \cO_4 \cO\>_\Omega &= f_{34\cO(b)}\<\cO_3 \cO_4 \cO\>^{(b)}.
\ee
(Sums over $a,b$ are implicit.)
Following the notation of \cite{Kravchuk:2018htv} (see also appendix~\ref{app:notation}), we use the subscript $\Omega$ to distinguish physical correlators from conformally-invariant structures.

Thus, when $\cO_1,\cO_2$ are not scalars, $\mathbb{O}_{\De,J}^\pm$ gets generalized to have an additional $\SO(d-2)$ representation label $\l$ and structure label $a$: $\mathbb{O}_{\De,J,\l(a)}^\pm$. It has residues
\be
\mathbb{O}_{\De,J,\l(a)}^\pm &\sim \frac{1}{\De-\De_i^\pm (J,\l)} \mathbb{O}_{i,J,\l(a)},
\ee
which for integer $J$ and signature $\pm =(-1)^J$ become light-transforms of local operators:
\be
\label{eq:becomelighttransformsoflocal}
\mathbb{O}_{i,J,\l(a)}^{(-1)^J} &= f_{12\cO^\dag_{i,J,\l}(a)} \wL[\cO_{i,J,\l}],\qquad J\in \Z_{\geq 0}.
\ee

Let $\cO_3,\cO_4$ be primary operators (not necessarily scalars). Three-point functions containing $\mathbb{O}_{i,J,\l(a)}^\pm$ are given by
\be
\<\cO_4 \mathbb{O}_{\De,J,\l(a)}^\pm(x,z) \cO_3\>_\Omega &= -C_{ab}^\pm(\De,J,\l) \<0|\cO_4 \wL[\cO](x,z) \cO_3|0\>^{(b)},\nn\\
\<\Omega|\cO_4 \mathbb{O}_{i,J,\l(a)}^\pm(x,z) \cO_3|\Omega\>_\Omega &= -\Res_{\De=\De_i^\pm(J,\l)} C_{ab}^\pm(\De,J,\l) \<0|\cO_4 \wL[\cO](x,z) \cO_3|0\>^{(b)}.
\label{eq:matrixelementintermsofstructures}
\ee
(We suppress spin indices on $\cO_3,\cO_4$ and only indicate the $x,z$ dependence of $\mathbb{O}$.)
The coefficients $C_{ab}^\pm(\De,J,\l)$ are given by the generalized Lorentzian inversion formula
\be
C_{ab}^\pm(\De,J,\l) &= \frac{-1}{2\pi i}  \int_{\substack{4>1\\2>3}} \frac{d^d x_1\cdots d^d x_4}{\vol(\tl \SO(d,2))} \<\O|[\cO_4, \cO_1] [\cO_2,\cO_3]|\O\>\nn\\
&\qquad\qquad   \x \tsym_2^{-1} \tsym_4^{-1}\frac{\p{\tsym_2\<0|\cO_2 \wL[\cO^\dagger]\cO_1|0\>^{(a)}}^{-1}\p{\tsym_4\<0|\cO_4 \wL[\cO]  \cO_3|0\>^{(b)}}^{-1}}{\<\wL[\cO]\wL[\cO^\dagger]\>^{-1}}\nn\\
&\quad \pm \frac{-1}{2\pi i}  \int_{\substack{4>2\\1>3}} \frac{d^d x_1\cdots d^d x_4}{\vol(\tl \SO(d,2))} \<\O|[\cO_4, \cO_2] [\cO_1,\cO_3]|\O\>\nn\\
&\qquad\qquad   \x \tsym_1^{-1} \tsym_4^{-1}\frac{\p{\tsym_1\<0|\bar{\cO}^\dag_2 \wL[\cO^\dagger](\bar x, -\bar z)\bar{\cO}^\dag_1|0\>^{(a)}}^{-1}\p{\tsym_4\<0|\cO_4 \wL[\cO]  \cO_3|0\>^{(b)}}^{-1}}{\<\wL[\cO]\wL[\cO^\dagger]\>^{-1}}
.
\label{eq:notsoobvious}
\ee
A cartoon diagram for the first integral on the right hand side is given in (\ref{eq:blockforlorentzinv}). Let us describe the ingredients in (\ref{eq:notsoobvious}) in detail. Again, we use the shorthand notation that $\cO_i$ is at position $x_i$. The integral is over a Lorentzian configuration where $4>1$, $2>3$, and all other pairs of points are spacelike separated. In terms of cross-ratios, this is the same as the integration region $0<z,\bar z < 1$ in (\ref{eq:simonsformula}).

The object in the second line of (\ref{eq:notsoobvious}) is schematic notation for a conformal block obtained by merging the two three-point structures $\p{\tsym_2\<0|\cO_2 \wL[\cO^\dagger]\cO_1|0\>^{(a)}}^{-1}$ and $\p{\tsym_4\<0|\cO_4 \wL[\cO]  \cO_3|0\>^{(b)}}^{-1}$, using the two-point structure $\<\wL[\cO]\wL[\cO^\dagger]\>^{-1}$. (It is not simply a ratio of three-point and two-point structures.) The precise merging procedure is described in \cite{Kravchuk:2018htv} --- it is essentially the usual procedure of summing over products of descendent three-point functions to obtain a conformal block, generalized to continuous spin. We will see some examples below. 
Pictorially, the block is 
\be
\input{BlockInLorentzianInv.tikz} \ .
\ee

The three-point structures making up the conformal block are defined by
\be
\p{\p{\tsym_2\<0|\cO_2 \wL[\cO^\dagger]\cO_1|0\>^{(a)}}^{-1}, \tsym_2\<0|\cO_2 \wL[\cO^\dagger]\cO_1|0\>^{(c)}}_L &= \de_a^c,\nn\\
\p{\p{\tsym_4\<0|\cO_4 \wL[\cO]  \cO_3|0\>^{(b)}}^{-1}, \tsym_4\<0|\cO_4 \wL[\cO]  \cO_3|0\>^{(d)}}_L &= \de_b^d,
\label{eq:threeptpairingsgeneralized}
\ee
where $\cT_i$ is translation to the next Minkowski patch discussed in section~\ref{eq:reviewlorentziancyl}.
Here, $(\cdot,\cdot)_L$ is a conformally-invariant pairing defined by using (\ref{eq:finitedimpairing}) for $\cO_1,\cO_2$ and (\ref{eq:lorentzianpairingbasic}) for $\cO$:
\be
\label{eq:lorentzianpairing}
&\p{\<\cO_1\cO_2 \cO\>, \<\tl\cO_1^\dag \tl \cO_2^\dag \cO^{\mathrm{S}\dag}\>}_L \nn\\
&\equiv \int_{\substack{2<1 \\ x\approx 1,2}} \frac{d^dx_1 d^dx_2 d^d x D^{d-2} z}{\vol(\tl{\SO}(d,2))} \<\cO_1(x_1)\cO_2(x_2) \cO(x,z)\> \<\tl\cO_1^\dag(x_1) \tl \cO_2^\dag(x_2) \cO^{\mathrm{S}\dag}(x,z)\> \nn\\
&= \frac{1}{2^{2d-2}\vol(\SO(d-2))} \frac{\<\cO_1(e^0)\cO_2(0) \cO(\oo,z)\> \<\tl\cO_1^\dag(e^0) \tl \cO_2^\dag(0) \cO^{\mathrm{S}\dag}(\oo,z)\>}{(-2z\.e^0)^{2-d}}.
\ee
The notation $1/\vol(\tl{\SO}(d,2))$, means that the integral should be gauge-fixed using the Fadeev-Popov procedure.
To obtain the last line, we used $\tl{\SO}(d,2)$ transformations to gauge-fix $x_2=0,x_1=e^0,x=\oo$, where $e^0$ is a unit-vector in the time direction. Finite-dimensional Lorentz indices are implicitly contracted between the two three-point structures.

The two-point structure in the denominator of (\ref{eq:notsoobvious}) is defined by
\be
\label{eq:twoptdefinition}
\p{\<\wL[\cO]\wL[\cO^\dagger]\>^{-1}, \<\wL[\cO]\wL[\cO^\dagger]\>}_L &= 1.
\ee
Here, $\<\wL[\cO]\wL[\cO^\dag]\>$ is the double light-transform of a time-ordered two-point structure. Even though the light-transform of an operator annihilates the vacuum, the light-transform of a time-ordered structure is delta-function supported. After light-transforming again, we obtain a two-point structure that is nonzero at separated points. These details are explained in \cite{Kravchuk:2018htv}. The Lorentzian two-point pairing is given by
\be\label{eq:2ptpairingL}
&\frac{(\<\cO\cO^\dagger\>,\<\cO^\mathrm{S}\cO^{\mathrm{S}\dagger}\>)_L}{\vol(\SO(1,1))^2}
\nn\\
&\equiv \int_{x_1\approx x_2} \frac{d^dx_1 d^dx_2 D^{d-2}z_1 D^{d-2} z_2}{\vol(\tl\SO(d,2))}\<\cO^a(x_1,z_1)\cO^{b\dagger}(x_2,z_2)\>\<\cO^\mathrm{S}_b(x_2,z_2)\cO^{\mathrm{S}\dagger}_a(x_1,z_1)\>,
\nn\\
&=\frac{\<\cO^a(0,z_1)\cO^{b\dagger}(\oo,z_2)\>\<\cO^\mathrm{S}_b(\oo,z_2)\cO^{\mathrm{S}\dagger}_a(0,z_1)\>}{2^{2d-2}\vol(\SO(d-2))}\frac{1}{(-2 z_1 \.z_2)^{2-d}}.
\ee
In the last line, we gauge-fixed $x_1=0,x_2=\oo$. 

The last line of (\ref{eq:notsoobvious}) includes a three-point structure that has been acted on by a combination of $\mathsf{CRT}$ and Hermitian conjugation,
\be
\bar{\cO_i}^\dag &\equiv \p{(\mathsf{CRT})\cO_i(\mathsf{CRT})}^\dag,\nn\\
\bar x &= (-x^0, -x^1, x^2,\cdots, x^{d-1}),\nn\\
\bar z &= (-z^0, -z^1, z^2,\cdots, z^{d-1}).
\ee
The role of this term is to ensure that $\mathbb{O}^\pm$ has the correct signature $\pm 1$. We give more details on this term in appendix~\ref{app:moreonanalyticcont}.

\subsection{Harmonic analysis on the celestial sphere}
\label{sec:harmoniccelestial}

Consider a product of light-transforms of local operators, placed at the same spacetime point
\be
\label{eq:productoflighttransforms}
\wL[\cO_1](x,z_1)\wL[\cO_2](x,z_2).
\ee
For simplicity, we take $\cO_1,\cO_2$ to be traceless symmetric tensors.
Each light-transformed operator has dimension $1-J_i$, and thus the product (\ref{eq:productoflighttransforms}) transforms like an operator with dimension $(1-J_1)+(1-J_2)=1-(J_1+J_2-1)$ located at $x$.

We would like to additionally decompose (\ref{eq:productoflighttransforms}) into irreducible representations of the Lorentz group that fixes $x$. To do so, we can use harmonic analysis \cite{Dobrev:1977qv} (or ``conglomeration" \cite{Fitzpatrick:2011dm}) for  $\SO(d-1,1)$, treating it as a Euclidean conformal group in $d-2$ dimensions. Harmonic analysis for $\SO(d+1,1)$ was reviewed in \cite{Karateev:2018oml}. In this section, we collect some of the needed ingredients from \cite{Karateev:2018oml}, replacing $d\to d-2$.

The $\SO(d-1,1)$ representations that will appear are $d{-}2$-dimensional operator representations $\cP_{\de,\l}$ with scaling dimension $\de$ and finite-dimensional $\SO(d-2)$-representation $\l$. We write $\cP_{\de}$ when $\l$ is trivial. We can think of the null vectors $z_i \in \R^{d-1,1}$ as embedding-space coordinates for the celestial sphere $S^{d-2}$. In this language, for example, we have a celestial three-point structure
\be
\label{eq:celestialspherethreept}
\<\cP_{\de_1}(z_1)\cP_{\de_2}(z_2)\cP_{\de_3}(z_3)\> &= \frac{1}{z_{12}^{\frac{\de_1+\de_2-\de_3}{2}}z_{23}^{\frac{\de_2+\de_3-\de_1}{2}} z_{13}^{\frac{\de_3+\de_1-\de_2}{2}}}, \nn\\
z_{ij} &\equiv -2 z_i\.z_j.
\ee
Here, $\cP_\de$ are not physical operators --- they label representations of $\SO(d-1,1)$, and (\ref{eq:celestialspherethreept}) denotes the unique three-point structure (up to normalization) for the given representations. We will also use the notation \cite{Kravchuk:2018htv}
\be
\tl \cP_{\de,\l} \equiv \cP_{2-d-\de,\l^R},\qquad
\tl \cP_{\de,\l}^\dag \equiv \cP_{2-d-\de,\l^*},
\ee
where $\l^R$ is the reflected representation and $\l^*$ is the dual representation to $\l$.\footnote{In odd dimensions, $\l^R=\l$. In even dimensions, $\l^R$ is given by swapping the spinor Dynkin labels of $\l$.}

We will be particularly interested in principal series representations of $\SO(d-1,1)$, which have $\de\in \frac{d-2}{2}+i\R$.  
Their significance is that they furnish a complete set of irreducible representations for decomposing objects that transform under $\SO(d-1,1)$.\footnote{When $d=3$, we can also have discrete-series representations appearing. We comment on the role of such representations in section~\ref{sec:contour}.}  For example, consider a function $f(z_1,z_2)$ that transforms like a product of scalar operators with dimensions $\de_1,\de_2$ on $S^{d-2}$. It can be decomposed into traceless-symmetric-tensor principal series representations, i.e.\ representations where $\l$ is the spin-$j$ traceless symmetric tensor representation of $\SO(d-2)$. We denote these by $\cP_{\de,j}$.

Let us define the ``partial wave" 
\be
\label{eq:celestialpartialwave}
W_{\de,j}(z) &\equiv \a_{\de,j} \int D^{d-2} z_1 D^{d-2}z_2 \<\tl\cP_{\de_1}^\dag(z_1) \tl \cP_{\de_2}^\dag(z_2)\cP_{\de,j}(z)\>f(z_1,z_2)
\\ &= \ \input{celestialPartialWave.tikz} \ ,
\ee
where
\be
\label{eq:defofalpha}
\a_{\de,j} &\equiv \frac{
\mu^{(d-2)}(\de,j) S_E^{(d-2)}(\cP_{\de_1}\cP_{\de_2}[\tl \cP_{\de,j}^\dag])
}{
(
\<\cP_{\de_1}\cP_{\de_2} \tl \cP_{\de,j}^\dag\>,
\<\tl \cP_{\de_1}^\dag \tl \cP_{\de_2}^\dag \cP_{\de,j}\>
)
}
\ee
and
\be
t_{\de,j}(z_1,z_2,z) &= \alpha_{\de,j} \<\tl\cP_{\de_1}^\dag(z_1) \tl \cP_{\de_2}^\dag(z_2)\cP_{\de,j}(z)\>.
\ee
The integration measure in (\ref{eq:celestialpartialwave}) is given by (\ref{eq:spheremebeddingmeasure}).
The quantities in (\ref{eq:defofalpha}) are the Plancherel measure  $\mu^{(d-2)}(\de,j)$ for $\SO(d-1,1)$, a shadow transform factor $S_E^{(d-2)}(\cP_{\de_1}\cP_{\de_2}[\tl \cP_{\de,j}])$, and a three-point pairing $(
\<\cP_{\de_1}\cP_{\de_2} \tl \cP_{\de,j}^\dag\>,
\<\tl \cP_{\de_1}^\dag \tl \cP_{\de_2}^\dag \cP_{\de,j}\>
)$. Explicit definitions and formulas for all of these quantities are available in \cite{Karateev:2018oml}. We will not need them here, since these factors will ultimately cancel. The only formula we will need is the ``bubble" integral \cite{Karateev:2018oml} 
\be
\input{celestialBubble.tikz} \ \times \vol \SO(1,1),
\ee which is
\be
\label{eq:bubbleformulacelestial}
&\a_{\de,j} \int D^{d-2} z_1 D^{d-2} z_2 \<\tl\cP_{\de_1}^\dag(z_1) \tl \cP_{\de_2}^\dag(z_2)\cP_{\de,j}(z)\> \<\cP_{\de_1}(z_1) \cP_{\de_2}(z_2) \cP_{\de,j}^\dag(z') \>\nn\\
 &\qquad\qquad= \<\cP_{\de,j}(z)\cP_{\de,j}^\dag(z')\>\vol\SO(1,1).
\ee
Here $\<\cP_{\de,j}(z)\cP_{\de,j}^\dag(z')\>$ is a two-point structure on the celestial sphere.\footnote{Specifically, it is the two-point structure used to obtain the shadow factor $S_E^{(d-2)}$ in the definition of $\a_{\de,j}$.} The infinite factor $\vol\SO(1,1)$ will cancel in all calculations below.
In the notation of section~\ref{sec:lightrayopesummary}, we have
\be
\label{eq:explicitequationfortdelta}
t_\de(z_1,z_2,z) &= t_{\de,0}(z_1,z_2,z).
\ee

The function $f(z_1,z_2)$ can be expanded in partial waves \cite{Dobrev:1977qv,Karateev:2018oml}
\be
\label{eq:theexpansion}
f(z_1,z_2) &= \sum_{j=0}^\oo \int_{\frac{d-2}{2}-i\oo}^{\frac{d-2}{2}+i\oo} \frac{d\de}{2\pi i} \cC_{\de,j}(z_1,z_2,\ptl_{z_2}) W_{\de,j}(z_2).
\ee
The differential operator $\cC_{\de,j}(z_1,z_2,\ptl_{z})$ is defined by
\be
\label{eq:celestialspherediffop}
\cC_{\de,j}(z_1,z_2,\ptl_{z_2}) \<\cP_{\de,j}(z_2) \cP_{\de,j}^\dag(z')\> &= \<\cP_{\de_1}(z_1)\cP_{\de_2}(z_2)  \cP_{\de,j}^\dag(z')\>,
\ee
This is simply the $d{-}2$-dimensional version of the usual differential operator appearing in an OPE of conformal primaries. Thus, (\ref{eq:theexpansion}) takes the form of an OPE in $d-2$ dimensions, where we have a contour integral over the principal series $\de\in \frac{d-2}{2}+i\R$ instead of a sum over $\de$. The contour can sometimes be deformed to give a sum, as we will see below.

Several objects above carry indices, and we are leaving the contraction of indices between dual objects implicit. For example, $\cP_{\de,j}(z)$ carries $j$ traceless-symmetric indices for the tangent bundle of $S^{d-2}$, and consequently $W_{\de,j}(z)$ does too. The differential operator $\cC_{\de,j}$ also carries these indices, and they are contracted in (\ref{eq:theexpansion}).

When $f(z_1,z_2)$ transforms like a product of more general operators in representations $\cP_{\de_1,\l_1}$ and $\cP_{\de_2,\l_2}$, then there can be multiple celestial three-point structures
\be
\<\tl\cP_{\de_1,\l_1}^\dag(z_1) \tl \cP_{\de_2,\l_2}^\dag(z_2)\cP_{\de,\l}(z)\>^{(\a)},
\ee
labeled by an index $\a$.
Consequently, the partial wave $W_{\de,\l}^{(\a)}(z)$ and differential operator $\cC_{\de,\l,\a}$ carry additional structure labels, and we have a more general expansion 
\be
\label{eq:theexpansionagain}
f(z_1,z_2) &= \sum_{\l,\a} \int_{\frac{d-2}{2}-i\oo}^{\frac{d-2}{2}+i\oo} \frac{d\de}{2\pi i} \cC_{\de,\l,\a}(z_1,z_2,\ptl_{z_2}) W_{\de,\l}^{(\a)}(z_2).
\ee

\subsection{Light-ray OPE from the Lorentzian inversion formula}
\label{sec:focusonirrep}

Applying (\ref{eq:celestialpartialwave}) and (\ref{eq:theexpansion}) to the product (\ref{eq:productoflighttransforms}), we have
\be
\label{eq:partialwaveofLL}
\wL[\cO_1](x,z_1)\wL[\cO_2](x,z_2) &= \sum_{j=0}^\oo \int_{\frac{d-2}{2}-i\oo}^{\frac{d-2}{2}+i\oo} \frac{d\de}{2\pi i} \cC_{\de,j}(z_1,z_2,\ptl_{z_2}) \mathbb{W}_{\de,j}(x,z_2),
\ee
where the partial waves are given by
\be
\mathbb{W}_{\de,j}(x,z) &= 
\a_{\de,j}
\int D^{d-2} z_1 D^{d-2}z_2 \<\tl\cP_{\de_1}^\dag(z_1) \tl \cP_{\de_2}^\dag(z_2)\cP_{\de,j}(z)\>\wL[\cO_1](x,z_1)\wL[\cO_2](x,z_2)
\nn\\
&=\int d^d x_1 d^d x_2 D^{d-2}z_1 D^{d-2}z_2 L_{\de,j}(x_1,z_1,x_2,z_2;x,z) \cO_1(x_1,z_1)\cO_2(x_2,z_2),
\ee
and the kernel $L_{\de,j}$ is given by
\be
\label{eq:doublelightkernel}
&L_{\de,j}(x_1,z_1,x_2,z_2;x,z) \nn
\\ & \quad \equiv
\a_{\de,j}
\<\tl\cP_{\de_1}^\dag(z_1) \tl \cP_{\de_2}^\dag(z_2)\cP_{\de,j}(z)\> \nn
\\ &\qquad\x
\int_{-\oo}^\oo d\a_1 d\a_2 
\,(-\a_1)^{-\de_1-J_1-1} (-\a_2)^{-\de_2-J_2-1}\de^{(d)}\p{x-\frac{z_1}{\a_1}-x_1} \de^{(d)}\p{x-\frac{z_2}{\a_2}-x_2}.
\ee
Here, we have defined $\de_i=\De_i-1$. We are taking $\cO_1,\cO_2$ to be traceless symmetric tensors for simplicity, so that the partial wave expansion (\ref{eq:partialwaveofLL}) only includes traceless symmetric tensors of spin-$j$ on the celestial sphere. We remove this restriction in section~\ref{sec:generalizationandcelestialmap}.
We can represent the kernel $L_{\de,j}$ pictorially as: 
\be
\input{LKernelDefWithj.tikz}\ .
\ee The blue solid and red dashed lines represent the Minkowski and celestial coordinates, respectively. We will not need to plug in the definition of $L_{\de,j}$ until the very end of our computation, accordingly we will omit the blue and red lines until necessary.

Note that $L_{\de,j}$ is nonvanishing for all $j\in \Z_{\geq 0}$. By contrast, the object $\mathbb{O}^\pm_{\De,J,j(a)}(x,z)$ is only defined when $j$ is such that operators with weights $(\De,J,j)$ can appear in the $\cO_1\x\cO_2$ OPE. For fixed $\cO_1,\cO_2$, this condition restricts $j$ to a finite set. For example, if $\cO_1,\cO_2$ are scalars, then only operators with $j=0$ (i.e.\ traceless symmetric tensors of $\SO(d-1,1)$) can appear in $\cO_1\x\cO_2$. See section~\ref{sec:generalizationandcelestialmap} for the rule that determines the allowed values of $j$ in the local operator OPE. Thus, the partial waves $\mathbb{W}_{\de,j}$ for higher values of $j$ cannot be related to $\mathbb{O}_{\De,J,j}^\pm$ itself. We show in \cite{Volume3} that they are related to derivatives of $\mathbb{O}_{\De,J,j}^\pm$.

\subsubsection{Matrix elements of $\mathbb{W}_{\de,j}$}

\begin{figure}[t]
	\centering
	\begin{tikzpicture}
	\draw[gray] (2-4-2.2,-2.5+4-2.2) -- (2-4,-2.5+4) -- (2-4+2.5,-2.5+4-2.5) -- (2-4,0.5-4) -- (2-4-2.2,0.5-4+2.2);
	\draw[gray] (2-4-2.2+8,-2.5+4-2.2) -- (2-4-2.2+8-0.3,-2.5+4-2.2-0.3) -- (2-4-2.2+8,0.5-4+2.2);
	\draw[blue,thick] (-3.3,-1) -- (-3.3+4,-1+4);
	\draw[purple,thick] (-3.3,-1) -- (-3.3-0.9,-1+0.9);
	\draw[purple,thick] (-3.3-0.9+8,-1+0.9) -- (-3.3+4,-1+4);
	\draw[dashed] (-4.2,-3.8) -- (-4.2,3.5);
	\draw[dashed] (3.8,-3.8) -- (3.8,3.5);	
	\draw[fill=black] (-2,0.3) circle (0.05);
	\node[below] at (-2+0.1,0.3) {$1$};
	\draw[fill=black] (-3.3,-1) circle (0.05);
	\node[below] at (-3.3,-1) {$x$};
	\draw[fill=black] (-3.3+4,-1+4) circle (0.05);
	\node[above] at (-3.3+4,-1+4) {$x^+$};
	\draw[fill=black] (-2+4,-2.3+4) circle (0.05);
	\node[above] at (-2+4+0.2,-2.3+4-0.05) {$2$};
	\draw[fill=black] (2,0.5) circle (0.05);
	\node[right] at (2,0.5) {$3$};
	\draw[fill=black] (2-4,0.5-4) circle (0.05);
	\node[below] at (2-4,0.5-4) {$3^-$};
	\draw[fill=black] (2-4,-2.5+4) circle (0.05);
	\node[above] at (2-4,-2.5+4) {$4$};
	\end{tikzpicture}
\caption{We study a configuration where $4>x>3^-$. Points $1$ and $2$ are integrated over distinct null lines from $x$ to $x^+$ (blue and purple). The diamond formed by the past null cone of $4$ and future null cone of $3^-$ is indicated in gray.}
	\label{fig:overallsetupmatrixeltsofW}
\end{figure}
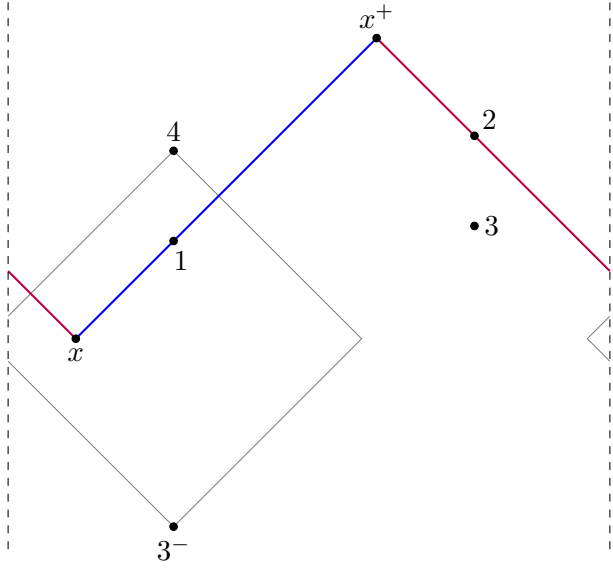

To determine $\mathbb{W}_{\de,j}(x,z)$, it suffices to study its matrix elements in states created by local primary operators $\cO_3$ and $\cO_4$:
\be
&\<\Omega|\cO_4 \mathbb{W}_{\de,j}(x,z) \cO_3 |\Omega\>\nn\\
&=
\a_{\de,j}
\int D^{d-2} z_1 D^{d-2}z_2 \<\tl\cP_{\de_1}^\dag(z_1) \tl \cP_{\de_2}^\dag(z_2)\cP_{\de,j}(z)\>\<\Omega|\cO_4 \wL[\cO_1](x,z_1)\wL[\cO_2](x,z_2)\cO_3 |\Omega\>.
\label{eq:matrixeltofp}
\ee
As usual, $\cO_i$ is at point $x_i$ unless otherwise specified.
Without loss of generality, let us assume the causal relationships $4>x>3^-$ (figure~\ref{fig:overallsetupmatrixeltsofW}). Other causal relationships can be obtained by analytic continuation in $x,x_3,x_4$.

Because $\wL[\cO_i]$ annihilates the vacuum, we can write (\ref{eq:matrixeltofp}) as the integral of a  double-commutator
\be
&=
\a_{\de,j}
\int D^{d-2} z_1 D^{d-2}z_2 \<\tl\cP_{\de_1}^\dag(z_1) \tl \cP_{\de_2}^\dag(z_2)\cP_{\de,j}(z)\>\<\Omega|\big[\cO_4, \wL[\cO_1](x,z_1)\big]\big[\wL[\cO_2](x,z_2),\cO_3\big] |\Omega\>
\nn\\
&=
\int d^dx_1 d^dx_2 D^{d-2} z_1 D^{d-2}z_2 L_{\de,j}(x_1,z_1,x_2,z_2;x,z)\th(4>1)\th(2>3)\nn\\
&\qquad\qquad\qquad\qquad\qquad\qquad \x \<\Omega|\big[\cO_4,\cO_1(x_1,z_1)\big]\big[\cO_2(x_2,z_2),\cO_3\big] |\Omega\>.
\label{eq:doublelightdoublecommutator}
\ee
In the last line, we introduced $\th$-functions $\th(4>1)\th(2>3)$ that remove the regions where $x_1$ is spacelike from $x_4$ and $x_2$ is spacelike from $x_3$. They are redundant because commutators vanish at spacelike separation. However, they will play an important role later, so we include them. Pictorially, we have
\be
\input{Lcontractedg.tikz} \ = \ \input{LcontracteddDisc.tikz} \ .
\ee To avoid visual clutter, we will omit theta functions in our diagrams.

The fact that (\ref{eq:doublelightdoublecommutator}) is the integral of a double commutator suggests that we should relate it to the Lorentzian inversion formula. In fact, the proof of the generalized Lorentzian inversion formula in \cite{Kravchuk:2018htv} proceeds from an expression similar to (\ref{eq:doublelightdoublecommutator}). We now follow the steps of that derivation.

First note that conformal invariance implies
\be
\label{eq:confinvariance}
\<\Omega|\cO_4 \mathbb{W}_{\de,j}(x,z) \cO_3 |\Omega\> &= A_{b}(\de,j) \<0|\cO_4 \wL[\cO](x,z)\cO_3|0\>^{(b)},
\ee
where $\cO$ has quantum numbers $(\De,J,\l)=(\de+1,J_1+J_2-1,j)$, $\<0|\cO_4\cO\cO_3|0\>^{(b)}$ is a basis of structures for the given representations, and $A_b(\de,j)$ are coefficients we would like to determine. A sum over $b$ is implicit. In terms of diagrams, that is
\be
\input{LcontracteddDisc.tikz} = A_b (\de,j) \times \ \input{L034str.tikz}\ .
\ee
Following \cite{Kravchuk:2018htv}, it is useful to act on both sides with $\cT_4$ (equivalently relabel $x_4\to \cT_4 x_4 = x_4^+$), giving
\be
\cT_4\<\Omega|\cO_4 \mathbb{W}_{\de,j}(x,z) \cO_3 |\Omega\> &= A_b(\de,j) \cT_4\<0|\cO_4 \wL[\cO](x,z)\cO_3|0\>^{(b)}.
\label{eq:afteractingwitht4}
\ee
Note that $\cT_4$ simply acts on three-point structures by multiplication by a phase. Nevertheless, it is useful to keep the abstract notation in (\ref{eq:afteractingwitht4}).
This relabeling turns the causal relationship $4>x>3^-$ into $3>4$ and $3\approx x$ and $4\approx x$ (see figure~\ref{fig:afterrelabelingpoints}). Here $i\approx j$ means $x_i$ is spacelike from $x_j$, see appendix~\ref{app:notation}. We write these relationships compactly as $(3>4)\approx x$.  Our Lorentzian pairing (\ref{eq:lorentzianpairing}) is defined for this type of causal relationship.
Thus, to isolate $A_b(\de,j)$, we can take the Lorentzian pairing of both sides with a dual structure
\be \label{eq:AdeltaFromPairing Diagram}
A_b(\de,j) 
&= \mkern-54mu \input{AdeltaFromPairing.tikz}\mkern-60mu .
\ee This gives
\be
A_b(\de,j) 
&= \p{\p{\cT_4\<0|\cO_4 \wL[\cO](x,z)\cO_3|0\>^{(b)}}^{-1},\cT_4\<\Omega|\cO_4 \mathbb{W}_{\de,j}(x,z) \cO_3 |\Omega\>}_L \nn\\
&= \int_{\substack{4>1\\ 2>3}} \frac{d^d x_1 d^d x_2 d^d x_3 d^d x_4 D^{d-2} z_1 D^{d-2} z_2}{\vol\tl\SO(d,2)}  \<\Omega|\big[\cO_4,\cO_1(x_1,z_1)\big]\big[\cO_2(x_2,z_2),\cO_3\big] |\Omega\>\nn\\
&\qquad \x \cT_2^{-1} \cT_4^{-1} \Bigg[\int  d^d x D^{d-2}z \p{\cT_4\<0|\cO_4 \wL[\cO](x,z)\cO_3|0\>^{(b)}}^{-1}\nn\\
&\qquad\qquad\qquad\qquad\qquad \x (\cT_2 L_{\de,j})(x_1,z_1,x_2,z_2;x,z)\th(4^+>1)\th(2^+>3)\Bigg]. 
\label{eq:exprfora}
\ee
Here, we have plugged in (\ref{eq:lorentzianpairing}) and (\ref{eq:doublelightdoublecommutator}). We then changed variables $x_4\to x_4^{-}$, acted with $\cT_2^{-1} \cT_2$ in the last line, and used $\cT_2 \th(2>3) = \th(2^+>3)$. Again, these relabelings are for the purposes of later applying the Lorentzian pairing (\ref{eq:lorentzianpairing}).

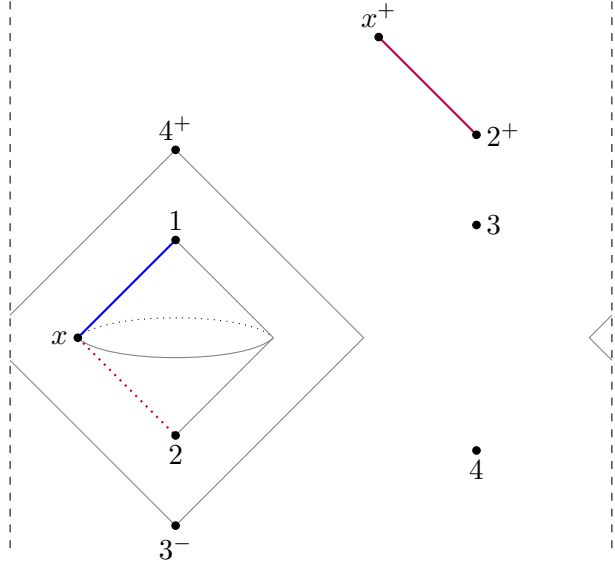
\begin{figure}[t]
	\centering
	\begin{tikzpicture}
	\draw[gray] (2-4-2.2,-2.5+4-2.2) -- (2-4,-2.5+4) -- (2-4+2.5,-2.5+4-2.5) -- (2-4,0.5-4) -- (2-4-2.2,0.5-4+2.2);
	\draw[gray] (2-4-2.2+8,-2.5+4-2.2) -- (2-4-2.2+8-0.3,-2.5+4-2.2-0.3) -- (2-4-2.2+8,0.5-4+2.2);
	\draw[gray] (-2,0.3) -- (-2+1.3,0.3-1.3);
	\draw[gray] (-2,-2.3) -- (-2+1.3,0.3-1.3);
	\draw[gray] (-3.3,-1) to[out=-45,in=-135,distance=0.5cm] (-2+1.3,0.3-1.3);
	\draw[dotted] (-3.3,-1) to[out=+45,in=+135,distance=0.5cm] (-2+1.3,0.3-1.3);
	\draw[blue,thick] (-3.3,-1) -- (-2,0.3);
	\draw[purple,thick,dotted] (-3.3,-1) -- (-2,-2.3);
	\draw[purple,thick] (-3.3+4,-1+4) -- (-2+4,-2.3+4);
	\draw[dashed] (-4.2,-3.8) -- (-4.2,3.5);
	\draw[dashed] (3.8,-3.8) -- (3.8,3.5);	
	\draw[fill=black] (-2,0.3) circle (0.05);
	\node[above] at (-2,0.3) {$1$};
	\draw[fill=black] (-3.3,-1) circle (0.05);
	\node[left] at (-3.3,-1) {$x$};
	\draw[fill=black] (-3.3+4,-1+4) circle (0.05);
	\node[above] at (-3.3+4,-1+4) {$x^+$};
	\draw[fill=black] (-2,-2.3) circle (0.05);
	\node[below] at (-2,-2.3) {$2$};
	\draw[fill=black] (-2+4,-2.3+4) circle (0.05);
	\node[right] at (-2+4,-2.3+4) {$2^+$};
	\draw[fill=black] (2,0.5) circle (0.05);
	\node[right] at (2,0.5) {$3$};
	\draw[fill=black] (2-4,0.5-4) circle (0.05);
	\node[below] at (2-4,0.5-4) {$3^-$};
	\draw[fill=black] (2,-2.5) circle (0.05);
	\node[below] at (2,-2.5) {$4$};
	\draw[fill=black] (2-4,-2.5+4) circle (0.05);
	\node[above] at (2-4,-2.5+4) {$4^+$};
	\end{tikzpicture}
\caption{
After relabeling $2\to 2^+$ and $4\to 4^+$, we have $4^+>1\gtrsim x \gtrsim 2 > 3^-$, where ``$i\gtrsim j$" means $i$ is on the future null cone of $j$. Let us imagine that $1,2,3,4$ are fixed and ask where $x$ can be.
We see that $x$ is spacelike from $3,4$ and $3>4$, equivalently $(3>4)\approx x$. Furthermore, $x$ is constrained to lie on the $S^{d-2}$ given by the intersection of the past lightcone of $1$ and future lightcone of $2$. We show lightlike segments between $x$ and $1$ (solid blue) and between $2^+$ and $x^+$ (solid purple), which are subsets of the light-transform contours from figure~\ref{fig:overallsetupmatrixeltsofW}. The image of the solid purple segment under $\cT^{-1}$ is shown in dotted purple.
}
	\label{fig:afterrelabelingpoints}
\end{figure}

\begin{figure}[t]
	\centering
	\begin{tikzpicture}
	\draw[gray] (-2,0.3) -- (-2+1.3,0.3-1.3);
	\draw[gray] (-2,-2.3) -- (-2+1.3,0.3-1.3);
	\draw[gray] (-3.3,-1) to[out=-45,in=-135,distance=0.5cm] (-2+1.3,0.3-1.3);
	\draw[dotted] (-3.3,-1) to[out=+45,in=+135,distance=0.5cm] (-2+1.3,0.3-1.3);
	\draw[blue,thick] (-3.3,-1) -- (-2,0.3);
	\draw[purple,thick,dotted] (-3.3,-1) -- (-2,-2.3);
	\draw[dashed] (-4.2,-3.2) -- (-4.2,1.2);
	\draw[dashed] (3.8,-3.2) -- (3.8,1.2);	
	\draw[
    decoration={markings,mark=at position 1 with {\arrow[scale=2]{>}}},
    postaction={decorate},
    shorten >=0.4pt,
    gray,dashed
    ]
    (2,0.5) -- (2,0.5-1.0);
    \draw[
    decoration={markings,mark=at position 1 with {\arrow[scale=2]{>}}},
    postaction={decorate},
    shorten >=0.4pt,
    gray,dashed
    ]
    (2,-2.5) -- (2,-2.5+1.0);	
	\draw[fill=black] (-2,0.3) circle (0.05);
	\node[above] at (-2,0.3) {$1$};
	\draw[fill=black] (-3.3,-1) circle (0.05);
	\node[left] at (-3.3,-1) {$x$};
	\draw[fill=black] (-2,-2.3) circle (0.05);
	\node[below] at (-2,-2.3) {$2$};
	\draw[fill=black] (2,-2.5+1.5) circle (0.05);
	\node[right] at (2,-2.5+1.5) {$x'$};
	\draw[fill=black] (2,0.5) circle (0.05);
	\node[above] at (2,0.5) {$3$};
	\draw[fill=black] (2,-2.5) circle (0.05);
	\node[below] at (2,-2.5) {$4$};
	\end{tikzpicture}
\caption{
To compute the block appearing in (\ref{eq:exprfora}), we take the limit $3,4\to x'$ inside the integral over $x,z$. Note that we have $(1>2)\approx x'$.
}
	\label{fig:afterrelabelingpointstakelimit}
\end{figure}
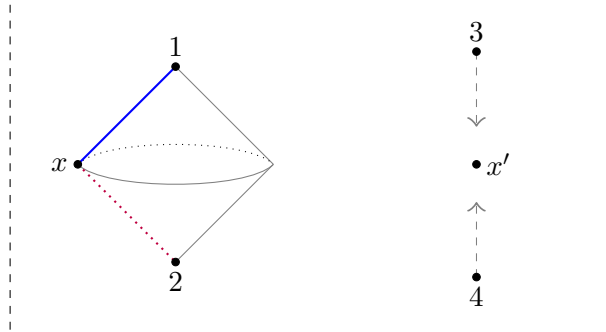

The bracketed quantity in (\ref{eq:exprfora}) is the object obtained by cutting the pairing (\ref{eq:AdeltaFromPairing Diagram}) on the lines labeled $1$, $2$, $3$, and $4$.
Because of the factors $\cT_2^{-1}\cT_4^{-1}$ outside the brackets, the configuration of points inside the brackets (figure~\ref{fig:afterrelabelingpoints}) is obtained from figure~\ref{fig:overallsetupmatrixeltsofW} by relabeling $2\to 2^+$ and $4\to 4^+$. Note that the bracketed quantity is a conformally-invariant function of $x_1,x_2,x_3,x_4$ that is an eigenfunction of the conformal Casimirs acting simultaneously on points $1,2$ (or equivalently $3,4$). Hence it is a linear combination of conformal blocks. To compute it, we can follow the computation in appendix H of \cite{Kravchuk:2018htv}. The kernel $\cT_2 L_{\de,j}$ forces $x$ to lie on the past lightcone of $x_1$ and the future lightcone of $x_2$ (see figure~\ref{fig:afterrelabelingpoints}). Thus, as $x_1\to x_2$ (equivalently $x_3\to x_4$) the integration point $x$ is forced to stay away from $x_3,x_4$. This means we can compute the integral by taking an OPE-type limit $x_3,x_4\to x'$ inside the integrand (figure~\ref{fig:afterrelabelingpointstakelimit}).

First, we must write the $34$ three-point structure as a linear operator $B(x_3,x_4,\ptl_{x'},\ptl_{z'})$ acting on a two-point function\footnote{Although we have written $B$ as a differential operator in $z'$, it must actually be an integral operator when $J$ is not an integer. See \cite{Kravchuk:2018htv} for an explicit expression.}
\be
\p{\cT_4\<0|\cO_4 \wL[\cO](x,z)\cO_3|0\>^{(b)}}^{-1} &= B(x_3,x_4,\ptl_{x'},\ptl_{z'}) \<\cO^\mathrm{F}(x',z') \cO^{\mathrm{F}\dag}(x,z)\>.
\label{eq:withtwopt}
\ee
Here, $\cO^\mathrm{F\dag}$ has the weights of something that can be paired with $\wL[\cO]$, namely $(J+d-1,\De-d+1,j)$ where $\De=\de+1$ and $J=J_1+J_2-1$.  We must also replace 
\be
\label{eq:replacetheta}
\th(4^+>1)\th(2^+>3) \to \th(x'^+>1)\th(2^+>x'),
\ee
 since we are taking the limit $x_3,x_4\to x'$. The factorization (\ref{eq:withtwopt}) of our three-point structure into a linear operator $B$ and a two-point structure is possible for generic dimensions and spins of the operator $\cO^\mathrm{F}$. However, it becomes invalid at special values of the quantum numbers of $\cO^\mathrm{F}$, where the operator $B$ develops singularities. It turns out that this special case occurs precisely when $j$ is a ``higher transverse spin" --- i.e., when $j$ is larger than the maximum transverse spin $j_\mathrm{max}$ that can appear in the usual OPE of the local operators $\cO_1\x\cO_2$. For this reason, the derivation that follows is only valid for $j\leq j_\mathrm{max}$; see \cite{Volume3} for details and the general case.
 
Because of the restriction $1>2$, (\ref{eq:replacetheta}) is equivalent to $\th((1>2)\approx x')$.
 The two-point function in (\ref{eq:withtwopt}) is then integrated against the $12$ three-point structure, giving a Lorentzian shadow transform
\be
\int_{x\approx x'} d^d x D^{d-2}z \<\cO^{\mathrm{F}}(x',z') \cO^{\mathrm{F}\dag}(x,z)\> \cT_2 L_{\de,j}(x_1,z_1,x_2,z_2;x,z)
&=\bS [\cT_2 L_{\de,j}](x_1,z_1,x_2,z_2;x',z').
 \label{eq:shadowtransformedstruct}
\ee
The result is the conformal block
\be
\label{eq:blockinnumerator}
&B(x_3,x_4,\ptl_{x'},\ptl_{z'}) \bS [\cT_2 L_{\de,j}](x_1,z_1,x_2,z_2;x',z') \th((1>2)\approx x') \nn\\
&=
\frac{
\p{\bS[\cT_2 L_{\de,j}]\th((1>2)\approx x')}
\p{\cT_4\<0|\cO_4 \wL[\cO](x,z)\cO_3|0\>^{(b)}}^{-1} 
}{\<\cO^{\mathrm{F}} \cO^{\mathrm{F}\dag}\>}.
\ee 
In the second line, we use the notation for a conformal block where the three-point structures in the numerator should be merged using the two-point function in the denominator. The precise meaning of this notation is the first line of (\ref{eq:blockinnumerator}). Pictorially, the block can be represented as 
\be
\input{BlockWithL.tikz}\ .
\ee

\subsubsection{Relating to the inversion formula}
\label{sec:relatingtolif}

After writing the quantity in brackets in (\ref{eq:exprfora}) as a conformal block, our formula for $A_b(\de,j)$ looks extremely similar to the Lorentzian inversion formula (\ref{eq:notsoobvious}). There are two main differences. Firstly, our formula for $A_b(\de,j)$ contains the three-point structure $\bS [\cT_2 L_{\de,j}]\th((1>2)\approx x')$ instead of $(\cT_2\<0|\cO_2 \wL[\cO^\dag] \cO_1|0\>^{(a)})^{-1}$. We need to express the former as a linear combination of the latter, and this is achieved by pairing with $\cT_2\<0|\cO_2 \wL[\cO^\dag] \cO_1|0\>^{(a)}$.

The second difference is that (\ref{eq:exprfora}) involves an integral only over the double-commutator $\<\Omega[\cO_4,\cO_1][\cO_2,\cO_3]|\Omega\>$, corresponding to the ``$t$-channel" term in (\ref{eq:notsoobvious}). It does not include a contribution from the ``$u$-channel" term. This is accounted for by averaging over even and odd spins.

In summary, comparing (\ref{eq:blockinnumerator}) and (\ref{eq:notsoobvious}), we find
\be
\label{eq:formulaforab}
A_b(\de,j) &= -2\pi i\x \frac 1 2\p{C^{+}_{ab}(\de+1,J_1+J_2-1,j) + C^{-}_{ab}(\de+1,J_1+J_2-1,j)}\nn\\
&\quad \x \frac{\<\wL[\cO]\wL[\cO^\dag]\>^{-1}}{\<\cO^\mathrm{F} \cO^{\mathrm{F\dag}}\>} 
 \p{
\bS[\cT_2 L_{\de,j}]\th((1>2)\approx x'),
\cT_2\<0|\cO_2 \wL[\cO^\dag] \cO_1|0\>^{(a)}
}_L.
\ee
Note that in this formula, the ratio of two-point structures $\frac{\<\wL[\cO]\wL[\cO^\dag]\>^{-1}}{\<\cO^\mathrm{F} \cO^{\mathrm{F\dag}}\>}$ is simply a number --- it does not refer to the formation of a conformal block.
The three-point pairing can be simplified further by rewriting it as a two-point pairing:
\be
\input{SofLKernelPairedWithStr.tikz} 
= \left( \input{LKernelPairedWithStrTwoPt.tikz}\, , \, \input{twoPtFnWithDagger.tikz} \right)_L\ .
\ee
In full detail, we have
\be
&\p{
\bS[\cT_2 L_{\de,j}]\th((1>2)\approx x'),
\cT_2\<0|\cO_2 \wL[\cO^\dag] \cO_1|0\>^{(a)}
}_L\nn\\
&= \int_{\substack{(1>2)\approx x' \\ x\approx x'}} \frac{d^d x_1 d^d x_2 d^d x' d^d x D^{d-2} z_1  D^{d-2} z_2 D^{d-2} z' D^{d-2} z}{\vol \tl \SO(d,2)}
\<\cO^{\mathrm{F}}(x',z') \cO^{\mathrm{F}\dag}(x,z)\>\nn\\
&\qquad\qquad \x \cT_2L_{\de,j}(x_1,z_1,x_2,z_2;x,z)
\cT_2\<0|\cO_2(x_2,z_2) \wL[\cO^\dag](x',z') \cO_1(x_1,z_1)|0\>^{(a)} \nn\\
&= 
\frac{1}{\vol\SO(1,1)^2}\bigg(\<\cO^{\mathrm{F}} \cO^{\mathrm{F}\dag}\>,
\int_{\substack{1\approx x' \\ 2 > x'}} d^d x_1 d^d x_2 D^{d-2} z_1 D^{d-2} z_2 L_{\de,j}(x_1,z_1,x_2,z_2;x,z) \nn\\
&\qquad\qquad \qquad\qquad\qquad\qquad\qquad \x 
\<0|\cO_2(x_2,z_2) \wL[\cO^\dag](x',z') \cO_1(x_1,z_1)|0\>^{(a)}
\bigg)_L.
\label{eq:simplifyingthreeptpairing}
\ee

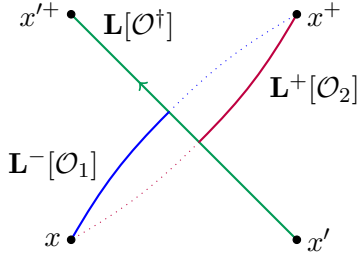
\begin{figure}[t]
	\centering
\begin{tikzpicture}
\begin{scope}[ForestGreen,thick,decoration={
    markings,
    mark=at position 0.7 with {\arrow{>}}}
    ]
	\draw[postaction={decorate}] (0,0) -- (-3,3);
\end{scope}	
	\draw[blue,thick] (-3,0) to[out=60,in=-135] (-1.7,1.7);
	\draw[blue,dotted] (-1.7,1.7) to[out=45,in=-150] (0,3);
	\draw[purple,thick] (-1.3,1.3) to[out=45,in=-120] (0,3);
	\draw[purple,dotted] (-1.3,1.3) to[out=-135,in=30] (-3,0);
	\draw[fill=black] (-3,0) circle (0.05);
	\node[left] at (-3,0) {$x$};
	\draw[fill=black] (-3+3,0+3) circle (0.05);
	\node[right] at (-3+3,0+3) {$x^+$};
	\draw[fill=black] (0,0) circle (0.05);
	\node[right] at (0,0) {$x'$};
	\draw[fill=black] (-3,3) circle (0.05);
	\node[left] at (-3,3) {$x'^+$};
	\node[left] at (-2.5,1.0) {$\wL^-[\cO_1]$};
	\node[right] at (-0.5,2.0) {$\wL^+[\cO_2]$};
	\node[above] at (-2.1,2.5) {$\wL[\cO^\dag]$};
\end{tikzpicture}
	\caption{
	Integration contours for the triple light-transform $\<0|\wL^+[\cO_2]\wL[\cO^\dag]\wL^-[\cO_1]|0\>$.  $\cO^\dag$ is integrated along a complete null line from $x'$ to $x'^+$ (solid green). $\cO_1$ is integrated along a half null line spacelike from $x'$ (solid blue), and $\cO_2$ is integrated along a half null line in the future of $x'$ (solid purple).
	}
	\label{eq:triplelighttransformcontours}
\end{figure}

In the last equality, we made the change of variables $x_2\to \cT_2^{-1} x_2 = x_2^-$ and recognized the integrals over $x,z,x',z'$ as a Lorentzian two-point pairing (\ref{eq:2ptpairingL}). The infinite factors $\vol\SO(1,1)^2$ will cancel shortly. Plugging in the definition of $L_{\de,j}$ (\ref{eq:doublelightkernel}), we have
\be
A_b(\de,j)&=-\pi i \p{C^{+}_{ab}(\de+1,J_1+J_2-1,j) + C^{-}_{ab}(\de+1,J_1+J_2-1,j)} \frac{\p{\<\wL[\cO]\wL[\cO^\dag]\>^{-1}, Q^{(a)}_{\de,j}}_L}{\vol\SO(1,1)^2},
\ee
where
\be
Q^{(a)}_{\de,j}(x,z;x',z') &= \a_{\de,j} \int D^{d-2}z_1 D^{d-2} z_2
\<\tl\cP_{\de_1}^\dag(z_1) \tl \cP_{\de_2}^\dag(z_2)\cP_{\de,j}(z)\> \nn\\
&\qquad\qquad\qquad  \x
\<0|\wL^+[\cO_2](x,z_2) \wL[\cO^\dag](x',z') \wL^-[\cO_1](x,z_1)|0\>^{(a)}.
\ee
Here, $\wL^-[\cO_1]$ indicates that the light-transform contour should be restricted to $x_1$ spacelike from $x'$, and $\wL^+[\cO_2]$ indicates that the light-transform contour should be restricted to $x_2$ in the future of $x'$ (figure~\ref{eq:triplelighttransformcontours}).

Thus our task reduces to expressing $Q_{\de,j}$ as a multiple of $\<\wL[\cO]\wL[\cO^\dag]\>$. To do so, it suffices to set $x=\oo$ and $x'=0$. Lorentz invariance and homogeneity in $z$'s guarantee
\be
\label{eq:triple light transform}
\frac{\<0|\wL^+[\cO_2](\oo,z_1) \wL[\cO^\dag](0,z') \wL^-[\cO_1](\oo,z_2)|0\>^{(a)}}{\vol\SO(1,1)} &= q^{(a)}_{\de,j} \<\cP_{\de_1}(z_1) \cP_{\de_2}(z_2) \cP_{\de,j}^\dag(z')\>,
\ee
for some constant $q^{(a)}_{\de,j}$. With hindsight, we have included a factor $\vol \SO(1,1)^{-1}$ on the left-hand side so that $q^{(a)}_{\de,j}$ is finite. Applying the bubble formula (\ref{eq:bubbleformulacelestial}), we find
\be
\frac{1}{\vol \SO(1,1)^2} Q^{(a)}_{\de,j}(\oo,z,0,z') &= q^{(a)}_{\de,j}\<\cP_{\de,j}(z)\cP_{\de,j}^\dag(z')\>.
\ee
Meanwhile, Lorentz invariance and homogeneity imply
\be
\<\wL[\cO](\oo,z) \wL[\cO^\dag](0,z')\> &= r_{\de,j} \<\cP_{\de,j}(z) \cP^\dag_{\de,j}(z')\>.
\ee
So that
\be
A_b(\de,j) &= -\pi i\p{C^{+}_{ab}(\de+1,J_1+J_2-1,j) + C^{-}_{ab}(\de+1,J_1+J_2-1,j)}\frac{q^{(a)}_{\de,j}}{r_{\de,j}}.
\ee
As explained above, this formula is valid when $j\leq j_\mathrm{max}$, where $j_\mathrm{max}$ is the maximum transverse spin that can appear in the local operator OPE $\cO_1\x\cO_2$.
Finally, combining (\ref{eq:partialwaveofLL}), (\ref{eq:confinvariance}), (\ref{eq:matrixelementintermsofstructures}), and writing $\de=\De-1$, we have
\be
\label{eq:summarysofar}
&\wL[\cO_1](x,z_1) \wL[\cO_2](x,z_2)
\nn\\
&= \pi i \sum_{j=0}^{j_\mathrm{max}} \int_{\frac d 2 - i\oo}^{\frac d 2 + i\oo} \frac{q_{\De-1,j}^{(a)}}{r_{\De-1,j}} \cC_{\de,j}(z_1,z_2,\ptl_{z_2})\p{\mathbb{O}^+_{\De,J_1+J_2-1,j(a)}(x,z_2)+\mathbb{O}^-_{\De,J_1+J_2-1,j(a)}(x,z_2)} \nn\\
&\quad+ \textrm{higher transverse spin}.
\ee
 The differential operator $\cC_{\de,j}$ is defined by (\ref{eq:celestialspherediffop}).
 
\subsubsection{Example: scalar $\cO_1,\cO_2$}
\label{sec:scalarexample}

As an example, consider the case where $\cO_1=\f_1$ and $\cO_2=\f_2$ are scalars.\footnote{As discussed in \cite{Kologlu:2019bco}, the product of light-transforms at coincident points may not be well-defined in this case. In this section, we ignore these issues and assume the product is well-defined.} We have $J=J_1+J_2-1=-1$. Furthermore, $j_\mathrm{max}=0$ since only traceless symmetric tensors of $\SO(d-1,1)$ can appear in the $\f_1\x\f_2$ OPE. 

Let us compute $q_{\de,0}$ and $r_{\de,0}$. The unique Wightman structure for two scalars and a spin $J=-1$ operator is
\be
\label{eq:scalarscalarspinJstruct}
\<0|\f_2(x_2)\cO(x_0,z_0) \f_1(x_1)|0\> &= \frac{(2V_{0,12})^{-1}}{x_{12}^{\De_1+\De_2-\De+1} x_{20}^{\De_2+\De-\De_1-1} x_{01}^{\De+\De_1-\De_2-1}},
\ee
The light-transform of $\cO$ is given by (\ref{eq:lighttransformsimplekinematics}) with the relabeling $(1,2,3)\to(2,1,0)$, and $J=-1$. In embedding-space language, we find
\be
\<0|\f_2(X_2)\wL[\cO](X_0,Z_0) \f_1(X_1)|0\>
&= \frac{ L(\f_1\f_2[\cO])(2V_{0,12})^{1-\De}}{
(X_{12})^{\frac{\De_1+\De_2-3+\De}{2}} (X_{10})^{\frac{\De_1-\De_2+3-\De}{2}} (-X_{20})^{\frac{\De_2-\De_1+3-\De}{2}}},
\label{eq:lighttransformstructureone}
\ee
where
\be
X_{ij} &\equiv -2 X_i\.X_j,\nn\\
V_{k,ij} &\equiv \frac{(Z_k\.X_i)(X_k\.X_j)-(Z_k\.X_j)(X_k\.X_i)}{X_i\.X_j}.
\ee

We should now specialize $X_0=(1,0,0)$ and compute the remaining light transforms $\wL^-[\f_1](X_\oo,Z_1)$ and $\wL^+[\f_2](X_\oo,Z_2)$. We set
\be
X_1&=Z_1 - \a_1 X_\oo=(0,0,z_1)-\a_1(0,1,0),\nn\\
X_2&=Z_2 - \a_2 X_\oo=(0,0,z_2)-\a_2(0,1,0),\nn\\
X_0 &= (1,0,0),\nn\\
Z_0 &= (0,0,z_0),
\ee
and integrate
\be
&\frac{\<0|\wL^+[\f_2](X_\oo,Z_2)\wL[\cO](X_0,Z_0) \wL^-[\f_1](X_\oo,Z_1)|0\>}{\vol\SO(1,1)} \nn\\
&=\frac{ L(\f_1\f_2[\cO]) }{\vol\SO(1,1)}\int_0^\oo d\a_2\int_{-\oo}^0 d\a_1  \frac{(-2\a_2 z_0\.z_1 + 2\a_1 z_0\.z_2)^{1-\De}}{(-2z_1\.z_2)^{\frac{\De_1+\De_2-3+\De}{2}} (-\a_1)^{\frac{\De_1-\De_2+3-\De}{2}} \a_2^{\frac{\De_2-\De_1+3-\De}{2}}} \nn\\
&= \frac{ L(\f_1\f_2[\cO]) }{\vol\SO(1,1)} \p{\int_0^\oo \frac{d\a_2}{\a_2}}   \frac{\G(\tfrac{\De-1+\De_1-\De_2}{2})\G(\tfrac{\De-1+\De_2-\De_1}{2})}{\G(\De-1)}\<\cP_{\de_1}(z_1)\cP_{\de_2}(z_2)\cP_\de(z_0)\> \nn\\
&= -\frac{2\pi i}{\De-2}\<\cP_{\de_1}(z_1)\cP_{\de_2}(z_2)\cP_\de(z_0)\>,
\ee
where $\de_i=\De_i-1$, $\de=\De-1$, and the celestial three-point structure $\<\cP_{\de_1}(z_1)\cP_{\de_2}(z_2)\cP_\de(z_0)\>$ is defined in (\ref{eq:celestialspherethreept}). To get the third line, we integrated over $\a_1$. The infinite factor $\vol\SO(1,1)$ cancels against the unbounded integral over $\a_2$. Alternatively, we could have used $\SO(1,1)$-gauge invariance to fix $\a_2=1$.

Thus, we find
\be
q_{\De-1,0} &=  -\frac{2\pi i}{\De-2}.
\ee
Meanwhile, the quantity $r_{\De-1,0}$ was computed in \cite{Kravchuk:2018htv} to be
\be
r_{\De-1,0} &= \left.-\frac{2\pi i}{\De+J-1}\right|_{J=-1} = -\frac{2\pi i}{\De-2}.
\ee
The ratio $q_{\De-1,0}/r_{\De-1,0}$ is simply $1$! We find
\be
\label{eq:scalar light transform ope}
\wL[\f_1](x,z_1) \wL[\f_2](x,z_2) &= \pi i  
\int_{\frac{d}{2}-i\oo}^{\frac{d}{2}+i\oo}\frac{d\De}{2\pi i}
\cC_{\De-1,0}(z_1,z_2,\ptl_{z_2}) \p{\mathbb{O}^+_{\De,-1}(x,z_2)+\mathbb{O}^-_{\De,-1}(x,z_2)}\nn\\
&\quad+\textrm{higher transverse spin}.
\ee

\subsubsection{Generalization and map to celestial structures}
\label{sec:generalizationandcelestialmap}

Let us summarize our result so far in slightly different language. In addition, we will generalize to the case where $\cO_1,\cO_2$ are not necessarily traceless symmetric tensors. Suppose $\cO_i$ have weights $(\De_i,J_i,\l_i)$, where the $\l_i$ are $\SO(d-2)$ representations. The light-transforms $\wL[\cO_i](\oo)$ transform as tensors in the representation $\l_i$ on the celestial sphere. To describe them, we can use the notation of appendix~\ref{sec:orthogonalreps}. We write $\wL[\cO_i](\oo,z,\vec w)$, where $\vec w = w_1,\dots,w_{n-1} \in \C^d$ is a collection of null polarization vectors orthogonal to $z$, encoding rows in the Young diagram of $\l_i$. The light-ray operators appearing in the OPE may also have nontrivial $\l$. In what follows, $\cO$ stands for the representation with weights $(\De,J,\l) = (\de+1,J_1+J_2-1,\l)$.

Lorentz-invariance guarantees that there exists an $\SO(d-1,1)$-invariant differential operator $\cD^{(a)}_{\de,\l}(z_1,\vec w_1,z_2,\vec w_2,\ptl_{z_2},\ptl_{\vec w_2})$ on the celestial sphere such that
\be
\label{eq:defofe}
&\cD^{(a)}_{\de,\l}(z_1,\vec w_1,z_2,\vec w_2,\ptl_{z_2},\ptl_{\vec w_2}) \<\wL[\cO](\oo,z_2,\vec w_2) \wL[\cO^\dag](0,z_0,\vec w_0)\>\nn\\
&=
\frac{\<0|\wL^+[\cO_2](\oo,z_2,\vec w_2) \wL[\cO^\dag](0,z_0,\vec w_0) \wL^-[\cO_1](\oo,z_1,\vec w_1)|0\>^{(a)} }{\vol\SO(1,1)}
.
\ee
In the notation of section~\ref{sec:relatingtolif}, when $\l$ is the spin-$j$ representation of $\SO(d-2)$, we have $\cD^{(a)}_{\de,j}=(q^{(a)}_{\de,j}/r_{\de,j})\cC_{\de,j}$.
The derivation of section~\ref{sec:relatingtolif} generalizes straightforwardly to give\footnote{As we discuss in section~\ref{sec:commutativity}, only the term with signature $(-1)^{J_1+J_2}$ contributes at $z_1\not\propto z_2$.}
\be
\label{eq:finalanswerforproductingeneral}
&\wL[\cO_1](x,z_1,\vec w_1) \wL[\cO_2](x,z_2,\vec w_2)\nn\\
&= \pi i\sum_{\l\in \Lambda_{12}} \int_{\frac{d}{2}-i\oo}^{\frac{d}{2}+i\oo}\frac{d\De}{2\pi i}\cD^{(a)}_{\de,\l}(z_1,\vec w_1,z_2,\vec w_2,\ptl_{z_2},\ptl_{\vec w_2})\nn\\
&\quad \qquad\qquad\x \p{\mathbb{O}^+_{\De,J_1+J_2-1,\l(a)}(x,z_2,\vec w_2) + \mathbb{O}^-_{\De,J_1+J_2-1,\l(a)}(x,z_2,\vec w_2)} \nn\\
&\quad+\textrm{higher transverse spin}.
\ee
Here, $\l$ ranges over the set $\Lambda_{12}$ of $\SO(d-2)$ representations that can appear in the $\cO_1\x\cO_2$ OPE and are also allowed by selection rules on the celestial sphere..

A simple rule to determine the set $\Lambda_{12}$ is as follows. Let $\rho_i=(J_i,\l_i)$ be the Lorentz irreps of $\cO_1$ and $\cO_2$. We have that the set $\L'_{12}$ of $\l$ appearing in $\cO_1\times\cO_2$ OPE is given by
\be
\Lambda'_{12} &= \mathrm{Res}^{\SO(d-1,1)}_{\SO(d-2)} \rho_1\otimes \rho_2,
\ee
where $\mathrm{Res}^G_H$ denotes restriction of a representation of group $G$ to its subgroup $H$. One can derive this rule by considering the three-point structure $\<\cO_1(x_1)\cO(x_0,z)\cO_2(x_2)\>$ as a function of $x_1,x_2,x_0$, and $z$. It furthermore carries indices for $\rho_1,\rho_2$ and $\l$ which we have suppressed. Using conformal invariance, we can fix $x_1,x_0,x_2$ to lie on a line in the time direction and $z$ to be $(1,1,0,\dots)$. The stabilizer group of this configuration is $\SO(d-2)$, and the correlator must be invariant under this stabilizer group. This leads to (\ref{eq:simplerlambdarule}). 

The set $\L''_{12}$ of $\l$ that are allowed from celestial sphere selection rules is
\be
\L''_{12} &= \{\l\,\,|\,\,(\l^\dagger\otimes \l_1\otimes \l_2)^{\SO(d-3)}\neq 0\},
\ee
where $\l^\dagger$ is the dual reflected of $\l$. This rule is just the $(d-2)$-dimensional version of the rule described in~\cite{Kravchuk:2016qvl}. Finally, the set $\L_{12}$ is just
\be
	\label{eq:simplerlambdarule}
	\L_{12}=\L'_{12}\cap \L''_{12}.
\ee

\begin{figure}[t]
	\centering
\begin{tikzpicture}
	\draw [gray] (-3,0) -- (0,3) -- (3,0);
	\draw [gray] (-3,0) to[out=-45,in=-135,distance=1.0cm] (3,0);
	\draw [gray,dashed] (-3,0) to[out=45,in=135,distance=1.0cm] (3,0);
	\draw [orange,thick] (-1.5,1.5) to[out=-45,in=-135,distance=0.5cm] (1.5,1.5);
	\draw [orange,thick,dashed] (-1.5,1.5) to[out=45,in=135,distance=0.5cm] (1.5,1.5);
	\draw [gray] (-3,3) to[out=-45,in=-135,distance=1.0cm] (3,3);
	\draw [gray,dashed] (-3,3) to[out=45,in=135,distance=1.0cm] (3,3);
	\draw[gray] (0,0) -- (3,3);
	\draw[gray] (0,0) -- (-3,3);
	\draw[fill=black] (-3,0) circle (0.05);
	\node [left] at (-3,0) {$\oo$};
	\draw[fill=black] (0,0) circle (0.05);
	\node [below] at (0,0) {$0$};
	\draw[ForestGreen,->,thick] (-1.5,1.5) -- (-1.5-0.75,1.5+0.75);
	\draw[blue,->,thick] (-1,1.69) -- (-0.5,2.35);
	\draw[purple,->,thick] (-0.75,1.28) -- (-0.39,2.13);
	\draw[fill=black] (-0.75,1.28) circle (0.05);
	\node[below] at (-0.55,1.28) {{\color{purple}$\cO_2$}};
	\draw[fill=black] (-1,1.69) circle (0.05);
	\node[above] at (-1-0.35,1.69+0.1) {{\color{blue}$\cO_1$}};
	\draw[fill=black] (-1.5,1.5) circle (0.05);
	\node[left] at (-1.55,1.5) {{\color{ForestGreen}$\cO^\dag$}};
\end{tikzpicture}
	\caption{The celestial locus configuration appearing in (\ref{eq:celestialmap}) and (\ref{eq:celestialstructmap}). The operators $\cO_1,\cO_2,$ and $\cO^\dag$ are placed on the celestial sphere  (orange) that is the intersection of the future null cones of $0$ and $\oo$. The arrows indicate the directions of the polarization vectors of each operator (which are inherited from their original light-transform contours).}
	\label{fig:celestiallocus}
\end{figure}
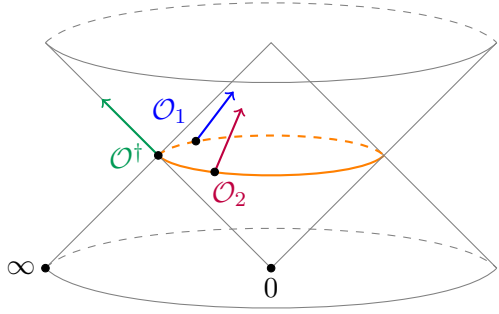

Equation (\ref{eq:defofe}) essentially defines a map from a three-point structure $\<\cO_1\cO_2\cO^\dag\>^{(a)}$ in $d$-dimensions to a differential operator $\cD_{\de,\l}^{(a)}$ in $d-2$ dimensions. We saw in section (\ref{sec:scalarexample}) that when $\cO_1,\cO_2$ are scalars, this map is surprisingly simple: it takes the standard Wightman structure (\ref{eq:scalarscalarspinJstruct}) to the standard differential operator $\cC_{\de,0}$. In fact, this map turns out to be simple in general. We claim that $\cD^{(a)}_{\de,\l}$ is determined by 
\be
&\cD^{(a)}_{\de,\l}(z_1,\vec w_1,z_2,\vec w_2,\ptl_{z_2},\ptl_{\vec w_2}) \p{\left.(-2H_{20})\<\cO(X_2,Z_2,\vec W_2)\cO^\dag (X_0,Z_0,\vec W_0)\>\right|_\mathrm{celestial}}\nn\\
&\quad=
\left.X_{12}(-2V_{0,21}) \<0|\cO_2(X_2,Z_2,\vec W_2) \cO^\dag(X_0,Z_0,\vec W_0) \cO_1(X_1,Z_1,\vec W_1)|0\>^{(a)}\right|_\mathrm{celestial}.
\label{eq:celestialmap}
\ee
Here, we use embedding-space language, as explained in appendix~\ref{sec:orthogonalreps}. The objects $V_{i,jk}$ and $H_{ij}$ are defined in appendix~\ref{sec:triple light transform}, see also \cite{Costa:2011mg}.
The two-point and three-point structures above are each specialized to the ``celestial" locus 
\be
\left.\phantom{\frac 1 2}f(X_i,Z_i,\vec W_i)\right|_\mathrm{celestial} &\equiv \left.\phantom{\frac 1 2}f(X_i,Z_i,\vec W_i)\right|_{
\substack{
Z_0=-(1,0,0) \\
Z_1=-(0,1,0) \\
Z_2=-(0,1,0) \\
X_i=(0,0,z_i) \\
W_{i,j}=(0,0,w_{i,j})
}}.
\label{eq:celestialstructmap}
\ee
This corresponds to placing all three operators on the celestial sphere given by the intersection of the future lightcone of the origin and the future lightcone of spatial infinity (figure~\ref{fig:celestiallocus}). It is easy to check that the three-point function on the right-hand side of (\ref{eq:celestialmap}), after restricting to the celestial locus, has homogenity $-\de_i=1-\De_i$ in $z_i$, and hence transforms like a three-point function of operators with dimensions $\de_i$ in $d-2$ dimensions. Similarly, the two-point function on the left-hand side transforms correctly in $d-2$ dimensions.

For example, when $\cO_1=\f_1,\cO_2=\f_2$ are scalars and $\cO$ is a traceless symmetric tensor with dimension $\De$, one can check from (\ref{eq:scalarscalarspinJstruct}) that 
\be
\left.(-2H_{20})\<\cO(X_2,Z_2)\cO^\dag (X_0,Z_0)\>\right|_\mathrm{celestial}
 &= \<\cP_\de(z_2) \cP_\de(z_0)\>,\nn\\
\left.X_{12}(-2V_{0,21}) \<0|\f_2(X_2) \cO^\dag(X_0,Z_0) \f_1(X_1)|0\>\right|_\mathrm{celestial}
&= \<\cP_{\de_1}(z_1) \cP_{\de_2}(z_2) \cP_{\de}(z_0)\>,
\ee
which easily gives $\cD_{\de,0}=\cC_{\de,0}$.

We have checked that (\ref{eq:celestialmap}) is equivalent to (\ref{eq:defofe}) for arbitrary traceless symmetric tensor representations by explicit calculation, see appendix~\ref{sec:triple light transform}. It can also be justified by examining the limit as $z_1\to z_2$ in (\ref{eq:defofe}). It would be nice to prove (\ref{eq:celestialmap}) more directly.

One important caveat to this discussion is that it only applies for separated points, i.e.\ when $z_1$ is not proportional to $z_2$. As we will see in section~\ref{sec:contactterms}, this map has to be modified in some special cases if one wishes to study $z_1\propto z_2$ contact terms.

\section{Commutativity}

\subsection{Light-ray OPE for the commutator}

\label{sec:commutativity}

In \cite{Kologlu:2019bco}, we argued on general grounds that $\wL[\cO_1](x,z_1,\vec w_1)$ and  $\wL[\cO_2](x,z_2,\vec w_2)$ commute, given certain conditions on $J_1$ and $J_2$. Our derivation of the light-ray OPE does not assume commutativity. In fact, even when commutativity holds, it is obscured in our derivation, since $\wL[\cO_1]$ and $\wL[\cO_2]$ are treated differently. For example, to obtain a double-commutator, we subtract the action of $\wL[\cO_1]$ on the future vacuum and $\wL[\cO_2]$ on the past vacuum.

It is instructive to see how commutativity appears from the point of view of the light-ray OPE. This will lead to nontrivial consistency conditions on the space of light-ray operators. In the remainder of this section, we assume the light-ray operators $\wL[\cO_1]$ and $\wL[\cO_2]$ are not coincident $z_1 \not\propto z_2$. We discuss how our arguments should be modified for coincident lightrays in section~\ref{sec:contactterms}.

We derived an expression for $\wL[\cO_1]\wL[\cO_2]$ in (\ref{eq:finalanswerforproductingeneral}). We can obtain an expression for the reverse ordering $\wL[\cO_2]\wL[\cO_1]$ by applying Rindler and Hermitian conjugation to both sides. Using (\ref{eq:rindlerhermitianlightlocal}) and (\ref{eq:rindlerhermitianlightnonlocal}), we find
\be
\label{eq:lightray product flipped ordering}
&\wL[\cO_2](x,z_2,\vec w_2)\wL[\cO_1](x,z_1,\vec w_1) \nn\\
&= \pi i\sum_{\l\in \Lambda_{12}} \int_{\frac{d}{2}-i\oo}^{\frac{d}{2}+i\oo}\frac{d\De}{2\pi i}\cD^{(a)}_{\de,\l}(z_1,\vec w_1,z_2,\vec w_2,\ptl_{z_2},\ptl_{\vec w_2})\nn\\
&\quad \qquad\qquad\x \p{(-1)^{J_1+J_2} \mathbb{O}^+_{\De,J_1+J_2-1,\l(a)}(x,z_2,\vec w_2) + (-1)^{J_1+J_2-1}  \mathbb{O}^-_{\De,J_1+J_2-1,\l(a)}(x,z_2,\vec w_2)} \nn\\
&\quad + \textrm{higher transverse spin}
\ee
Taking the difference with (\ref{eq:finalanswerforproductingeneral}), we get the commutator
\be
&[\wL[\cO_1](x,z_1,\vec w_1), \wL[\cO_2](x,z_2,\vec w_2)] \nn\\
&= 2\pi i\sum_{\l\in \Lambda_{12}} \int_{\frac{d}{2}-i\oo}^{\frac{d}{2}+i\oo}\frac{d\De}{2\pi i}\cD^{(a)}_{\de,\l}(z_1,\vec w_1,z_2,\vec w_2,\ptl_{z_2},\ptl_{\vec w_2})
\mathbb{O}^{(-1)^{J_1+J_2-1}}_{\De,J_1+J_2-1,\l(a)}(x,z_2,\vec w_2) + \dots\nn\\
&= -2\pi i \sum_i \cD^{(a)}_{\de_i,\l_i}(z_1,\vec w_1,z_2,\vec w_2,\ptl_{z_2},\ptl_{\vec w_2})
\mathbb{O}^{(-1)^{J_1+J_2-1}}_{i,J_1+J_2-1,\l(a)}(x,z_2,\vec w_2) + \dots,
\label{eq:expressionforcommutator}
\ee
where ``$\dots$" represents contributions with higher transverse spin $\l\notin \Lambda_{12}$. We explain while these terms vanish in~\cite{Volume3}, and here we focus just on the low-transverse spin part.
In the last line, we have assumed that the behavior of the integrand at large $\De$ is such that we can deform the $\De$-contour to pick up poles on the positive real axis, obtaining a sum over Regge trajectories $i$. For more detail on deforming the $\De$ contour, see section~\ref{sec:contour}.

The operators on the right-hand side of (\ref{eq:expressionforcommutator}) have spin $J=J_1+J_2-1$ and signature $(-1)^J$. For example, when $J_1\equiv J_2 \mod 2$, the commutator is given by a sum of light-ray operators with odd $J$ and odd signature. This is easy to understand from symmetries: the light-transforms $\wL[\cO_i]$ have signature $(-1)^{J_i}$, and the commutator introduces an additional $-1$, since Hermitian conjugation reverses operator ordering.

These quantum numbers are exactly the ones needed for $\mathbb{O}^{(-1)^{J_1+J_2-1}}_{i,J_1+J_2-1,\l_i}$ to be the light-transform of a local operator. Let us assume this is the case (we return to this assumption in section~\ref{sec:finishing}).  Using (\ref{eq:becomelighttransformsoflocal}), we have
\be\label{eq:commutatorislocal}
&[\wL[\cO_1](x,z_1,\vec w_1), \wL[\cO_2](x,z_2,\vec w_2)]\nn\\
 &= -2\pi i \sum_{i} \cD^{(a)}_{\de_i,\l_i}(z_1,\vec w_1,z_2,\vec w_2,\ptl_{z_2},\ptl_{\vec w_2}) f_{12\cO_{i}^\dag(a)} \wL[\cO_{i}](x,z_2,\vec w_2)+\dots,
\ee
where each $\cO_i$ has quantum numbers $(\De,J,\l)=(\de_i+1,J_1+J_2-1,\l_i)$.

There are now two slightly different cases. In the first case, the local operators that would appear in the right hand side of~\eqref{eq:commutatorislocal} are not allowed to appear in the Euclidean OPE.\footnote{This includes the cases when $J_1+J_2-1$ is negative, i.e.\ $J_1=J_2=0$.} In other words, $f_{12\cO_{i}^\dag(a)}$ are zero by selection rules. In this case we immediately find that the commutator is identically zero.

The second case is when $ f_{12\cO_{i}^\dag(a)}$ are not forbidden by Euclidean selection rules. To see that the commutator vanishes in this case, recall that the differential operator $\cD^{(a)}_{\de,\l}$ is nonzero only if the three-point structure $\<\cdots\>^{(a)}$ survives the map to celestial structures (\ref{eq:celestialmap}). However, the structure $\<\cdots\>^{(a)}$ cannot survive this map if it also appears in a three-point function of local operators, modulo a small subtlety to be discussed below. The reason is that $V_{0,21}|_\mathrm{celestial}=0$, so the right-hand side of (\ref{eq:celestialmap}) vanishes unless $\<\cdots\>^{(a)}$ contains a pole $V_{0,21}^{-1}$ that can cancel this zero. Such poles are not allowed in three-point functions of local operators (which must be polynomial in polarization vectors $z_i$). 
It follows that
\be
f_{12\cO_{i}^\dag(a)} \cD^{(a)}_{\de_i,\l_i} &= 0
\ee
for any local operator $\cO_i$. Hence, the commutator $[\wL[\cO_1],\wL[\cO_2]]$ vanishes again.

There is a small subtlety in the above argument, which is due to the fact the statements about the map to celestial structures are correct for separated points only. As we will show in section~\ref{sec:contactterms}, it does sometimes happen that tensor structures appearing in three-point functions of local operators map to contact terms on the celestial sphere.

The above argument was somewhat abstract, so let us give a concrete example. Consider the case $\cO_1=\cO_2=T$, where $T$ is the stress tensor in a 3d CFT. The commutator $[\wL[T],\wL[T]]$ is a sum of spin-3 light-ray operators on odd-signature Regge trajectories. By our assumption above, such operators are light-transforms of local spin-3 operators. However, the $T\x T$ OPE does not contain any spin-3 operators, due to selection rules and Ward identities \cite{Kravchuk:2016qvl,Dymarsky:2017yzx}. (Odd-spin operators appearing in $T\x T$ have spins $5,7,\dots$.) Thus, the commutator $[\wL[T],\wL[T]]$ must vanish. No contact terms arise in this case.

\subsection{Finishing the argument with conformal Regge theory}
\label{sec:finishing}

The key step in the above argument was the assumption that
\be
\label{eq:thingwewant}
\mathbb{O}^{(-1)^{J_1+J_2-1}}_{i,J_1+J_2-1,\l(a)} &= f_{12\cO_i^\dag(a)} \wL[\cO_{i}],
\ee
where $\cO_i$ is a local operator of spin $J_1+J_2-1$. As discussed in section~\ref{sec:moregeneralrepresentations}, this is true by construction in the case when $f_{12\cO_i^\dag(a)}$ is allowed to be non-zero by selection rules of the Euclidean OPE.\footnote{Saying that $f_{12\cO_i^\dag(a)}=0$ even thought it is allowed by Euclidean OPE amounts to saying that there is no corresponding pole in $\mathbb{O}_{\De,J,\l(a)}^\pm$ and hence no $\mathbb{O}^{(-1)^{J_1+J_2-1}}_{i,J_1+J_2-1,\l(a)}$ in the first place.} More precisely, this is true under the condition $J_1+J_2-1>J_0$, which comes from the fact that the Lorentzian inversion formula is only guaranteed to reproduce Euclidean OPE data for spins larger than $J_0$. We return to this condition later in this section.

We are also interested in the case where $f_{12\cO_i^\dag(a)}$ is forbidden by the selection rules of the Euclidean OPE. In this case, there is nothing that we can write in the right-hand side of~\eqref{eq:thingwewant} and so we would like to argue that in this case
\be
\label{eq:thingweactuallywant}
	\mathbb{O}^{(-1)^{J_1+J_2-1}}_{i,J_1+J_2-1,\l(a)} &= 0.
\ee

We can argue for (\ref{eq:thingweactuallywant}) using conformal Regge theory and boundedness in the Regge limit. 
Let us first review some aspects of conformal Regge theory, using a four-point function of scalars for simplicity. We follow the presentation of \cite{Kravchuk:2018htv}. One starts with a four-point function in a Euclidean partial wave expansion
\be
\label{eq:euclideanpartialwavedecomposition}
\<\f_1\f_2\f_3\f_4\> &= \sum_{J=0}^\oo \oint \frac{d\De}{2\pi i} C(\De,J) (\cF_{\De,J}(x_i) + \cH_{\De,J}(x_i)).
\ee
Here, we've split each partial wave into a piece $\cF_{\De,J}(x_i)$ that dies at large positive $J$ and a piece $\cH_{\De,J}(x_i)$ that dies at large negative $J$. For simplicity, we only keep track of $\cF_{\De,J}$. The sum runs over nonnegative integer $J$ because these are the allowed spins in the Euclidean OPE.

\begin{figure}[t]
	\centering
	{\tikzset{tikzfig/.append style={scale=1.3}}
		\input{ReggeContour.tikz}
		~
		\input{ReggeContourT.tikz}
	}
	\caption{Deformation of Regge contour in Sommerfeld-Watson transform. Left: even spins in the case of scalar four-point function. Right: odd spins in the case of $\<TT\cO_3\cO_4\>$.}
	\label{fig:reggecontours}
\end{figure}
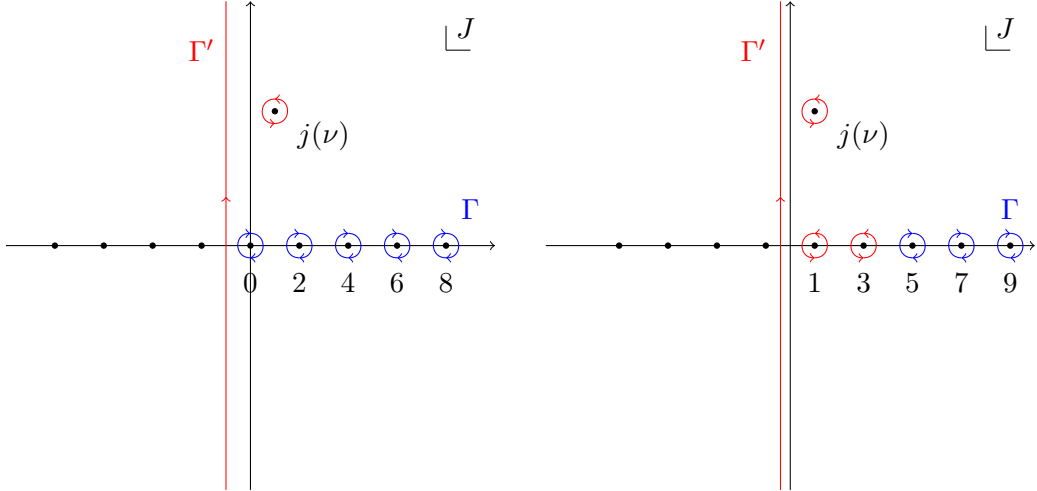

The key step is the Sommerfeld-Watson transform: we rewrite the sum over $J$ as a contour integral
\be
\sum_{J=0}^\oo \oint \frac{d\De}{2\pi i} C(\De,J) \cF_{\De,J}(x_i)
&= -\oint_\G dJ \oint \frac{d\De}{2\pi i} \p{\frac{C^+(\De,J)}{1-e^{-i\pi J}}+\frac{C^-(\De,J)}{1+ e^{-i\pi J}}} \cF_{\De,J}(x_i),
\label{eq:sommerfeldwatsonintegrand}
\ee
where $\G$ encircles all nonnegative integers clockwise. We now deform the contour $\G\to \G'$ towards the imaginary $J$ axis (left panel of figure~\ref{fig:reggecontours}). When we do, we pick up any poles or branch cuts in the integrand that were not encircled by the original contour $\G$.  We refer to such singularities as ``Regge poles". In figure~\ref{fig:reggecontours} we show a single Regge pole at $J=j(\nu)$. The behavior of the correlator in the Regge limit is determined by the Regge poles. 
If the Regge growth exponent is $J_0$, then all Regge poles must have real part less or equal to $J_0$.

Let us now consider what happens in spinning four-point functions when we have non-trivial selection rules. For concreteness, we will focus on the case $\cO_1=\cO_2=T$ and study matrix elements of $\mathbb{O}^{(-1)^{J_1+J_2-1}}_{i,J_1+J_2-1,\l(a)}=\mathbb{O}^{-}_{i,3,(a)}$ (i.e.\ $\l=0$) between generic states created by $\cO_3$ and $\cO_4$. These matrix elements show up as residues of the poles of the function $C^-_{ab}(\De,J=3,\l=0)$ which appears in the partial wave expansion of $\<TT\cO_3\cO_4\>$.  Note that there are no local spin-3 (or spin-1) operators in $T\times T$ OPE allowed by selection rules. In order to prove~\eqref{eq:thingweactuallywant} we must show that this function does not have physical poles.

To see this, imagine applying conformal Regge theory to $\<TT\cO_3\cO_4\>$. We will arrive at the generalization of~\eqref{eq:sommerfeldwatsonintegrand}, where the factor\footnote{One might argue that in this case we should use a different factor in the Sommerfeld-Watson transform. However, the factor $\frac{1}{1+e^{-i\pi J}}$ is the unique factor which has the same residue at all sufficiently large odd $J$ and an appropriate behavior at infinity in the complex plane.}
\be
	\frac{1}{1+e^{-i\pi J}}
\ee
will create poles for all odd $J$, including $J=1$ and $J=3$. However, since $J=1,3$ are not allowed in the Euclidean OPE, the contour $\G$ must not circle these poles, see right panel of figure~\ref{fig:reggecontours}. This implies that these poles will be picked up by $\G'$. If $J_0<3$, we must conclude that the residue of $J=3$ pole vanishes, and so 
\be
	C^-_{ab}(\De,J=3,\l=0)=0.
\ee

This straightforwardly generalizes to other situations, and we conclude that~\eqref{eq:thingwewant} holds provided $J_1+J_2-1>J_0$. If this condition is satisfied, the arguments in the previous section show that $[\wL[\cO_1](x,z_1),\wL[\cO_2](x,z_2)]$ vanishes for $z_1\not\propto z_2$. This is precisely the same result as obtained in~\cite{Kologlu:2019bco}, where it was shown that $J_1+J_2-1>J_0$ is a necessary condition for the product $\wL[\cO_1](x,z_1)\wL[\cO_2](x,z_2)$ to be well-defined and commutative.

\subsection{Superconvergence in $\nu$-space}
\label{sec:superconvergenceinnuspace}

We have seen that when $J_1+J_2-1>J_0$, the commutator (\ref{eq:expressionforcommutator}) vanishes. This follows from the analysis of \cite{Kologlu:2019bco}, or alternatively from the arguments of sections~\ref{sec:commutativity} and~\ref{sec:finishing} using the light-ray OPE and conformal Regge theory.\footnote{More precisely, those arguments applied to the case where the null directions $z_1$ and $z_2$ are not coincident $z_1\not\propto z_2$. For coincident null directions, there can be contact terms. In that case, the discussion in this section would need to be modified by subtracting those contact terms before passing to $\nu$-space. In the case of ANEC operators, contact terms are absent.} From (\ref{eq:expressionforcommutator}), commutativity is equivalent to the statement that
\be
\<\Omega|\cO_4 \mathbb{O}^{(-1)^{J_1+J_2-1}}_{\frac{d}2+i\nu,J_1+J_2-1,\l(a)}(x,z_2,\vec w_2) \cO_3 |\Omega\> &= 0,\qquad \textrm{if $J_1+J_2-1>J_0$},
\ee
where we have written $\De=\frac d 2 + i\nu$, and the above conditions hold for all $\nu\in \R$. Using (\ref{eq:matrixelementintermsofstructures}), we can also write this as
\be
\label{eq:superconvergenceintermsofC}
C_{ab}^{(-1)^{J_1+J_2-1}}\p{\frac d 2 + i \nu, J_1 + J_2 - 1, \l} &= 0,\qquad \textrm{if $J_1+J_2-1>J_0$}.
\ee
What constraints do these conditions imply on CFT data?

For simplicity, let us specialize to the case where $\cO_1,\cO_2,\cO_3,\cO_4$ are scalars, so that $\l=\bullet$ and the labels $a,b$ are trivial. We further assume that the $\cO_i$ have equal dimensions $\De_\f$, though we do not assume they are identical, since otherwise the superconvergence sum rules are trivial. Recall that $C^\pm(\De,J)$ is computed by plugging the physical four-point function $g(z,\bar z)$ into the Lorentzian inversion formula (\ref{eq:simonsformula}) and performing the integral. The four-point function has an expansion in $t$-channel conformal blocks that converges exponentially inside the square $z,\bar z \in (0,1)$ \cite{Pappadopulo:2012jk}:
\be
\label{eq:tchannelforfourpointfunction}
g(z,\bar z) &= \p{\frac{z\bar z}{(1-z)(1-\bar z)}}^{\De_\f}\sum_{\De',J'} p^t_{\De',J'} G_{\De',J'}(1-z,1-\bar z). \nn\\
\mathrm{dDisc}_t[g](z,\bar z) &= \p{\frac{z\bar z}{(1-z)(1-\bar z)}}^{\De_\f}\sum_{\De',J'} 2\sin^2\p{\pi \frac{\De'-2\De_\f}{2}} p^t_{\De',J'} G_{\De',J'}(1-z,1-\bar z).
\ee
On the second line, we have written an expansion for $\mathrm{dDisc}[g]$.
Because $\mathrm{dDisc}$ inserts positive, bounded factors $2\sin^2\p{\pi \frac{\De'-2\De_\f}{2}}$ into the $t$-channel block expansion, the $t$-channel block expansion for $\mathrm{dDisc}[g]$  converges exponentially inside the square as well.\footnote{The analysis of  \cite{Pappadopulo:2012jk} relied on positivity of $p^t_{\Delta',J'} =f_{14 \cO_{\Delta',J'}^\dagger}  f_{\cO_{\Delta',J'} 23} $ which is not guaranteed for the OPE of non-identical scalar operators at hand. Using the simple inequality $| f_{14 \cO^\dagger}  f_{\cO 23} | \leq {1 \over 2} \left(| f_{14 \cO^\dagger}|^2 +  | f_{\cO 23} |^2\right)$ we can readily apply the results of \cite{Pappadopulo:2012jk} to our case.} The four-point function has a similar expansion in $u$-channel blocks with coefficients $p^u_{\De',J'}$.

Inserting (\ref{eq:tchannelforfourpointfunction}) into the Lorentzian inversion formula, we obtain an expression for $C^\pm(\De,J)$ as a sum
\be
\label{eq:sumforc}
C^\pm(\De,J) &= \sum_{\De',J'} (p^t_{\De',J'}\pm p^u_{\De',J'})\cB(\De,J; \De',J'),
\ee
where $\cB(\De,J;\De',J')$ is the Lorentzian inversion of a single $t$-channel block. The function $\cB(\De,J;\De',J')$ was computed in $d=2$ and $d=4$ dimensions in \cite{Liu:2018jhs}. We expect the sum (\ref{eq:sumforc}) to converge whenever $\De= \frac{d}{2}+i\nu$ is on the principal series and $J>J_0$ is larger than the Regge intercept. We argue for this using the Fubini-Tonelli theorem in appendix~\ref{app:swapping}.

Plugging (\ref{eq:sumforc}) into (\ref{eq:superconvergenceintermsofC}), we obtain an infinite set of sum rules\footnote{We expect that $J_0<-1$ is not true in most interesting theories. Here, we have this condition because we specialized to scalar operators for simplicity.}
\be
\label{eq:scalarsuperconvergencerule}
0 &= \sum_{\De',J'} (p^t_{\De',J'}-p^u_{\De',J'})\cB\p{\frac d 2 + i \nu,-1; \De',J'},\qquad \textrm{if $J_0 < -1$}.
\ee
As we will see in section~\ref{sec:schanvstchan}, these are precisely the superconvergence sum rules of \cite{Kologlu:2019bco}, written as a function of a different variable $\nu$. In $\nu$-space, we have a clear argument that the sum is convergent.
Equation (\ref{eq:scalarsuperconvergencerule}) and its generalization to spinning correlators may be a good starting point for analyzing contributions of stringy states to superconvergence sum rules in holographic theories.

We give more details on the relationship between $\nu$-space sum rules and the sum rules from \cite{Kologlu:2019bco} in section~\ref{sec:schanvstchan}.

\section{The celestial block expansion}
\label{sec:celestialblocks}

\subsection{Celestial blocks}

For the purpose of computing event shapes, we would like to apply the light-ray OPE inside momentum eigenstates. Matrix elements of individual light-ray operators $\mathbb{O}_{\De,J}$ in momentum eigenstates are proportional to the one-point event shape (\ref{eq:onepointeventshape}). To apply the OPE (\ref{eq:summarysofar}), we must understand how to apply the differential operator $\cC_{\de,0}(z_1,z_2,\ptl_{z_2})$ to these matrix elements:
\be
\label{eq:defcelestialblock}
\cC_{\de,0}(z_1,z_2,\ptl_{z_2})\<\f(p)|\mathbb{O}_{\De,J}(\oo,z_2)|\f(p)\> &\,\propto\, \cC_{\de,0}(z_1,z_2,\ptl_{z_2})(-2 z_2\.p)^{-\de}.
\ee
We call the resulting objects ``celestial blocks" because they capture the full contribution of a light-ray operator and its $z$-derivatives to an event shape.

The right-hand side of (\ref{eq:defcelestialblock}) is fixed by Lorentz-invariance and homogeneity to have the form
\be
\label{eq:lorentzandhomogeneity}
\cC_{\de,0}(z_1,z_2,\ptl_{z_2})(-2 z_2\.p)^{-\de} &= \frac{(-p^2)^{\frac{\de_1+\de_2-\de}{2}}}{(-2z_1\.p)^{\de_1}(-2z_2\.p)^{\de_2}} f(\z),
\ee
where the cross-ratio $\z$ is given by
\be
\label{eq:zetadef}
\z &= \frac{(-2z_1\.z_2)(-p^2)}{(-2z_1\.p)(-2z_2\.p)}.
\ee
Furthermore, it is an eigenvector of the quadratic Casimir of the Lorentz group acting simultaneously on $z_1,z_2$, or equivalently acting on $p$. Specifically, it is killed by the differential operator
\be
-\frac 1 2 \p{p_\mu\pdr{}{p^\nu} - p_\nu\pdr{}{p^\mu}}\p{p^\mu\pdr{}{p_\nu} - p^\nu\pdr{}{p_\mu}} - \de(\de-d+2).
\ee
This gives the Casimir differential equation
\be
0 &=
4 (1-\zeta)
   \zeta ^2 f''(\zeta )
-2 \zeta  \left(2 \left(\delta _1+\delta _2+1\right) \zeta +d-2 \left(\delta _1+\delta
   _2+2\right)\right) f'(\zeta )
   \nn\\
&\quad
+\left(\left(\delta -\delta _1-\delta _2\right)
   \left(d-\delta -\delta _1-\delta _2-2\right)-4 \delta _1 \delta _2 \zeta \right)
   f(\zeta ).
\ee

Meanwhile, from the definition of $\cC_{\de,0}$, we see that
\be
\label{eq:celestialminiblock}
\cC_{\de,0}(z_1,z_2,\ptl_{z_2})(-2 z_2\.p)^{-\de}
&= 
(-2z_1\.z_2)^{\frac{\de-\de_1-\de_2}{2}}(-2 z_2\.p)^{-\de} + \dots,
\ee
where ``$\dots$" represent higher-order terms in the separation between $z_1$ and $z_2$ on the celestial sphere.
In terms of $f(\z)$, this becomes
\be
\label{eq:theboundarycondition}
f(\z) &= \z^{\frac{\de-\de_1-\de_2}{2}}(1+O(\z)).
\ee
The solution to the Casimir equation with boundary condition (\ref{eq:theboundarycondition}) is
\be
\label{eq:celestialblockfunction}
f_{\De}^{\De_1,\De_2}(\z) &= \z^{\frac{\De-\De_1-\De_2+1}{2}}{}_2F_1\p{\frac{\De-1+\De_1-\De_2}{2},\frac{\De-1-\De_1+\De_2}{2},\De+1-\frac{d}{2},\z},
\ee
where we have written $\de_i=\De_i-1$ for future convenience.

Essentially the same function has appeared previously in the literature as the conformal block for a two-point function of local operators in the presence of a spherical codimension-1 boundary \cite{McAvity:1995zd,Liendo:2012hy}. The reason is that the momentum $p$ breaks $\SO(d-1,1)$ in a similar way to a boundary in a $d-2$ dimensions. To see this, consider an embedding space coordinate $X\in \R^{d-1,1}$ for a $d-2$-dimensional CFT. A spherical codimension-1 boundary is specified by $P\.X=0$, for some spacelike $P\in \R^{d-1,1}$ \cite{Gadde:2016fbj}. The vector $P$ breaks the symmetry from $\SO(d-1,1)$ to $\SO(d-2,1)$. In our case, we have a timelike vector $p$ that breaks the symmetry from $\SO(d-1,1)$ to $\SO(d-1)$. However, the Casimir equation is the same in both cases, and the only difference is a minus sign in our definition of the cross-ratio $\z$.

Now, we can finally write the light-ray OPE for a two-point event shape. For simplicity, we consider the case where the sink, source, and detectors are all scalars. In this case, the higher transverse spin terms in the light-ray OPE cannot contribute, because there is no conformally-invariant three-point structure between two scalars and a non-traceless-symmetric tensor. From the OPE (\ref{eq:summarysofar}), we have
\be
&\<\f_4(p)|\wL[\f_1](\oo,z_1) \wL[\f_2](\oo,z_2)|\f_3(p)\> \nn\\
&= \pi i  
\int_{\frac{d}{2}-i\oo}^{\frac{d}{2}+i\oo}\frac{d\De}{2\pi i}
\cC_{\De-1,0}(z_1,z_2,\ptl_{z_2}) \<\f_4(p)|\mathbb{O}^+_{\De,-1}(\oo,z_2)+\mathbb{O}^-_{\De,-1}(\oo,z_2)|\f_3(p)\> \nn\\
&= -\pi i  
\int_{\frac{d}{2}-i\oo}^{\frac{d}{2}+i\oo}\frac{d\De}{2\pi i}
 (C^+(\De,-1) + C^-(\De,-1))\cC_{\De-1,0}(z_1,z_2,\ptl_{z_2})\<\f_4(p)|\wL[\cO](\oo,z_2)|\f_3(p)\>,
\ee
where $\<0|\f_4\cO\f_3|0\>$ is the standard Wightman structure (\ref{eq:scalarscalarspinJstruct}) with $2\to 4$ and $1\to 3$. Plugging in the expression (\ref{eq:onepointeventshape}) for the light transform and Fourier transform (with appropriate relabelings), and using (\ref{eq:lorentzandhomogeneity}) we find 
\be\label{eq:scalareventshapecrossratio}
\<\f_4(p)|\wL[\f_1](\oo,z_1) \wL[\f_2](\oo,z_2)|\f_3(p)\>
&= 
\frac{(-p^2)^{\frac{\De_1+\De_2+\De_3+\De_4-4-d}{2}} \th(p) }{(-2z_1\.p)^{\De_1-1}(-2z_2\.p)^{\De_2-1}}
\cG_{\f_1\f_2}(\z),
\ee
where 
\be
\label{eq:predictionforg}
\cG_{\f_1\f_2}(\z) &=  2^{d+4-\De_{3}-\De_{4}}\pi ^{\frac{d}{2}+3} e^{i \pi  \frac{\De_{4}-\De_{3}}{2}}
  \nn\\
 &\quad \x \int_{\frac{d}{2}-i\oo}^{\frac{d}{2}+i\oo}\frac{d\De}{2\pi i} \frac{\Gamma (\Delta -2) \p{C^+(\De,-1) + C^-(\De,-1)}}{\Gamma (\frac{\Delta-1 +\De_{3}-\De_{4}}{2} ) \Gamma (\frac{\Delta-1 -\De_{3}+\De_{4}}{2}) \Gamma (\frac{\De_{3}+\De_{4}-\Delta-1}{2}) \Gamma (\frac{\Delta-1 +\De_{3}+\De_{4}-d}{2})}f^{\De_1,\De_2}_\De(\z).
\ee

In the special case where the sink and source are the same $\f_3=\f_4=\f$, it is natural to define an expectation value by dividing by a zero-point event shape:
\be
\<\f(p)|\f(p)\> &\equiv \int d^d x e^{ip\.x} \<0|\f(0) \f(x)|0\>
 = \frac{2^{d+1-2\De_\f}\pi^{\frac{d+2}{2}}}{\G(\De_\f-\tfrac{d-2}{2}) \G(\De_\f)}(-p^2)^{\frac{2\De_\f-d}{2}} \th(p).
\ee
We find
\be
\label{eq:scalareventshape}
\frac{\<\f(p)|\wL[\f_1](\oo,z_1) \wL[\f_2](\oo,z_2)|\f(p)\>}{\<\f(p)|\f(p)\>}
&=
\frac{\G(\De_\f-\tfrac{d-2}{2}) \G(\De_\f)}{2^{d+1-2\De_\f}\pi^{\frac{d+2}{2}}} \frac{(-p^2)^{\frac{\De_1+\De_2}{2}-2}\cG_{\f_1\f_2}(\z)}{(-2z_1\.p)^{\De_1-1}(-2z_2\.p)^{\De_2-1}}.
\ee

\subsection{Contour deformation in $\De$ and spurious poles}
\label{sec:contour}

In (\ref{eq:predictionforg}), the celestial block expansion of $\cG_{\f_1\f_2}(\zeta)$ takes the form of an integral over the principal series $\De\in \frac d 2 + i\R$. When $\zeta<1$, the celestial block $f_\De^{\De_1,\De_2}(\zeta)$ is exponentially damped in the right-half $\De$-plane, so we can deform the contour into this region and pick up poles in the integrand.

The coefficient function $C^\pm(\De,J)$ contains poles of the form
\be
C^\pm(\De,J) &\ni -\frac{p_{i}^\pm(J)}{\De-\De_{i}^\pm(J)},
\ee
where $p_{i}^\pm(J)$ are products of OPE coefficients analytically-continued in $J$, and $\De_{i}^\pm(J)$ are dimensions analytically-continued in $J$.\footnote{We comment on the possibility of non-simple poles or branch-cuts in $\De$ below.} When we deform the $\De$-contour, we pick up contributions from these poles. They are interpreted as light-ray operators in the light-ray OPE.

In general, $C^\pm(\De,J)$ can also contain ``spurious" poles at $\Delta = d+J+n$ for integer $n$, originating from poles in the conformal block $G_{J+d-1,\De-d+1}(z,\bar z)$ in the Lorentzian inversion formula (\ref{eq:simonsformula}). In the usual conformal block expansion, these spurious poles cancel with poles in $G_{\De,J}(z,\bar z)$ that are encountered when deforming the $\De$-contour from the principal series to the positive real axis \cite{Dobrev:1977qv,Cornalba:2007fs,Caron-Huot:2017vep,Simmons-Duffin:2017nub}. However, the celestial block $f_\De^{\De_1,\De_2}(\zeta)$ does not have poles in $\De$ to the right of the principal series.\footnote{Assuming $|\De_1-\De_2|$ is not too large. See \cite{Kravchuk:2018htv,Karateev:2018oml} for examples of how to treat the case where $|\De_1-\De_2|$ is large.} Thus, it is not clear how spurious poles in $C^\pm(\De,J)$ could cancel.

Remarkably, it turns out that when we set $J=-1$, spurious poles in $C^\pm(\De,J)$ are absent. This can be seen as follows. Note that the following combination of conformal blocks is free of poles to the right of $\De=\frac d 2$ \cite{Caron-Huot:2017vep}:
\be
G_{J+d-1,\De-d+1}(z,\bar z) - r_{\De,J} G_{\De,J}(z,\bar z),
\ee
where
\be
\label{eq:rdeltaj}
r_{\De,J} &=
\frac{
\Gamma (J+\frac{d-2}{2})
\Gamma(J+\frac{d}{2})
}{
\Gamma (J+1)
\Gamma (J+d-2)
}
\frac{
\Gamma (\Delta -1)
\Gamma (\Delta-d +2)
}{
\Gamma (\Delta -\frac{d}{2})
\Gamma(\Delta-\frac{d-2}{2})
}
\nn\\
&\quad\x
\frac{
\Gamma (J-\Delta +d)
\Gamma (\frac{-d-J+\Delta -\Delta _1+\Delta _2+2}{2})
\Gamma (\frac{-d-J+\Delta +\Delta _3-\Delta _4+2}{2})
}{
\Gamma (\Delta-J-d+2)
\Gamma(\frac{d+J-\Delta -\Delta _1+\Delta _2}{2})
\Gamma (\frac{d+J-\Delta +\Delta _3-\Delta _4}{2})
}.
\ee
Suppose first that $d\neq 4$. Setting $J=-1$, the factor $\G(J+1)^{-1}$ in (\ref{eq:rdeltaj}) ensures that $r_{\De,-1}=0$, so that $G_{J+d-1,\De-d+1}|_{J=-1}$ is free of poles to the right of $\De=\frac d 2$.  In the special case $d=4$, we have \cite{DO1,DO2}
\be
G_{\De,-1} &= \frac{z\bar z}{z-\bar z}(k_{\De-1}(z)k_{\De-1}(\bar z) - k_{\De-1}(z)k_{\De-1}(\bar z)) = 0,
\ee
so that $G_{J+d-1,\De-d+1}|_{J=-1}$ is again free of poles to the right of $\De=\frac d 2$.\footnote{Note that the case $d=2$ is not relevant to our discussion, since there is no transverse space $\R^{d-2}$ in which to consider the light-ray OPE.}

Let us also comment on the case $d=3$. There, the Lorentz group is $\SL(2,\R)$, whose harmonic analysis is slightly different than for the higher-dimensional Lorentz groups. In particular, the Plancherel measure for $\SL(2,\R)$ has support on discrete series representations in addition to principal series representations. In this case, we expect the contribution of discrete series representations to be cancelled by poles in $C^\pm(\De,J)$, in the same way as occurs in the four-point function of fermions in the SYK model \cite{Maldacena:2016hyu}.

The end result is that spurious poles and discrete state contributions are absent in the celestial block expansion for all $d>2$. Deforming the $\De$-contour, we obtain $\cG_{\f_1\f_2}(\zeta)$ as a sum of contributions from light-ray operators when $\z<1$
\be
\label{eq:cgassum}
\cG_{\f_1\f_2}(\z) &=  2^{d+4-\De_{3}-\De_{4}}\pi ^{\frac{d}{2}+3} e^{i \pi  \frac{\De_{4}-\De_{3}}{2}}
  \nn\\
 &\quad \x \sum_i \frac{\Gamma (\Delta_i -2) \p{p_{\De_i}^+ + p_{\De_i}^-}}{\Gamma (\frac{\Delta_i-1 +\De_{3}-\De_{4}}{2} ) \Gamma (\frac{\Delta_i-1 -\De_{3}+\De_{4}}{2}) \Gamma (\frac{\De_{3}+\De_{4}-\Delta_i-1}{2}) \Gamma (\frac{\Delta_i-1 +\De_{3}+\De_{4}-d}{2})}f^{\De_1,\De_2}_{\De_i}(\z) \nn\\
 &(\textrm{when $\z<1$}).
\ee
Here, $i$ labels Regge trajectories and we have abbreviated $\De_i=\De_i(J=-1)$ and $p^\pm_{\De_i} = p^\pm_{i}(J=-1)$.
When $\z=1$, the celestial block $f_{\De}^{\De_1,\De_2}(\z)$ is no longer exponentially damped at large positive $\De$, so (\ref{eq:cgassum}) does not apply. We will see examples of how to treat the case $\z=1$ in section~\ref{sec:treelevel}.

We expect that the above analysis extends to the more general light-ray OPE $\wL[\cO_1]\wL[\cO_2]$, where $\cO_1$ and $\cO_2$ have general spins $J_1$ and $J_2$. In this case, the contour integral over $\De$ in (\ref{eq:finalanswerforproductingeneral}) should become (in schematic notation)
\be
\label{eq:finalanswerforproductingeneralsum}
\wL[\cO_1]
\wL[\cO_2]
&= -\pi i\sum_{i,\l} \cD^{(a)}_{\De_i-1,\l}
\p{\mathbb{O}^+_{i,J_1+J_2-1,\l(a)}
 + \mathbb{O}^-_{i,J_1+J_2-1,\l(a)}
} + \dots.
\ee

Let us return to the assumption that $C^\pm(\De,J)$ (more generally $\mathbb{O}^\pm_{\De,J,\l(a)}$) has only simple poles in $\De$. This is known to be true when the signature and spin are such that $C^\pm(\De,J)$ describes light-transforms of local operators, i.e.\ when $J\in \Z_{\geq 0}$ and $\pm1=(-1)^J$. However, for more general values of $J$, the singularity structure of $C^\pm(\De,J)$ as a function of $\De$ is not known. In the presence of other types of singularities like higher poles and branch cuts, one can define light-ray operators $\mathbb{O}^\pm_{i,J}$ in terms of discontinuities across those singularities, and then suitable generalizations of (\ref{eq:cgassum}) and (\ref{eq:finalanswerforproductingeneralsum}) apply.

\subsection{No contribution from disconnected terms}
\label{sec:disconnected}

Consider an event shape of identical scalars
\be
\label{eq:identicalscalareventshape}
\<\f(p)|\wL[\f]\wL[\f]|\f(p)\>.
\ee
The four-point function of $\phi$'s can be split into connected and disconnected pieces
\be
\label{eq:connectedanddisconnected}
&\<\f(x_1)\f(x_2)\f(x_3)\f(x_4)\>\nn\\
&= \<\f(x_1)\f(x_2)\f(x_3)\f(x_4)\>_c\nn\\
&\quad + \<\f(x_1)\f(x_2)\>\<\f(x_3)\f(x_4)\> + \<\f(x_1)\f(x_3)\>\<\f(x_2)\f(x_4)\> + \<\f(x_1)\f(x_4)\>\<\f(x_3)\f(x_2)\>.
\ee
After taking the light-transforms to compute (\ref{eq:identicalscalareventshape}), the disconnected terms in (\ref{eq:connectedanddisconnected}) must drop out. The reason is that the light-transform of a Wightman two-point function vanishes, since the light-transformed operator annihilates the vacuum.

Despite the simplicity of this argument, vanishing of disconnected contributions in the celestial block expansion is slightly nontrivial. The mechanism is similar to the vanishing of spurious poles discussed in section~\ref{sec:contour}. Note that the contribution of disconnected terms to $C^+(\De,J)$ is given by the OPE coefficient function of Mean Field Theory (MFT). This is \cite{Fitzpatrick:2011dm,Karateev:2018oml}
\be
&C^\mathrm{MFT}(\De,J) \nn\\
&= \frac{2^{d+1-2\De} \G(J+\tfrac d 2)\G(\tfrac{d+1+J-\De}{2})\G(\De-1)\G(\tfrac{\De+J}{2})\G(\tfrac d 2 - \De_\f)^2 \G(\tfrac{J-\De}{2}+\De_\f)\G(\tfrac{\De+J-d}{2}+\De_\f)}{\G(J+1)\G(\tfrac{d+J-\De}{2})\G(\De-\tfrac d 2)\G(\tfrac{\De+J-1}{2}) \G(\tfrac{2d+J-\De-2\De_\f}{2})\G(\tfrac{d+J+\De-2\De_\f}{2})\G(\De_\f)^2} .
\label{eq:mftcontribution}
\ee
Due to the factor $\G(J+1)^{-1}$, this function vanishes at $J=-1$. Thus, we have
\be
C^\pm(\De,J=-1) &= C^\pm_c(\De,J=-1),
\ee
where the subscript ${}_c$ indicates the contribution of the connected term alone. Consequently, disconnected terms don't contribute to the celestial block expansion (\ref{eq:predictionforg}), as expected.

\subsection{Relationship to $t$-channel blocks and superconvergence}
\label{sec:schanvstchan}

In \cite{Kologlu:2019bco}, we introduced an alternative expansion for event shapes in terms of $t$-channel event-shape blocks $G^t_{\De',J'}(p,z_1,z_2)$.  We computed $G^t_{\De',J'}(p,z_1,z_2)$ by inserting a projector onto an individual conformal multiplet $\cO_{\De',J'}$ between $\wL[\cO_1]$ and $\wL[\cO_2]$. An alternative way to obtain it is to first find the contribution of the $t$-channel four-point block $G_{\De',J'}(1-z,1-\bar z)$ in the Lorentzian inversion formula and then plug this into the celestial block expansion (\ref{eq:predictionforg}). 

For example, in the case of scalars $\cO_i=\f_i$ with dimensions $\De_i$, we claim that
\be\label{eq:scalareventshapecrossratiotchannelblock}
G^{t}_{\De',J'}(p,z_1,z_2)
&= 
\frac{(-p^2)^{\frac{\De_1+\De_2+\De_3+\De_4-4-d}{2}} \th(p) }{(-2z_1\.p)^{\De_1-1}(-2z_2\.p)^{\De_2-1}}
\cG_{\De',J'}^t(\z),
\ee
where 
\be
\label{eq:predictionforgtchannel}
&\cG_{\De',J'}^t(\z)=  2^{d+4-\De_{3}-\De_{4}}\pi ^{\frac{d}{2}+3} e^{i \pi  \frac{\De_{4}-\De_{3}}{2}}
  \nn\\
 &\quad \x \int_{\frac{d}{2}-i\oo}^{\frac{d}{2}+i\oo}\frac{d\De}{2\pi i} \frac{2\Gamma (\Delta -2) \cB(\De,-1;\De',J')}{\Gamma (\frac{\Delta-1 +\De_{3}-\De_{4}}{2} ) \Gamma (\frac{\Delta-1 -\De_{3}+\De_{4}}{2}) \Gamma (\frac{\De_{3}+\De_{4}-\Delta-1}{2}) \Gamma (\frac{\Delta-1 +\De_{3}+\De_{4}-d}{2})}f^{\De_1,\De_2}_\De(\z).
\ee
Here $\cB(\De,J;\De',J')$ is the Lorentzian inversion of a single $t$-channel block (defined near (\ref{eq:sumforc})) and $G^t_{\De',J'}(p,z_1,z_2)$ are the functions defined in (5.160) in \cite{Kologlu:2019bco}.
We have verified this identity numerically for some special cases in $d=2$ and $d=4$ using formulas for $\cB^\pm$ from \cite{Liu:2018jhs}.

One property of event-shape $t$-channel blocks is that they are regular in the limit $z_1\to z_2$. This is consistent with (\ref{eq:predictionforgtchannel}) because the Lorentzian inversion of a single $t$-channel block contains double and single poles at double-trace values of $\De$, and no other singularities in $\De$ \cite{Caron-Huot:2017vep,Liu:2018jhs}. Thus, when we deform the $\De$-contour in (\ref{eq:predictionforgtchannel}) to pick up poles, we obtain only double-trace celestial blocks, which are indeed regular near $\z=0$.

Equation (\ref{eq:predictionforgtchannel}) lets us clarify the relationship between (\ref{eq:scalarsuperconvergencerule}) and the superconvergence sum rules written in \cite{Kologlu:2019bco}. Equation~(\ref{eq:scalarsuperconvergencerule}) is a superconvergence sum rule written in $\nu$-space, obtained by decomposing the commutator (\ref{eq:expressionforcommutator}) into celestial conformal partial waves. By contrast, the sum rules of \cite{Kologlu:2019bco} are obtained by decomposing the commutator into $t$-channel conformal multiplets (each of which is a finite sum of spherical harmonics on the celestial sphere). To go from (\ref{eq:scalarsuperconvergencerule}) to the formulas of \cite{Kologlu:2019bco}, we can integrate (\ref{eq:scalarsuperconvergencerule}) against celestial blocks.

\section{Contact terms}
\label{sec:contactterms}

In addition to giving a convergent expansion for the product
\be
\wL[\cO_1](x,z_1,\vec w_1)\wL[\cO_2](x,z_2,\vec w_1)
\ee
for $z_1\not\propto z_2$, the OPE expansion~\eqref{eq:finalanswerforproductingeneral} can also capture contact terms in the limit $z_1\to z_2$, such as those studied in~\cite{Cordova:2018ygx}. A complete description of possible contact terms is beyond the scope of this work. Instead, in this section, we will study two specific examples and explain how \eqref{eq:finalanswerforproductingeneral}, suitably interpreted, can be used to determine contact terms at $z_1\propto z_2$. The contact terms in both examples ultimately arise for the same reason: we must be careful about the distributional interpretation of the integrand in~\eqref{eq:finalanswerforproductingeneral}. In particular, we must ensure that this distribution is analytic in $\De$.

\subsection{Charge detector commutator}
\label{sec:chargecommutator}

Our first example concerns contact terms in the OPE of charge detectors,\footnote{Note that a sufficient condition for the charge-charge correlator to exist is $J_0 < 1$. Therefore, we expect that we encounter divergences in gauge theories both in the weak and strong coupling perturbative expansion. On the other hand, we expect that it exists in the critical $O(N)$ model.}
\be
\wL[J^a](x,z_1)\wL[J^b](x,z_2),
\ee
where $J^a$ is a current for a global symmetry group $G$, and $a$ is an adjoint index for $G$. From~\cite{Cordova:2018ygx}, the commutator should contain a contact term
\be\label{eq:JLRcommutator}
[\wL[J^a](x,z_1),\wL[J^b](x,z_2)]=i f^{abc} \de^{d-2}(z_1,z_2)\wL[J^c](x,z_1),
\ee
where $f^{abc}$ are the structure constants of $G$, and $\de^{d-2}(z_1,z_2)$ is a delta-function on the null cone. To see this, note that
\be
	\int D^{d-2}z\, \wL[J^a](x,z)=Q^a,
\ee
and we should have
\be
	[Q^a,J^b(x,z)]=i f^{abc} J^c(x,z).
\ee
Requiring that $[\wL[J^a](x,z_1),\wL[J^b](x,z_2)]$ vanishes for $z_1\not\propto z_2$ we arrive at~\eqref{eq:JLRcommutator}. Vanishing of this commutator for $z_1\not\propto z_2$ was justified in~\cite{Kologlu:2019bco} if $J_0<1$. This also follows from the arguments of section~\ref{sec:commutativity}.

We would now like to argue for~\eqref{eq:JLRcommutator} using the light-ray OPE. Recall that the commutator is a sum of light-transforms of local operators with spin $J_1+J_2-1=1$. Thus, we must understand three-point structures
\be\label{eq:JJO}
\<J^a(x_1,z_1)J^b(x_2,z_2)\cO^c_{\De}(x_3,z_3)\>^{(a)}
\ee
where $\cO^c_\De$ is a local spin-1 operator in the adjoint representation of $G$, with dimension $\De$. There exist two tensor structures
\be
	&\<J(x_1,z_1)J(x_2,z_2)\cO_{\De}(x_3,z_3)\>^{(1)}=\frac{V_1H_{23}+V_2H_{13}+(\De+d-2)V_3H_{12}+(\De+3)V_1V_2V_3}{X_{12}^{\frac{2d-\De-1}{2}}X_{13}^{\frac{\De+1}{2}}X_{23}^{\frac{\De+1}{2}}},\label{eq:JJnormalstruct}\\
	&\<J(x_1,z_1)J(x_2,z_2)\cO_{\De}(x_3,z_3)\>^{(2)}=\nn\\
	&\frac{(\De-2d+3)((\De+1)(V_1H_{23}+V_2H_{13})-(\De-2d+1)V_3H_{12})+(\De-d+1)(\De+3)V_3^{-1}H_{23}H_{13}}{X_{12}^{\frac{2d-\De-1}{2}}X_{13}^{\frac{\De+1}{2}}X_{23}^{\frac{\De+1}{2}}}.\label{eq:JJspecialstructure}
\ee
Here we used the convention $V_1=V_{1,23}$ and its cyclic permutations, and $H_{ij},V_{i,jk},X_{ij}$ are defined in appendix~\ref{sec:triple light transform}, see also \cite{Costa:2011mg}. The second structure cannot appear in the local three-point function~\eqref{eq:JJO} for generic $\De$ because of the term involving $V_{3}^{-1}$. However, when $\cO=J$ and $\De=d-1$, the term with $V_3^{-1}$ vanishes, and this structure is allowed.\footnote{We thank Simon Caron-Huot for pointing out this interpretation of the second structure at $\De=d-1$.} Moreover, at $\De=d-1$ Ward identities fix the coefficient $\l_2$ of the second structure as
\be
	\l_2^{abc}=\frac{C_J f^{abc}}{(d^2-4)\vol S^{d-1}},
\ee
where $C_J$ is defined by
\be
	\<J^a(x_1,z_1)J^b(x_2,z_2)\>=C_J\frac{H_{12}\de^{ab}}{X_{12}^{d}}.
\ee
We will now argue that the second structure survives the map to celestial structures even at $\De=d-1$ as a contact term.

According to the results of section~\ref{sec:generalizationandcelestialmap}, na\"ively, when $\De=d-1$ the structure~\eqref{eq:JJspecialstructure} does not survive the map to celestial structures because it does not contain factors of $V_3^{-1}$. However, this is only true for $z_1\not\propto z_2$. When $z_1\propto z_2$, this claim must be modified. It should be possible to see this directly by performing a more careful analysis of the map to celestial structures. However, we can also use the following indirect argument. According to the results of section~\ref{sec:generalizationandcelestialmap}, for generic $\De$ the structure~\eqref{eq:JJspecialstructure} gets mapped to the following OPE contribution
\be
\label{eq:equationwithA}
	\wL[J^a](x_0,\vec y_1)\wL[J^b](x_0,\vec y_2)\ni i\pi\frac{(\De+3)(\De-d+1)\l_2^{abc}}{C_J}(|\vec y_{12}|^{(\De-1)-2(d-2)}+\cdots)\wL[\cO^c](x_0,\vec y_2).
\ee
This result is obtained using~\eqref{eq:finalanswerforproductingeneral} and~\eqref{eq:celestialmap}. Here, we put $x_0$ at past null infinity and used transverse coordinates $\vec y_i$ to parametrize the detectors. The factor $(\De-d+1)$ appears because only the term with $V_{3}^{-1}$ in~\eqref{eq:JJspecialstructure} contributes. We can now take the limit $\De\to d-1$ in this expression, taking into account that
\be
	(\De-d+1)|\vec y_{12}|^{(\De-1)-2(d-2)}\to (\vol S^{d-3})\de^{d-2}(\vec y_1-\vec y_2),
\ee
while the higher-order terms in the parenthesis in~\eqref{eq:equationwithA} are less singular and go to zero. We then find
\be
	\wL[J^a](x_0,\vec y_1)\wL[J^b](x_0,\vec y_2)&\ni i\pi\frac{(d+2)\l_2^{abc}\vol S^{d-3}}{C_J}\de^{d-2}(\vec y_1-\vec y_2)\wL[J^c](x_0,\vec y_2)\nn\\
	&=\frac{i f^{abc}}{2}\de^{d-2}(\vec y_1-\vec y_2)\wL[J^c](x_0,\vec y_2).
\ee
It follows from the discussion in~\ref{sec:commutativity} that this is the only term that survives after taking the commutator,\footnote{This is assuming that the first structure~\eqref{eq:JJnormalstruct} does not produce contact terms under the map to celestial structures.} and so we find
\be
	[\wL[J^a](x_0,\vec y_1),\wL[J^b](x_0,\vec y_2)]=i f^{abc}\de^{d-2}(\vec y_1-\vec y_2)\wL[J^c](x_0,\vec y_2),
\ee
as expected.

We expect that it should be possible to generalize this discussion to other commutators considered in~\cite{Cordova:2018ygx}. The main difficulty in this generalization is that the operators considered in~\cite{Cordova:2018ygx} are descendants of light transforms~\cite{Kologlu:2019bco}. We expect that the light-ray OPE can be generalized to OPE of these descendants; we briefly discuss this direction in section~\ref{sec:discussion}.

\subsection{Contact terms in energy correlators in $\cN=4$ SYM}
\label{sec:contacttermsEE}

Our second example concerns the celestial block expansion~\eqref{eq:predictionforg}. For simplicity, we will specialize to $\De_i=2$, which is relevant for the case of energy-energy correlator in $\cN=4$ SYM studied in the next section, see ~\eqref{eq:eecorrelatorF} and ~\eqref{eq:celestiablockforenergy}.

We will focus on the function 
\be\label{eq:fhatdefinition}
	\hat f_\De(\z)&=\frac{4 \pi^4 \Gamma(\Delta-2) }{ \Gamma(\frac{\Delta-1 }{2})^3 \Gamma(\frac{3-\Delta }{ 2})} f^{4,4}_\Delta (\z)\nn\\
	&=\frac{4 \pi^4 \Gamma(\Delta-2) }{ \Gamma(\frac{\Delta-1 }{2})^3 \Gamma(\frac{3-\Delta }{ 2})} \z^{\frac{\Delta -7 }{ 2} } \  _2 F_1( \tfrac{\Delta -1}{2}, \tfrac{\Delta -1}{2}, \Delta -1, \z)
\ee
that multiplies $C^+(\De,-1)$ under the integral in~\eqref{eq:celestiablockforenergy}. Na\"ively, this function vanishes at $\De=3+2n$ due to the $\G$-function in the denominator. However, at the same time the factor $\z^{\frac{\De-7}{2}}$ becomes singular as $\z^{n-2}$ if $n=0,1$. To interpret~\eqref{eq:celestiablockforenergy} in a distributional sense and simultaneously treat it as the integral of an analytic function, we must ensure that we make sense of $\hat f_\De(\z)$ as a distribution that is analytic in $\De$. This distribution must be defined for $\zeta\in[0,1]$.

For $\mathrm{Re}\,\De>5$, $\hat f_\De(z)$ is integrable near $\z=0$ and thus uniquely defines a distribution analytic in $\De$. Therefore, for all other $\De$ the distribution $\hat f_\De(z)$ must be defined by analytic continuation in $\De$. For example,
\be
	\hat f_5(\z)=\lim_{\De\to 5}\hat f_\De(\z)=\lim_{\De\to 5}{8 \pi^4} \tfrac{\De-5}{2}\z^{\frac{\Delta -7 }{ 2}}=8\pi^4\de(\z),
\ee
and similarly
\be
	\hat f_3(\z)=4\pi^4\de'(\z)-2\pi^4\de(\z).
\ee
The other values of $\De$ that give negative integer powers of $\zeta$ are $\De=1-2n$ for $n\geq 0$. In these cases, we find $\hat f_{\De}(\z)=0$, due to higher-order zeros coming from $\G(\tfrac{\De-1}{2})^3$ in the denominator. For other values of $\De$, the exponent of $\z$, even if large and negative, is non-integer, and analytic continuation in $\De$ defines a distribution even though there is no zero coming from the $\G$-functions.

As we will see in the next section, the function relevant for scalar event shapes in $\cN=4$ SYM is $\z^2 \hat f_\De(\z)$. Since we only a found delta function and its first derivative in $\hat f_\De(\z)$, this means that there are no contact terms in the scalar event shapes. Alternatively, by repeating the above analysis for $\z^2 \hat f_\De(\z)$ we find that it stops being integrable at $\De=1$, at which point the $\G(\tfrac{\De-1}{2})^3$ factor in denominator kicks in, and we do not get interesting distributions.

We will also need a slight refinement of the result for $\hat f_\De(\z)$ near $\De=5$. Near this point, the only term non-integrable in $\z$ comes from the leading term of the ${}_2F_1$, so we can write
\be
	\hat f_\De(\z)\sim \frac{4 \pi^4 \Gamma(\Delta-2) }{ \Gamma(\frac{\Delta-1 }{2})^3 \Gamma(\frac{3-\Delta }{ 2})} \z^{\frac{\Delta -7 }{ 2} }.
\ee
Furthermore,
\be
	\frac{4 \pi^4 \Gamma(\Delta-2) }{ \Gamma(\frac{\Delta-1 }{2})^3 \Gamma(\frac{3-\Delta }{ 2})}=4\pi^4(\De-5)+2\pi^4(\De-5)^2-\pi^4(\De-5)^3+O((\De-5)^4),
\ee
and
\be
	\z^{\frac{\Delta -7 }{ 2} }=\frac{2}{\De-5}\de(\z)+\left[\frac{1}{\z}\right]_0 +\frac{\De-5}{2}\left[\frac{\log \z}{\z}\right]_0+ O((\De-5)^2),
\ee
so 
\be\label{eq:f5expansion}
	\hat f_\De(\z)\sim& 8\pi^4\de(\z)+4\pi^4\p{\de(\z)+\left[\frac{1}{\z}\right]_0}(\De-5)\nn\\
	&-2\pi^4\p{\de(\z)-\left[\frac{1}{\z}\right]_0-\left[\frac{\log \z}{\z}\right]_0}(\De-5)^2+
	O((\De-5)^3).
\ee
Here the distribution $[\z^{-1}]_0$ is in principle defined by the Laurent expansion in which it appears. Otherwise, one can define it as the unique distribution which agrees with $\z^{-1}$ on test functions which vanish at $\z=0$ and for which
\be\label{eq:[]_0definition}
	\int_0^1d\z \left[\frac{1}{\z}\right]_0=0.
\ee
Similar comments apply to $[\z^{-1}\log \z]_0$. It is straightforward to obtain subleading terms in~\eqref{eq:f5expansion}. In section~\ref{sec:nequalsfour} we will see that the contact terms we just described are necessary to satisfy the Ward identities for the energy-energy correlator.

\section{Event shapes in $\mathcal{N}=4$ SYM}
\label{sec:nequalsfour}

In this section, we apply the machinery derived above to scalar half-BPS operators in $\cN= 4$ SYM. We re-derive some previous results both at weak and strong coupling and make further predictions. The basic operators of interest are
\be
\cO^{IJ} &= \Tr\p{\Phi^I \Phi^J - \frac 1 6 \de^{IJ} \Phi^K \Phi^K},
\ee
which transform as traceless symmetric tensors of $\SO(6)$, i.e.\ in the $\mathbf{20}'$ representation. These operators are part of a supermultiplet that also contains R-symmetry conserved currents, supersymmetric currents, and the stress tensor, among other operators. 

We will study a scalar event shape, where the detectors, source, and sink are all built from $\cO^{IJ}$'s.
Superconformal Ward identities relate the four-point function of $\mathbf{20}'$ scalars to four-point functions of other operators in the stress tensor multiplet \cite{Eden:1999gh,Howe:1999hz}. These relations were explicitly worked out in \cite{Belitsky:2014zha, Korchemsky:2015ssa}. In particular they imply a simple relation between scalar event shapes and energy-energy correlators which we review below.

The structure of the section is as follows. We first review basic properties of the four-point function of $\mathbf{20}'$ operators and define the scalar event shape of interest. We then explain its relation to the energy-energy correlator which is the main subject of our interest. 
In sections~\ref{sec:treelevel},~\ref{sec:oneloop},~\ref{sec:twoloops}, we apply the light-ray OPE at weak coupling at tree-level, 1-loop, and 2-loops (at leading and subleading twist), finding agreement with previous results and completing them with contact term contributions. In section~\ref{sec:finitecoupling}, we use known OPE data to make a new prediction for the the small-angle limit at $3$ and $4$-loops. In section~\ref{sec:strongcoupling}, we apply the OPE at strong coupling, again finding agreement with previous results.

\subsection{Review: event shapes in $\cN=4$ SYM}

The scalar event shape of interest is built from $\cO^{IJ}$'s, where the $R$-symmetry indices are contracted with particular polarizations. Following the conventions of \cite{Belitsky:2013xxa}, we treat the in- and out-states differently from the detectors. 
For the in- and out-states, we contract $\cO^{IJ}$ with null polarization vectors $Y_I \in \C^{6}$,
\be
\cO(x,Y) &= \left( \frac{N_c^2 - 1}{2\pi^4} \right)^{-1/2} \cO^{IJ}(x) Y_I Y_J.
\ee
The two-point function of $\cO(x,Y)$ is given by
\be
\<\cO(x,\bar Y)\cO(0,Y)\> &=\left( \frac{N_c^2 - 1}{2\pi^4} \right)^{-1}\frac{N_c^2-1}{32\pi^4} \frac{(Y\.\bar Y)^2 }{ x^4} = \frac{(Y\.\bar Y)^2}{16 x^4}.
\ee
For the detectors, we contract the $R$-symmetry indices of $\cO^{IJ}$ with traceless symmetric tensors $S_{IJ}$,
\be
\cO(x,S) &= 2\cO^{IJ}(x) S_{IJ}.
\ee
Obviously, $\cO(x,Y)$ and $\cO(x,S)$ encode the same thing, with the $R$-symmetry indices treated slightly differently.

Let us denote $\cO(z)\equiv \frac 1 2 \wL[\cO](\oo,z)$, where $z$ is a future-pointing null vector.\footnote{The factor of $\frac 1 2$ is for consistency with the definitions of \cite{Belitsky:2013xxa}.} Our scalar event shape is defined by
\be
\label{eq:defscflow}
\<\cO(z_1,S_1) \cO(z_2,S_2)\>_{p,Y} &\equiv \s_\mathrm{tot}^{-1}(p,Y) \int d^d x e^{-ip\. x} \<\Omega|\cO(x,\bar Y) \cO(z_1,S_1) \cO(z_2,S_2) \cO(0,Y) |\Omega\>, \\
\s_\mathrm{tot}(p,Y) &\equiv \int d^d x e^{-ip\. x} \<\Omega|\cO(x,\bar Y) \cO(0,Y) |\Omega\> = 2\pi^3 \frac{(Y\.\bar Y)^2}{16} \th(p) .
\ee
This event shape is sometimes called ``scalar flow," by analogy with energy flow observables that measure the flow of energy at null infinity.

Following \cite{Belitsky:2013xxa}, let us choose the $R$-symmetry structures
\be
Y_0 &= (1,0,1,0,i,i),  \nn \\
S_0 &= \mathrm{diag}(1,-1,0,0,0,0),  \nn\\
S'_0 &= \mathrm{diag}(0,0,1,-1,0,0).
\label{eq:specialrstructs}
\ee
With this choice, we have $\< \cO(x,\bar Y_0) \cO(0, Y_0) \> = \frac{1 }{ x^4}$. Moreover, only the $\mathbf{105}$ representation of $\SO(6)$ contributes to the $\cO(n_1,S)\x \cO(n_2,S')$ OPE.  Finally, superconformal Ward identities relate the event shape with these choices to energy correlators
\be\label{eq:eescalarrelation}
\<\cE(z_1)\cE(z_2)\>_{p,Y_0} &= \frac{16 (-p^2)^2}{(-2z_1\.z_2)^2} \<\cO(z_1,S_0)\cO(z_2,S'_0)\>_{p,Y_0} + \text{protected contact terms}.
\ee
(The energy correlators are independent of $Y$.) In~\cite{Belitsky:2014zha}, this relation was derived while ignoring contact terms at $z_1\propto z_2$. We will find that consistency with the OPE requires correcting this relation by contact terms. We expect that these contact terms come from the protected part of the $\mathbf{20}'$ four-point function. We discuss them in more detail below.

Using (\ref{eq:scalareventshape}), the scalar event shape can be written
\be
\label{eq:definitionoff}
\<\cO(z_1,S_0)\cO(z_2,S_0')\>_{p,Y_0} &= \p{\frac 1 2}^2\frac{1}{2\pi^3} \frac{\mathcal{G}_{\cO \cO}(\zeta)}{(-2z_1 \. p ) (-2z_2 \. p )},
\ee
where the factor $(\tfrac 1 2)^2$ in (\ref{eq:definitionoff}) comes from $\cO(z)\equiv \frac 1 2 \wL[\cO](\oo,z)$.
The function $\cG_{\cO\cO}(\z)$ has a celestial block expansion given by (\ref{eq:predictionforg}):
\be
\cG_{\cO\cO}(\z)
&= 
\int_{2 - i \infty}^{2+i \infty} \frac{d \Delta }{ 2 \pi i} C^+(\Delta, -1) \frac{16 \pi^5 \Gamma(\Delta-2) }{ \Gamma(\frac{\Delta-1 }{2})^3 \Gamma(\frac{3-\Delta }{ 2})} f_\Delta^{2,2} (\z).
\label{eq:celestialblockforscalar}
\ee
Here, $C^+(\De,-1)$ encodes the OPE data of the $\<\cO\cO\cO\cO\>$ four-point function, analytically continued to spin $J=-1$. We discuss this four-point function in section~\ref{sec:20'fourpt}. Since the $\mathbf{105}$ representation appears in the symmetrized tensor square of the $\mathbf{20}'$ representation, the OPE contains only even spin operators. This is the reason for the absence of $C^{-}(\Delta,-1)$ in (\ref{eq:celestialblockforscalar}).

The superconformal Ward identity (\ref{eq:eescalarrelation}) lets us obtain a celestial block expansion for the energy-energy correlator in terms of OPE data for the scalar four-point function. Let us define the function $\cF_\cE(\zeta)$ by
\be
\label{eq:eecorrelatorF}
\<\cE(z_1)\cE(z_2)\>_{p,Y_0} &= \frac{4\vol S^{2}}{2\pi^3} \frac{(-p^2)^4}{(-2z_1\.p)^3(-2z_2\.p)^3}\cF_\cE(\z).
\ee
Here we include the factor $4\vol S^2=16\pi$ because it simplifies the Ward identities discussed below. The relation (\ref{eq:eescalarrelation}) implies that $\cF_\cE$ has the celestial block expansion
\be
\label{eq:celestiablockforenergy}
\cF_\cE(\z) &=  \int_{2 - i \infty}^{2+i \infty} \frac{d \Delta }{ 2 \pi i} C^+(\Delta, -1) \frac{4 \pi^4 \Gamma(\Delta-2) }{ \Gamma(\frac{\Delta-1 }{2})^3 \Gamma(\frac{3-\Delta }{ 2})} f_\Delta^{4,4} (\z) + \xi(\z),
\ee
where
\be
f_\Delta^{4,4}(\z) &= \z^{\frac{\Delta -7 }{ 2} } \  _2 F_1\p{ \frac{\Delta -1 }{ 2}, \frac{\Delta -1 }{ 2}, \Delta -1, \z}, \\
\xi (\z) &\equiv \frac{1}{4}(2\de(\z)-\de'(\z)). \label{eq:univcontactterm}
\ee
Here,  $C^+(\Delta,-1)$ is the same function that enters (\ref{eq:celestialblockforscalar}). The function $\xi(\z)$ represents the protected coupling-independent contact terms mentioned in (\ref{eq:eescalarrelation}). Below, we fix $\xi(\z)$ by requiring consistency with Ward identities and check that it is indeed independent of the coupling (at one and two loops, and at strong coupling). Its effect is to remove the contribution of short multiplets from $C^+(\De,-1)$ in (\ref{eq:celestiablockforenergy}). It would be interesting to derive the presence of $\xi(\z)$ from first principles along the lines of \cite{Belitsky:2014zha}. 

For $0<\z\leq 1$, the superconformal Ward identity relating scalar flow and the energy-energy correlator takes the simple form
\be
\label{eq:fgrelationship}
\cF_\cE(\z) &= \frac{\cG_{\cO\cO}(\z)}{4\pi \z^2},\qquad (0<\z\leq 1).
\ee
However, the celestial block expansion (\ref{eq:celestiablockforenergy}) also captures contact terms at $\z=0$ that are not captured by (\ref{eq:fgrelationship}).

When evaluating the celestial block expansion for $\zeta<1$, we will find it convenient to close the $\De$-contour to the right as discussed in section \ref{sec:contour} and write the event shape as a sum over Regge trajectories, see (\ref{eq:cgassum}). For example, we have
\be
\cF_{\cE}(\z) &= \sum_i p_{\Delta_i}  \frac{4 \pi^4 \Gamma(\Delta_i -2) }{ \Gamma(\frac{\Delta_i -1 }{2})^3 \Gamma(\frac{3-\Delta_i }{ 2})} f^{4,4}_{\Delta_i} (\z) + \xi(\z),\qquad
(\z<1).
\label{eq:finCoupl}
\ee

Before computing $\cF_\cE(\z)$, let us comment on some of its properties.
First, $\cF_\cE(\z)$ is constrained by Ward identities. By integrating $\cE(z_1)$ over the celestial sphere with the appropriate weight, we can produce different components of the translation generators $P^\mu$. In the energy correlator~\eqref{eq:eecorrelatorF}, these must evaluate to $p^\mu$, which leads to the Ward identities
\be
	\int_0^1 d\z\,\cF_{\cE}(\z)&=\half,\label{eq:energyWI}\\
	\int_0^1 d\z(2\z-1) \cF_{\cE} (\z)&=0.\label{eq:momentumWI}
\ee
Since (\ref{eq:energyWI}),(\ref{eq:momentumWI}) are sensitive to the values of $\cF_\cE(\z)$ at arbitrary angle $\z$ they can be used as a nontrivial consistency check on the computations of $\cF_{\cE}(\z)$. 

Finally, note that $\cF_{\cE}(\zeta)$ has a weak-coupling expansion
\be
\cF_{\cE}(\zeta) &= \cF_{\cE}^{(0)}(\zeta) + a \cF_{\cE}^{(1)}(\zeta)  + \dots ,
\qquad
a \equiv \frac{g_{\mathrm{YM}}^2 N_c}{4\pi^2}.
\ee
 $\cF_{\cE}(\zeta)$ is explicitly known up to two-loop order \cite{Belitsky:2013ofa}, and as a two-fold integral at three loops \cite{Henn:2019gkr}.\footnote{The quantity ${\rm EEC}(\z)$ computed in \cite{Belitsky:2013ofa,Henn:2019gkr} is equal to our $\cF_{\cE}(\z)$.}  It is also easily computable at strong coupling, reproducing the result of Hofman and Maldacena \cite{Hofman:2008ar}.
 
\subsection{Review: four-point function of $\mathbf{20}'$ operators}
\label{sec:20'fourpt}

The main ingredient in computing $\cF_{\cE}(\zeta)$ is the four-point function of $\mathbf{20}'$ operators that enters in the definition of the scalar event shape (\ref{eq:defscflow}), specialized to the $R$-symmetry structures (\ref{eq:specialrstructs}). This is
\be
\label{eq:fourpointneq4}
\<\cO(x_4,\bar Y_0) \cO(x_1,S_0) \cO(x_2,S'_0)\cO(x_3,Y_0) \>
&= \frac{G^{(105)}(u,v) }{ x_{12}^4 x_{34}^4}.
\ee

It will be convenient to write $G^{(105)}(u,v)$ in two different ways. Firstly, we have
\be
\label{eq:fourpointneq4g105}
G^{(105)}(u,v)&=
 \frac{c}{2(2\pi)^4}
 \p{u^2 + \frac{u^2 }{ v^2} }  + \frac{1}{(2\pi)^4} \frac{u^2 }{v} \p{\frac 1 2 + u \Phi(u,v) },
\ee
where the central charge $c$ is given by
\be
c &= \frac{N_c^2-1}{4}.
\ee
Here, the function $\Phi(u,v)= \Phi(v,u) = \frac{1 }{ v}\Phi\left( \frac{u }{ v} , \frac{1 }{ v} \right)$ encodes the dependence of the correlator on the coupling $a$ (it is zero for $a=0$). 
It is known explicitly
up to three loops \cite{Drummond:2013nda}.  The integrand for $G^{(105)}(u,v)$ is known up to ten loops in the planar limit \cite{Eden:2011we,Bourjaily:2016evz}. 

The other way of writing $G^{(105)}(u,v)$ is to organize it into the contribution of short and long supermultiplets in the superconformal block expansion,
\be
G^{(105)}(u,v) = 
\frac{c}{2(2\pi)^4}
u^2 G^{(\mathrm{short})}(u,v) + \cH(u,v),
\ee
where $G^{(\mathrm{short})}(u,v)$ encodes the contribution from protected operators and was computed in \cite{Beem:2016wfs}. $\cH(u,v)$ encodes the contribution of long multiplets and can be written in terms of superconformal blocks as follows
\be
\label{eq:superOPE}
\cH(u,v) = \sum_{\Delta}\sum_{\text{even}~J} a_{\Delta,J} g_{\Delta+4, J}(u,v) =  \sum_{\Delta}\sum_{\text{even}~J} p_{\Delta, J} g_{\Delta, J}(u,v)
\ee
where $g_{\Delta, J}(u,v)$ are the usual conformal blocks and $a_{\Delta, J}$ is the three-point coupling to a given superconformal primary, see e.g.\ \cite{Beem:2016wfs}.\footnote{Note that \cite{Alday:2017vkk} used a different conformal block normalization.} We will use $p_{\Delta, J}$ to denote the three-point coupling to a given conformal primary. Note that only even spin operators enter in the OPE decomposition of $G^{(105)}(u,v)$.

Because of the factor $\Gamma(\frac{3-\Delta }{ 2})^{-1}$ in (\ref{eq:finCoupl}), most protected operators from $G^{\mathrm{(short)}}(u,v)$ will not contribute to $\cF_\cE(\z)$. However, operators with dimensions $\De=3$ and $\De=5$ can contribute contact terms at $\z=0$, in accordance with the discussion in section~\ref{sec:contacttermsEE}.

\subsection{Tree level}
\label{sec:treelevel}

To get the tree-level correlator we set $\Phi(u,v) = 0$ in (\ref{eq:fourpointneq4}). 
Recall from section~\ref{sec:disconnected} that
\be
	C^+(\De,J=-1)=C_c(\De,J=-1),
\ee
where $C_c(\De,-1)$ corresponds to the connected part of the four-point function. Written in terms of cross-ratios, the connected tree-level correlator takes the form
\be\label{eq:4ptN4SYMtreelevel}
\<\cO(x_4,\bar Y_0) \cO(x_1,S_0) \cO(x_2,S'_0)\cO(x_3,Y_0) \>_{c}^{\mathrm{tree}} = \frac{1}{2(2\pi)^4} \frac{1}{x_{14}^2 x_{23}^2 x_{13}^2 x_{24}^2} &= \frac{1}{x_{12}^4 x_{34}^4}\p{\frac{1}{2(2\pi)^4} \frac{u^2}{v}}.
\ee
Plugging into the inversion formula, we have
\be
C_{c}^{\mathrm{tree}}(\De,J) &= 2\frac{\ka_{\De+J}}{4} \int_0^1 \int_0^1 \frac{dz}{z^2} \frac{d\bar z}{\bar z^2} \frac{(z-\bar z)^2}{(z\bar z)^2} G_{J+3,\De-3}(z,\bar z) \mathrm{dDisc}\left[\frac{1}{2(2\pi)^4}\frac{(z\bar z)^2}{(1-z)(1-\bar z)}\right],
\ee
where the factor of $2$ in front comes from the fact that the $t$- and $u$-channel terms in the inversion formula are equal.
$\mathrm{dDisc}\frac{1}{1-\bar z}$ is delta-function supported near $\bar z=1$. To regulate it, we will replace 
\be
\frac{z\bar z}{(1-z)(1-\bar z)} &\to \frac{(z\bar z)^{1+\de}}{((1-z)(1-\bar z))^{1+\de}}.
\ee
Recall that \cite{DO1,DO2}
\be
G_{J+3,\De-3}(z,\bar z) &=  \frac{z\bar z}{z-\bar z}(k_{\De+J}(z) k_{4+J-\De}(\bar z)-k_{\De+J}(\bar z) k_{4+J-\De}(z)),\nn\\
k_\beta(z) &= z^{\beta/2} {}_2F_1\p{\frac{\beta}{2},\frac{\beta}{2},\beta,z}.
\ee
Doing the integral and removing the regulator $\delta \to 0$ leads to the result
\be
\label{eq:treelevelC}
& C^{\mathrm{tree}}_c(\De,J)\nn\\
&= \frac{\G(\tfrac{\De+J}{2})^4}{\G(\De+J-1)\G(\De+J)}\frac{1}{2(2\pi)^4} \p{ \frac{\G(\De+J)}{\G(\tfrac{\De+J}{2})^2}I_1(\tfrac{J+4-\De}{2},-1)-I_1(\tfrac{\De+J}{2},-1)\frac{\G(J+4-\De)}{\G(\tfrac{J+4-\De}{2})^2}}, 
\ee
where
\be
I_\a(h,p)&\equiv \int_0^1 \frac{dz}{z(1-z)} z^{\a} \p{\frac{z}{1-z}}^p k_{2h}(z)\nn\\
&= \frac{\G(\a+p+h)\G(-p)}{\G(\a+h)}{}_3F_2(h,h,\a+p+h;2h,\a+h;1).
\ee
We can now compute the energy-energy correlator by plugging (\ref{eq:treelevelC}) at $J=-1$ into (\ref{eq:celestiablockforenergy}). The result is
\be
&C^{\mathrm{tree}}_c(\Delta, -1) \frac{4 \pi^4 \Gamma(\Delta-2) }{ \Gamma(\frac{\Delta-1 }{2})^3 \Gamma(\frac{3-\Delta }{ 2})}  \nn\\
&=\frac{1}{8} \p{\frac{1}{\G(\tfrac{\De-1}{2})}\frac{I_1(\tfrac{3-\De}{2},-1)}{\G(\tfrac{3-\De}{2}) }-I_1(\tfrac{\De-1}{2},-1)\frac{\G(\tfrac{\De-1}{2})}{\G(\De-1)}\frac{\G(3-\De)}{\G(\tfrac{3-\De}{2})^3}} \nn\\
&= 
\frac{1}{8}
\frac{\G(\tfrac{\De-1}{2})}{\G(\De-1)\G(\tfrac{3-\De}{2})} \int_0^1 \frac{dz}{z} \p{
\frac{\G(\De-1)}{\G(\tfrac{\De-1}{2})^2} k_{3-\De}(z)
- \frac{\G(3-\De)}{\G(\tfrac{3-\De}{2})^2} k_{\De-1}(z) } \nn\\
&=\frac{1 }{ 8} \frac{\Gamma(\frac{\Delta - 1 }{ 2})^2 }{ \Gamma(\Delta-2)} \ .
\label{eq:nicecombination}
\ee
This expression is free of poles to the right of the principal series, so by closing the contour in~\eqref{eq:celestiablockforenergy} to the right we conclude that $\cF_{\cE}(\z) = 0$ for $0<\z<1$. This ignores the possibility of contact terms discussed in section~\ref{sec:contacttermsEE}, which we now address.

Let us start by studying contact terms at $\z=0$. As explained in section~\ref{sec:contacttermsEE}, apart from the protected contact term $\xi(\z)$ in (\ref{eq:univcontactterm}), the energy correlator $\cF_\cE(x)$ may receive contact terms from the integral~\eqref{eq:celestiablockforenergy}. Indeed, when $\z=0$, the distribution $\hat f_\De(\z)$ does not vanish at $\De=3,5$, and we in fact have
\be
	\cF_{\cE}^{(0)}(\z) &=-(4\pi^4\de'(\z)-2\pi^4\de(\z))\mathrm{res}_{\De=3}\, C^\text{tree}_c(\Delta, -1) -8\pi^4\de(\z)\mathrm{res}_{\De=5}\, C^\text{tree}_c(\Delta, -1) + \xi(\z) \nn\\
	&  =\frac{1}{4} \de(\z)  \ , \qquad
(\z<1).
\ee

Let us now analyze contact terms at $\z=1$. When $\z=1$, we should worry about the convergence of the integral when closing the contour, since there is no suppression coming from $\z^{\frac{\De-7}{2}}$ in the celestial block.
To probe possible delta-function terms localized at $\z=1$ let us consider moments of the energy flow 
\be
\int_0^1 d \z \ \z^N \cF_{\cE}^{(0)}(\z) = \frac{\delta_{N,0} }{ 4} +  \int_{C_0- i \infty}^{C_0+i \infty} \frac{d \Delta }{ 2 \pi i} C^{\mathrm{tree}}_c(\Delta, J=-1) \frac{4 \pi^4 \Gamma(\Delta-2) }{ \Gamma(\frac{\Delta-1 }{2})^3 \Gamma(\frac{3-\Delta }{ 2})}  \int_0^1 d \z \ \z^N f_\Delta^{4,4} (\z) , \.
\ee
where we deformed the integration contour to ${\rm Re}[\Delta]=C_0>5$ so that the integral $ \int_0^1 d \z \ \z^N f_\Delta^{4,4} (\z)$ converges for $N \geq 0$.
We find that at large $|\De|\gg 1$ the integrand behaves as
\be
\int_0^1 d \z \ \z^N \cF_{\cE}^{(0)}(\z) =\frac{\delta_{N,0} }{ 4} + \int_{2 - i \infty}^{2+i \infty} \frac{d \Delta }{ 2 \pi i} \frac{1 }{ 2 \Delta} = \frac{1 +\delta_{N,0}}{ 4},
\ee
where we evaluated the $\Delta$ integral using the principal value prescription. If we subtract off this leading behavior, then the contour deformation in $\De$ becomes legitimate and we get $0$ for the remainder. This implies that $\cF_\cE^{(0)}(\z) \ni \frac{1 }{ 4} \delta(1 - \z )$, in agreement with the straightforward scattering amplitude evaluation, see e.g. \cite{Belitsky:2013xxa}. More generally, we see that distributional terms supported at $\z=1$ are encoded in the large-$\Delta$ behavior of $C^+(\Delta, J=-1)$.

To summarize, the energy-energy correlator at tree-level is given by
\be
	\cF^{(0)}_{\cE}(\z)=\frac{1}{4}(\de(\z)+\de(1-\z)).
\ee
Note that this is the unique expression with delta functions at $\z=0$ and $\z=1$ that satisfies both Ward identities~\eqref{eq:energyWI} and~\eqref{eq:momentumWI}.

\subsection{One loop}
\label{sec:oneloop}

To study perturbative corrections, let us briefly discuss how they are encoded in $C^+(\De,J)$. Non-perturbatively, we have poles of the form
\be
	C^+(\De,J)\sim -\frac{a_i(a)}{\De-\De_i(a)},
\ee
where $a_i(a)$ and $\De_i(a)$ are, respectively, the product of OPE coefficients and scaling dimension of an exchanged operator. 

We furthermore have expansions
\be
	a_i(a)&=a_i^{(0)}+a \ a_i^{(1)} +a^2 a_i^{(2)}+\cdots,\\
	\De_i(a)&=\De_i^{(0)}+a \ \g_i^{(1)} +a^2 \g_i^{(2)}+\cdots,
\ee
and thus
\be
	C^+(\De,J)\sim -\frac{a_i^{(0)}}{\De-\De_i^{(0)}}+a\p{
		-\frac{a_i^{(1)}}{\De-\De_i^{(0)}}
		-\frac{a_i^{(0)}\g_i^{(1)}}{(\De-\De_i^{(0)})^2}
	}+\cdots.
\ee
Suppose now that there is a degeneracy at tree level, i.e.\ $\De^{(0)}_i=\De^{(0)}_*$. Then we have
\be
	C^+(\De,J)&\sim -\frac{\sum_i a_i^{(0)}}{\De-\De_*^{(0)}}+a\p{
		-\frac{\sum_i a_i^{(1)}}{\De-\De_*^{(0)}}
		-\frac{\sum_i a_i^{(0)}\g_i^{(1)}}{(\De-\De_*^{(0)})^2}
	}+\cdots\nn\\
	&\sim -\frac{\<a_*^{(0)}\>}{\De-\De_*^{(0)}}+a\p{
		-\frac{\<a_*^{(1)}\>}{\De-\De_*^{(0)}}
		-\frac{\<a_*^{(0)}\g_*^{(1)}\>}{(\De-\De_*^{(0)})^2}
	}+\cdots,
\ee
where we introduced the notation $\<\cdots\>$ (used extensively below) representing the total contribution of operators that are degenerate at tree level. Below, the subscript ${}_*$ will be replaced by a label referring to the degenerate group of operators. The contribution of these poles to~\eqref{eq:celestiablockforenergy} now becomes
\be\label{eq:eventshape-one-loop}
	\cF_\cE^{(1)}(\z)\ni \<a_*^{(1)}\>\left[\frac{4 \pi^4 \Gamma(\Delta-2) }{ \Gamma(\frac{\Delta-1 }{2})^3 \Gamma(\frac{3-\Delta }{ 2})} f_\Delta^{4,4} (\z)\right]_{\De=\De_*^{(0)}}
	+\<a_*^{(0)}\g_*^{(1)}\>\left[\ptl_\De \frac{4 \pi^4 \Gamma(\Delta-2) }{ \Gamma(\frac{\Delta-1 }{2})^3 \Gamma(\frac{3-\Delta }{ 2})} f_\Delta^{4,4} (\z)\right]_{\De=\De_*^{(0)}}.
\ee

In this section, we will not compute $C^+(\De,-1)$, but rather use the known OPE data $\<a_*^{(1)}\>$ and $\<a_*^{(0)}\g_*^{(1)}\>$, analytically continued to $J=-1$. The complete OPE data for the one-loop correlator was written down in~\cite{Henriksson:2017eej}. Recall from section~\ref{sec:20'fourpt} that the contribution of long multiplets, which are the ones that receive loop corrections, is given by
\be
\label{eq:handphi}
{\cal H}(u,v) =
\frac{c}{2(2\pi)^4}
u^2 \p{1 + \frac{1 }{ v^2} - G^{\mathrm{(short)}}(u,v) }  + \frac{1}{(2\pi)^4} \frac{u^2 }{v} \p{\frac 1 2 + u \Phi(u,v) }.
\ee
At tree level, this can be decomposed into superconformal blocks (\ref{eq:superOPE}) as follows, see (2.21) in \cite{Henriksson:2017eej},
\be
\label{eq:treelevel}
\la a_{\tau=2 , J}^{(0)} \ra &= \frac{1 }{ (2 \pi)^4} \frac{\Gamma(J+3)^2 }{ \Gamma(2J+5)} \ , \nn \\
\la a_{\tau, J}^{(0)} \ra &=\frac{c }{ (2 \pi)^4} \frac{\Gamma(\frac{\tau }{ 2}+1)^2 \Gamma(\frac{\tau }{ 2}+J+2)^2 }{ \Gamma(\tau+1)\Gamma(\tau+2J+3) } \left( (\tau+J+2)(J+1) + \frac{(-1)^{\tau/2} }{ c} \right) , ~~~ \tau = 4,6,8, ... \ ,
\ee
where we used twist $\tau = \Delta - J$ and even spin $J\geq 0$ to label the operators. 

Note that for $\tau>2$ there are degeneracies in the spectrum, so $\<\cdots\>$ notation is necessary. One can check that (\ref{eq:treelevel}) indeed correctly reproduces (\ref{eq:handphi}) upon setting $\Phi(u,v)$ to zero. 

In perturbation theory, we write
\be
	\Phi(u,v) = a \ \Phi^{(1)}(u,v)+a^2 \Phi^{(2)}(u,v) + \cdots,
\ee
and similarly for $\cH(u,v)$.
At one loop we have
\be
{\cal H}^{(1)}(u,v) &=\frac{1 }{ (2 \pi)^4} \frac{u^3 }{ v} \Phi^{(1)}(u,v) , \nn \\
\Phi^{(1)}(u,v) &=- \frac{1 }{ 4}\frac{1 }{ z - \bar z} \left( 2 {\rm Li}_2(z) - 2 {\rm Li}_2(\bar z) + \log z \bar z \log \frac{1- z }{ 1 - \bar z} \right) \ .
\ee
Analogously to~\eqref{eq:eventshape-one-loop}, the OPE data enters as
\be
\delta {\cal H}(u,v) = \sum_{\tau=2,4,... ;\ {\rm even} \ J}^{\infty} \left( \la a_{\tau, J}^{(1)}\ra G_{\tau+ 4+J, J} + \la a_{\tau,J}^{(0)} \gamma_{\tau, J}^{(1)} \ra \pa_{\tau} G_{\tau+4+J,J}  \right),
\ee
where for convenience we labeled the superconformal primaries by twist $\tau = \Delta - J$ instead of the dimension (as we did in (\ref{eq:superOPE})). 

The result of the one-loop decomposition for anomalous dimensions is, see (A.24-A.25) in \cite{Henriksson:2017eej},  
\be
 \la \gamma_{2,J} \ra &\equiv \frac{\la a_{2 , J}^{(0)} \gamma_{2,J}^{(1)} \ra }{ \la a_{2 , J}^{(0)} \ra } = 2 S_1 (J+2) , \nn \\
 \la \gamma_{\tau,J} \ra &\equiv  \frac{\la a_{\tau , J}^{(0)} \gamma_{\tau,J}^{(1)} \ra }{ \la a_{\tau , J}^{(0)} \ra } = - \frac{2 }{ c} \frac{ [\eta+1] S_1 (\frac{\tau }{ 2}) + [\eta-1] S_1(\frac{\tau }{ 2} + J + 1) 
  }{
    (\tau+J+2)(J+1) + \frac{\eta }{ c} },
\ee
where following \cite{Henriksson:2017eej} we introduced $\eta = (-1)^{\tau/2}$ and
\be
S_k(N) = \sum_{i=1}^N \frac{1 }{ i^k}.
\ee

We can concisely write the OPE coefficients at one loop by defining
\be
\la a_{\tau, J}^{(1)} \ra \equiv \la \alpha_{\tau, J} \ra \la a_{\tau, J}^{(0)} \ra + \frac{1 }{ 2} \pa_J \la a_{\tau , J}^{(0)} \gamma_{\tau,J}^{(1)} \ra .
\ee
The coefficients $\<\alpha_{\tau,J}\>$ are
\be
\la \alpha_{2,J}\ra &= - \zeta_2 , \nn \\
\la \alpha_{\tau,J}\ra &= - \frac{2 }{ c}  \frac{1 }{ \left( (\tau+J+2)(L+1) + \frac{(-1)^{\tau/2} }{ c} \right)} \Bigg( \frac{1 - \eta }{ 2} \zeta_2 + (1+\eta) S_1 (\tfrac{\tau}{2})^2 - \frac{1 + \eta }{ 2} S_2 (\tfrac{\tau}{2})  \nn \\
&\qquad  - (1+\eta) S_1 (\tfrac{\tau}{2}) S_1(\tau) + [(2 \eta -1 ) S_1 (\tfrac{\tau}{2}) + (1-\eta) S_1(\tau)  ] S_1 (\tfrac{\tau}{2} + J + 1) \Bigg),
\ee

Note that for superconformal primaries of twist $\tau$ and spin $J$ we should set $\De^{(0)}_*=4+\tau+J$ in~\eqref{eq:eventshape-one-loop}. Here the shift by $4$ is due to the form of the superconformal block in~\eqref{eq:superOPE}. This means that for twist $\tau=2n$, $n\geq 1$, and spin $J=-1$ we have to use $\De_*^{(0)}=3+2n$. In this case, the first term in~\eqref{eq:eventshape-one-loop} vanishes for $\z\neq 0$ due to the factor $\G(\tfrac{\De-3}{2})^{-1}$. Thus, the only relevant term is the one proportional to $\la a_{\tau , -1}^{(0)} \gamma_{\tau,-1}^{(1)} \ra$ for which we get
\be
\la a_{\tau , -1}^{(0)} \gamma_{\tau,-1}^{(1)} \ra = (-1)^{\tau/2+1} \frac{\Gamma(1+\frac{\tau }{ 2})^4 S_{1}(\frac{\tau }{ 2}) }{ 4 \pi^4 \Gamma(1+\tau)^2} \ .
\ee
From this we conclude that for $0< \z < 1$
\be
\cF_{\cE}^{(1)}(\z) &=2 \pi^4  \sum_{n=1}^{\infty} (-1)^{n+1} \la a_{\tau=2n , -1}^{(0)} \gamma_{\tau=2n,-1}^{(1)} \ra \frac{1 }{ r_{n+1} }  f_{3+2n}^{4,4}(\z) \nn \\
&=\sum_{n=1}^{\infty} \frac{(n!)^2 }{ 2 (2n)!} S_1(n) f_{3+2n}^{4,4}(z) =  - \frac{1 }{ 4} \frac{ \log (1- \z) }{ \z^2 (1 - \z)} \ ,
\ee
where 
\be
\label{eq:definitionofhrh}
	r_h=\frac{\G(h)^2}{\G(2h-1)}.
\ee
Again our results are in perfect agreement with the direct evaluation performed in \cite{Belitsky:2013xxa}. 

Let us now analyze the contact terms at $\z=0$ and $\z=1$ in $\cF_\cE(\z)$. First, let us fix these contact terms using the result for $0<\z<1$
\be
\label{eq:awayfromzerooneoneloop}
	\cF_\cE^{(1)}(\z)=- \frac{1}{4} \frac{\log(1-\z)}{\z^2(1-\z)} \qquad (0<\z<1),
\ee
together with Ward identities.
We will then check that we reproduce the same contact terms at $\z=0$ using the light-ray OPE.
We can rewrite (\ref{eq:awayfromzerooneoneloop}) as
\be
\label{eq:regdefinitiononeloop}
	\cF_\cE^{(1)}(\z)=\frac{1}{4\z}-\frac{1}{4}\frac{\log(1-\z)}{1-\z}+\cF_\cE^{(1),\text{reg}}(\z),
\ee
where $\cF_\cE^{(1),\text{reg}}(\z)$ is integrable near $0$ and $1$, and so has an unambiguous distributional interpretation. We then only need to interpret the first two terms. The most general expression we can write is\footnote{We assume that there are no derivatives of delta-functions. We verify this at $\z=0$ using the OPE.}
\be
	\cF_\cE^{(1)}(\z)=c_0^{(1)}\de(\z)+c_1^{(1)}\de(1-\z)+\frac{1}{4}\left[\frac{1}{\z}\right]_0-
	\frac{1}{4}\left[\frac{\log(1-\z)}{1-\z}\right]_1+\cF_\cE^{(1),\text{reg}}(\z),
\ee
where $[\cdots]_0$ is defined near~\eqref{eq:[]_0definition}, and the definition of $[\cdots]_1$ is analogous with $\z\to 1-\z$. Ward identities~\eqref{eq:energyWI} and~\eqref{eq:momentumWI} require
\be
	\int_0^1 d\z\, \cF_\cE^{(1)}(\z)=\int_0^1 d\z(2\z-1) \cF_\cE^{(1)}(\z)=0,
\ee
from which we find 
\be
	c_0^{(1)}=-\frac{1}{4},\qquad c_1^{(1)}=-\frac{\z_2}{4}.
\ee

We would now like to reproduce the distributional piece near $\z=0$
\be
	\cF_\cE^{(1)}(\z)=-\frac{1}{4}\de(\z)+\frac{1}{4}\left[\frac{1}{\z}\right]_0+\text{regular}
\ee
from the OPE. From the discussion in section~\ref{sec:contacttermsEE} together with~\eqref{eq:eventshape-one-loop}, this piece is given by
\be
	\cF_\cE^{(1)}(\z)&\ni \<a^{(1)}_{\tau=2,-1}\>\hat f_{5}(\z)+\<a^{(0)}_{\tau=2,-1}\g^{(1)}_{\tau=2,-1}\>\ptl_\De\hat f_{\De}(\z)\vert_{\De=5}\nn\\
	&=-\frac{1}{16\pi^4}\times 8\pi^4\de(\z)
	+\frac{1}{16\pi^4}\times 4\pi^4\p{\de(\z)+\left[\frac{1}{\z}\right]_0}\nn\\
	&=-\frac{1}{4}\de(\z)+\frac{1}{4}\left[\frac{1}{\z}\right]_0,
\ee
where we used~\eqref{eq:f5expansion}. This is precisely the expected result. 

To summarize, the full one-loop energy-energy correlator takes the form
\be
\label{eq:fulloneloop}
\cF_\cE^{(1)}(\z)=-\frac{1 }{ 4}\de(\z) - \frac{\zeta_2 }{ 4} \de(1-\z)+\frac{1}{4}\left[\frac{1}{\z}\right]_0-
	\frac{1}{4}\left[\frac{\log(1-\z)}{1-\z}\right]_1+\cF_\cE^{(1),\text{reg}}(\z) ,
\ee
where $\cF_\cE^{(1),\text{reg}}(\z)$ is defined via (\ref{eq:regdefinitiononeloop}). The distributional part at $\z=1$ is in agreement with the one obtained in \cite{Belitsky:2013xxa}. We also derive this $\z=1$ contact term from a different point of view in appendix~\ref{app:z1contacts}.

\subsection{Two loops}
\label{sec:twoloops}

Next, we would like to perform a similar analysis for the two-loop result \cite{Eden:2000mv,Bianchi:2000hn}. In this case, we must expand both the three-point coefficients and the anomalous dimensions up to second order. We have
\be
{\cal H}^{(2)}(u,v) =  \sum_{\tau=2,4,... ;\ \mathrm{even}~J}^{\infty} &\left( \la a_{\tau, J}^{(2)}\ra G_{\tau + 4, J} + \la a_{\tau,J}^{(1)} \gamma_{\tau, J}^{(1)}+a_{\tau,J}^{(0)}\g_{\tau,J}^{(2)} \ra   \pa_{\tau} G_{\tau+4,J}  + \frac{1 }{ 2} \la a_{\tau,J}^{(0)} (\gamma_{\tau, J}^{(1)} )^2 \ra  \pa_{\tau}^2 G_{\tau+4,J}  \right) ,
\ee
and a similar extension of~\eqref{eq:eventshape-one-loop} for the celestial block expansion (\ref{eq:finCoupl}). The explicit expression for $\cH^{(2)}$ is \cite{Alday:2010zy}
\be
\label{eq:twoloopsfull}
 {\cal H}^{(2)}(u,v) &= \frac{1 }{ (2 \pi)^4} \frac{u^3 }{ v} \left( \frac{1 }{ 2} (1+u+v) \left[ \Phi^{(1)}(u,v)  \right]^2
+2 \left[  \Phi^{(2)}(u,v) + \Phi^{(2)}\left(v, u \right) + \frac{1 }{ v}  \Phi^{(2)}\left(\frac{u }{ v} , \frac{1 }{ v} \right)   \right] \right) ,
\nn\\
\Phi^{(2)}(z,\bar z) &=\frac{1 }{ 16} \frac{1 }{ z - \bar z} \Big( \,  6 ({\rm Li}_4(z) - {\rm Li}_4(\bar z)) - 3 \log ( z \bar z) ( {\rm Li}_3(z) - {\rm Li}_3(\bar z) ) \nn \\
& \qquad\qquad\qquad + \frac{1 }{ 2} \log^2 ( z \bar z)  ( {\rm Li}_2(z) - {\rm Li}_2(\bar z) ) \Big) \ .
\ee

A complete OPE expansion of this result is not available in the literature (as far as we know). Otherwise, we could simply evaluate the OPE data at $J=-1$, plug into the celestial OPE formula, and read off the answer for the energy-energy correlator. Some parts of the OPE expansion were obtained in \cite{Dolan:2004iy}, whose results we use below. For simplicity we focus on the term that involves $\la a_{\tau,J}^{(1)} (\gamma_{\tau, J}^{(1)} )^2 \ra$, which on the celestial sphere maps to terms containing $\log^2 \zeta$.

Below, it will be useful to explicitly write the small-$z$ expansion of $ \pa_{\tau}^2 G_{\tau+4,J}$, which takes the form
\be
\label{eq:blockexpansion}
\pa_{\tau}^2 G_{\tau+4,J} &= (z \bar z)^{2+\frac{\tau }{ 2}} \log^2 z \left( \frac{1 }{ 4} \tilde g_{\tau+4, J} + \frac{1 }{ 4} (z \bar z) \tilde g_{\tau+4,J}^{\mathrm{sub}} + ... \right) \ , \nn \\
\tilde g_{\tau, J} = g_{\tau/2 , J} &=  \bar z^J \ _2 F_1 \p{\frac{\tau }{ 2}+J, \frac{\tau }{ 2}+J, \tau + 2 J, \bar z} \ , \nn \\
\tilde g^{\mathrm{sub}}_{\tau , J} (\bar z) &=\tilde g_{\tau+4, J-2}(\bar z) +\frac{\tau - 2 }{ 4} \tilde g_{\tau +2, J-1}(\bar z) - \delta_{J,0} \tilde g_{\tau +2, - 2}(\bar z) \ ,
\ee
where we only kept the terms containing $\log^2 z$.

\subsubsection{Leading twist}

The leading-twist contribution to $\cH^{(2)}$ takes the form $(z \bar z)^3 \log^2 z f_{3}(\bar z)$, where
\be
\label{eq:leadingtwist}
f_{3}(\bar z) =\frac{1 }{ 16} \frac{1 }{ (2 \pi)^4} \frac{ 1 }{ (1- \bar z) \bar z^2} \left( \log^2 [1-\bar z] + 2 \bar z \ {\rm Li}_2(\bar z)\right).
\ee
Since there is no tree-level degeneracy for twist-two operators, this is equal to 
\be
\label{eq:matchingtwisttwo}
f_3(\bar z) &= \frac{1 }{ 2} \sum_{{\rm even} \ J} a_{2,J}^{(0)} \left( \gamma_{2, J}^{(1)} \right)^2   \left[ \frac{1 }{ 4} \tilde g_{6, J}(\bar z) \right] .
\ee
Indeed one can check that (\ref{eq:matchingtwisttwo}) reproduces (\ref{eq:leadingtwist}) . 

\subsubsection{Subleading twist}

Knowing $\la a_{\tau,J}^{(1)} (\gamma_{\tau, J}^{(1)} )^2 \ra$ at $\tau=2$ allows us to compute the $\z^2\log\z$ piece in $\cF^{(2)}(\z)$. A really nontrivial check would be to reproduce the $\zeta^3 \log \zeta$ term. Indeed the two-loop result of \cite{Belitsky:2013ofa} contains both rational and transcendental pieces ($\pi^2$) at this order. The latter should come from the analytically continued $\la \gamma^2 \ra \neq \la \gamma \ra^2$, due to the degeneracy of twist $4$ operators. 

We can compute the required OPE data from the piece $(z \bar z)^4 \log^2 z f_{4}(\bar z)\in \cH^{(2)}$, where
\be
\label{eq:subleadingtwist}
f_{4}(\bar z) = - \frac{1 }{ 16} \frac{1 }{ (2 \pi)^4} \frac{1 }{ (1-\bar z) \bar z^4} \left( 2 \bar z^2 + \bar z( \bar z - 2) \log [1 - \bar z] - (2 + \bar z^2) \log^2 [1 - \bar z] - 2 \bar z (1+\bar z)  {\rm Li}_2(\bar z)  \right).
\ee
This receives contributions from descendants of twist-two operators as well as from the subleading twist-four Regge trajectory. The subleading trajectory has tree-level degeneracies that we have not resolved, and therefore we cannot simply compute the result using our one-loop analysis.

The function (\ref{eq:subleadingtwist}) has decomposition
\be
\label{eq:decomposition}
f_{4}(\bar z) &=\frac{1 }{ 2}  \sum_{{\rm even} \ J}  a_{2,J}^{(0)} \left(\gamma_{2, J}^{(1)} \right)^2 \left[ \frac{1 }{ 4}  \tilde g^{\mathrm{sub}}_{8, J} (\bar z) \right]+ \frac{1 }{ 2}  \sum_{{\rm even} \ J}  \la a_{4,J}^{(0)} [ \gamma_{4, J}^{(1)} ]^2 \ra \left[ \frac{1 }{ 4} \tilde g_{8, J}(\bar z) \right] .
\ee
Using 
\be
\label{eq:identityB}
	\tilde g_{\tau, J} (\bar z) = \bar z^{-\frac{\tau }{ 2}} \tilde g_{0, \tau+J}(\bar z)
\ee
and (\ref{eq:blockexpansion}) it is easy to compute the contribution of descendants of twist $2$ operators. After that we are left with the contribution of twist-four primaries
\be
\tilde f_{4}(\bar z) = \frac{1 }{ 2 (2 \pi)^4 \bar z^4}  \left( - \frac{9 }{ 2} - \frac{1 }{ 4} \frac{\bar z^2 }{ 1 - \bar z} + \frac{(\bar z -2 ) ( 18 - \frac{\bar z^2 }{ 1 - \bar z}  ) }{ 8 \bar z} \log (1- \bar z)+ \frac{1 }{ 8} (1 + \frac{\bar z^2 }{ 1 - \bar z}  ) \log^2 (1 -\bar z) \right)  , 
\ee
which admits the decomposition (\ref{eq:decomposition}) with the second term only. From this we find\footnote{To solve this decomposition problem, one can use the methods of \cite{Dolan:2004iy}.}

\be
 \la a_{4,J}^{(0)} [ \gamma_{4, J}^{(1)} ]^2 \ra &= \frac{1 }{ 4 \pi^4}  \frac{2^{-8-2J} \sqrt \pi }{ 3 \Gamma ( J + 7/2) } \left( - \frac{6 (11+7 J + J^2) \Gamma(3+J) }{ 4+J}  \right. \nn \\
&\quad\left.\phantom{\frac{1}{1}} -\Gamma(4+J) \p{\pi^2 +6 (1 - 2S_1(3+J)) S_1(3+J) + 3 S_2(2+\tfrac{J}{2}) - 3S_2(\tfrac{5+J}{2})}\right) .
\ee
Evaluating at $J=-1$, we finally get
\be
 \la a_{4,-1}^{(0)} [ \gamma_{4, -1}^{(1)} ]^2 \ra = - \frac{2 }{ 9} ( \pi^2 - 11) \frac{1 }{ 2 (2 \pi)^4} .
\ee
Note the appearance of the transcendental quantity $\pi^2$ which is absent for even integer $J$.

\subsubsection{Two-loop energy correlator}

Expanding (\ref{eq:finCoupl}) to the second order we get
\be
\label{eq:expandingtosecondorder}
\cF_{\cE}^{(2)}(\z) &=2 \pi^4  \sum_{n=1}^{\infty} \bigg(\left( \la a_{\tau =2n,-1}^{(0)} \gamma_{\tau=2n,-1}^{(2)} \ra  + \la a_{\tau = 2n,-1}^{(1)} \gamma_{\tau=2n, -1}^{(1)} \ra  \right)\frac{(-1)^{n+1} }{ r_{n+1}} f_{3+2n}^{4,4}(\z)  \nn \\
&\qquad\qquad\qquad+ \la a_{\tau = 2n,-1}^{(0)} [\gamma_{\tau =2n, -1}^{(1)}]^2 \ra (-1)^{n+1} \frac{1 }{ 2} \pa_n \left[ \frac{f_{3+2n}^{4,4}(\z) }{ r_{n+1}} \right]  \bigg).
\ee
Here, we used the fact that corrections to three-point coefficients alone do not contribute to scalar flow, due to the vanishing of the prefactor in (\ref{eq:finCoupl}) at tree-level twists.

Since we do not have degeneracies at twist two, we can fully predict the $n=1$ term in (\ref{eq:expandingtosecondorder}). For $n=2$, corresponding to twist-four operators, we only computed the term $\la a_{4,-1}^{(0)} [\gamma_{4, -1}^{(1)}]^2 \ra$. The only missing element in the twist two sector is the two-loop anomalous dimension. It takes the following form (see e.g. formula (5.29) in \cite{Dolan:2004iy})
\be
\gamma_{2, J}^{(2)} = 2 S_{-2,1}(J+2) - 2 S_1 (J+2) ( S_2(J+2) + S_{-2}(J+2) ) - ( S_3(J+2) + S_{-3}(J+2) ) \ ,
\ee
where $S_{-2,1}(N) = \sum_{n=1}^{N} \frac{(-1)^n }{ n^2} S_1 (n)$ is an example of a nested harmonic sum.
The relevant analytic continuation from even spins to $J=-1$ gives
\be
\gamma_{2, -1}^{(2),+} =2 S_{-2,1}^{+}(1)+ \frac{\pi^2 }{ 3} - 6 + \frac{3 }{ 2} \zeta_3 \  = -4 + \frac{\pi^2 }{ 3} - \zeta_3 ,
\ee
where we used standard methods \cite{Kotikov:2005gr} to perform the analytic continuation.

Plugging everything back, we get the following prediction for the small-angle expansion of the scalar flow observable at two loops
\be
\cF_\cE^{(2)}(\z) &=\frac{1 }{ 4 \zeta} \left(1 + \frac{\pi^2 - 5 }{ 6} \z + \dots \right) \log \z  +  \frac{1 }{ 4 \z} \left( - \frac{1}{ 2} \zeta_3  +\frac{\pi^2 }{ 6} - 3 \right) + \dots \ .
\ee
This coincides with the expansion of the result in \cite{Belitsky:2013ofa}. In principle, by performing the OPE decomposition of the small $z$ expansion of the two-loop result (\ref{eq:twoloopsfull}) further and evaluating it at $J=-1$, we can predict higher order terms in the small-angle (small $\zeta$) expansion of the scalar event shape. 

\subsubsection{Contact terms}
\label{sec:twoloopscontact}

\begin{figure}[t]
	\centering
	\includegraphics[scale=0.7]{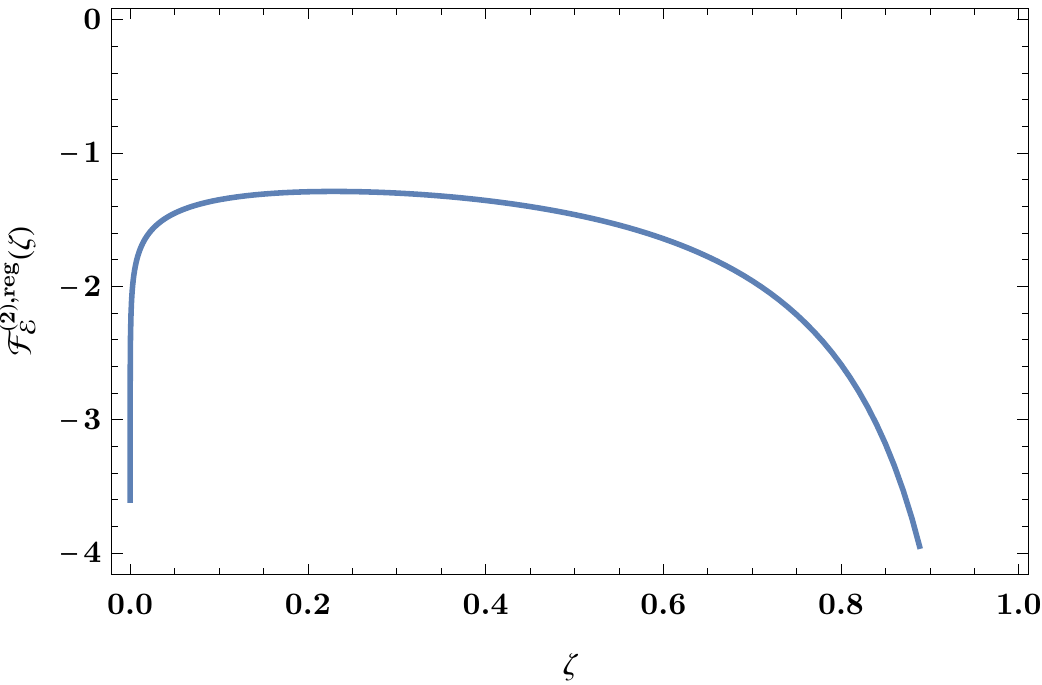}~
	\includegraphics[scale=0.7]{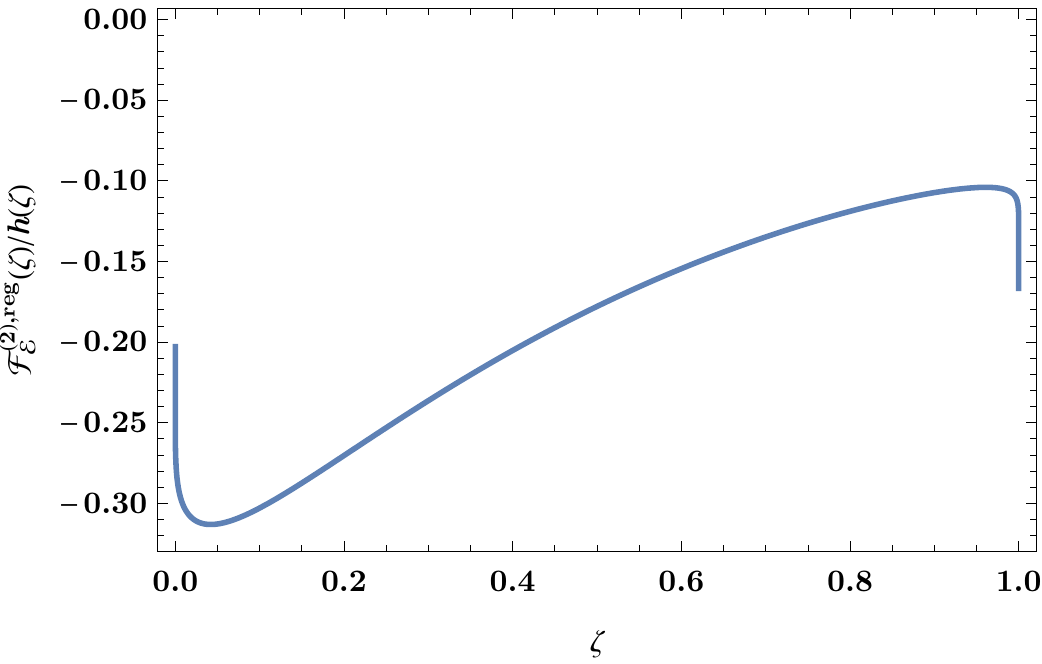}
	\caption{Integrable part $\cF_\cE^{(2),\text{reg}}(\z)$ of the two-loop energy correlator. Left: $\cF_\cE^{(2),\text{reg}}(\z)$ as a function of $\z$. Right: $\cF_\cE^{(2),\text{reg}}(\z)/h(\z)$ as a function of $\z$, where $h(\z)=(1-\log\z)(1-\log(1-\z))^3$.}
	\label{fig:twoloopFreg}
\end{figure}

Let us also check that we reproduce the correct $\z=0$ contact terms in $\cF^{(2)}_\cE(\z)$. Firstly, as in the one-loop example, we can use the Ward identities to fix the contact terms in the two-loop result of~\cite{Belitsky:2013ofa}. We have
\be
	\cF_\cE^{(2)}(\z)=&\frac{1}{4}\frac{1}{\z}\p{\frac{\pi^2}{6}-\half\z_3-3}+\frac{1}{4}\frac{\log\z}{\z}
	+\frac{\z_3}{8}\frac{1}{1-\z}+\frac{\pi^2}{16}\frac{\log (1-\z)}{1-\z}+\frac{1}{8}\frac{\log^3(1-\z)}{1-\z}\nn\\
	&+\cF_\cE^{(2),\text{reg}}(\z),
\ee
where $\cF_\cE^{(2),\text{reg}}(\z)$ is integrable both at $\z=0$ and $\z=1$. We show the plot of $\cF_\cE^{(2),\text{reg}}(\z)$ in figure~\ref{fig:twoloopFreg}. It only has integrable $\log^k$-type singularities at the endpoints. To demonstrate this, we show also the ratio $\cF_\cE^{(2),\text{reg}}(\z)/h(\z)$ with $h(\z)=(1-\log\z)(1-\log(1-\z))^3$. This ratio is finite, but approaches its limits near $\z=0,1$ in a non-analytic way due to $1/\log^k$ type non-analyticities. 

As before, we make an ansatz for the distribution by writing
\be
\label{eq:twoloopcomplete}
	\cF_\cE^{(2)}(\z)=&\frac{1}{4}\left[\frac{1}{\z}\right]_0\p{\frac{\pi^2}{6}-\half\z_3-3}+\frac{1}{4}\left[\frac{\log\z}{\z}\right]_0
	+\frac{\z_3}{8}\left[\frac{1}{1-\z}\right]_1+\frac{\pi^2}{16}\left[\frac{\log (1-\z)}{1-\z}\right]_1+\frac{1}{8}\left[\frac{\log^3(1-\z)}{1-\z}\right]_1\nn\\
	&+c_0^{(2)}\de(\z)+c_1^{(2)}\de(1-\z)+\cF_\cE^{(2),\text{reg}}(\z),
\ee
where $[\z^{-1}\log^k\z]_0$ is defined by the Taylor expansion of $\z^{-1+\e}$ in $\e$ to the appropriate order, and similarly for $[(1-\z)^{-1}\log^k(1-\z)]_1$. The Ward identities~\eqref{eq:energyWI} and~\eqref{eq:momentumWI} require that
\be
	0=&c_0^{(2)}+c_1^{(2)}+\int_0^1 d\z\,\cF_\cE^{(2),\text{reg}}(\z),\\
	0=&-c_0^{(2)}+c_1^{(2)}-\half(\z_3+1)+\frac{5\pi^2}{24}+\int_0^1 d\z(2\z-1)\cF_\cE^{(2),\text{reg}}(\z).
\ee
The explicit expression for $\cF_\cE^{(2),\text{reg}}(\z)$ follows easily from the definition and the results of~\cite{Belitsky:2013ofa}. Due to its complexity, we computed the above integrals numerically,
\be
	\int_0^1 d\z\,\cF_\cE^{(2),\text{reg}}(\z)&=-2.6133007151791604187079457\dots,\\
	\int_0^1 d\z(2\z-1)\cF_\cE^{(2),\text{reg}}(\z)&=-1.047646501079170962972713\dots,
\ee
from which we can determine
\be
	c_0^{(2)}&=1.26039667304023767931294\dots,\label{eq:c02numerics} \nn \\
	c_1^{(2)}&=1.35290404213892273939500\dots.
\ee
Using Mathematica's \texttt{FindIntegerNullVector} we found that to the available precision these numbers are given by
\be
\label{eq:twoloopcontacterms}
	c_0^{(2)}&=\frac{11\pi^4}{1440}-\frac{\pi^2}{8}+\frac{7}{4}, \nn \\
	c_1^{(2)}&=\frac{\pi^4}{72}.
\ee
To summarize, the distributional piece of $\cF^{(2)}_\cE(\z)$ near $\z=0$ is
\be\label{eq:ccF2fromWI}
	\cF^{(2)}_\cE(\z)=\p{\frac{11\pi^4}{1440}-\frac{\pi^2}{8}+\frac{7}{4}}\de(z)+\frac{1}{4}\left[\frac{1}{\z}\right]_0\p{\frac{\pi^2}{6}-\half\z_3-3}+\frac{1}{4}\left[\frac{\log\z}{\z}\right]_0+\cdots \ .
\ee

As at one loop, from the OPE point of view these pieces are determined completely by twist-two OPE data. In particular, we have
\be
	\cF^{(2)}_\cE(\z)=&\<a^{(2)}_{\tau=2,-1}\>\hat f_5(\z)+\<a^{(1)}_{\tau=2,-1}\g^{(1)}_{\tau=2,-1}+a^{(0)}_{\tau=2,-1}\g^{(2)}_{\tau=2,-1}\>\ptl_\De\hat f_\De(\z)\vert_{\De=5}\nn\\
	&+\half \<a^{(0)}_{\tau=2,-1}(\g^{(1)}_{\tau=2,-1})^2\>\ptl_\De^2\hat f_\De(\z)\vert_{\De=5}+\cdots.
\ee
All OPE data in this equation except for $\<a^{(2)}_{\tau=2,-1}\>$ has been described above. We give $\<a^{(2)}_{\tau=2,-1}\>$ in the next section in equation~\eqref{eq:a3loop}. Using these results and~\eqref{eq:f5expansion} we precisely reproduce~\eqref{eq:ccF2fromWI}. A calculation in appendix~\ref{app:z1contacts} also reproduces the value of $c_1^{(2)}$ in~\eqref{eq:twoloopcontacterms}. Note that this is non-trivial consistency check of the result~\cite{Belitsky:2013ofa}, since in order to fix the contact terms we used Ward identities which involve integrals of the even shape over $\z$, not just the $\z\to 0$ and $\z\to 1$ limits.

To summarize, the full two loop energy-energy correlator is given by (\ref{eq:twoloopcomplete}), where $c_0^{(2)}$ and $c_1^{(2)}$ are given by (\ref{eq:twoloopcontacterms}). This completes the $0 < \z < 1$ result of \cite{Belitsky:2013ofa}. We checked numerically that the complete two-loop energy-energy correlator satisfies Ward identities~\eqref{eq:energyWI} and~\eqref{eq:momentumWI}. This check was also performed in \cite{BacktoBackGrisha}.

\subsection{Three loops}
\label{sec:threeloopsN4}

Recently the three loop the energy-energy correlator have been computed in~\cite{Henn:2019gkr}. The authors have verified that the leading $\z$ asymptotic of their result agrees with our prediction (see section~\ref{sec:finitecoupling}).\footnote{This was also independently verified in \cite{LanceFuture} based on the two-loop result \cite{Belitsky:2013ofa} and the energy Ward identity~\eqref{eq:energyWI}.}  In this section we extend this check to contact terms at $\z=0$, similarly to what we did at the two-loop level above. Namely, we will use the results of~\cite{Henn:2019gkr} and Ward identities to fix the contact terms at $\z=0$ and $\z=1$, and then compare to the $\z=0$ contact terms predicted by the light-ray OPE. This provides a highly non-trivial consistency check of the results of~\cite{Henn:2019gkr}, since the Ward identities involve integrals of $\cF_{\cE}^{(3)}(\z)$ over $\z$.

\begin{figure}[t]
	\centering
	\includegraphics[scale=0.7]{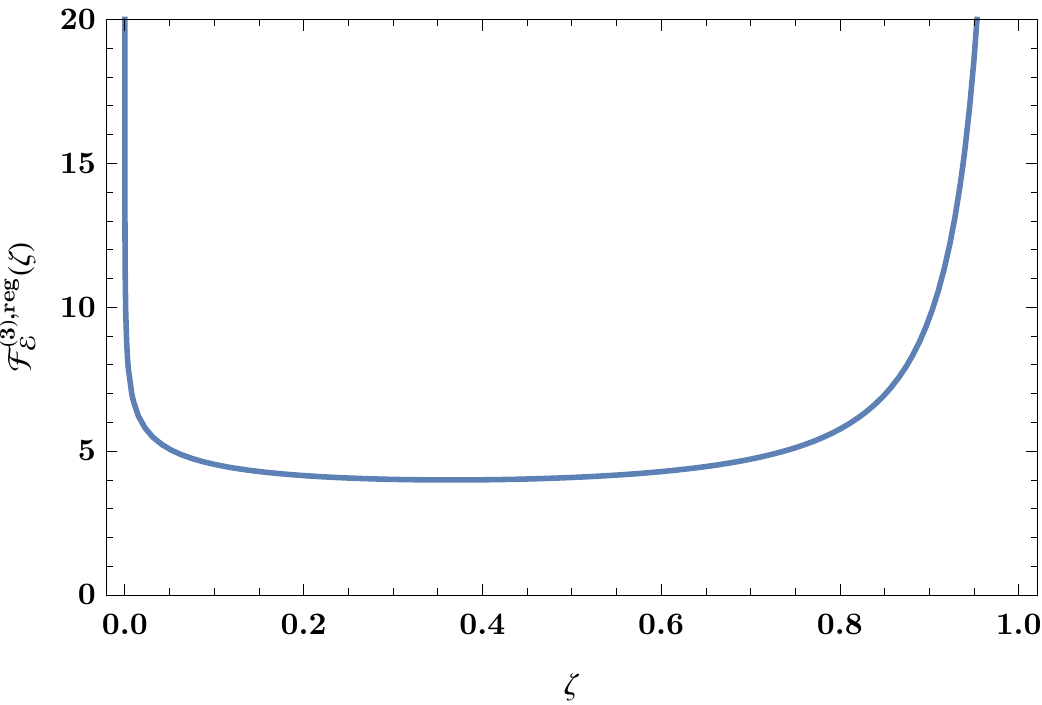}~
	\includegraphics[scale=0.7]{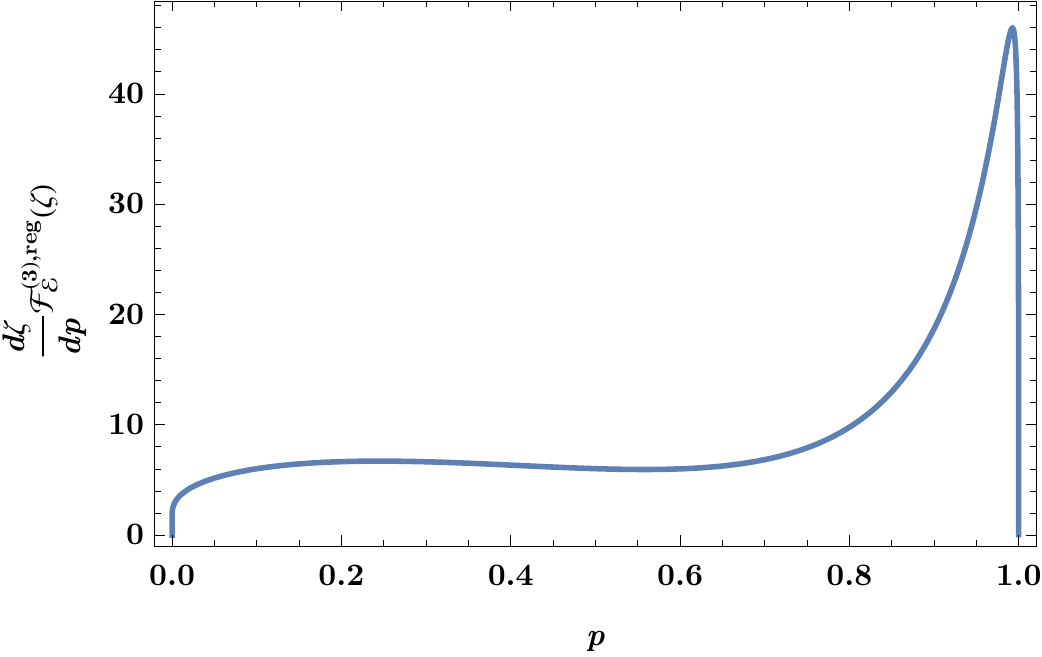}
	\caption{Integrable part $\cF_\cE^{(3),\text{reg}}(\z)$ of the two-loop energy correlator. Left: $\cF_\cE^{(3),\text{reg}}(\z)$ as a function of $\z$. Right: $\frac{\ptl\z}{\ptl p}\cF_\cE^{(3),\text{reg}}(\z)$ as a function of $p$.}
	\label{fig:threeloopFreg}
\end{figure}

We proceed as before, by writing
\be\label{eq:3loopansatz}
	4\cF_{\cE}^{(3)}(\z)=&4c_0^{(3)}\de(\z)
	+\tfrac{1}{2}\left[ \tfrac{\log^2\z}{\z} \right]_0
	+\p{\tfrac{\pi^2}{3}-\z_3-5}\left[ \tfrac{\log\z}{\z} \right]_0
	+\p{17-\tfrac{4\pi^2}{3}+\tfrac{5\pi^4}{144}-\z_3+\tfrac{3}{2}\z_5}\left[\tfrac{1}{\z}\right]_0\nn\\
	&+4c_1^{(3)}\de(y)
	-\tfrac{1}{8}\left[\tfrac{\log^5y}{y} \right]_1
	-\tfrac{\pi^2}{6}\left[\tfrac{\log^3y}{y} \right]_1
	-\tfrac{11\z_3}{4}\left[\tfrac{\log^2y}{y} \right]_1
	-\tfrac{61\pi^4}{720}\left[\tfrac{\log y}{y} \right]_1
	-\p{\tfrac{7}{2}\z_5+\tfrac{\pi^2}{3} \z_3}\left[\tfrac{1}{y} \right]_1\nn\\
	&+4\cF_{\cE}^{(3),\text{reg}}(\z),
\ee
where $y=1-\z$ and $\cF_{\cE}^{(3),\text{reg}}(\z)$ is integrable at $\z=0$ and $\z=1$. We show the plot of $\cF_{\cE}^{(3),\text{reg}}(\z)$ in the left panel of figure~\ref{fig:threeloopFreg}. Again, it only has integrable $\log^k$ singularities. In order to perform numerical integration of these singularities we change the variable from $\z\in[0,1]$ to $p\in[0,1]$ defined as
\be
	\z=\frac{(1-p)^2+(1-p)^3}{\log^2p}\p{1+\frac{p^5(1-p)}{\log^5(1-p)}}.
\ee 
This change of variables is designed so that the Jacobian $\tfrac{\ptl\z}{\ptl p}$ has appropriate $1/\log^k$ behavior to cancel $\log^k$ singularities of $\cF_{\cE}^{(3),\text{reg}}(\z)$ near $\z=0,1$. We show the plot of the resulting function $\tfrac{\ptl\z}{\ptl p}\cF_{\cE}^{(3),\text{reg}}(\z)$ in the right panel of figure~\ref{fig:threeloopFreg}.

The singular part, except from the delta functions (and distributional interpretation of other pieces), can be obtained from the results of~\cite{Henn:2019gkr}. We can fix the coefficients $c_i^{(3)}$ by requiring that the Ward identities~\eqref{eq:energyWI} and~\eqref{eq:momentumWI} are satisfied. We find the equations
\be
	0&=c_0^{(3)}+c_1^{(3)}+\int d\z\,\cF_\cE^{(3),\text{reg}}(\z),\\
	0&=-c_0^{(3)}+c_1^{(3)}+
	4-\frac{4\pi^2}{3}-\frac{\pi^4}{40}+\frac{11\z_3}{4}+\frac{\pi^2\z_3}{6}+\frac{5\z_5}{2}+
	\int d\z(2\z-1)\cF_\cE^{(3),\text{reg}}(\z).	
\ee
Integrating the result of~\cite{Henn:2019gkr} numerically we find
\be\label{eq:numerical3loopintegrals}
	\int d\z\,\cF_\cE^{(3),\text{reg}}(\z)&\approx 9.53135,\nn\\
	\int d\z(2\z-1)\cF_\cE^{(3),\text{reg}}(\z)&\approx 4.84686.
\ee
In~\cite{Henn:2019gkr}, $\cF^{(3)}_\cE(\z)$ contains a piece expressed as a double integral, and the integrals above are therefore effectively triple integrals. Because of this, it is non-trivial to control the numerical errors, and we have not attempted to get an a priori error estimate for~\eqref{eq:numerical3loopintegrals}. Based on the agreement with the light-ray OPE below, we expect that the errors in the numbers above are in the last digit.

Using this data we find
\be\label{eq:3loopc1c2numeric}
	c_0^{(3)}&\approx -4.20195,\nn\\
	c_1^{(3)}&\approx -5.32939.
\ee
Using the same methods as above, and the OPE data described in section~\ref{sec:finitecoupling}, we find the light-ray OPE prediction for $c_0^{(3)}$,
\be\label{eq:c03}
	c_0^{(3)}&=-\frac{49}{4}+\pi^2-\frac{\pi^4}{576}-\frac{109\pi^6}{30240}+\frac{5\z_3}{4}-\frac{7}{24}\pi^2\z_3+\frac{3\z_3^2}{16}+\frac{27\z_5}{8}\nn\\
	&=-4.2019873198181\cdots.
\ee
This agrees well with~\eqref{eq:3loopc1c2numeric}, and based on the accuracy of the agreement, we expect for $c_1^{(3)}$
\be\label{eq:c13prediction}
	c_1^{(3)}&\approx -5.3294(1).
\ee
We show in appendix~\ref{app:z1contacts} that $c_1^{(3)}$ is given by
\be\label{eq:c13}
	c_1^{(3)}= - \frac{197 \pi^6 }{ 40320} - \frac{7 \zeta_3^2}{ 16}=-5.329425268\cdots,
\ee
which precisely agrees with~\eqref{eq:c13prediction}.\footnote{In deriving ~\eqref{eq:c13} we used the three-loop result for the so-called coefficient function $H(a)$ \cite{BacktoBackGrisha}.}  This numerical check was also done in \cite{BacktoBackGrisha}.

To summarize, the complete three-loop energy-energy correlator, including contact terms, is given by~\eqref{eq:3loopansatz}, where $c_0^{(3)}$ and $c_1^{(3)}$ are given by~\eqref{eq:c03} and~\eqref{eq:c13}, while $\cF_\cE^{(3)}(\z)$ follows from its definition and results of~\cite{Henn:2019gkr}. We checked numerically that the complete three-loop energy-energy correlator satisfies Ward identities~\eqref{eq:energyWI} and~\eqref{eq:momentumWI}. 

\subsection{Four loops in the planar limit and finite coupling}
\label{sec:finitecoupling}

\begin{figure}[t]
	\centering
	\includegraphics[scale=0.52]{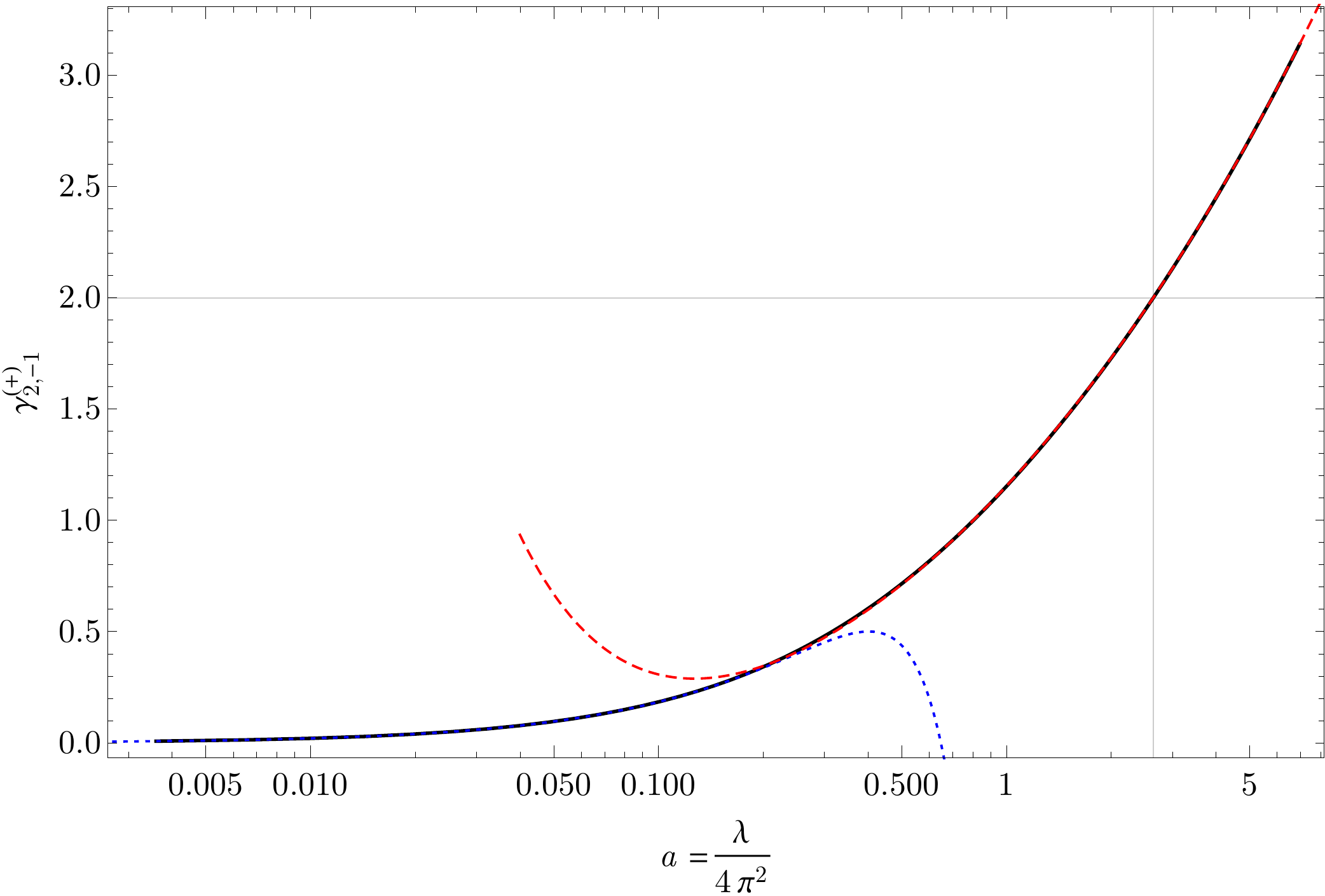}
	\caption{$\gamma^{(+)}_{2,-1}$ as a function of the coupling constant $a = \frac{\lambda }{ 4 \pi^2}$. The plot was kindly made for us by Nikolay Gromov. The actual numerics was done for $J=-1+10^{-5}$. The blue dotted line represents a four-loop weak coupling approximation to $\gamma^{(+)}_{2,-1} $, the red dashed line corresponds to the first four terms at the strong coupling expansion \cite{Gromov:2014bva}. The solid line was obtained using the quantum spectral curve technique \cite{Gromov:2014bva}. The curve intersects $\gamma^{(+)}_{2,-1}=2$ at $a\approx 2.645$. At this point the small angle expansion of the EEC becomes regular and dominated by the twist four double trace operators .}
	\label{fig:Kolyaplot}
\end{figure}

Using known results for the OPE data of twist-2 operators, we can make new predictions for the leading small-angle asymptotics of the energy-energy correlator.  At finite coupling the contribution of twist-two operators takes the form 
\be
\label{eq:fincoupltw2}
\cF_{\cE}^\text{twist-two}(\z) &= a^{(+)}_{2,-1} \frac{4 \pi^4 \Gamma\p{3+\gamma^{(+)}_{2,-1}} }{ \Gamma\p{2 + \frac{\gamma^{(+)}_{2,-1} }{2}}^3 \Gamma\p{-1 - \frac{\gamma^{(+)}_{2,-1}}{ 2}}} f_{5 +\gamma^{(+)}_{2,-1} }^{4,4}(\z) \ , 
\ee
where by $(+)$ we indicate analytic continuation from even spin. Note that $\gamma^{(+)}_{2,-1}$ can be computed at any 't Hooft coupling using integrability methods \cite{Gromov:2013pga,Gromov:2014bva}. At small angles we have $f_{5 +\gamma^{(+)}_{2,-1} }^{4,4}(\z) \approx \z^{\frac{\gamma^{(+)}_{2,-1}}{ 2}  - 1}$. Therefore, at weak coupling (\ref{eq:fincoupltw2}) controls the small angle $\z \to 0$ expansion of the EEC. When the coupling becomes large, operators with twist two at tree level become very heavy and the leading small-angle asymptotic is controlled by the approximately twist-four double trace operators. This transition happens at $a\approx 2.645$, see figure~\ref{fig:Kolyaplot}.

At finite coupling there is no contact term coming from~\eqref{eq:fincoupltw2}, since the anomalous dimension of twist-two operators is finite. The term $\xi(\z)$ in~\eqref{eq:finCoupl} is completely canceled by a contribution of a protected operator. This cancellation is the same as at strong coupling and is described in the next section. In summary, the event shape at finite coupling is integrable near $\z=0$ and the contact terms only appear at weak coupling through the expansion~\eqref{eq:f5expansion}.

Using (\ref{eq:fincoupltw2}) we can easily make a planar four-loop prediction for the leading asymptotic of $\cF(\z)$.\footnote{Starting from the four loops there are non-planar corrections to the correlator \cite{Eden:2012tu}.} The relevant OPE data takes the form
\be
\gamma_{2,-1}^{(+)} &= 2 a + \p{-4 + \frac{\pi^2 }{ 3} - \zeta_3} a^2 + \left( 16 - \frac{4 }{ 3} \pi^2 + \frac{\pi^4 }{ 120} - 3 \zeta_3 + 3 \zeta_5 \right) a^3
\nn \\
&\quad+ \left( - 80 + \frac{\pi^2 }{ 6}[48 - 13 \zeta_3+\zeta_5]   - \frac{1 }{ 720} \pi^4 [46+5 \zeta_3]
\right.
\nn\\
&\quad\quad\quad\left.  - \frac{107 \pi^6}{ 15120} + 14 \zeta_3 +\frac{9 }{ 2} \zeta_3^2+16 \zeta_5 - \frac{69 }{ 8} \zeta_7 \right) a^4  + \dots,
\\
\label{eq:a3loop}
\frac{a_{2,-1}^{(+)} }{ a_{2,-1}^{(0)}} &= 1 - 2 a + \left( 12 - \frac{2 \pi^2 }{ 3} + \frac{11 \pi^4 }{ 360} +\frac{1 }{ 2} \zeta_3 \right) a^2 \nn\\
&\quad+\left( - 80 + \pi^2 (6 - \frac{7 }{ 6} \zeta_3 ) - \frac{\pi^4 }{ 24} - \frac{109 }{ 7560} \pi^6  + 6 \zeta_3 + \frac{3 }{ 4} \zeta_3^2 + 12 \zeta_5 \right) a^3 + a_4 a^4 + \dots \ ,
\ee
where for our normalization of the four-point function the tree-level three-point function is $a_{2,-1}^{(0)} = \frac{1 }{ 32 \pi^4}$. Up to three loops, the results can be found in \cite{Eden:2012rr}, where the three-loop correction to the structure constant was first explicitly computed.\footnote{The currently available online version (arXiv v1) of \cite{Eden:2012rr} contains a typo. The corrected version of the formula can be found for example in \cite{Alday:2013cwa} which we used in our computation.} For the four-loop anomalous dimensions, we combined the results of \cite{Kotikov:2007cy} and \cite{Bajnok:2008qj}. To analytically continue in spin, we used the HPL package \cite{Maitre:2005uu} together with the supplement developed in \cite{Gromov:2015vua}.\footnote{In the papers cited above, the anomalous dimension and three-point coupling of ${\rm Tr}[Z D^J Z]$ operator are computed. These operators transform in the $\mathbf{20'}$ representation and their dimensions and couplings are related to the anomalous dimension $\gamma_{\tau=2,J}$ and $a_{\tau=2,J}$ of the superconformal primaries that appear in (\ref{eq:superOPE}) by a spin shift $J \to J+2$, see e.g.\ \cite{Dolan:2001tt}. Therefore, the formulas of \cite{Eden:2012rr} should be evaluated at $J=1$ for our purposes. 
}

Plugging these results into (\ref{eq:fincoupltw2}), we easily obtain the leading small-angle expansion of the energy-energy correlator up to four loops. Due to the factor 
\be
\frac{1 }{ \Gamma\p{-1 - \frac{\gamma^{(+)}_{2,-1}}{ 2}} },
\ee
only the three-loop correction to three-point coefficients is needed to compute the four-loop result for $0<\z<1$. At $\z=0$, $\z=1$, there are contact terms that depend on additional data at four loops (discussed below). The first two terms in the expansion in the coupling reproduce the two-loop computation of \cite{Belitsky:2013ofa}. The three- and four-loop predictions are new. Our three-loop prediction was recently independently confirmed in \cite{Henn:2019gkr}. 

In more detail, we can write the following expression for the planar four-loop energy-energy correlator\footnote{By planar we mean that it was obtained from the planar four-loop correlation function. Starting from four loops there are corrections to the energy correlator suppressed by $\frac{1 }{ c}$.}
\be
&\cF_{\cE}^{(4), \text{pl}}(\z)= c_0^{(4)} \delta(\z)  + \frac{1 }{ 24} \left[ \frac{\log^3 \z }{ \z} \right]_0 + \left( - \frac{7 }{ 8} + \frac{1 }{ 16} \pi^2 - {3 \over 16} \z_3 \right) \left[  {\log^2 \z \over \z} \right]_0  \nn \\
&+ {1 \over 4} \left( 31 - {17  \over 6} \pi^2 +{1 \over 15}\pi^4 - {1 \over 6} \pi^2 \zeta_3 +{1 \over 4} \zeta_3^2 + 3 \zeta_5 \right)  \left[  {\log \z \over \z} \right]_0 \nn \\
&+{1 \over 4} \left(-111+{65\over 6} \pi^2  - {3  \over 16}\pi^4 - {389  \over 30240}\pi^6 + 10 \zeta_3 - 2 \pi^2 \zeta_3 - {3 \over 160} \pi^4 \zeta_3 + 3 \zeta_3^2 + 20 \zeta_5 + {1 \over 12} \pi^2 \zeta_5 - {69 \over 16} \zeta_7  \right) \left[  {1 \over \z} \right]_0 \nn \\
&+ c_1^{(4)} \delta(y)+ {1 \over 192} \left[ {\log^7 y \over y} \right]_1 +{5 \over 384} \pi^2 \left[ {\log^5 y \over y} \right]_1  + {95 \over 192} \zeta_3  \left[ {\log^4 y \over y} \right]_1 + {29 \over 1920} \pi^4 \left[ {\log^3 y \over y} \right]_1 \nn \\
&+\left( {67 \over 192} \pi^2 \z_3 + {69 \over 16} \z_5 \right) \left[ {\log^2 y \over y} \right]_1 + \left( {367  \over 48384} \pi^4 + {97 \over 32} \z_3^2 \right) \left[ {\log y \over y} \right]_1 \nn \\
&+\left( {187 \over 5760} \pi^4 \z_3 + {95 \over 192} \pi^2 \z_5 + {785 \over 128} \z_7 \right)  \left[ {1 \over y} \right]_1 +\cF_\cE^{(4),\text{reg}}(\z) ,
\ee
where $\cF_\cE^{(4),\text{reg}}(\z)$ is integrable at $\z=0,1$. We also included the leading terms in the $\z\to 1$ limit, which we obtained using results of~\cite{Collins:1981uk,Belitsky:2013ofa,BacktoBackGrisha} as described in appendix~\ref{app:z1contacts} (recall $y=1-\z$).\footnote{Here we again made use of the three-loop result for $H(a)$~\cite{BacktoBackGrisha}.} The contact term coefficients $c_0^{(4)}$ and $c_1^{(4)}$ are equal to 
\be
c_0^{(4)} &={1 \over 4} \left( - 209 + {37 \over 2} \pi^2  - {23 \over 80} \pi^4 - {389 \over 30240} \pi^6 + 20 \z_3 - {8 \over 3} \pi^2 \z_3 - {3 \over 160} \pi^4 \z_3  \right.  \   \nn \\
&\left.  \qquad\quad+{11 \over 2} \z_3^2 + 14 \z_5 + {1 \over 12} \pi^2 \z_5 - {69 \over 16} \z_7 + a_4 \right) \ , \nn \\ 
c_1^{(4)} &=\tfrac{17}{144}\pi^2\z_3^2+{7 \over 2} \z_3\z_5+{1 \over 4}H_4  \ ,
\ee
where $a_4$ is a four-loop correction to the three-point function at $J=-1$, see ~\eqref{eq:a3loop}, and $H_4$ is a four-loop correction to the coefficient function, see appendix ~\ref{app:z1contacts}, which are presently unknown. The Ward identities ~\eqref{eq:energyWI}, ~\eqref{eq:momentumWI} thus take the form
\be\label{eq:fourloopWI}
	c_0^{(4)}+c_1^{(4)} + \int_0^1 d\z\,\cF_\cE^{(4),\text{reg}}(\z)&=0, \nn \\
	-c_0^{(4)}+c_1^{(4)} +\int_0^1 d\z(2\z-1) \cF_\cE^{(4),\text{reg}}(\z) 
	&={45 \over 2} - {245  \over 24} \pi^2  - {13 \over 240} \pi^4 - {151 \over 17280} \pi^6 + {39 \over 2} \zeta_3 + {37 \over 16} \pi^2 \z_3 \nn \\
	&\quad+{107 \over 1440} \pi^4 \z_3  - {119 \over 16} \z_3^2 + {35 \over 4} \z_5 + {91 \over 96} \pi^2 \z_5 +{923 \over 64} \z_7\nn\\
	&\approx - 9.784125919 \dots \ .
\ee
As was the case at three loops, these identities provide a nontrivial test for any future four-loop computation. Because we explicitly isolated all the distributional terms it is particularly suited for numerical tests. Alternatively, given a four-loop result for $\cF_\cE^{(4), \text{pl}}$, one can use~\eqref{eq:fourloopWI} to predict $a_4$ and $H_4$. These values can then be used to predict leading five-loop asymptotics at $\z\to 0$ and $\z\to 1$.

\subsection{Strong coupling in the planar limit}
\label{sec:strongcoupling}

The four-point function at strong-coupling is simple enough that we can directly compute $C^{+,\text{sugra}}(\De,J)$ and use the celestial block expansion to obtain the full scalar flow observable as a function of $\zeta$. The four-point function is \cite{Arutyunov:2000py}
\be
\Phi^{\mathrm{(sugra)}}(u,v) = u v \bar D_{2422}(u,v) .
\ee
For a review of $\bar D$-functions see e.g. \cite{Dolan:2001tt}.

As explained in \cite{Alday:2017vkk}, remarkably the tree-level supergravity answer is fixed by the protected half-BPS data and is given by 
\be
{\rm dDisc}[G^{(105)}(z,\bar z)] &= {\rm dDisc}\left[{z \bar z \over (1-z) (1 - \bar z)}\right] f(z, \bar z) ,
\ee
where $f(z, \bar z)$ is regular at $z, \bar z =1$ and is symmetric under permutations of $z$ and $\bar z$. The relation to the ${\cal G} (z,\bar z)$ used in \cite{Alday:2017vkk} is $G^{(105)}(z,\bar z) =c {1 \over 2} {1 \over (2 \pi)^4} (z \bar z)^2 {\cal G}(z,\bar z)$.
Thus,
\be
C^{+,\text{sugra}}(\Delta, J) &= 2 {\kappa_{\Delta + J} \over 4}  \int_0^1 {d z \over z^2} \int_0^1 {d \bar z \over \bar z^2} {z - \bar z \over z \bar z} \nn \\
&\left.
\quad\x(k_{\Delta+J}(z) k_{4+J-\Delta}(\bar z) - k_{\Delta+J}(\bar z) k_{4+J-\Delta}(z)) {2 (\sin \pi \delta)^2 z^{1+ \delta} \bar z^{1+ \delta} \over (1- z)^{1+ \delta} (1-\bar z)^{1+\delta}} f(z, \bar z)
\right|_{\de\to 0},
\ee
where we have regulated the integral by introducing $\de$ in the same way as we did in section~\ref{sec:treelevel}. 

To isolate the contribution that survives as $\de\to 0$, we rewrite $z - \bar z = (1-\bar z) - (1-z)$. By the symmetry of the integral under the exchange of $z$ and $\bar{z}$, each of the terms produces an identical contribution, giving a factor of $2$. We can rewrite the integral as
\be
C^{+,\text{sugra}}(\Delta, J) &=  {\kappa_{\Delta + J} \over (2 \pi)^4}  (\sin \pi \delta)^2 \int_0^1 {d z \over z^2} \int_0^1 {d \bar z \over \bar z^2}  \left({1 \over 2} D (D-2) {z \log z \over 1 - z} \right) \nn \\
&\quad\left.\x\left(k_{\Delta+J}(z) k_{4+J-\Delta}(\bar z) - k_{\Delta+J}(\bar z) k_{4+J-\Delta}(z) \right) {\bar z^{1+ \delta} \over (1-\bar z)^{1+\delta}}\right|_{\de\to 0}\, ,
\ee
where we set $\bar z =1$ in $f(z,\bar z)$ since it does not affect the $\delta = 0$ result, and used $f(z,1) =  {1 \over 2} {1 \over (2 \pi)^4}  \left(- {1 \over 2} D (D-2) {z \log z \over 1 - z} \right) $, see \cite{Alday:2017vkk}. We have also introduced the differential operator
\be
D &= z^2 \partial_z (1-z) \partial_z \, ,
\ee 
which is the Casimir operator of which $k_{\beta}(z)$ is an eigenfunction with eigenvalue ${\beta (\beta - 2) \over 4}$. Doing the integrals, we get
\be
C^{+,\text{sugra}}(\Delta,J) = {\Gamma({\Delta + J \over 2})^2 \over 64 \pi^4 \Gamma(\Delta + J - 1)} \left( - \tilde I (4+J-\Delta)  + {\Gamma(4+J-\Delta) \Gamma({\Delta + J \over 2})^2 \over \Gamma(\Delta + J) \Gamma({4+J-\Delta \over 2})^2 } \tilde I (\Delta+J) \right),
\ee
where
\be
\tilde I(\beta) &=\int_0^1 {d z \over z^2} k_{\beta}(z) D (D-2) {z \log z \over 1 - z} \nn \\
&= {1 \over 16} (\beta+2) \beta (\beta-2) (\beta - 4) \int_0^1 {d z \over z^2} k_{\beta}(z) {z \log z \over 1 - z} + {\beta \over 16} {(\beta^2 - 2 \beta -10)\Gamma(1+\beta) \over \Gamma(1 + {\beta \over 2})^2} \ ,
\ee
where the second term in the second line comes from boundary terms when we integrate by parts. Its contribution to $C^{+,\text{sugra}}(\Delta,-1)$ is equal to zero.

Specializing to $J=-1$, we find
\be\label{eq:Csugraresult}
C^{+,\text{sugra}}(\Delta, -1)  {4 \pi^4 \Gamma(\Delta-2) \over \Gamma({\Delta-1 \over2})^3 \Gamma({3-\Delta \over 2})} = - \pi {(\Delta+1)(\Delta-1)(\Delta-5) \Gamma({\Delta -1 \over 2})^2 \over 256 \Gamma(\Delta -3) \cos {\pi \Delta \over 2} } \ .
\ee
This provides the data needed to compute $\cF^{\mathrm{(sugra)}}$ using the celestial block expansion.
Formula (\ref{eq:celestiablockforenergy}) gives an integral which we can evaluate by residues when $0<\z<1$,
\be\label{eq:Fsugra}
\cF^{\mathrm{(sugra)}}_{\cE}(\z) &=\int_{2 - i \infty}^{2+i \infty} {d \Delta \over 2 \pi i} C(\Delta, -1) {4 \pi^4 \Gamma(\Delta-2) \over \Gamma({\Delta-1 \over2})^3 \Gamma({3-\Delta \over 2})} f_\Delta^{4,4} (\z) \nn \\
&=\sum_{n=0}^{\infty} {(-1)^n \over 8}  (n+1) (n+2) (n+3) (n+4) \ r_{n+3} f_{7+2n}^{4,4}(\z)\nn\\
 &= {1 \over 2} , \qquad 0 < \z < 1 .
\ee
where $r_h$ was defined in (\ref{eq:definitionofhrh}). This answer coincides with the one obtained in \cite{Hofman:2008ar}. Alternatively, we could have directly continued the known OPE decomposition of the correlation function to $J=-1$. Indeed, in the one-loop example above the sum above is equal to 
\be
\label{eq:OPEsugra}
\cF_{\cE}^{\mathrm{(sugra)}}(\z) &=2 \pi^4  \sum_{n=0}^{\infty} (-1)^{n+1} \la a_{\tau=4+2n , -1}^{(0)} \gamma_{\tau=4+2n,-1}^{\mathrm{(sugra)}} \ra {1 \over r_{n+3} }  f_{7+2n}^{4,4}(\z) ,
\ee
where the sum goes over the Regge trajectories of double trace operators with scaling dimension $\Delta(J)= 4 + 2n + J + \gamma^{\mathrm{(sugra)}}_{\tau=4+2n, J}$. Note that in our normalization $a_{\tau=4+2n , -1}^{(0)} \sim O(c)$, see ~\eqref{eq:treelevel}, whereas $\gamma^{\mathrm{(sugra)}}_{\tau=4+2n, -1} \sim O({1 \over c})$. After an appropriate overall rescaling related to the normalization of the conformal blocks the coefficients in the celestial block expansion (\ref{eq:Fsugra}) and (\ref{eq:OPEsugra}) coincide with the analytic continuation of the OPE data worked out in \cite{Alday:2017vkk} to $J=-1$. 

The result ~\eqref{eq:Fsugra} already satisfies Ward identities~\eqref{eq:energyWI} and~\eqref{eq:momentumWI}, so we do not need to add any distributional terms at $\z=0$ or $\z=1$. Let us now check this using the light-ray OPE. Using ~\eqref{eq:Csugraresult} and formulas from \ref{sec:contacttermsEE} we find for the distributional terms at $\zeta = 0$
\be
	&-\mathrm{res}_{\De=3} C(\De,-1) \hat f_3(\z)-\mathrm{res}_{\De=5} C(\De,-1) \hat f_5(\z)+\xi(\z)\nn\\
	&=\frac{1}{16\pi^4}\p{4\pi^4\de'(\z)-2\pi^4\de(\z)}-\frac{3}{64\pi^4}8\pi^4\de(\z)+\xi(\z)\nn\\
	&=\frac{1}{4}\p{\de'(\z)-2\de(\z)}+\xi(\z)=0.
\ee
Similarly, to probe distributional terms at $\z=1$ we consider $\int_0^1 d \zeta\, \zeta^N \cF^{\mathrm{(sugra)}}(\z)$ and evaluate the integral over $\Delta$. The result is that distributional terms are absent. 

To summarize, the complete strong coupling result takes the form
\be
\cF_{\cE}^{(\text{sugra})}(\z)&={1 \over 2} \ .
\ee

\subsection{Comments on supergravity at one loop}

Recently, the function $G^{(105)}(u,v)$ was also computed at strong coupling to the ${1 \over N^4}$ order \cite{Aprile:2017bgs}, see also \cite{Aprile:2017qoy,Alday:2017xua,Alday:2017vkk}.  It corresponds to a one-loop computation in supergravity. It is therefore natural to ask if we can use it to compute the corresponding correction to the two-point energy correlator.  As discussed in \cite{Kologlu:2019bco} the existence of the two-point energy correlator is guaranteed in the non-perturbative theory as well as in the planar theory. This, however, does not have to be the case in ${1 \over N^2}$ perturbation theory. Indeed, in this case the Regge behavior of the correlation function becomes more and more singular and the condition for the existence of the energy correlator $J_0<3$ can be violated (here $J_0$ is the Regge intercept of the correlator).   

At infinite 't Hooft coupling and order ${1 \over N^4}$ we have $J_0 = 3$ and thus the energy correlator becomes ill-defined. In other words, to compute it we have to first re-sum ${1 \over N^2}$ corrections before doing the light transforms and taking the coincident limit, see \cite{Kologlu:2019bco}. It is very easy to see the manifestation of the problem at the level of the OPE as well. If we are to try to evaluate corrections to the spectrum at $J=-1$ as we did above in section (\ref{sec:twoloops}) we find a pole in $\la a_{\tau,-1} [\gamma_{\tau, -1}]^2 \ra$, see e.g. (3.15) in \cite{Alday:2017vkk}. It is an interesting question how to compute subleading large $N$ corrections to the energy correlator. We leave this question for the future.

\subsection{Multi-point event shapes}

It is also interesting to consider higher-point event shapes. To our knowledge, the only higher-point event shapes available in the literature are the ones due to Hofman and Maldacena \cite{Hofman:2008ar} for planar ${\cal N}=4$ SYM at strong coupling. In principle, higher-point event shapes can be computed via repeated light-ray OPEs, in the same way that correlation functions of local operators can be computed by repeated local OPEs. (Alternatively, we can use the $t$-channel block decomposition introduced in \cite{Kologlu:2019bco}.) Although we have not developed the formalism for taking OPEs of completely general light-ray operators in this work, it is reasonable to conjecture that the light-ray OPE closes on the light-ray operators of \cite{Kravchuk:2018htv}. This is already enough information to make nontrivial predictions about the small-angle limit of multi-point event shapes.

As a simplest nontrivial example, consider a three-point event shape of null-integrated scalars. We assume that the Regge behavior of the theory is such that the event shape exists, and the null-integrated scalars commute. By taking consecutive OPEs, we have
\be
 \wL[\f_1](\vec y_1) \wL[\f_2](\vec y_2) \wL[\f_3](\vec y_3) 
&= \sum_i \cC_{\De_i-1}(\vec y_{12},\ptl_{\vec y_2}) \mathbb{O}^+_{i,-1}(\vec y_2) \wL[\f_3](\vec y_3) \nn\\
&= \sum_{i,j} \cC_{\De_i-1}(\vec y_{12},\ptl_{\vec y_2}) \cC_{\De_j-1}(\vec y_{23},\ptl_{\vec y_3}) \mathbb{O}^+_{j,-2}(\vec y_3),
\ee
where for simplicity we have ignored transverse spins in the second OPE and we are dropping overall constants. We have also abused notation and written the light-ray operators as a function of the transverse position $\vec y$, as opposed to $x,z$ used in most of this work.

Inserting the above expression into an event shape, we obtain a sum of multi-point celestial blocks (which would be interesting to compute explicitly). In the limit $|\vec y_{12}|\ll |\vec y_{23}| \ll 1$, the product of operators is dominated by the lightest-dimension terms in each OPE
\be
\lim_{\vec y_{23}\to 0}\lim_{\vec y_{12}\to 0}\wL[\f_1](\vec y_1) \wL[\f_2](\vec y_2) \wL[\f_3](\vec y_3) 
&\propto |\vec y_{12}|^{\De^+_{-1} - \De_1 - \De_2 + 1} |\vec y_{23}|^{\De^+_{-2} - \De^+_{-1} - \De_3 + 1} \mathbb{O}^+_{\mathrm{lightest},-2}(\vec y_1),
\ee
where $\De^+_{-1}$ and $\De^+_{-2}$ represent the lightest dimensions at spin $-1$ and $-2$.

Similarly, we can take repeated OPE limits of an arbitrary number of scalar light-ray operators (assuming their products exist). This leads to a very simple formula for the multi-collinear limit of scalar event shapes
\be
\lim_{\th_{1k}\to 0} \cdots  \lim_{\th_{12}\to 0} \langle \wL[\phi_1](\infty, z_1) \cdots \wL[\phi_k](\infty, z_k) \rangle_{\psi(p)} \propto |\theta_{1k}|^{\De^{+}_{1-k}- \De^{+}_{2-k} - (\Delta_k -1) } \cdots |\theta_{12}|^{\De^+_{-1} - \Delta_1 - \Delta_2 + 1} ,
\ee  
where we have suppressed subleading terms and an overall proportionality constant that does not depend on relative angles.

Of course, a more physically interesting case is to consider multi-point energy correlators. A difference compared to the scalar case is that the OPE of ANEC operators contains light-ray operators transforming nontrivially under $\SO(d-2)$ (except for $d=3$), see (\ref{eq:finalanswerforproductingeneral}) and \cite{Hofman:2008ar}. Let us ignore this for the moment. Repeated OPEs give 
\be
\label{eq:multipointenergyshape}
\lim_{\th_{1k}\to 0} \cdots \lim_{\th_{12}\to 0} \<\cE(z_1)\cdots \cE(z_k)\>_{\psi(p)} &\propto |\th_{1k}|^{\tau_{k+1}^+ - \tau_k^+ + 2 - d} \cdots |\th_{12}|^{\tau_3^+ + 4 - 2d}.
\ee
Here $\tau^+_J$ represents the leading twist at spin $J$. 
When operators transform non-trivially under $\SO(d-2)$, the overall scaling with respect to the corresponding small angle will not change --- it will still be controlled by the minimal twist \cite{Hofman:2008ar}.

A fascinating property of repeated ANEC OPEs is that alternating steps are controlled by local operators. Specifically, after a single OPE, we obtain light-ray operators with even signature and spin $3$. After taking an additional OPE with an ANEC operator, we obtain light-ray operators with even signature and spin $4$. These are the quantum numbers of a light-transformed local operator. We expect that arguments like the ones in sections~\ref{sec:commutativity} and~\ref{sec:finishing} establish that the resulting operator is indeed the light-transform of a local spin-4 operator. Thus, the structure of the light-ray OPE is\footnote{In writing (\ref{eq:nonlocallocalOPE}) we assumed that the nonlocal spin-3 operators that appear in the OPE of two ANEC operators commute with the ANEC operator. This is consitent with the fact that $[\cE(z_1) \cE(z_2), \cE(z_3)] = 0$.}
\be
\label{eq:nonlocallocalOPE}
\wL[\textrm{local}] \x \wL[\textrm{local}] &\sim \textrm{(nonlocal)} \nn\\
\textrm{(nonlocal)} \x \wL[\textrm{local}] & \sim \wL[\textrm{local}].
\ee
We have already determined the form of the first line above. To understand OPEs for multi-point event shapes, it suffices to understand the second line.

\section{Discussion and future directions}
\label{sec:discussion}

\subsection{Generalizations}

In this work, we derived an OPE for a product of null-integrated operators on the same null plane. There are several possible generalizations that would be interesting to consider.

One possibility is to derive OPEs of more general continuous-spin light-ray operators \cite{Kravchuk:2018htv}. Such an OPE would enable repeated OPEs in multi-point event shapes. For example, a three-point energy correlator could be computed by applying the OPE in this paper to merge two ANEC operators into nontrivial light-ray operators, followed by a generalized OPE with the remaining ANEC operator to produce additional light-ray operators. From symmetries, the low transverse-spin terms in a multi-point OPE of $n$ ANEC operators will produce light-ray operators with spin $n+1$. The average null energy condition implies positivity of the leading light-ray operator in this product, which is presumably the lowest-twist light-ray operator with spin $n+1$.\footnote{We thank Clay C\'ordova for discussions on this point.} This gives an alternative derivation of the higher-even-spin ANEC \cite{Hartman:2016lgu} that additionally includes the case of odd spins, but is not as general as the continuous spin version in \cite{Kravchuk:2018htv}.

A possible application of repeated OPEs for multi-point event shapes is to set up a bootstrap program for event shapes similar to the bootstrap program for four-point functions of local operators \cite{Ferrara:1973yt,Polyakov:1974gs}.\footnote{We discuss a different kind of bootstrap program for event shapes in the next subsection.} Specifically, one could demand that the light-ray OPE is associative and use this condition to study the space of possible event shapes abstractly. One can also consider mixed light-ray and $t$-channel OPEs of the type discussed in \cite{Kologlu:2019bco}. With sufficient positivity conditions, perhaps one could apply numerical bootstrap techniques \cite{Rattazzi:2008pe,ElShowk:2012ht,Poland:2018epd}. Even without deriving the details of the generalized light-ray OPE, it is reasonable to conjecture that it closes on the light-ray operators of \cite{Kravchuk:2018htv}, and thus multi-point event shapes should admit an expansion in multi-point celestial blocks (which would be interesting to compute).

It would also be interesting to study OPEs of other types of null-integrated operators, such as those studied in \cite{Casini:2017roe,Cordova:2018ygx}. As explained in \cite{Kologlu:2019bco}, these can be viewed as descendants of light-transformed operators $\wL[\cO]$. Consider two such descendants inserted at the same point, say $x=0$,
\be
	(P^{k_1}\wL[\cO_1])(0,z_1)(P^{k_2}\wL[\cO_2])(0,z_2),
\ee
where we denoted the descendants schematically by $P^{k_i}\wL[\cO_i]$ and suppressed polarizations associated to $P$. Acting on this with $K^{k_1+k_2+1}$ we get $0$, and so we must conclude that this product has an expansion in terms of descendants of light-ray operators at level at most $k_1+k_2$. A conformally-invariant way to think about descendants $P^{k_i}\wL[\cO_i]$ is in terms of weight-shifting operators~\cite{Kologlu:2019bco,Karateev:2017jgd}. It is likely that the derivation of the light-ray OPE in this paper can be dressed appropriately with weight-shifting operators using methods described in~\cite{Karateev:2017jgd,Kravchuk:2018htv}.

Another generalization is to allow null-integrated operators to be on different null planes that approach each other. It should still be possible to relate matrix elements of such a product to the Lorentzian inversion formula.

In \cite{Kologlu:2019bco}, we introduced shock amplitudes, which describe the flat-space limit of the bulk dual of a null-integrated operator. In theories with bounded Regge growth, it should be possible to analytically continue shock amplitudes in spin, giving a vast generalization of the amplitudes usually considered. This work suggests a simple way to partially achieve this generalization: one can take coincident limits of shock particles to produce other types of shocks with different (integer) spin. For example, a coincident limit of shock gravitons can produce the spin-3 ``stringy" shock studied by Hofman and Maldacena \cite{Hofman:2008ar}.

A more speculative possible direction is to derive a nonperturbative OPE for amplitudes, describing a convergent expansion around the collinear limit. Such an OPE expansion exists in planar $\cN=4$ \cite{Alday:2010ku,Basso:2013vsa,Basso:2013aha,Basso:2014koa,Basso:2014jfa,Basso:2014nra}, relying on special properties of the theory like amplitude-Wilson-loop duality and integrability; it would be nice to generalize to a generic CFT. (Presumably, this would also require finding a good nonperturbative definition of an amplitude in a generic CFT.) Perhaps the conformal basis \cite{Pasterski:2016qvg,Pasterski:2017kqt} could be helpful for this. The soft limit of an external particle should correspond to the insertion of a null-integrated operator, so perhaps the hypothetical amplitudes OPE would be related to the light-ray OPE in this limit.

\subsection{More applications to event shapes}

It would be interesting to understand whether the light-ray OPE can be applied to asymptotically-free theories like QCD. The small angle behavior of the EEC in QCD was analyzed in \cite{Konishi:1979cb}. A more general factorization formula describing the collinear limit $\z\to 0$ and applicable to any weakly coupled gauge theory was derived in \cite{LanceFuture}. The energy-energy correlator (EEC) in QCD was recently computed at 2 loops (NLO) for arbitrary $\zeta$ \cite{Dixon:2018qgp, Luo:2019nig}. The light-ray OPE gives a way to resum large logarithms using symmetries as opposed to RG equations. The celestial block expansion is ultimately a consequence of Lorentz symmetry, which is still present when conformal symmetry is broken. Thus, event shapes in any theory should admit a celestial block expansion. However, when dilatation symmetry is broken, the selection rule $J=J_1+J_2-1$ will no longer hold. Thus, we expect the celestial block expansion in asymptotically-free theories to involve light-ray operators with other spins.\footnote{We thank Ian Moult for discussions on this point.}

In \cite{Hofman:2008ar}, it was shown how to relate the EEC to spin-3 moments of PDFs. Because these spin-3 moments compute matrix elements of spin-3 light-ray operators, it is natural to guess that spin-$J$ moments  of PDFs for general $J\in \C$ compute matrix elements of general spin-$J$ light-ray operators.\footnote{We thank Juan Maldacena and Aneesh Manohar for making this suggestion, and Ian Moult and Cyuan Han Chang for discussions.} It would be interesting to derive this connection directly.

The celestial block expansion suggests a way of ``perturbatively bootstrapping" the EEC in the same sense as the perturbative bootstrap for amplitudes and Wilson loops in $\cN=4$ SYM \cite{Dixon:2011pw, Dixon:2013eka,Dixon:2014iba,Drummond:2014ffa,Dixon:2014xca,Golden:2014pua,Caron-Huot:2016owq}. The idea of the perturbative bootstrap is to guess a basis of functions for the answer at some loop order (for example, by guessing the symbol alphabet). One then imposes consistency conditions to fix the coefficients in this basis. In the case of amplitudes in $\cN=4$, this program has been wildly successful, for example resulting in expressions for the $6$-point gluon amplitude up to 7 loops \cite{Caron-Huot:2019vjl}. There, consistency with the OPE for amplitudes \cite{Alday:2010ku,Basso:2013vsa,Basso:2013aha,Basso:2014koa,Basso:2014jfa,Basso:2014nra} and data from integrability provide powerful constraints. The celestial block expansion can provide analogous constraints for the EEC. Furthermore, in \cite{Kologlu:2019bco}, we gave a different expansion for the EEC in terms of ``$t$-channel blocks." OPE data from integrability can be used in either channel to make predictions that could help bootstrap the EEC.

An important ingredient in the perturbative bootstrap is the presence of contact terms in perturbative event shapes at $\z=0$ and $\z=1$. Because of Ward identities, the coefficients of contact terms serve as a check on the entire event shape. The light-ray OPE gives a systematic way to compute contact terms at $\z=0$. Furthermore, it provides a connection between the $\z=0$ contact term at $L$ loops and the leading non-contact term as $\z\to 0$ at $L+1$ loops.\footnote{Meanwhile, the back-to-back expansion (\ref{eq:btb}) provides a description of contact terms and leading non-contact terms at $\z=1$, given knowledge of the hard function $H(a)$ and cusp/collinear anomalous dimensions. Given this, one could imagine a poor-man's version of the perturbative bootstrap, where one uses contact terms at $L$ loops to predict leading non-contact terms at $L+1$ loops, fits the leading non-contact terms to a simple ansatz, integrates the ansatz to obtain contact terms at $L+1$ loops, and repeats.}

It would also be interesting to understand event shapes in $\cN=4$ SYM in a systematic expansion in $1/\lambda$ and $1/N$. The leading $1/\lambda$ corrections to energy-energy correlators were computed in \cite{Hofman:2008ar}, see also \cite{Goncalves:2014ffa}. They take the form of a finite sum of the $t$-channel event-shape blocks defined in \cite{Kologlu:2019bco}. This suggests that $t$-channel blocks could be simple ingredients for setting up a perturbative expansion in $1/\lambda$. One advantage of the $t$-channel expansion is the absence of contributions from double-trace operators in the planar limit. (By contrast, the light-ray OPE discussed in this paper gets contributions from both single- and double-trace operators.) The extreme simplicity of the $1/\lambda$ corrections in \cite{Hofman:2008ar} stems from the fact that the string shockwave $S$-matrix, expanded to leading order in $\alpha'$, only mixes adjacent levels on the string worldsheet, see e.g. \cite{Kologlu:2019bco}.

The problem of developing a $1/N$ expansion at large $\l$ is conceptually interesting because the condition  $J_0 < 3$ for the event shape to be well-defined is violated in na\"ive $1/N$ perturbation theory. To study $1/N$ corrections, it will be necessary to re-sum the four-point function in the Regge regime.

\subsection{Other applications and future directions}

Null-integrated operators arise naturally in information-theoretic quantities in quantum field theory. 
For example, the full modular Hamiltonian in the vacuum state of a region bounded by a cut $v=f(\vec y)$ of the null plane $u=0$ is \cite{Casini:2017roe}
\be
H_f &= 2\pi(K-P_f),\nn\\
P_f &= \int d^{d-2}\vec y f(\vec y)\int_{-\oo}^\oo dv T_{vv}(u=0,v,\vec y) = \int d^{d-2} \vec y f(\vec y) \wL[T](\vec y),
\ee
where $K$ is the generator of a boost in the $u$-$v$ plane. Here, we have abused notation and written $\wL[T]$ as a function of the transverse position $\vec y$, instead of the usual arguments $x,z$.

The vacuum modular flow operator is $U_f(s) = e^{-i s H_f}$. It is interesting to ask how $U_f$ changes as we deform the cut $f(\vec y)\to f(\vec y)+\de f(\vec y)$. Because the ANEC operator $\wL[T]$ appears in the modular Hamiltonian, we can use the algebra of $K$ and $P_f$ together with the light-ray OPE to do perturbation theory in $\de f(\vec y)$:
\be
\label{eq:perturbationtheoryformodularoperator}
U_{f+\de f}(s) U_{f}(-s) 
&= \exp\p{-\frac{i}{2\pi}(e^{2\pi s}-1) (H_{f+\de f}-H_{f})} \nn\\
&= \exp \p{it\int d^{d-2} \vec y\,\de f(\vec y) \wL[T](\vec y)} \nn\\
&= 1 + it\int d^{d-2} \vec y\,\de f(\vec y) \wL[T](\vec y) 
\nn\\
&\quad -(-i\pi)
\frac{t^2}{2} \sum_i \int d^{d-2} \vec y_1 d^{d-2} \vec y_2 \de f(\vec y_1) \de f(\vec y_2)  \cC_{\De_i-1}(\vec y_{12},\ptl_{\vec y_2}) \mathbb{O}_{i,J=3}(\vec y) + \dots,
\ee
where $t=e^{2\pi s}-1$. The expression (\ref{eq:perturbationtheoryformodularoperator}) gives a direct connection between the spectrum of a CFT and the shape dependence of the vacuum modular flow operator. It may be useful for understanding aspects of the quantum null energy condition (QNEC) \cite{Bousso:2015mna,Bousso:2015wca,Balakrishnan:2017bjg,Ceyhan:2018zfg}. Furthermore, it would be interesting to see whether it (or other manifestations of the light-ray OPE) has implications for bulk locality in holographic theories.

It would also be interesting to study the light-ray OPE for strongly-coupled theories like the 3d Ising model. With enough CFT data, it may be possible to compute event shapes and study modular flow quantitatively in this theory.

Particle colliders like the LHC have given us a wealth of data on event shapes in the Standard Model. In principle, it should be possible to measure event shapes in condensed matter systems using a tabletop collider. One must prepare a material in a state described by a QFT, excite it at a point, and measure the pattern of excitations on the boundary of the material. Several quantum critical points have both Euclidean and Lorentzian avatars in the laboratory. Traditionally, the most precise measurements are available for the Euclidean avatars, in the form of scaling dimensions of low-dimension operators. Event shapes for these systems could reveal intrinsically Lorentzian dynamics that would otherwise remain deeply hidden in the Euclidean measurements. 

Finally, it could be interesting to study event shapes in gravitational theories in an asymptotically flat spacetime, see e.g. \cite{Strominger:2017zoo} and references therein.\footnote{The same comment applies to electromagnetism.} In this case, physical measurements are performed at the future null infinity $\mathscr{I}^+$. As in a particle collider experiment, one can measure energy flux through the celestial sphere created in a gravitational collision. In addition to energy carried away by matter fields, there is a contribution due to gravity waves $\cE(\vec n) \sim \int_{\mathscr{I}^+} \textrm{News}^2$ which is quadratic in the so-called news tensor. In a gravitational theory, however, it is also natural to consider light-ray operators that are linear in the metric, similar to the ones measured in the current gravitational wave experiments. One such example is a memory light-ray operator $\cM(\vec n) \sim \int_{\mathscr{I}^+} \textrm{News}$ which measures the memory effect on the celestial sphere. As in the main body of the paper, we can consider multi-point gravitational event shapes and possibly study the corresponding light-ray OPE. One appealing feature of these observables is that they are IR safe --- in other words all IR divergencies that arise in the computations of scattering amplitudes should cancel in the event shapes. BMS symmetry \cite{Strominger:2013jfa} and familiar soft theorems \cite{He:2014laa} should become statements that relate different gravitational event shapes.\footnote{For example, an integral of the energy flux operator over the celestial sphere is related to the insertion of the memory operator \cite{Strominger:2014pwa}.}

\section*{Acknowledgements}	

We thank Mikhail Alfimov, Cyuan Han Chang, Clay C\'ordova, Lance Dixon, Claude Duhr, Tom Faulkner, Tom Hartman, Johannes Henn, Gregory Korchemsky,  Adam Levine, Juan Maldacena, Ian Moult, Gavin Salam, Amit Sever, Emery Sokatchev, and Kai Yan for discussions. We thank Lance Dixon, Ian Moult, and Hua Xing Zhu for sharing a draft of their work before publication \cite{LanceFuture}. We also thank Gregory Korchemsky for sharing a draft of his work before publication \cite{BacktoBackGrisha}. We thank Nikolay Gromov for producing and sharing with us figure~ \ref{fig:Kolyaplot}. DSD is supported by Simons Foundation grant 488657 (Simons Collaboration on the Nonperturbative Bootstrap), a Sloan Research Fellowship, and a DOE Early Career Award under grant No.\ DE-SC0019085. PK is supported by DOE grant No.\ DE-SC0009988. This research was supported in part by the National Science Foundation under Grant No.\ NSF PHY-1748958.

\pagebreak
	
\appendix

\section{Notation}
\label{app:notation}

In this appendix, we summarize some of our notation. Many of our conventions are taken from \cite{Kravchuk:2018htv}.

It is useful to distinguish between physical correlation functions and conformally invariant structures. A correlation function in the state $|\Omega\>$ represents a physical correlation function in a CFT. For example,
\be
\<\Omega|\cO_1\cdots\cO_n|\Omega\>
\ee
is a Wightman $n$-point function in a physical theory, and
\be
\<\cO_1\cdots\cO_n\>_\Omega
\ee
is a time-ordered $n$-point function in a physical theory.

Two- or three-point functions in the fictitious state $|0\>$ represent conformally-invariant functions that are fixed by conformal invariance. If conformal symmetry allows a finite set of possible tensor structures, then we index the possibilities by a label $(a)$, $(b)$, etc.. For example,
\be
\<0|\cO_1\cO_2\cO_3|0\>^{(a)}
\ee
represents a conformally-invariant tensor structure for the representations of $\cO_1,\cO_2,\cO_3$, and $a$ runs over the possible solutions to the conformal Ward identities. The above structure has an $i\e$ prescription appropriate for a Wightman function. Meanwhile,
\be
\<\cO_1\cO_2\cO_3\>^{(a)}
\ee
represents a conformally-invariant structure with the $i\e$ prescription of a time-ordered correlator.

Primary operators are labeled by weights $(\De,\rho)$ with respect to the conformal group $\tl\SO(d,2)$. Here, $\De\in\C$ and $\rho$ is an irreducible representation of $\SO(d-1,1)$. ($\De$ is constrained to be real and sufficiently positive for local operators in unitary theories.)  The weights of $\rho$ can be futher decomposed into $\rho=(J,\l)$, where $J$ is a positive integer for local operators, but in general $J\in \C$ can be continuous in Lorentzian signature. Here, $\l$ is a finite-dimensional representation of $\SO(d-2)$.  We can think of $J$ as the length of the first row of the Young diagram of $\rho$, while $\l$ encodes the remaining rows.  Altogether, we specify a conformal representation by the triplet $(\De,J,\l)$.

We often use the symbol $\cO$ to stand for the conformal representation with quantum numbers $(\De,J,\l)$. We use $\phi$ to represent a scalar operator with quantum numbers $(\De_\f,0,\bullet)$, where $\bullet$ is the trivial representation. (An exception is in section~\ref{sec:nequalsfour}, where $\cO^{IJ}$ refers to a $\mathbf{20'}$ operator in $\cN=4$ SYM.)

If $\cO$ is a local operator, then $\rho$ is a finite-dimensional representation. In this case, we define shadow and Hermitian conjugate representations as follows 
\be
\tl \cO &: (d-\De,\rho^R), \nn\\
\cO^\dag &: (\De,(\rho^R)^*),
\ee
where $\rho^R$ denotes the reflection of $\rho$ and $(\rho^R)^*$ is the dual of $\rho^R$. 

For continuous-spin operators, $\rho=(J,\l)$ is infinite-dimensional. The light transform turns a local operator into a continuous spin operator
\be
\wL[\cO] &: (1-J,1-\Delta,\lambda) \ .
\ee 
To define a conformally-invariant pairing for continuous spin operators we define
\be
\cO^{S} &: (d-\De,2-d-J,\l),\nn \\
\cO^{S\dagger} &: (d-\De,2-d-J,\l^*).
\ee 
Similarly, we define $\cO^F$ as an operator that can be paired with $\wL[\cO]$ (upon Hermitian conjugation)
\be
\cO^F &: (J+d-1,\Delta - d + 1, \lambda), \nn\\
\cO^{F\dag} &: (J+d-1,\Delta - d + 1, \lambda^*).
\ee

To describe the causal relation between two points we use the following symbols:  
\begin{itemize}
\item $x \approx y$ if $x$ and $y$ are space-like;
\item $x > y$ ($x < y$) if $x$ lies in the future (past) light-cone of $y$; 
\item $x \gtrsim y$ ($x \lesssim y$) if $x$ is on the future (past) null cone of $y$.
\end{itemize}

In section~\ref{sec:schannel}, we extensively use Euclidean and Lorentzian pairings between the 2-, 3- and 4-point functions. These are described in detail in appendix C and D of \cite{Kravchuk:2018htv} correspondingly. 

\section{Representations of orthogonal groups}
\label{sec:orthogonalreps}

\subsection{General index-free notation}
\label{sec:indexfree}

A finite-dimensional representation of $\SO(d)$ is labeled by a sequence $\bm_d=(m_{d,1},\dots,m_{d,n})$ such that
\be
&m_{d,1} \geq m_{d,2} \geq \dots \geq m_{d,n-1}\geq |m_{d,n}| && d=2n \\
&m_{d,1}\geq m_{d,2} \geq \dots \geq m_{d,n} \geq 0 && d=2n+1
\ee
The $m_{d,i}$ are either all integers (in the case of tensor representations) or all half-integers. When they are integers, we can think of them as lengths of rows of a Young diagram. See \cite{Kravchuk:2017dzd} for a recent review.

A spin-$J$ traceless symmetric tensor has labels $\mathbf{m}_d=(J,0,\dots,0)$, corresponding to a single-row Young diagram with length $J$. More generally, an object in the representation $\bm_d$ is a tensor with indices
\be
f^{\mu_1\cdots \mu_{m_{d,1}}\, \nu_1 \cdots \nu_{m_{d,2}}\, \cdots\, \rho_{1}\cdots \rho_{m_{d,n}}}.
\ee
For a given Young diagram, we can choose to make either symmetry of the rows manifest or antisymmetry of the columns manifest. We choose to make symmetry of the rows manifest. Thus, $f$ is symmetric in each of its $n$ groups of indices
\be
f^{\mu_1\cdots \mu_{m_{d,1}}\, \nu_1 \cdots \nu_{m_{d,2}}\, \cdots\, \rho_{1}\cdots \rho_{m_{d,n}}} &= f^{(\mu_1\cdots \mu_{m_{d,1}}) (\nu_1 \cdots \nu_{m_{d,2}}) \cdots (\rho_{1}\cdots \rho_{m_{d,n}})}.
\ee
Furthermore, it is traceless in all pairs of indices. Antisymmetrization of columns of the Young diagram is reflected in the fact that if we try to symmetrize too many indices, we get zero. For example,
\be
\label{eq:antisymmetrizationcondition}
f^{(\mu_1\cdots \mu_{m_{d,1}} \nu_1)\nu_2 \cdots \nu_{m_{d,2}}\, \cdots\, \rho_{1}\cdots \rho_{m_{d,n}}} &= 0.
\ee

It is useful to encode the tensor $f$ using index-free notation. We introduce polarization vectors $z_1,\dots,z_n\in \C^d$ for each row of the Young diagram and contract them with the corresponding indices to form a polynomial
\be
f(z_1,\dots,z_n) &\equiv f^{\mu_1\cdots \mu_{m_{d,1}}\, \nu_1 \cdots \nu_{m_{d,2}}\, \cdots\, \rho_{1}\cdots \rho_{m_{d,n}}} z_{1\mu_1}\cdots z_{1\mu_{m_{d,1}}} z_{2\nu_1}\cdots z_{2\nu_{m_{d,2}}}\cdots z_{n\rho_1}\cdots z_{n\rho_{m_{d,n}}}.
\ee
By construction, $f(z_i)$ is homogeneous in each polarization vector
\be
\label{eq:itishomogeneous}
f(\a_1 z_1,\cdots, \a_{n} z_{n}) &= \a_1^{m_{d,1}} \cdots \a_{n}^{m_{d,n}} f(z_1,\cdots, z_{n})\qquad (\a_i\in \C).
\ee
Because $f$ is traceless, we can impose the conditions
\be
\label{eq:orthogonalityconditions}
z_i^2 = 0,\quad z_i\.z_j = 0.
\ee
These conditions mean that shifting $f$ by anything proportional to $\de^{\mu\nu}$ leads to the same polynomial $f(z_i)$. The traceless tensor $f$ can thus be recovered from the polynomial $f(z_i)$ by choosing any tensor leading to the correct polynomial and subtracting traces.

In index-free notation, the antisymmetrization condition (\ref{eq:antisymmetrizationcondition}) becomes
\be
f(z_1,z_2+\b z_1,z_3,\dots,z_{n}) &= f(z_1,z_2,z_3,\dots, z_{n}).
\ee
In other words, $f$ is gauge-invariant under shifts $z_2\to z_2+\b z_1$. (Note that this gauge-redundancy is consistent with the orthogonality conditions (\ref{eq:orthogonalityconditions}).) More general antisymmetrization conditions show that $f$ is invariant under the gauge redundancies
\be
\label{eq:generalgaugeredundancy}
z_2 &\to z_2 + \# z_1 \nn\\
z_3 &\to z_3 + \# z_2 + \# z_1 \nn\\
\vdots &  \nn\\
z_{n} &\to z_{n} + \# z_{n-1} +\cdots+ \# z_1.
\ee

Finally, in even dimensions, the tensor $f$ can satisfy
\be
&\e^{\mu_1\cdots\rho_1}{}_{\mu_0\cdots\rho_0}f^{\mu_0\mu_2\cdots \mu_{m_{d,1}}\, \cdots\, \rho_{0}\rho_2\cdots \rho_{m_{d,n}}} z_{1\mu_1}\cdots z_{1\mu_{m_{d,1}}} z_{2\nu_1}\cdots z_{2\nu_{m_{d,2}}}\cdots z_{n\rho_1}\cdots z_{n\rho_{m_{d,n}}}\nn\\
&= \pm p_n f(z_1,\dots,z_n)
\ee
where $p_n$ is a constant depending only on $n$. This is equivalent to imposing an (anti-)self-duality condition on the polarization vectors
\be
\label{eq:dualitycondition}
\e^{\mu_1\cdots\rho_1}{}_{\mu_0\cdots\rho_0} z_{1\mu_1}\cdots z_{n\rho_1} &= \pm p_d n! z_{[1\mu_0}\cdots z_{n\rho_0]}.
\ee

To summarize, the representation $\bm_d$ is equivalent to the space of homogeneous polynomials of polarization vectors $z_1,\dots, z_{n} \in \C^d$ with degrees $m_{d,1},\dots, m_{d,n}$, satisfying the orthogonality conditions (\ref{eq:orthogonalityconditions}), duality condition (\ref{eq:dualitycondition}) in even dimensions, and subject to gauge-redundancy (\ref{eq:generalgaugeredundancy}).

We have essentially arrived at the Borel-Weil theorem, specialized to orthogonal groups. The theorem states that each irreducible finite-dimensional representation of a reductive Lie group $G$ is equivalent to the space of global sections of a holomorphic line bundle on the flag manifold $G/B$, where $B\subset G$ is a Borel subgroup.  In the case $G=\SO(d)$, the flag manifold $G/B$ is the projectivization of the space of vectors $z_1,\dots, z_n$ satisfying the above conditions and gauge-redundancies. A section of a line bundle on this space is a homogeneous polynomial of the polarization vectors.

It is sometimes useful to use mixed index-free notation, where only some of the polarization vectors are contracted. For example, we could consider
\be
\label{eq:mixed}
f^{\nu_1 \cdots \nu_{m_{d,2}}\, \cdots\, \rho_{1}\cdots \rho_{m_{d,n}}}(z_1) &\equiv f^{\mu_1\cdots \mu_{m_{d,1}}\, \nu_1 \cdots \nu_{m_{d,2}}\, \cdots\, \rho_{1}\cdots \rho_{m_{d,n}}} z_{1\mu_1}\cdots z_{1\mu_{m_{d,1}}}.
\ee
The object $f^{\nu_1 \cdots \nu_{m_{d,2}}\, \cdots\, \rho_{1}\cdots \rho_{m_{d,n}}}(z_1)$ is a tensor on the null cone $z_1^2 = 0$. Its indices satisfy all the symmetry conditions appropriate for the Young diagram $(m_{d,2},\dots,m_{d,n})$ obtained by discarding the first row of the Young diagram $(m_{d,1},m_{d,2},\dots,m_{d,n})$. Furthermore, antisymmetry conditions like (\ref{eq:antisymmetrizationcondition}) mean that if we contract any of the indices of (\ref{eq:mixed}) with $z_1$, the result is zero. We say that (\ref{eq:mixed}) is ``transverse".

\subsection{Poincare patches}

We can think of the polarization vector $z_1$ as an embedding-space coordinate in $d-2$ dimensions. It is natural to ask what the function $f(z_1,\dots,z_n)$ looks like in flat coordinates. Let us write the metric on $\C^d$ as
\be
z\.z &= -z^+ z^- +  z^\perp\.z^\perp,
\ee
where $ z^\perp\in \C^{d-2}$. For generic $z_1$, we can use homogeneity to set
\be
z_1 &= (z_1^+,z_1^-,z_1^\perp) = (1, (y^{\perp})^{2}, y^\perp),\qquad  y^\perp \in \C^{d-2}.
\ee
Using the gauge redundancies (\ref{eq:generalgaugeredundancy}), we can set $z_2^+=\cdots =z_n^+=0$. The orthogonality conditions (\ref{eq:orthogonalityconditions}) then imply that the other $z_i$ take the form
\be
z_i &= (0,2 z_i^\perp\. y^\perp,  z_i^\perp),\qquad z_i^\perp\in \C^{d-2},
\ee
where
\be
  z_i^\perp\. z_j^\perp=0\quad (i,j=2,\dots,n).
\ee
Thus, we obtain a function
\be
f^\downarrow( y^\perp;  z_2^\perp,\dots, z_n^\perp) &\equiv \left.f(z_1,\dots,z_n)\right|_{\substack{
z_1=(1,\vec y^{\perp 2}, y^\perp) \\
z_i = (0,2z_i^\perp\. y^\perp,  z_i^\perp).
}}
\ee

The function $f^\downarrow$ is not homogeneous in $y^\perp$, but it is a homogeneous polynomial in the remaining arguments $ z_2^\perp,\dots,  z_n^\perp \in \C^{d-2}$. Furthermore, the $ z_2^\perp,\dots,  z_n^\perp$ are subject to the same orthogonality and gauge redundancies as before, except now in 2-fewer dimensions. Thus, $f^\downarrow$ is equivalent to a tensor field on $\C^{d-2}$, transforming in the $\SO(d-2)$ representation $(m_{d,2},\dots,m_{d,n})$
\be
f^\downarrow( y^\perp; z_2^\perp,\dots,z_n^\perp) &= f^{\downarrow\a_1\cdots \a_{m_{d,2}}\,\cdots\,\b_1\cdots \b_{m_{d,n}}}( y^\perp) z^\perp_{2\a_1}\cdots  z^\perp_{2\a_{m_{d,2}}} \cdots  z^\perp_{n\b_1}\cdots  z^\perp_{n\b_{m_{d,n}}},
\ee
where $\a_i,\b_i$ are vector indices in $d-2$-dimensions. This is the usual procedure of restricting to a Poincare patch in the embedding formalism.

The function $f(z_1,\dots,z_n)$ can easily be recovered from $f^\downarrow( y^\perp; z_2^\perp,\dots,z_n^\perp)$ by imposing the correct homogeneity and gauge redundancy
\be
f(z_1,\dots,z_n) &= (f^\downarrow)^\uparrow(z_1,\dots,z_n) \nn\\
&= (z_1^+)^{m_1} f^\downarrow\p{\frac{ z^\perp_1}{z_1^+};  z^\perp_2-\frac{z_2^+}{z_1^+} z^\perp_1,\dots, z^\perp_n-\frac{z_n^+}{z_1^+} z^\perp_1}.
\ee
This is the usual procedure of lifting to the embedding space.

If we like, restriction to a Poincare patch can be iterated again to obtain a tensor field on $\C^{d-2} \x \C^{d-4}$ with indices valued in the $\SO(d-4)$ representation $(m_{d,3},\dots,m_{d,n})$,
\be
&f^{\downarrow\downarrow}( y^\perp, x^{\perp\perp};z^{\perp\perp}_3,\dots,z^{\perp\perp}_n), \nn\\
&= f^{\downarrow\downarrow\a_1\cdots \a_{m_{d,3}}\,\cdots,\b_1\cdots \b_{m_{d,n}}}(y^\perp,x^{\perp\perp})  z^{\perp\perp}_{3\a_1}\cdots z^{\perp\perp}_{3\a_{m_{d,3}}} \cdots z^{\perp\perp}_{n\b_1}z^{\perp\perp}_{n\b_{m_{d,n}}}
\qquad x^{\perp\perp},z^{\perp\perp}_j\in \C^{d-4}.
\ee
Here, $\a_i,\b_i$ are vector indices in $d-4$ dimensions.
Similarly, we can obtain $f^{\downarrow\downarrow\downarrow}$ which is a tensor field on $\C^{d-2}\x \C^{d-4} \x \C^{d-6}$, etc..  All of these functions can be lifted back to the original homogeneous polynomial $f(z_1,\dots,z_n)$.

\subsection{Application to CFT}

Most of the above constructions still work when some of the weights $m_{d,i}$ become continuous. We can now no longer demand that $f$ is a polynomial in the polarization vectors with continuous weights. However, we can still demand that $f$ is a homogeneous function. Such homogeneous functions yield infinite-dimensional representations of $\SO(d)$.\footnote{An index-free formalism for CFT operators in general tensor representations was introduced in \cite{Costa:2014rya}. That formalism introduces fermionic polarization vectors, and essentially differs from the one here by privileging the columns of Young tableaux instead of the rows.}

We are interested in studying infinite-dimensional representations of $\tl\SO(d,2)$, corresponding to operators in CFT. These are labeled by a weight $\bm_{d+2}=(-\De,m_{d,1},\dots,m_{d,n})$, where $\De$ is not necessarily a negative integer. To describe light-ray operators, we must additionally allow $m_{d,1}=J$ to be non-integer. We often use the notation
\be
\mathbf{m}_{d+2} &= (-\De,J,\l),\nn\\
\l&=(m_{d,2},\dots, m_{d,n}),
\ee
where $\l$ are weights of a finite-dimensional representation of $\SO(d-2)$. When $J$ is an integer satisfying $J\geq m_{d,2}$, we can also define the finite-dimensional representation of $\SO(d-1,1)$ 
\be
\rho &= (J,m_{d,2},\dots,m_{d,n}).
\ee

The elements of the representation with weights $\bm_{d+2}$ are homogeneous functions of the kind described in section~\ref{sec:indexfree}. Here, we simply introduce some specialized notation for the case at hand. The functions are
\be
\cO(X,Z,W_1,\dots,W_{n-1}),\qquad X,Z\in\R^{d,2},\quad W_i\in \C^{d+2},
\ee
where the vectors $X,Z,W_i$ are null and mutually orthogonal. Furthermore, they satisfy gauge redundancies
\be \label{eq:polarization gauge redundancy}
Z &\sim Z + \# X \nn\\
W_1 &\sim W_1 + \# Z + \# X \nn\\
\vdots \nn\\
W_{n-1} &\sim W_{n-1} + \# W_{n-2} +\dots +\# X.
\ee
The homogeneity condition is
\be
\cO(\a X, \b Z, \a_1 W_1,\dots,\a_{n-1} W_{n-1}) &= \a^{-\De} \b^J \a_1^{m_{d,2}} \cdots \a_{n-1}^{m_{d,n}} \cO(X,Z,W_1,\dots,W_{n-1}).
\ee
Furthermore, $\cO$ is constrained to be a polynomial in the  $W_i$'s (but not in $X,Z$).

The restriction of $\cO$ to a Poincare patch is given by
\be
\cO^\downarrow(x,z,w_1,\dots,w_{n-1}) &= \left.\cO(X,Z,W_1,\dots,W_{n-1})\right|_{\substack{
X=(1,x^2,x)  \\
Z=(0,2x\.z,z) \\
W_i=(0,2x\.w_i,w_i)
}}.
\ee
Here, $z,w_i$ are mutually orthogonal null vectors, subject to the gauge redundancies
\be
w_1 &\sim w_1 + \# z \nn\\
w_2 &\sim w_2 + \# w_1 + \# z \nn\\
\vdots \nn\\
w_{n-1} &\sim w_{n-1} + \# w_{n-2} +\cdots+ \# z.
\ee
The function $\cO^\downarrow$ satisfies the homogeneity condition
\be
\cO^\downarrow(x,\b z,\a_1 w_1,\dots,\a_{n-1} w_{n-1}) &= \b^J \a_1^{m_{d,2}} \cdots \a_{n-1}^{m_{d,n}}\cO^\downarrow(x,z,w_1,\dots,w_{n-1}).
\ee

The transverse coordinates $\vec y$ discussed in section~\ref{sec:boostselection} come about when we do an additional restriction to a Poincare patch in the $z$ variable:
\be
\cO^{\downarrow\downarrow}(x,\vec y;\vec w_1,\dots,\vec w_{n-1}) &= \left.\cO^\downarrow(x,z,w_1,\dots,w_{n-1})\right|_{
\substack{
z=(1,\vec y^2,\vec y)\\
w_i = (0,2\vec y\.\vec w_i, \vec w_i)
}
},
\ee
where
\be
 x\in \R^{d-1,1},\quad \vec y\in \R^{d-2},\quad \vec w_i\in \C^{d-2}.
\ee
We can equivalently think of $\cO^{\downarrow\downarrow}(x,\vec y)$ as a tensor field on $\R^{d-1,1}\x\R^{d-2}$ transforming in the $\SO(d-2)$ representation $\l$. When $\cO$ is a traceless symmetric tensor (i.e.\ $\l$ is trivial), we have
\be
\int_{-\oo}^\oo \cO_{v\cdots v}(u=0,v,\vec y) &\propto \wL[\cO]^{\downarrow\downarrow}(-\oo z_0, \vec y),
\ee
where $z_0=(1,1,0,\dots,0)$ is a null vector in the $v$ direction.

We almost always abuse notation and drop the $\downarrow$ superscripts, relying on the arguments of $\cO$ to distinguish between the embedding-space function and its restrictions to Poincare patches.
We also often use mixed index-free notation, where we strip off the $w_i$'s to obtain a tensor operator
\be
\cO(x,z,w_1,\dots,w_{n-1}) &= \cO^{\mu_1\cdots \mu_{m_{d,2}}\,\cdots\,\nu_1\cdots\nu_{m_{d,n}}}(x,z) w_{1\mu_1}\cdots w_{1\mu_{m_{d,2}}}\cdots w_{n{-}1\nu_1} \cdots w_{n{-}1\nu_{m_{d,n}}}.
\ee
The tensor $\cO^{\mu_1\cdots \mu_{m_{d,2}}\,\cdots\,\nu_1\cdots\nu_{m_{d,n}}}(x,z)$ has indices symmetrized using the Young tableau $\l=(m_{d,2},\dots,m_{d,n})$, and furthermore all its indices are transverse to $z$.
Finally, we often suppress tensor indices and simply write $\cO(x,z)$, where it is understood that $\cO$ can carry indices transverse to $z$.

All of these different formalisms for representing $\cO$ are equivalent, and they are convenient for different purposes. For example, to define the celestial map in section~\ref{sec:generalizationandcelestialmap}, it is convenient to use embedding-space operators $\cO(X,Z,W_1,\dots,W_{n-1})$. To define the Lorentzian pairings (\ref{eq:lorentzianpairing}) and (\ref{eq:2ptpairingL}), it is convenient to use the object $\cO^{\mu_1\cdots \mu_{m_{d,2}}\,\cdots\,\nu_1\cdots\nu_{m_{d,n}}}(x,z)$ which caries a finite set of indices transverse to $z$. We move freely between the different formalisms as needed.

\section{More on analytic continuation and even/odd spin}
\label{app:moreonanalyticcont}

In this section, we give more detail on the relationship between $\mathsf{CRT}$ and the generalized Lorentzian inversion formula. In particular, we explain how to go from the formula in \cite{Kravchuk:2018htv} to the formula (\ref{eq:notsoobvious}) in the main text. 

The formula derived in \cite{Kravchuk:2018htv} is
\be
C_{ab}^\pm(\De,J,\l) &= -\frac{1}{2\pi i}  \int_{\substack{4>1\\2>3}} \frac{d^d x_1\cdots d^d x_4}{\vol(\tl \SO(d,2))} \<\O|[\cO_4, \cO_1] [\cO_2,\cO_3]|\O\>\nn\\
&\qquad\qquad \qquad  \x \tsym_2^{-1} \tsym_4^{-1}\frac{\p{\tsym_2\<\cO_1 \cO_2 \wL[\cO^\dagger]\>^{(a)}}^{-1}\p{\tsym_4\<\cO_4 \cO_3 \wL[\cO]\>^{(b)}}^{-1}}{\<\wL[\cO]\wL[\cO^\dagger]\>^{-1}}\nn\\
&\quad +(1\leftrightarrow 2).
\label{eq:obviousgeneralization}
\ee
It involves light-transforms of time-ordered structures $\<\cO_1\cO_2 \wL[\cO^\dag]\>^{(a)}$ and $\<\cO_3\cO_4 \wL[\cO]\>^{(b)}$.\footnote{By a ``time-ordered structure," we mean a conformally-invariant function of positions, with the $i\e$ prescription appropriate for a time-ordered correlator. By a ``Wightman structure," we mean a conformally-invariant function of positions, with the $i\e$ prescription appropriate for a Wightman function with the given ordering.}
 Time-ordered structures only make sense for integer $J$ (see appendix~A of \cite{Kravchuk:2018htv}), so we must give a prescription for how to analytically continue (\ref{eq:obviousgeneralization}) in $J$. Such a prescription was described in \cite{Kravchuk:2018htv}.\footnote{It is as follows: we should first compute $\<\cO_1\cO_2 \wL[\cO^\dag]\>^{(a)}$ for general nonnegative integer $J$ (where $J$ is the spin of $\cO$). The result is no longer a time-ordered structure (e.g.\ it has $\th$-functions of positions). It can then analytically continued from even or odd $J$, depending on whether we are computing $C^+_{ab}(\De,J,\l)$ or $C^-_{ab}(\De,J,\l)$. The analytic continuations are fixed by demanding that they are well-behaved in the right-half $J$-plane.} However, for our purposes, it will be helpful to phrase it in a different way. In particular, this requires clarifying the role of the $\pm$ sign in the definition of $\cO^\pm_{\De,J,\l(a)}$.

Note that there are two terms in the Lorentzian inversion formula. The $t$-channel term written explicitly in (\ref{eq:obviousgeneralization}) depends on
\be
\label{eq:pairofstructs}
\cT_2\<\cO_1\cO_2 \wL[\cO^\dag](x_0,z_0)\>^{(a)} &= \cT_2\<0|\cO_2 \wL[\cO^\dag](x_0,z_0) \cO_1|0\>^{(a)} & ((1>2)\approx 0), \\
\cT_4\<\cO_3\cO_4 \wL[\cO](x_0,z_0)\>^{(b)} &= \cT_4\<0|\cO_4 \wL[\cO](x_0,z_0) \cO_3|0\>^{(b)} & ((3>4) \approx 0).
\label{eq:pairofstructsagain}
\ee
On the right, we indicate the causal relationship between points for which the structure is needed. We also give light-transformed Wightman structures that equal the light-transformed time-ordered structures when those causal relationships hold.
Meanwhile, the $u$-channel term $(1\leftrightarrow 2)$ depends on 
\be
\label{eq:uchannelstruct}
\cT_1\<\cO_1\cO_2 \wL[\cO^\dag](x_0,z_0)\>^{(a)} &= \cT_1\<0|\cO_1 \wL[\cO^\dag](x_0,z_0) \cO_2|0\>^{(a)} & ((2>1)\approx 0),
\ee
instead of (\ref{eq:pairofstructs}).

We see from (\ref{eq:pairofstructs}) and (\ref{eq:uchannelstruct}) that the Lorentzian inversion formula actually depends on a pair of Wightman structures
\be
\<0|\cO_2 \cO^\dag(x_0,z_0) \cO_1|0\>^{(a)},\quad
\<0|\cO_1 \cO^\dag(x_0,z_0) \cO_2|0\>^{(a)}.
\label{eq:wightmanstructuresneeded}
\ee
It is easy to separately analytically continue each Wightman structure in spin. However, we should take care to preserve the correct relationship between the structures.
Let us describe this relationship when $J$ is an integer, and then generalize to non-integer $J$.

The simplest way to relate the structures (\ref{eq:wightmanstructuresneeded}) for integer $J$ is to demand that they are equal when all operators are spacelike separated. Unfortunately, this type of relationship does not generalize to non-integer $J$ due to branch cuts in the spacelike region \cite{Kravchuk:2018htv}.

A different way to state the relationship between the structures (\ref{eq:wightmanstructuresneeded}) for integer $J$ is to say how they transform under a combination of $\mathsf{CRT}$ and Hermitian conjugation. Recall that $\mathsf{CRT}$ is an anti-unitary symmetry that takes $x=(x^0,x^1,x^2,\dots,x^{d-1})$ to its Rindler reflection $\bar x=(-x^0,-x^1,x^2,\dots,x^{d-1})$. Its action on a local operator is given by
\be
\label{eq:localcrt}
(\mathsf{CRT}) \cO_\mathrm{local}^\a(x) (\mathsf{CRT}) &=\p{(e^{-i\pi \cM^{01}})^\a{}_\b \cO_\mathrm{local}^\b(\bar x)}^\dag,
\ee
where $\a,\b$ are indices for the Lorentz representation of $\cO$, and and $\cM^{01}$ is the generator of a boost in the $01$ plane. (We assume $\cO_\mathrm{local}$ is bosonic, for simplicity.) In general, we define the ``Rindler conjugate" of any (not necessarily local) operator $\cO$ by
\be
\bar \cO &\equiv (\mathsf{CRT})\cO(\mathsf{CRT}).
\ee
Note that Rindler conjugation preserves operator ordering, since it is simply conjugation by a symmetry.

If we combine Rindler conjugation with Hermitian conjugation, we obtain a linear map that reverses operator ordering
\be
\label{eq:rindlerhermitian}
\cO &\to \bar \cO^\dag.
\ee
For local operators, this is equivalent to a rotation by $\pi$ in the plane spanned by $x^1$ and Euclidean time $ix^0$,
\be
\label{eq:rindlerplushermitianlocal}
\bar{\cO_\mathrm{local}^\a(x)}^\dag &= (e^{-i\pi \cM^{01}})^\a{}_\b \cO_\mathrm{local}^\b(\bar x).
\ee
(One way to understand why this reverses operator ordering is that such a rotation reverses all the $i\e$'s.) However, for non-local operators, (\ref{eq:rindlerhermitian}) cannot be described in terms of a Euclidean rotation.  We call the eigenvalue of an operator under (\ref{eq:rindlerplushermitianlocal}) its ``signature."

Let $z_0=(1,1,0,\dots,0)$ be a null vector satisfying $\bar z_0=-z_0$. Given a local operator $\cO_\mathrm{local}$ with dimension $\De$ and spin-$J$, it is easy to check using (\ref{eq:localcrt}) that $\wL[\cO_\mathrm{local}](-\oo z_0,z_0)$ has signature $(-1)^J$,
\be
\label{eq:rindlerhermitianlightlocal}
\bar{\wL[\cO_\mathrm{local}](-\oo z_0,z_0)}^\dag &= (-1)^J \wL[\cO_\mathrm{local}](-\oo z_0,z_0).
\ee
However, more general light-ray operators can have a signature that is not necessarily related to $J$, and this is what the superscript $\pm$ encodes:
\be
\label{eq:rindlerhermitianlightnonlocal}
\bar{\mathbb{O}_{\De,J}^\pm(-\oo z_0,z_0)}^\dag &= \pm \mathbb{O}_{\De,J}^\pm(-\oo z_0,z_0).
\ee

Let us understand how signature is encoded in the inversion formula.
Since (\ref{eq:rindlerhermitian}) acts as a complexified Lorentz transformation (\ref{eq:rindlerplushermitianlocal}) on local operators, it is an operator-order-reversing ``symmetry" of three-point functions of local operators. Let $\cO_1,\cO_2$ be any local operators. We have
\be
\<0|\cO_1 \cO_\mathrm{local}^\dag(x,z) \cO_2|0\> 
&= \<0|\bar{\cO}_2^\dag\, \bar{\cO_\mathrm{local}^\dag (x,z)}^\dag\, \bar{\cO}_1^\dag |0\> \nn\\
&= \<0|\bar{\cO}_2^\dag\, \cO_\mathrm{local}^\dag (\bar x,\bar z)\, \bar{\cO}_1^\dag |0\>\nn\\
&= (-1)^J \<0|\bar{\cO}_2^\dag\, \cO_\mathrm{local}^\dag (\bar x,-\bar z)\,  \bar{\cO}_1^\dag |0\>.
\ee
In the last line, we used that $\cO_\mathrm{local}^\dag(x,z)$ is a degree-$J$ polynomial in $z$ to give it a future-pointing polarization vector $-\bar z$. Here, $\bar{\cO}_{1,2}^\dag$ are given by (\ref{eq:localcrt}).

The natural generalization to non-integer $J$ is that the Wightman structures (\ref{eq:wightmanstructuresneeded}) should be related by
\be
\label{eq:rindlerhermitiannoninteger}
\<0|\cO_1 \cO^\dag(x,z) \cO_2|0\>^{(a)}
 &= \pm \<0|\bar{\cO}_2^\dag\, \cO^\dag (\bar x,-\bar z)\,  \bar{\cO}_1^\dag |0\>^{(a)},
\ee
where $\pm$ indicates whether we have analytically continued from even or odd spin. Again, $\bar\cO_i^\dag$ is given by (\ref{eq:rindlerplushermitianlocal}). Plugging this in to (\ref{eq:obviousgeneralization}) gives equation~(\ref{eq:notsoobvious}).

\section{Checking the celestial map with triple light transforms}
\label{sec:triple light transform}

For symmetric traceless tensors $\cO_1$ and $\cO_2$, our OPE formula \eqref{eq:summarysofar} relies on the computation of the coefficient $q_{\de,j}^{(a)}$ defined by the triple light-transform in \eqref{eq:triple light transform}. For more general representations of $\cO_1$ and $\cO_2$, our formula \eqref{eq:finalanswerforproductingeneral} requires computation of the map defined by \eqref{eq:defofe}. We claim that this map is determined by the celestial map \eqref{eq:celestialmap}. In this appendix, we will prove the celestial map for operators in symmetric traceless tensor representations. We leave proving it for more general representations for the future.

Let $\cO_1$ and $\cO_2$ be symmetric traceless tensors of spins $J_1$ and $J_2$, and consider the three-point structures
\be
\langle0| \cO_1(X_1,Z_1) \cO_2(X_2,Z_2) \cO (X_0,Z_0) |0\rangle^{(a)} \, .
\ee
For simplicity, we consider the case with $\cO$ in a symmetric traceless tensor representation, $(\De,J=J_1+J_2-1,\l =\bullet)$, as well. Then, the relevant three-point structures were classified in~\cite{Costa:2011mg}. In embedding space, we can use the following basis of tensor structures;
\begin{align}
\langle0| \cO_1(X_1,Z_1) \cO_2(X_2,Z_2) \cO (X_0,Z_0) |0\rangle^{(a)} = \frac{\prod_i (-2V_i)^{m_i} \prod_{i<j} (-2H_{ij})^{n_{ij}}}{X_{12}^{\frac{\bar \tau_1+\bar \tau_2-\bar \tau_0}{2}}X_{20}^{\frac{\bar \tau_2+\bar \tau_0-\bar \tau_1}{2}}X_{01}^{\frac{\bar \tau_0+\bar \tau_1-\bar \tau_2}{2}}},
\end{align} where $i,j=0,1,2$, $\bar \tau_i = \Delta_i + J_i$ and the basis index $(a)$ is determined by six numbers $\{m_i, n_{ij}\}$ satisfying
\begin{align}
m_i + \sum_{j\ne i} n_{ij} = J_i.
\end{align} Recall that $X_{ij}\equiv-2 X_i \cdot X_j$. The building blocks for the structures are \cite{Costa:2011mg}
\begin{align}
X_{ij} &\equiv -2 X_i\.X_j,\\
V_{i,jk} &\equiv 
 \frac{Z_i\cdot X_j \; X_i\cdot X_k - Z_i \cdot X_k \;X_i \cdot X_j}{X_j \cdot X_k},
\\ H_{ij} & \equiv 
-2 \left( Z_i\cdot Z_j \; X_i \cdot X_j - Z_i\cdot X_j \; Z_j\cdot X_i\right).
\end{align} For brevity, we define $V_i \equiv V_{i,jk}$ for $\{i,j,k\}$ in cyclic order. We have shown in \cite{Kologlu:2019bco} that
\begin{align}
\langle0| \cO_2\wL[\cO] \cO_1  |0\rangle^{(a)} = (-2V_0)^{m_0} \prod_{i<j} (-2H_{ij})^{n_{ij}} \, \langle0| \cO'_2 \wL[\f]  \cO'_1 |0 \rangle^{(a')} \, .
\end{align} The new structure $\langle0| \cO'_2 \f \cO'_1 |0\rangle^{(a')}$ is the unique one that has
\begin{align}
n_{ij}'=0,\qquad m_0'=0, \qquad m_1'=m_1,\qquad m_2'=m_2\,.
\end{align} The new formal operators $\cO'_i$ have spin $J_i'=m_i$ and dimension $\De'_i = \De_i +J_i-m_i$. (Note that $\bar \tau_i = \bar \tau_i'$.) The formal scalar $\f$ has dimension $\De_\f=\bar \tau$. The light-transform of the structure $(a')$ is \cite{Kologlu:2019bco}
\begin{align}
&\langle0| \cO'_2\wL[\f]  \cO'_1 |0\rangle^{(a')} \nn
\\ &=L(\cO'_1 \cO'_2 [\f]) \frac{
		(-2 V_{0})^{1-\bar \tau}
		(-2V_{1})^{m_1}
		(-2V_{2})^{m_2}
	}{
		(-X_{02})^{\frac{\bar\tau^L+\bar\tau_2-\bar\tau_1}{2}}
		X_{01}^{\frac{\bar\tau^L+\bar\tau_1-\bar\tau_2}{2}}
		X_{12}^{\frac{\bar\tau_1+\bar\tau_2-\bar\tau^L}{2}}
	}
	f\p{-\frac{H_{01}}{2V_{0}V_{1}},-\frac{H_{02}}{2V_{0}V_{2}}}\qquad ((2>0)\approx 1),
\end{align}
where $\bar \tau^L = (1-J) + (1-\De) = 2-\bar \tau$, 
\begin{align}
L(\cO'_1 \cO'_2 [\f]) = -2\pi i \frac{\Gamma(\De_\f-1)}{\Gamma(\tfrac{\De_\f+\tau'_1-\tau_2'}{2})\Gamma(\tfrac{\De_\f-\tau'_1+\tau_2'}{2})}\, ,
\end{align} and
\begin{align}
f(x,y) = F_2(\bar\tau-1;-m_1,-m_2;\tfrac{1}{2}(\bar \tau+\tau'_1-\tau'_2),\tfrac{1}{2}(\bar \tau-\tau'_1+\tau'_2);x,y)\, .
\end{align} $F_2$ is the Appell hypergeometric function
\begin{align}
F_2 (\a;\b,\b';\g,\g';x,y) \equiv \sum_{k=0}^\oo \sum_{l=0}^\oo \frac{(\a)_{k+l}(\b)_k(\b')_l}{k!l!(\g)_k (\g')_l} x^k y^l\, .
\end{align}

Now, we'd like to specialize $X_0 = (1,0,0)$ and compute the remaining light transforms $\wL^-[\cO_1](X_\oo,Z_1)$ and $\wL^+[\cO_2](X_\oo,Z_2)$.
\be
&\frac{\<0|\wL^+[\cO_2](X_\oo,Z_2)\wL[\cO](X_0,Z_0) \wL^-[\cO_1](X_\oo,Z_1)|0\>^{(a)}}{\vol\SO(1,1)} \nn\\
&=\frac{1}{\vol\SO(1,1)}\int_0^\oo d\a_2\int_{-\oo}^0 d\a_1 (2V_0)^{m_0} \prod_{i<j} (-2H_{ij})^{n_{ij}} \, \langle \cO'_1 \wL[\f] \cO'_2 \rangle^{(a')} \nn
\\ &=\frac{1}{\vol\SO(1,1)}\int_0^\oo d\a_2\int_{-\oo}^0 d\a_1 \prod_{i<j} (-2H_{ij})^{n_{ij}} \, L(\cO'_1 \cO'_2 [\f]) \frac{
		(2 V_{0})^{1-\bar \tau+m_0}
		(2V_{1})^{m_1}
		(2V_{2})^{m_2}
	}{
		(-X_{02})^{\frac{\bar\tau^L+\bar\tau_1-\bar\tau_2}{2}}
		X_{01}^{\frac{\bar\tau^L+\bar\tau_2-\bar\tau_1}{2}}
		X_{12}^{\frac{\bar\tau_1+\bar\tau_2-\bar\tau^L}{2}}
	}\nn
\\ &\qquad \times
	f\p{-\frac{H_{01}}{2V_{0}V_{1}},-\frac{H_{02}}{2V_{0}V_{2}}}
\ee
Inside the integral, the light-transform instructs us to replace
\begin{align}
X_1 &\rightarrow Z_1 -\alpha_1 X_\infty = (0,-\alpha_1,z_1) \nn
\\ X_2 &\rightarrow Z_2 -\alpha_2 X_\infty = (0,-\alpha_2,z_2) \nn
\\ Z_{1,2} &\rightarrow -X_\infty = (0,-1,\vec 0) 
\end{align} where $Z_i = (0,0,z_i)$, and accordingly,
\begin{align}
V_1 &= -\frac{z_1\cdot z_2}{\alpha_2} \nn
\\ V_2 &= \frac{z_1\cdot z_2}{\alpha_1} \nn
\\ V_0 &= \frac{\alpha_2 z_1\cdot z_0-\alpha_1 z_2\cdot z_0}{2 z_1\cdot z_2} \nn 
\\H_{01} &=z_0\cdot z_1 \nn
\\ H_{02} &= z_0\cdot z_2 \nn
\\ H_{12} &=0. 
\end{align} 
Since $H_{12}=0$, only structures with $n_{12}=0$ will survive. In that case, the selection rule $J=J_1+J_2-1$ implies 
\begin{align}
m_0=m_1+m_2-1.
\end{align}
 Expanding the Appell $F_2$ sum, we evaluate the integral for each term;
\begin{align}
&\int_0^\oo d\a_2\int_{-\oo}^0 d\a_1 
\frac{
		(-2 H_{01})^{J_1-m_1+k}
		(-2 H_{02})^{J_2-m_2+l}
		(2 V_{0})^{1-\bar \tau+m_0-k-l}
		(2V_{1})^{m_1-k}
		(2V_{2})^{m_2-l}
	}{
		(-X_{02})^{\frac{\bar\tau^L+\bar\tau_1-\bar\tau_2}{2}}
		X_{01}^{\frac{\bar\tau^L+\bar\tau_2-\bar\tau_1}{2}}
		X_{12}^{\frac{\bar\tau_1+\bar\tau_2-\bar\tau^L}{2}} 
	} \nn
\\ &\quad= \frac{z_{01}^{k+J_1-m_1}z_{02}^{l+J_2-m_2}}{z_{12}^{\tfrac{\bar \tau_1+\bar \tau_2-\tau}{2} }}\int_0^\oo d\a_2\int_{-\oo}^0 d\a_1 \frac{\left(\alpha _2 z_{01}- \alpha _1 z_{02}\right){}^{1-\bar{\tau }-k-l+m_0}}{\alpha _2^{\tfrac{\bar{\tau }^L-\bar{\tau }_1+\bar{\tau }_2}{2}-k+m_1} \left(-\alpha _1\right){}^{\tfrac{\bar{\tau }^L+\bar{\tau }_1-\bar{\tau }_2}{2}-l+m_2}} \nn
\\ &\quad = \frac{\Gamma(\tfrac{ \de+ \de_{12}}{2} +J_1-m_1+k)\Gamma(\tfrac{ \de+ \de_{21}}{2} +J_2-m_2+l)}{\Gamma( \de+J-m_0+k+l) } \left( \int_0^{\oo} \frac{d\a_2}{\a_2}\right)
\<\cP_{\de_1}(z_1) \cP_{\de_2}(z_2) \cP_{\de}(z_0)\>
\end{align}
Combining with the remaining factors, we have
\be
&\frac{\<0|\wL^+[\cO_2](X_\oo,Z_2)\wL[\cO](X_0,Z_0) \wL^-[\cO_1](X_\oo,Z_1)|0\>^{(a)}}{\vol\SO(1,1)} 
= q_{\de,0}^{(a)} \<\cP_{\de_1}(z_1) \cP_{\de_2}(z_2) \cP_{\de}(z_0)\>  
\ee
with 
\be
q_{\de,0}^{(a)}&= -2\pi i\, \delta_{n_{12},0}\,  \frac{(\delta+J-m_0)_{m_0}}{(\tfrac{\delta+\delta_1-\de_2}{2} +J_1 -m_1)_{m_2}(\tfrac{\delta+\delta_{2}-\de_1}{2} +J_2 -m_2)_{m_1}} \nn
\\ &\quad \times \sum_{k,l=0}^{\infty} \frac{1}{k! \, l!} \frac{(-m_1)_k (-m_2)_l  (\delta+J)_{k+l}(\tfrac{\delta+\delta_{1}-\de_2}{2}+J_1-m_1)_k(\tfrac{\delta+\delta_{2}-\de_1}{2}+J_2-m_2)_l}{(\delta+J-m_0)_{k+l}(\tfrac{\delta+\delta_{1}-\de_2}{2}+J_1-m_1+m_2)_k(\tfrac{\delta+\delta_{2}-\de_1}{2}+J_2-m_2+m_1)_l}\, .
\ee
Quite remarkably, this sum completely simplifies, yielding a pair of Kronecker delta functions. Finally, we have
\be
q_{\de,0}^{(a)} &= -2\pi i \de_{n_{12},0}\frac{(\delta+J-m_0)_{m_0}}{(\tfrac{\delta+\delta_1-\de_2}{2} +J_1 -m_1)_{m_2}(\tfrac{\delta+\delta_{2}-\de_1}{2} +J_2 -m_2)_{m_1}}  \de_{m_1,0}\, \de_{m_2,0} \nn
\\ &= -2\pi i \frac{1}{\de+J} \de_{n_{12},0}\, \de_{m_1,0}\, \de_{m_2,0}\,. 
\ee
Recalling that
\be
r_{\de,0} &= -\frac{2\pi i}{\de+J},
\ee
the OPE differential on the celestial sphere is given by
\be
\cD^{(a)}_{\de,0} (z_1,z_2,\partial_{z_2}) = \frac{q^{(a)}_{\de,0}}{r_{\de,0}} \cC_{\de,0} = \de_{n_{12},0}\, \de_{m_1,0}\, \de_{m_2,0} \, \cC_{\de,0} \, .
\ee In other words, the differential is $\cC_{\de,0}$ if $(a)$ is the structure 
\be 
(a) = \{m_0,m_1,m_2,n_{01},n_{02},n_{12} \} = \{ -1,0,0,J_1,J_2,0 \}
\ee
is proportional to
\be
V_{0}^{-1} H_{01}^{J_1} H_{02}^{J_2}\, ,
\ee and zero otherwise. This precisely agrees with the celestial map \eqref{eq:celestialmap}.

\section{Swapping the integral and $t$-channel sum in the inversion formula}
\label{app:swapping}

We would like to argue that 
\be
\label{eq:sumforc appendix}
C^\pm(\De,J) &= \sum_{\De',J'} p_{\De',J'}\cB(\De,J; \De',J')
\ee
is a convergent sum, where $\cB(\De,J;\De',J')$ is the Lorentzian inversion of a single $t$-channel block, and we have $J>J_0$ and $\De=\frac d 2 + i\nu$. We can argue for this using the Fubini-Tonelli theorem. The theorem implies that we can exchange the sum over $\De',J'$ and the integral over $z,\bar z$ in the Lorentzian inversion formula if the result after replacing each term with its absolute value is finite:
\be
&\int_0^1\int_0^1 dz d\bar z \frac{|z-\bar z|^{d-2}}{(z\bar z)^d} |G_{J+d-1,\De-d+1}(z,\bar z)|
\nn\\
&\quad\quad \x\sum_{\De',J'} \left|p_{\De',J'} \mathrm{dDisc}_t\left[\p{\frac{z\bar z}{(1-z)(1-\bar z)}}^{\De_\f} G_{\De',J'}(1-z,1-\bar z)\right]\right|
 &< \oo.
\ee
Because $p_{\De',J'}$ is positive and $\mathrm{dDisc}_t[\dots]$ is as well, we can write this condition more simply as
\be
\int_0^1\int_0^1 dz d\bar z \frac{|z-\bar z|^{d-2}}{(z\bar z)^d} |G_{J+d-1,\De-d+1}(z,\bar z)| \mathrm{dDisc}_t[g](z,\bar z) &< \oo.
\label{eq:lorentzianinversionwithabsolutevalue}
\ee

Note that the Lorentzian inversion formula converges for $J>J_0$ and $\De=\frac d 2 + i\nu$ on the principal series \cite{Caron-Huot:2017vep,Simmons-Duffin:2017nub}. Thus, it suffices to bound the integral (\ref{eq:lorentzianinversionwithabsolutevalue}) by a constant times the Lorentzian inversion formula with $\De=\frac d 2$ (which is on the principal series). Specifically, we will argue that
\be
\label{eqtheratio}
\frac{|G_{J+d-1,\De-d+1}(z,\bar z)|}{G_{J+d-1,\frac d 2-d+1}(z,\bar z)} < \mathrm{const},\quad z,\bar z \in [0,1],\ \De=\frac d 2 + i \nu,
\ee
where the constant can depend on $\De$ and $J$ but is independent of $z,\bar z$. Because the functions in the numerator and denominator of (\ref{eqtheratio}) are smooth and nonzero in the interior of the square, it suffices to argue that their ratio is bounded in a neighborhood of the boundary of the square. By symmetry, it suffices to consider $z\leq \bar z$.

When $z\ll \bar z$, the ratio takes the form
\be
\frac{|G_{J+d-1,\De-d+1}(z,\bar z)|}{G_{J+d-1,\frac d 2-d+1}(z,\bar z)} &\sim
\frac{
\left|z^{\frac{J-\De+2d-2}{2}} k_{\De+J}(\bar z)\right|
}{
z^{\frac{J-\frac d 2+2d-2}{2}} k_{\frac d 2+J}(\bar z)
} = \frac{|k_{\De+J}(\bar z)|}{k_{\frac d 2+J}(\bar z)}, \qquad z\ll\bar z
\label{eq:smallzratio}
\ee
where $k_\b(x)$ is an $\SL_2$ block. The above ratio is equal to $1$ (and hence bounded) when $\bar z =0$. Since both $\SL_2$ blocks behave like $\log (1-\bar z)$ near $\bar z = 1$, their ratio is bounded near $\bar z = 1$ as well. Because the numerator and denominator are smooth and nonzero for $0<\bar z < 1$, the ratio (\ref{eq:smallzratio}) is bounded by a $\bar z$-independent constant.

In the Regge limit $z,\bar z \ll 1$ with $z/\bar z$ fixed, (\ref{eqtheratio}) is
\be
\frac{|G_{J+d-1,\De-d+1}(z,\bar z)|}{G_{J+d-1,\frac d 2-d+1}(z,\bar z)} 
&\sim
\frac{|C_{\De-d+1}(x)|}{C_{\frac d 2 - d + 1}(x)},\qquad z,\bar z \ll 1,
\ee
where $C_J(x)$ is a Gegenbauer function and $x=\frac{z+\bar z}{2\sqrt{z\bar z}}$ ranges from $1$ to $\oo$. Again, by examining the limits $x\to 1$ and $x\to \oo$, one finds that the above ratio is bounded.

The $\bar z \to 1$ limit of a conformal block can be studied by solving the Casimir equation. Again in this case, one finds that the numerator and denominator of (\ref{eqtheratio}) both behave as the same function of $1-\bar z$, times functions of $z$ whose ratios are bounded. This completes our argument.

\section{Contact terms at $\z=1$ in $\cN=4$ SYM}
\label{app:z1contacts}

In the main text we described how one can recover the contact terms in the energy-energy correlator $\cF_\cE(\zeta)$ in $\cN=4$ SYM at $\z=0$ and $\z=1$ using Ward identities~\eqref{eq:energyWI} and~\eqref{eq:momentumWI}. We were also able to recover the $\z=0$ contact terms using the light-ray OPE formula~\eqref{eq:eecorrelatorF}. In this appendix we explain how the $\z=1$ contact terms can be obtained by another independent argument.

In the back-to-back region the energy-energy correlator in $\cN=4$ SYM takes the following form \cite{Collins:1981uk,Belitsky:2013ofa,BacktoBackGrisha}
\be\label{eq:btb}
\cF_\cE(\z)\sim_{\z\to 1} \frac{H(a)}{8y}\int_0^\oo e^{-\tfrac{1}{2}\G_\text{cusp}(a)\log^2\p{\tfrac{b^2}{y b_0^2}}-\G_\text{coll}(a)\log\p{\tfrac{b^2}{y b_0^2}}} b J_0(b) db,
\ee
where $y=1-\z$, $b_0=2e^{-\gamma_E}$, $\G_\text{cusp}(a)$ is the cusp anomalous dimension  and $\G_\text{coll}(a)$ is the so-called collinear anomalous dimension. Both $\G_\text{cusp}(a)$ and $\G_\text{coll}(a)$ are known at any coupling $a$ from integrability \cite{Freyhult:2010kc}. Note that starting from four loops there are non-planar corrections to $\G_\text{cusp}(a)$ and $\G_\text{coll}(a)$ which we do not write here \cite{Boels:2017skl, Boels:2017fng,Henn:2019rmi}. 

At weak coupling $\G_\text{cusp}(a)$ is given by the following expansion ~\cite{Beisert:2006ez}
\be
\G_\text{cusp}(a)=a-\tfrac{1}{2}\z_2 a^2+{11 \over 8} \z_4 a^3 - \left( \frac{1 }{ 8} \z_3^2 + \frac{219 }{ 64} \z_6 \right) a^4 + \cdots,
\ee
$\G_\text{coll}(a)$ is the collinear anomalous dimension given by~\cite{Freyhult:2007pz,Dixon:2017nat}
\be
\G_\text{coll}(a)=-\tfrac{3}{2}\z_3 a^2+(\tfrac{1}{2}\z_2\z_3+\tfrac{5}{2}\z_5)a^3- \left( \frac{21 }{ 16} \z_3 \z_4  +  \frac{5 }{ 8} \z_2 \z_5+\frac{175 }{ 32}  \z_7 \right) a^4 +  \cdots,
\ee
and $H(a)$ is the so-called coefficient function given by\footnote{We thank Grisha Korchemsky for sharing the coefficient of $a^3$ in $H(a)$ with us.}~\cite{Henn:2019gkr,BacktoBackGrisha}
\be
H(a)=1-\z_2a+2\z_2^2 a^2+\p{
	-\tfrac{33}{8}\z_2^3-\tfrac{1}{4}\z_4\z_2-\tfrac{17}{12}\z_3^2+\tfrac{1}{64}\z_6
}a^3+H_4 a^4+\cdots.
\ee
The coefficient $H_4$ is at present unknown.

At finite $a$~\eqref{eq:btb} is integrable near $y=0$, and so does not have any contact terms. It is possible that even at finite coupling there is an extra contact term that has to be added to~\eqref{eq:btb}. We assume that this is not the case, and that there are no contact terms at $\z=1$ at finite coupling. Under this assumption, we can therefore obtain perturbative contact terms at $\z=1$ if we carefully expand~\eqref{eq:btb} in powers of $a$. Na\"ive expansion yields terms of the form $y^{-1}\log^k y$. In our conventions for the distributional part of $\cF_\cE(\z)$ we interpret these terms as $[y^{-1}\log y]_1$, which satisfy
\be
\int_0^1 d\z \left[\frac{\log^k(1-\z)}{1-\z}\right]_1=0.
\ee
Therefore, to determine the coefficient of $\de(y)=\de(1-\z)$ in~\eqref{eq:btb}, it suffices to integrate~\eqref{eq:btb} from $0$ to $1$, and expand the result as a power series in $a$.

The $y$ integral we need to perform is
\be
\cI_a(b)&=\int_0^1 dy y^{-1+2\G_\text{cusp}(a)\log\frac{b}{b_0}+\G_\text{coll}(a)}e^{-\tfrac{1}{2}\G_\text{cusp}(a)\log^2 y}\nn\\
&=e^{\frac{(2\G_\text{cusp}(a)\log\frac{b}{b_0}+\G_\text{coll}(a))^2}{2\G_\text{cusp}(a)}}\frac{\sqrt \pi \mathrm{erfc}\p{\tfrac{2\G_\text{cusp}(a)\log\frac{b}{b_0}+\G_\text{coll}(a)}{\sqrt{2\G_\text{cusp}(a)}}}}{\sqrt{2\G_\text{cusp}(a)}}.
\ee
This can be expanded in powers of $a$, with $b$-dependence entering as powers $\log \tfrac{b}{b_0}$. Note that na\"ively this function has an expansion in powers of $\sqrt{a}$. However, all non-integer powers of $a$ will go away after performing $b$-integral.

We now want to perform the $a$-expansion of the integral
\be
\int_0^\oo \cI_a(b)e^{-2\G_\text{coll}(a)\log\frac{b}{b_0}}e^{-2\G_\text{cusp}(a)\log^2\frac{b}{b_0}}b J_0(b) db.
\ee
The product
\be
\cI_a(b)e^{-2\G_\text{coll}(a)\log\frac{b}{b_0}}
\ee
can be expanded in $a$ with coefficients polynomial in $\log\frac{b}{b_0}$. This is legal since the integral still converges after the expansion. This means that it suffices to compute the integrals
\be
\int_0^\oo \log^k\tfrac{b}{b_0}e^{-2\G_\text{cusp}\log^2\frac{b}{b_0}}b J_0(b) db,
\ee
where we treat $\G_\text{cusp}$ as arbitrary parameter. It suffices only to compute this in the case $k=0,1$ since to get higher $k$ we can simply take derivatives with respect to $\G_\text{cusp}$. Let us consider the case $k=0$; $k=1$ is completely analogous. We first integrate by parts,
\be
&\int_0^\oo e^{-2\G_\text{cusp}\log^2\frac{b}{b_0}}b J_0(b) db=\int_0^\oo e^{-2\G_\text{cusp}\log^2\frac{b}{b_0}}d(b J_1(b))\nn\\
&=4\G_\text{cusp}\int_0^\oo \log\tfrac{b}{b_0} e^{-2\G_\text{cusp}\log^2\frac{b}{b_0}}J_1(b) db.
\ee
Now the integral converges even for $\G_\text{cusp}=0$, so we can expand the exponential since $\G_\text{cusp}\in O(a)$. This way, we reduce to integrals
\be
\int_0^\oo \log^k\tfrac{b}{b_0} J_1(b)db= \p{\ptl_\nu^k\int_0^\oo \p{\tfrac{b}{b_0}}^\nu J_1(b)db}_{\nu=0}=\p{\ptl_\nu^k\p{\tfrac{2}{b_0}}^\nu\frac{\G(1+\tfrac{\nu}{2})}{\G(1-\tfrac{\nu}{2})}}_{\nu=0}.
\ee
Using this algorithm we find that the coefficient $c_1$ in front of $\de(1-\z)$ is given by
\be
c_1&=\frac{H(a)}{8}\p{
	2
	-4[\G_\text{coll}(a)\G_\text{cusp}(a)\z_3+\tfrac{5}{3}\G_\text{cusp}(a)^3\z_3^2]
	+12\z_5[\G_\text{coll}(a)\G_\text{cusp}(a)^2+\tfrac{14}{3}\G_\text{cusp}(a)^4\z_3]
	+O(a^5)
}\nn\\
&=\frac{H(a)}{8}\p{
	2
	-\tfrac{2}{3}\z_3^2 a^3
	+(28\z_3\z_5+5\z_2\z_3^2)a^4+O(a^5)
}\nn\\
&=\tfrac{1}{4}
-\tfrac{1}{4}\z_2 a
+\tfrac{1}{2}\z_2^2 a^2
-\p{\tfrac{197\pi^6}{40320}+\tfrac{7\z_3^2}{16}}a^3
+\tfrac{1}{144}\p{17\pi^2\z_3^2+504\z_3\z_5+36H_4}a^4+O(a^5).
\ee

\bibliographystyle{JHEP}
\bibliography{refs}

\end{document}

%% file: harmonic_analysis.tikzstyles
\tikzstyle{small circle}=[shape=circle, fill=white, draw=black]
\tikzstyle{medium circle}=[fill=white, draw=black, shape=circle, minimum width=1cm, minimum height=1cm]
\tikzstyle{twoPoint}=[fill=white, draw=black, shape=circle, twopt]
\tikzstyle{threePoint}=[fill=white, draw=black, shape=circle, threept]
\tikzstyle{m vert rect}=[fill=white, draw=black, shape=rectangle, minimum height=2cm, minimum width=.5cm]
\tikzstyle{s box}=[fill=white, draw=black, shape=rectangle]
\tikzstyle{celestialThreePoint}=[fill=white, draw=red, shape=circle, threept]
\tikzstyle{large circle}=[fill=white, draw=black, shape=circle, minimum height=1.5cm, minimum width=1.5cm]
\tikzstyle{green circle of inversion}=[draw={black!20!green}, shape=circle, minimum size=2cm, dashed, tikzit draw={black!20!green}]
\tikzstyle{large green circle}=[draw={black!20!green}, shape=circle, dashed, minimum size=3cm, semithick]
\tikzstyle{small triangle}=[fill=white, draw=black, regular polygon, regular polygon sides=3, inner sep=1.5pt, rotate=90]
\tikzstyle{small blue triangle}=[fill=white, draw={black!20!blue}, regular polygon, regular polygon sides=3, inner sep=1.5pt, rotate=-90, tikzit draw=blue]
\tikzstyle{small green circle}=[draw={black!20!green}, shape=circle, tikzit draw={black!20!green}, dashed, minimum size=.5cm]
\tikzstyle{celestialTwoPoint}=[fill=red, draw=red, shape=circle, inner sep=1pt, minimum size=1pt]

\tikzstyle{arrow}=[->]
\tikzstyle{op}=[-, spinning]
\tikzstyle{dashedOp}=[-, scalar]
\tikzstyle{celestial}=[-, color=red, scalar, tikzit draw=red]
\tikzstyle{minkowski}=[-, spinning, color={black!20!blue}, tikzit draw=blue]

%% file: contour_integrals.tikzstyles
\tikzstyle{pole}=[fill=black, draw=black, shape=circle, inner sep=0pt, minimum size=2pt]

\tikzstyle{Arrow}=[->, draw=blue]
\tikzstyle{anti arrow}=[<-, draw=blue]
\tikzstyle{axis}=[->]
\tikzstyle{dashed_arrow}=[->, draw=red]
\tikzstyle{dashed_only}=[-, draw=red]
\tikzstyle{dashed_anti}=[<-, draw=red]

%% file: styles.tikzstyles
\tikzstyle{point}=[fill=black, draw=black, shape=circle, inner sep=0pt, minimum size=3pt]
\tikzstyle{lr_point}=[fill=red, draw=red, shape=circle, inner sep=0pt, minimum size=2pt]
\tikzstyle{sh_point}=[fill={rgb,255: red,128; green,128; blue,128}, draw={rgb,255: red,128; green,128; blue,128}, shape=circle, inner sep=0pt, minimum size=3pt]
\tikzstyle{sh_lr_point}=[fill={red!40}, draw={red!40}, shape=circle, inner sep=0pt, minimum size=2pt, tikzit fill={rgb,255: red,255; green,128; blue,0}, tikzit draw={rgb,255: red,255; green,128; blue,0}]

\tikzstyle{dashed_st}=[-, dashed]
\tikzstyle{arrow}=[->]
\tikzstyle{dashed_grey}=[-, draw=black, dotted]
\tikzstyle{blue_line}=[draw=blue, ->]
\tikzstyle{energy}=[draw={rgb,255: red,255; green,128; blue,0}, decoration={{snake,amplitude=1pt,segment length=6pt,post length=1pt}}, decorate, ->]
\tikzstyle{blue_line_0}=[-, draw=blue]
\tikzstyle{redline}=[-, draw=red]

%% file: figures/C_F_plot_3_anomalous.tikz
\begin{tikzpicture}
	\begin{pgfonlayer}{nodelayer}
		\node [style=none] (0) at (0, 5) {};
		\node [style=none] (1) at (0, -2) {};
		\node [style=none] (2) at (5, 0) {};
		\node [style=none] (3) at (-5, 0) {};
		\node [style=none] (4) at (0.5, 5) {$J$};
		\node [style=none] (5) at (5, -0.5) {$\Delta-\frac{d}{2}$};
		\node [style=point] (6) at (1.5, 2) {};
		\node [style=point] (7) at (3.775, 4) {};
		\node [style=none] (10) at (0, 1.275) {};
		\node [style=none] (11) at (4.8, 5) {};
		\node [style=point] (13) at (3.575, 2) {};
		\node [style=point] (14) at (5.85, 4) {};
		\node [style=none] (17) at (6.875, 5) {};
		\node [style=none] (20) at (2.3, 1) {};
		\node [style=none] (22) at (1.2, 0.275) {};
		\node [style=none] (23) at (-7, 2) {};
		\node [style=none] (24) at (-7, 4) {};
		\node [style=none] (25) at (7, 4) {};
		\node [style=none] (26) at (7, 2) {};
		\node [style=none] (27) at (0.5, 4) {4};
		\node [style=none] (28) at (0.5, 2) {2};
		\node [style=none] (30) at (1.5, 2.5) {$T$};
		\node [style=none] (31) at (-7, 3) {};
		\node [style=none] (32) at (7, 3) {};
		\node [style=none] (38) at (0.5, 3) {${\color{red}3}$};
		\node [style=none] (39) at (0, 4) {};
		\node [style=none] (40) at (0.175, 4) {};
		\node [style=none] (41) at (0, 3) {};
		\node [style=none] (42) at (0.175, 3) {};
		\node [style=none] (43) at (0, 2) {};
		\node [style=none] (44) at (0.175, 2) {};
		\node [style=none] (45) at (2.7, 3) {};
		\node [style=none] (46) at (3.25, 3.5) {};
		\node [style=none] (47) at (2.125, 2.5) {};
		\node [style=none] (48) at (0.75, 1.5) {};
		\node [style=none] (51) at (4.2, 2.5) {};
		\node [style=none] (52) at (4.775, 3) {};
		\node [style=none] (53) at (5.325, 3.5) {};
		\node [style=none] (54) at (2.95, 1.5) {};
		\node [style={sh_point}] (55) at (-1.5, 2) {};
		\node [style={sh_point}] (56) at (-3.775, 4) {};
		\node [style=none] (57) at (0, 1.275) {};
		\node [style=none] (58) at (-4.8, 5) {};
		\node [style=none] (59) at (-2.7, 3) {};
		\node [style=none] (60) at (-3.25, 3.5) {};
		\node [style=none] (61) at (-2.125, 2.5) {};
		\node [style=none] (62) at (-0.75, 1.5) {};
		\node [style={sh_point}] (63) at (-3.575, 2) {};
		\node [style={sh_point}] (64) at (-5.85, 4) {};
		\node [style=none] (65) at (-6.875, 5) {};
		\node [style=none] (66) at (-2.3, 1) {};
		\node [style=none] (67) at (-1.2, 0.275) {};
		\node [style=none] (68) at (-4.2, 2.5) {};
		\node [style=none] (69) at (-4.775, 3) {};
		\node [style=none] (70) at (-5.325, 3.5) {};
		\node [style=none] (71) at (-2.95, 1.5) {};
		\node [style=none] (72) at (2.7, 3.175) {};
		\node [style=none] (73) at (2.7, 2.825) {};
		\node [style=none] (74) at (2.875, 3) {};
		\node [style=none] (75) at (2.525, 3) {};
		\node [style=none] (77) at (4.775, 3.175) {};
		\node [style=none] (78) at (4.775, 2.825) {};
		\node [style=none] (79) at (4.95, 3) {};
		\node [style=none] (80) at (4.6, 3) {};
	\end{pgfonlayer}
	\begin{pgfonlayer}{edgelayer}
		\draw [style=arrow] (3.center) to (2.center);
		\draw [style=arrow] (1.center) to (0.center);
		\draw (17.center) to (14);
		\draw [style={dashed_st}] (20.center) to (22.center);
		\draw [style={dashed_grey}] (24.center) to (25.center);
		\draw [style={dashed_grey}] (26.center) to (23.center);
		\draw [style={dashed_grey}] (31.center) to (32.center);
		\draw (39.center) to (40.center);
		\draw (41.center) to (42.center);
		\draw (43.center) to (44.center);
		\draw [in=-150, out=0, looseness=0.75] (10.center) to (48.center);
		\draw [in=-145, out=30, looseness=0.75] (48.center) to (6);
		\draw (6) to (47.center);
		\draw (47.center) to (45.center);
		\draw (45.center) to (46.center);
		\draw (46.center) to (7);
		\draw (7) to (11.center);
		\draw (13) to (51.center);
		\draw (51.center) to (52.center);
		\draw (52.center) to (53.center);
		\draw (53.center) to (14);
		\draw (13) to (54.center);
		\draw (54.center) to (20.center);
		\draw [in=-30, out=180, looseness=0.75] (57.center) to (62.center);
		\draw [in=-35, out=150, looseness=0.75] (62.center) to (55);
		\draw (55) to (61.center);
		\draw (61.center) to (59.center);
		\draw (59.center) to (60.center);
		\draw (60.center) to (56);
		\draw (56) to (58.center);
		\draw (65.center) to (64);
		\draw [style={dashed_st}] (66.center) to (67.center);
		\draw (63) to (68.center);
		\draw (68.center) to (69.center);
		\draw (69.center) to (70.center);
		\draw (70.center) to (64);
		\draw (63) to (71.center);
		\draw (71.center) to (66.center);
		\draw [style=redline] (72.center) to (73.center);
		\draw [style=redline] (74.center) to (75.center);
		\draw [style=redline] (77.center) to (78.center);
		\draw [style=redline] (79.center) to (80.center);
	\end{pgfonlayer}
\end{tikzpicture}

%% file: figures/LKernelDef.tikz
\begin{tikzpicture}
	\begin{pgfonlayer}{nodelayer}
		\node [style=none] (0) at (-3, 0) {};
		\node [style=none] (1) at (-10, 2) {};
		\node [style=none] (2) at (-10, -2) {};
		\node [style=threePoint] (3) at (-7, 0) {$L_{\delta}$};
		\node [style=none] (4) at (-10.5, 2.225) {$1$};
		\node [style=none] (5) at (-10.5, -2.275) {$2$};
		\node [style=none] (6) at (-2, 0.125) {$\mathcal{O}^L$};
		\node [style=none] (13) at (0, 0) {$=$};
		\node [style=small blue triangle] (17) at (7.25, -0.5) {};
		\node [style=none] (18) at (1.5, 1.975) {$1$};
		\node [style=none] (19) at (1.5, -2.025) {$2$};
		\node [style=none] (20) at (13.5, 0.05) {$\mathcal{O}^L$};
		\node [style=none] (21) at (4.75, 2) {};
		\node [style=none] (22) at (4.75, -2) {};
		\node [style=none] (24) at (11, 0) {};
		\node [style=s box] (25) at (3.5, 1.975) {$\mathbf{L}$};
		\node [style=s box] (26) at (3.5, -2) {$\mathbf{L}$};
		\node [style=none] (29) at (2, 2) {};
		\node [style=none] (30) at (3, 2) {};
		\node [style=none] (33) at (2, -2) {};
		\node [style=none] (34) at (3, -2) {};
		\node [style=celestialThreePoint] (35) at (8.25, 0.675) {$t_\delta$};
		\node [style=none] (37) at (8, 0.75) {};
		\node [style=none] (38) at (8, 0.525) {};
		\node [style=none] (40) at (12.5, 0) {};
		\node [style=none] (41) at (4.75, 2) {};
		\node [style=none] (42) at (7.15, -0.35) {};
		\node [style=none] (43) at (4.75, -2) {};
		\node [style=none] (44) at (7.125, -0.725) {};
		\node [style=none] (45) at (7.5, -0.5) {};
		\node [style=none] (46) at (11, 0) {};
	\end{pgfonlayer}
	\begin{pgfonlayer}{edgelayer}
		\draw [style=op] (3) to (0.center);
		\draw [style=op] (1.center) to (3);
		\draw [style=op] (2.center) to (3);
		\draw [style=op] (29.center) to (30.center);
		\draw [style=op] (33.center) to (34.center);
		\draw [style=celestial, in=180, out=0] (35) to (24.center);
		\draw [style=celestial, in=165, out=0, looseness=1.50] (21.center) to (37.center);
		\draw [style=celestial, in=180, out=0] (22.center) to (38.center);
		\draw [style=op] (24.center) to (40.center);
		\draw [style=op] (25) to (21.center);
		\draw [style=op] (26) to (22.center);
		\draw [style=minkowski, in=120, out=0] (41.center) to (42.center);
		\draw [style=minkowski, in=-135, out=0] (43.center) to (44.center);
		\draw [style=minkowski, in=-180, out=0] (45.center) to (46.center);
	\end{pgfonlayer}
\end{tikzpicture}

%% file: figures/LcontracteddDiscCartoon.tikz
\begin{tikzpicture}
	\begin{pgfonlayer}{nodelayer}
		\node [style=none] (0) at (7.5, 0) {};
		\node [style=none] (4) at (7, 0.75) {$\mathcal{O}^L$};
		\node [style=none] (5) at (0.5, 2.25) {$1$};
		\node [style=none] (6) at (0.5, -2.25) {$2$};
		\node [style=none] (7) at (-5.5, 2) {};
		\node [style=none] (8) at (-5.5, -2) {};
		\node [style=none] (9) at (-6, 2) {$4$};
		\node [style=none] (10) at (-6, -2) {$3$};
		\node [style=threePoint] (14) at (3.5, 0) {$L_{\delta}$};
		\node [style=medium circle] (15) at (-2.5, 0) {$\mathrm{dDisc}[g]$};
	\end{pgfonlayer}
	\begin{pgfonlayer}{edgelayer}
		\draw [style=op, bend right, looseness=1.50] (15) to (14);
		\draw [style=op] (14) to (0.center);
		\draw [style=op] (15) to (7.center);
		\draw [style=op] (15) to (8.center);
		\draw [style=op, bend right=330, looseness=1.50] (15) to (14);
	\end{pgfonlayer}
\end{tikzpicture}

%% file: figures/L034str.tikz
\begin{tikzpicture}
	\begin{pgfonlayer}{nodelayer}
		\node [style=none] (0) at (0.75, 0) {};
		\node [style=none] (1) at (3, 0.75) {$\mathcal{O}^L$};
		\node [style=none] (4) at (-4.5, 2) {};
		\node [style=none] (5) at (-4.5, -2) {};
		\node [style=none] (6) at (-5, 2) {$4$};
		\node [style=none] (7) at (-5, -2) {$3$};
		\node [style=threePoint] (9) at (-1.5, 0) {$b$};
		\node [style=s box] (10) at (1, 0) {$\mathbf{L}$};
		\node [style=none] (11) at (1.25, 0) {};
		\node [style=none] (12) at (3.5, 0) {};
		\node [style=none] (13) at (-0.25, 0.725) {$\mathcal{O}$};
	\end{pgfonlayer}
	\begin{pgfonlayer}{edgelayer}
		\draw [style=op] (9) to (4.center);
		\draw [style=op] (9) to (5.center);
		\draw [style=op] (9) to (0.center);
		\draw [style=op] (11.center) to (12.center);
	\end{pgfonlayer}
\end{tikzpicture}

%% file: figures/L034strInv.tikz
\begin{tikzpicture}
	\begin{pgfonlayer}{nodelayer}
		\node [style=none] (0) at (0, 0) {};
		\node [style=none] (4) at (-3.725, 1.325) {};
		\node [style=none] (6) at (-5.25, 2) {$4$};
		\node [style=none] (7) at (-5.25, -2) {$3$};
		\node [style=s box] (10) at (0.25, 0) {$\mathbf{L}$};
		\node [style=none] (11) at (0.75, 0) {};
		\node [style=none] (12) at (2, 0) {};
		\node [style=none] (13) at (-0.9, 0.725) {$\mathcal{O}$};
		\node [style=large green circle] (14) at (-1, 0) {};
		\node [style=none] (15) at (3.5, 0) {};
		\node [style=none] (16) at (-4.75, 2) {};
		\node [style=none] (17) at (-4.75, -2) {};
		\node [style=none] (18) at (-3.725, -1.325) {};
		\node [style=threePoint] (19) at (-1.75, 0) {$a$};
		\node [style=none] (20) at (-4.75, -2) {};
		\node [style=none] (21) at (3, 0.75) {$\mathcal{O}^L$};
		\node [style=none] (22) at (1.75, 2) {\color{black!20!green}{$^{-1}$}};
	\end{pgfonlayer}
	\begin{pgfonlayer}{edgelayer}
		\draw [style=op] (11.center) to (12.center);
		\draw [style=op] (15.center) to (12.center);
		\draw [style=op] (16.center) to (4.center);
		\draw [style=op] (19) to (18.center);
		\draw [style=op] (20.center) to (18.center);
		\draw [style=op] (19) to (4.center);
		\draw [style=op] (19) to (0.center);
	\end{pgfonlayer}
\end{tikzpicture}

%% file: figures/greenCircleOfInversion.tikz
\begin{tikzpicture}
	\begin{pgfonlayer}{nodelayer}
		\node [style=small green circle] (0) at (0, 0) {};
		\node [style=none] (1) at (0.75, 0.5) {\color{black!20!green}{$^{-1}$}};
	\end{pgfonlayer}
\end{tikzpicture}

%% file: figures/AdeltaFromPairingCartoon.tikz
\begin{tikzpicture}
	\begin{pgfonlayer}{nodelayer}
		\node [style=none] (4) at (8.075, 0.75) {$\mathcal{O}^L$};
		\node [style=none] (5) at (3.575, 2.25) {$1$};
		\node [style=none] (6) at (3.575, -2.25) {$2$};
		\node [style=none] (9) at (-1.925, 3) {$4$};
		\node [style=none] (10) at (-1.925, -3) {$3$};
		\node [style=threePoint] (14) at (6.575, 0) {$L_{\delta}$};
		\node [style=none] (21) at (8.325, 0) {};
		\node [style=none] (22) at (13.25, -1.675) {};
		\node [style=medium circle] (28) at (0.575, 0) {$\mathrm{dDisc}[g]$};
		\node [style=none] (29) at (13.25, 1.675) {};
		\node [style=threePoint] (30) at (11.575, 0) {$b$};
		\node [style=large green circle] (31) at (11.575, 0) {};
		\node [style=none] (32) at (14.575, 2) {\color{black!20!green}{$^{-1}$}};
		\node [style=s box] (34) at (9.775, 0) {$\mathbf{L}$};
	\end{pgfonlayer}
	\begin{pgfonlayer}{edgelayer}
		\draw [style=op] (14) to (21.center);
		\draw [style=op, in=45, out=150, looseness=1.25] (28) to (29.center);
		\draw [style=op] (30) to (29.center);
		\draw [style=op, bend right, looseness=1.50] (28) to (14);
		\draw [style=op, bend right=330, looseness=1.50] (28) to (14);
		\draw [style=op, in=-45, out=-150, looseness=1.25] (28) to (22.center);
		\draw [style=op] (30) to (22.center);
		\draw [style=op] (34) to (21.center);
		\draw [style=op] (30) to (34);
	\end{pgfonlayer}
\end{tikzpicture}

%% file: figures/dDisc.tikz
\begin{tikzpicture}
	\begin{pgfonlayer}{nodelayer}
		\node [style=none] (1) at (3, 3) {};
		\node [style=none] (2) at (3, -3) {};
		\node [style=none] (5) at (3.5, 3) {$4$};
		\node [style=none] (6) at (3.5, -3) {$3$};
		\node [style=none] (7) at (-3, 3) {};
		\node [style=none] (8) at (-3, -3) {};
		\node [style=medium circle] (9) at (0, 0) {$\mathrm{dDisc}[g]$};
		\node [style=none] (10) at (-3.5, 3) {$1$};
		\node [style=none] (11) at (-3.5, -3) {$2$};
	\end{pgfonlayer}
	\begin{pgfonlayer}{edgelayer}
		\draw [style=op] (9) to (7.center);
		\draw [style=op] (9) to (8.center);
		\draw [style=op] (9) to (1.center);
		\draw [style=op] (9) to (2.center);
	\end{pgfonlayer}
\end{tikzpicture}

%% file: figures/BlockWithLCartoon.tikz
\begin{tikzpicture}
	\begin{pgfonlayer}{nodelayer}
		\node [style=none] (4) at (3.825, 0.75) {$\mathcal{O}^L$};
		\node [style=none] (5) at (-1.425, 3) {$1$};
		\node [style=none] (6) at (-1.425, -3) {$2$};
		\node [style=none] (9) at (11.5, 3) {$4$};
		\node [style=none] (10) at (11.5, -3) {$3$};
		\node [style=none] (21) at (4.75, 0) {};
		\node [style=none] (22) at (9.925, -1.925) {};
		\node [style=none] (29) at (9.925, 1.925) {};
		\node [style=threePoint] (30) at (8, 0) {$b$};
		\node [style=large green circle] (31) at (7.75, 0) {};
		\node [style=none] (32) at (10.75, 1.75) {\color{black!20!green}{$^{-1}$}};
		\node [style=s box] (34) at (6.2, 0) {$\mathbf{L}$};
		\node [style=none] (35) at (-1, 3) {};
		\node [style=threePoint] (36) at (2.075, 0) {$L_{\delta}$};
		\node [style=none] (37) at (-1, -3) {};
		\node [style=none] (38) at (11, 3) {};
		\node [style=none] (39) at (11, -3) {};
	\end{pgfonlayer}
	\begin{pgfonlayer}{edgelayer}
		\draw [style=op] (30) to (29.center);
		\draw [style=op] (30) to (22.center);
		\draw [style=op] (34) to (21.center);
		\draw [style=op] (30) to (34);
		\draw [style=op] (37.center) to (36);
		\draw [style=op] (36) to (21.center);
		\draw [style=op] (35.center) to (36);
		\draw [style=op] (38.center) to (29.center);
		\draw [style=op] (39.center) to (22.center);
	\end{pgfonlayer}
\end{tikzpicture}

%% file: figures/BlockInLorentzianInvCartoon.tikz
\begin{tikzpicture}
	\begin{pgfonlayer}{nodelayer}
		\node [style=none] (5) at (-1.675, 3) {$1$};
		\node [style=none] (6) at (-1.675, -3) {$2$};
		\node [style=none] (9) at (14.5, 3) {$4$};
		\node [style=none] (10) at (14.5, -3) {$3$};
		\node [style=none] (21) at (7.75, 0) {};
		\node [style=none] (22) at (12.925, -1.925) {};
		\node [style=none] (29) at (12.925, 1.925) {};
		\node [style=threePoint] (30) at (11, 0) {$b$};
		\node [style=large green circle] (31) at (10.75, 0) {};
		\node [style=none] (32) at (13.75, 1.75) {\color{black!20!green}{$^{-1}$}};
		\node [style=s box] (34) at (9.2, 0) {$\mathbf{L}$};
		\node [style=none] (38) at (14, 3) {};
		\node [style=none] (39) at (14, -3) {};
		\node [style=none] (44) at (5.25, 0) {};
		\node [style=none] (45) at (0.0750002, -1.925) {};
		\node [style=none] (46) at (0.0750002, 1.925) {};
		\node [style=threePoint] (47) at (2, 0) {$a$};
		\node [style=large green circle] (48) at (2.25, 0) {};
		\node [style=none] (49) at (5.25, 1.75) {\color{black!20!green}{$^{-1}$}};
		\node [style=s box] (50) at (3.8, 0) {$\mathbf{L}$};
		\node [style=none] (51) at (-1, 3) {};
		\node [style=none] (52) at (-1, -3) {};
		\node [style=none] (53) at (6.5, 0) {$\times$};
		\node [style=none] (54) at (7.25, 0.675) {$\mathcal{O}^L$};
	\end{pgfonlayer}
	\begin{pgfonlayer}{edgelayer}
		\draw [style=op] (30) to (29.center);
		\draw [style=op] (30) to (22.center);
		\draw [style=op] (34) to (21.center);
		\draw [style=op] (30) to (34);
		\draw [style=op] (38.center) to (29.center);
		\draw [style=op] (39.center) to (22.center);
		\draw [style=op] (47) to (46.center);
		\draw [style=op] (47) to (45.center);
		\draw [style=op] (50) to (44.center);
		\draw [style=op] (47) to (50);
		\draw [style=op] (51.center) to (46.center);
		\draw [style=op] (52.center) to (45.center);
		\draw [style=op] (53.center) to (44.center);
		\draw [style=op] (53.center) to (21.center);
	\end{pgfonlayer}
\end{tikzpicture}

%% file: figures/SofLKernel.tikz
\begin{tikzpicture}
	\begin{pgfonlayer}{nodelayer}
		\node [style=none] (5) at (-1.675, 3) {$1$};
		\node [style=none] (6) at (-1.675, -3) {$2$};
		\node [style=twoPoint] (44) at (5, 0) {};
		\node [style=threePoint] (47) at (2, 0) {$L_\delta$};
		\node [style=none] (51) at (-1, 3) {};
		\node [style=none] (52) at (-1, -3) {};
		\node [style=none] (53) at (6.5, 0) {};
		\node [style=none] (55) at (6.5, 0.75) {$\mathcal{O}^L$};
	\end{pgfonlayer}
	\begin{pgfonlayer}{edgelayer}
		\draw [style=op] (53.center) to (44);
		\draw [style=op] (51.center) to (47);
		\draw [style=op] (52.center) to (47);
		\draw [style=op] (47) to (44);
	\end{pgfonlayer}
\end{tikzpicture}

%% file: figures/aLofO12strInv.tikz
\begin{tikzpicture}
	\begin{pgfonlayer}{nodelayer}
		\node [style=none] (5) at (-1.675, 3) {$1$};
		\node [style=none] (6) at (-1.675, -3) {$2$};
		\node [style=none] (44) at (5.25, 0) {};
		\node [style=none] (45) at (0.0750002, -1.925) {};
		\node [style=none] (46) at (0.0750002, 1.925) {};
		\node [style=threePoint] (47) at (2, 0) {$a$};
		\node [style=large green circle] (48) at (2.25, 0) {};
		\node [style=none] (49) at (5.25, 1.75) {\color{black!20!green}{$^{-1}$}};
		\node [style=s box] (50) at (3.8, 0) {$\mathbf{L}$};
		\node [style=none] (51) at (-1, 3) {};
		\node [style=none] (52) at (-1, -3) {};
		\node [style=none] (53) at (6.5, 0) {};
		\node [style=none] (54) at (6.5, 0.75) {$\mathcal{O}^L$};
	\end{pgfonlayer}
	\begin{pgfonlayer}{edgelayer}
		\draw [style=op] (47) to (46.center);
		\draw [style=op] (47) to (45.center);
		\draw [style=op] (50) to (44.center);
		\draw [style=op] (47) to (50);
		\draw [style=op] (51.center) to (46.center);
		\draw [style=op] (52.center) to (45.center);
		\draw [style=op] (53.center) to (44.center);
	\end{pgfonlayer}
\end{tikzpicture}

%% file: figures/SofLKernelPairedWithStrCartoon.tikz
\begin{tikzpicture}
	\begin{pgfonlayer}{nodelayer}
		\node [style=none] (5) at (0, 3.975) {$1$};
		\node [style=none] (6) at (0, -0.025) {$2$};
		\node [style=twoPoint] (44) at (0, -2.525) {};
		\node [style=threePoint] (47) at (2.25, 1.975) {$L_\delta$};
		\node [style=threePoint] (53) at (-2.25, 1.975) {$a$};
		\node [style=none] (54) at (3.25, -0.525) {$\mathcal{O}^L$};
		\node [style=s box] (55) at (-3, -0.025) {$\mathbf{L}$};
	\end{pgfonlayer}
	\begin{pgfonlayer}{edgelayer}
		\draw [style=op, bend left=45, looseness=1.25] (47) to (44);
		\draw [style=op, bend right=45, looseness=1.25] (53) to (47);
		\draw [style=op, bend left=45, looseness=1.25] (53) to (47);
		\draw [style=op, bend right=15] (53) to (55);
		\draw [style=op, bend right] (55) to (44);
	\end{pgfonlayer}
\end{tikzpicture}

%% file: figures/LKernelPairedWithStrTwoPtCartoon.tikz
\begin{tikzpicture}
	\begin{pgfonlayer}{nodelayer}
		\node [style=none] (5) at (0, 1.975) {$1$};
		\node [style=none] (6) at (0, -2.025) {$2$};
		\node [style=threePoint] (47) at (2.25, -0.025) {$L_\delta$};
		\node [style=threePoint] (53) at (-2.25, -0.025) {$a$};
		\node [style=none] (54) at (4.25, 0.725) {$\mathcal{O}^L$};
		\node [style=s box] (55) at (-4, -0.025) {$\mathbf{L}$};
		\node [style=none] (56) at (-6, 0) {};
		\node [style=none] (57) at (5.5, 0) {};
		\node [style=none] (58) at (-5, 0.725) {$\mathcal{O}^L$};
	\end{pgfonlayer}
	\begin{pgfonlayer}{edgelayer}
		\draw [style=op, bend right=45, looseness=1.25] (53) to (47);
		\draw [style=op, bend left=45, looseness=1.25] (53) to (47);
		\draw [style=op] (47) to (57.center);
		\draw [style=op] (55) to (56.center);
		\draw [style=op] (53) to (55);
	\end{pgfonlayer}
\end{tikzpicture}

%% file: figures/twoPtFn.tikz
\begin{tikzpicture}
	\begin{pgfonlayer}{nodelayer}
		\node [style=twoPoint] (0) at (0, 0) {};
		\node [style=none] (1) at (2, 0) {};
		\node [style=none] (2) at (-2, 0) {};
		\node [style=none] (3) at (1.5, 0.75) {$\mathcal{O}^{L}$};
		\node [style=none] (4) at (-1, 0.75) {$\mathcal{O}^{L}$};
	\end{pgfonlayer}
	\begin{pgfonlayer}{edgelayer}
		\draw [style=op] (1.center) to (0);
		\draw [style=op] (2.center) to (0);
	\end{pgfonlayer}
\end{tikzpicture}

%% file: figures/tripleLofStr.tikz
\begin{tikzpicture}
	\begin{pgfonlayer}{nodelayer}
		\node [style=none] (5) at (0.25, 3.225) {$1$};
		\node [style=none] (6) at (0.25, -3.275) {$2$};
		\node [style=none] (47) at (7, 0) {};
		\node [style=threePoint] (53) at (-2.25, 0) {$a$};
		\node [style=none] (54) at (8, 0.725) {$\mathcal{O}^L$};
		\node [style=s box] (55) at (-4, -0.025) {$\mathbf{L}$};
		\node [style=none] (56) at (-6.5, 0) {};
		\node [style=none] (57) at (9.25, 0) {};
		\node [style=none] (58) at (-5.5, 0.725) {$\mathcal{O}^L$};
		\node [style=none, rotate=45] (59) at (0.175, 2.425) {};
		\node [style=none, rotate=-45] (61) at (0.175, -2.425) {};
		\node [style=celestialThreePoint] (62) at (4.5, 0.5) {$t_\delta$};
		\node [style=none, rotate=180] (63) at (3.275, -0.725) {};
		\node [style=small blue triangle] (64) at (3.425, -0.5) {};
		\node [style=none] (65) at (3.275, -0.275) {};
		\node [style=none] (66) at (3.7, -0.5) {};
		\node [style=s box, rotate=45] (67) at (-1, 1.25) {$\!\mathbf{L}^+\!$};
		\node [style=s box, rotate=-45] (68) at (-1, -1.25) {$\!\mathbf{L}^-\!$};
		\node [style=none] (69) at (-0.725, 1.525) {};
		\node [style=none] (70) at (-0.7, -1.55) {};
	\end{pgfonlayer}
	\begin{pgfonlayer}{edgelayer}
		\draw [style=op] (47.center) to (57.center);
		\draw [style=op] (55) to (56.center);
		\draw [style=op] (53) to (55);
		\draw [style=minkowski, in=225, out=-45] (61.center) to (63.center);
		\draw [style=celestial, in=165, out=45] (59.center) to (62);
		\draw [style=celestial, in=240, out=-45] (61.center) to (62);
		\draw [style=minkowski, in=150, out=30] (59.center) to (65.center);
		\draw [style=celestial, in=180, out=-15] (62) to (47.center);
		\draw [style=minkowski, in=180, out=0] (66.center) to (47.center);
		\draw [style=op] (53) to (67);
		\draw [style=op] (53) to (68);
		\draw [style=op] (69.center) to (59.center);
		\draw [style=op] (70.center) to (61.center);
	\end{pgfonlayer}
\end{tikzpicture}

%% file: figures/BlockInLorentzianInv.tikz
\begin{tikzpicture}
	\begin{pgfonlayer}{nodelayer}
		\node [style=none] (5) at (-1.925, 3) {$1$};
		\node [style=none] (6) at (-1.675, -3) {$2$};
		\node [style=none] (9) at (14.75, 3) {$4$};
		\node [style=none] (10) at (14.5, -3) {$3$};
		\node [style=none] (21) at (7.75, 0) {};
		\node [style=none] (22) at (12.925, -1.925) {};
		\node [style=none] (29) at (12.925, 1.925) {};
		\node [style=threePoint] (30) at (11, 0) {$b$};
		\node [style=large green circle] (31) at (10.75, 0) {};
		\node [style=none] (32) at (13.75, 1.75) {\color{black!20!green}{$^{-1}$}};
		\node [style=s box] (34) at (9.2, 0) {$\mathbf{L}$};
		\node [style=none] (38) at (14, 3) {};
		\node [style=none] (39) at (14, -3) {};
		\node [style=none] (44) at (5.25, 0) {};
		\node [style=none] (45) at (0.0750002, -1.925) {};
		\node [style=none] (46) at (0.0750002, 1.925) {};
		\node [style=threePoint] (47) at (2, 0) {$a$};
		\node [style=large green circle] (48) at (2.25, 0) {};
		\node [style=none] (49) at (5.25, 1.75) {\color{black!20!green}{$^{-1}$}};
		\node [style=s box] (50) at (3.8, 0) {$\mathbf{L}$};
		\node [style=none] (51) at (-1, 3) {};
		\node [style=none] (52) at (-1, -3) {};
		\node [style=none] (53) at (6.5, 0) {$\times$};
		\node [style=none] (54) at (7.25, 0.675) {$\mathcal{O}^L$};
	\end{pgfonlayer}
	\begin{pgfonlayer}{edgelayer}
		\draw [style=op] (30) to (29.center);
		\draw [style=op] (30) to (22.center);
		\draw [style=op] (34) to (21.center);
		\draw [style=op] (30) to (34);
		\draw [style=op] (38.center) to (29.center);
		\draw [style=op] (39.center) to (22.center);
		\draw [style=op] (47) to (46.center);
		\draw [style=op] (47) to (45.center);
		\draw [style=op] (50) to (44.center);
		\draw [style=op] (47) to (50);
		\draw [style=op] (51.center) to (46.center);
		\draw [style=op] (52.center) to (45.center);
		\draw [style=op] (53.center) to (44.center);
		\draw [style=op] (53.center) to (21.center);
	\end{pgfonlayer}
\end{tikzpicture}

%% file: figures/celestialPartialWave.tikz
\begin{tikzpicture}
	\begin{pgfonlayer}{nodelayer}
		\node [style=celestialThreePoint] (0) at (0, 0) {$t_{\delta,j}$};
		\node [style=celestialThreePoint] (1) at (4, 0) {$f$};
		\node [style=none] (2) at (-3, 0) {};
		\node [style=none] (3) at (6, 1.25) {};
		\node [style=none] (5) at (6, -1.25) {};
		\node [style=none] (6) at (5.75, 0.5) {\color{red}{.}};
		\node [style=none] (7) at (5.825, 0) {\color{red}{.}};
		\node [style=none] (8) at (5.75, -0.5) {\color{red}{.}};
		\node [style=none] (9) at (2, 1.75) {{$1$}};
		\node [style=none] (10) at (2, -1.75) {{$2$}};
		\node [style=none] (11) at (-2.5, 0.75) {$\mathcal{P}_{\delta,j}$};
	\end{pgfonlayer}
	\begin{pgfonlayer}{edgelayer}
		\draw [style=celestial] (0) to (2.center);
		\draw [style=celestial, in=405, out=135] (1) to (0);
		\draw [style=celestial, in=315, out=225] (1) to (0);
		\draw [style=celestial] (1) to (5.center);
		\draw [style=celestial] (1) to (3.center);
	\end{pgfonlayer}
\end{tikzpicture}

%% file: figures/celestialBubble.tikz
\begin{tikzpicture}
	\begin{pgfonlayer}{nodelayer}
		\node [style=celestialThreePoint] (0) at (-7, 0) {$t_{\delta,j}$};
		\node [style=celestialThreePoint] (1) at (-3, 0) {};
		\node [style=none] (2) at (-10, 0) {};
		\node [style=none] (9) at (-5, 2) {{$1$}};
		\node [style=none] (10) at (-5, -2) {{$2$}};
		\node [style=none] (11) at (-9.5, 0.75) {$\mathcal{P}_{\delta,j}$};
		\node [style=none] (12) at (0, 0) {};
		\node [style=none] (13) at (-0.5, 0.75) {$\mathcal{P}_{\delta,j}^\dagger$};
		\node [style=none] (14) at (2, 0) {$=$};
		\node [style=celestialTwoPoint] (15) at (6, 0) {};
		\node [style=none] (16) at (4, 0) {};
		\node [style=none] (17) at (8, 0) {};
		\node [style=none] (18) at (4.5, 0.75) {$\mathcal{P}_{\delta,j}$};
		\node [style=none] (19) at (7.5, 0.75) {$\mathcal{P}_{\delta,j}^\dagger$};
	\end{pgfonlayer}
	\begin{pgfonlayer}{edgelayer}
		\draw [style=celestial] (0) to (2.center);
		\draw [style=celestial, in=420, out=120] (1) to (0);
		\draw [style=celestial, in=300, out=240] (1) to (0);
		\draw [style=celestial] (1) to (12.center);
		\draw [style=celestial] (15) to (16.center);
		\draw [style=celestial] (15) to (17.center);
	\end{pgfonlayer}
\end{tikzpicture}

%% file: figures/LKernelDefWithj.tikz
\begin{tikzpicture}
	\begin{pgfonlayer}{nodelayer}
		\node [style=none] (0) at (-2.5, 0) {};
		\node [style=none] (1) at (-9.5, 2) {};
		\node [style=none] (2) at (-9.5, -2) {};
		\node [style=threePoint] (3) at (-6.5, 0) {$L_{\delta,j}$};
		\node [style=none] (4) at (-10, 2.225) {$1$};
		\node [style=none] (5) at (-10, -1.775) {$2$};
		\node [style=none] (6) at (-1.5, 0.125) {$\mathcal{O}^L$};
		\node [style=none] (13) at (0, 0) {$=$};
		\node [style=small blue triangle] (17) at (7.5, -0.5) {};
		\node [style=none] (18) at (1.5, 1.925) {$1$};
		\node [style=none] (19) at (1.5, -2.075) {$2$};
		\node [style=none] (20) at (13.5, 0.05) {$\mathcal{O}^L$};
		\node [style=none] (21) at (5, 2) {};
		\node [style=none] (22) at (5, -2) {};
		\node [style=none] (24) at (11, 0) {};
		\node [style=s box] (25) at (3.5, 1.975) {$\mathbf{L}$};
		\node [style=s box] (26) at (3.5, -2) {$\mathbf{L}$};
		\node [style=none] (29) at (2, 2) {};
		\node [style=none] (30) at (3, 2) {};
		\node [style=none] (33) at (2, -2) {};
		\node [style=none] (34) at (3, -2) {};
		\node [style=celestialThreePoint] (35) at (8.7, 0.75) {$t_{\delta,j}$};
		\node [style=none] (37) at (8.075, 1) {};
		\node [style=none] (38) at (8.15, 0.45) {};
		\node [style=none] (40) at (12.5, 0) {};
		\node [style=none] (41) at (5, 2) {};
		\node [style=none] (42) at (7.4, -0.35) {};
		\node [style=none] (43) at (5, -2) {};
		\node [style=none] (44) at (7.375, -0.725) {};
		\node [style=none] (45) at (7.75, -0.5) {};
		\node [style=none] (46) at (11, 0) {};
	\end{pgfonlayer}
	\begin{pgfonlayer}{edgelayer}
		\draw [style=op] (3) to (0.center);
		\draw [style=op] (1.center) to (3);
		\draw [style=op] (2.center) to (3);
		\draw [style=op] (29.center) to (30.center);
		\draw [style=op] (33.center) to (34.center);
		\draw [style=celestial, in=180, out=0] (35) to (24.center);
		\draw [style=celestial, in=165, out=0, looseness=1.50] (21.center) to (37.center);
		\draw [style=celestial, in=180, out=0] (22.center) to (38.center);
		\draw [style=op] (24.center) to (40.center);
		\draw [style=op] (25) to (21.center);
		\draw [style=op] (26) to (22.center);
		\draw [style=minkowski, in=120, out=0] (41.center) to (42.center);
		\draw [style=minkowski, in=-135, out=0] (43.center) to (44.center);
		\draw [style=minkowski, in=-180, out=0] (45.center) to (46.center);
	\end{pgfonlayer}
\end{tikzpicture}

%% file: figures/Lcontractedg.tikz
\begin{tikzpicture}
	\begin{pgfonlayer}{nodelayer}
		\node [style=none] (0) at (7.5, 0) {};
		\node [style=none] (4) at (7, 0.75) {$\mathcal{O}^L$};
		\node [style=none] (5) at (0.5, 2.25) {$1$};
		\node [style=none] (6) at (0.5, -2.25) {$2$};
		\node [style=none] (7) at (-5.5, 2) {};
		\node [style=none] (8) at (-5.5, -2) {};
		\node [style=none] (9) at (-6, 2) {$4$};
		\node [style=none] (10) at (-6, -2) {$3$};
		\node [style=threePoint] (14) at (3.5, 0) {$L_{\delta,j}$};
		\node [style=large circle] (15) at (-2.5, 0) {$g$};
	\end{pgfonlayer}
	\begin{pgfonlayer}{edgelayer}
		\draw [style=op, bend right, looseness=1.50] (15) to (14);
		\draw [style=op] (14) to (0.center);
		\draw [style=op] (15) to (7.center);
		\draw [style=op] (15) to (8.center);
		\draw [style=op, bend right=330, looseness=1.50] (15) to (14);
	\end{pgfonlayer}
\end{tikzpicture}

%% file: figures/LcontracteddDisc.tikz
\begin{tikzpicture}
	\begin{pgfonlayer}{nodelayer}
		\node [style=none] (0) at (7.5, 0) {};
		\node [style=none] (4) at (7, 0.75) {$\mathcal{O}^L$};
		\node [style=none] (5) at (0.5, 2.25) {$1$};
		\node [style=none] (6) at (0.5, -2.25) {$2$};
		\node [style=none] (7) at (-5.5, 2) {};
		\node [style=none] (8) at (-5.5, -2) {};
		\node [style=none] (9) at (-6, 2) {$4$};
		\node [style=none] (10) at (-6, -2) {$3$};
		\node [style=threePoint] (14) at (3.5, 0) {$L_{\delta,j}$};
		\node [style=medium circle] (15) at (-2.5, 0) {$\mathrm{dDisc}[g]$};
	\end{pgfonlayer}
	\begin{pgfonlayer}{edgelayer}
		\draw [style=op, bend right, looseness=1.50] (15) to (14);
		\draw [style=op] (14) to (0.center);
		\draw [style=op] (15) to (7.center);
		\draw [style=op] (15) to (8.center);
		\draw [style=op, bend right=330, looseness=1.50] (15) to (14);
	\end{pgfonlayer}
\end{tikzpicture}

%% file: figures/AdeltaFromPairing.tikz
\begin{tikzpicture}
	\begin{pgfonlayer}{nodelayer}
		\node [style=none] (4) at (7.55, 0.75) {$\mathcal{O}^L$};
		\node [style=none] (5) at (3.075, 2.25) {$1$};
		\node [style=none] (6) at (3.075, -2.25) {$2$};
		\node [style=none] (9) at (-1.925, 3.25) {$4$};
		\node [style=none] (10) at (-1.925, -3.25) {$3$};
		\node [style=threePoint] (14) at (5.825, 0) {$L_{\delta,j}$};
		\node [style=none] (21) at (7.825, 0) {};
		\node [style=none] (22) at (13.25, -1.925) {};
		\node [style=medium circle] (28) at (0.0749998, 0) {$\mathrm{dDisc}[g]$};
		\node [style=none] (29) at (13.25, 1.925) {};
		\node [style=threePoint] (30) at (11.325, 0) {$b$};
		\node [style=large green circle] (31) at (11.075, 0) {};
		\node [style=none] (32) at (14.075, 1.75) {\color{black!20!green}{$^{-1}$}};
		\node [style=s box] (34) at (9.525, 0) {$\mathbf{L}$};
	\end{pgfonlayer}
	\begin{pgfonlayer}{edgelayer}
		\draw [style=op] (14) to (21.center);
		\draw [style=op, in=45, out=150] (28) to (29.center);
		\draw [style=op] (30) to (29.center);
		\draw [style=op, bend right, looseness=1.50] (28) to (14);
		\draw [style=op, bend right=330, looseness=1.50] (28) to (14);
		\draw [style=op, in=-45, out=-150] (28) to (22.center);
		\draw [style=op] (30) to (22.center);
		\draw [style=op] (34) to (21.center);
		\draw [style=op] (30) to (34);
	\end{pgfonlayer}
\end{tikzpicture}

%% file: figures/BlockWithL.tikz
\begin{tikzpicture}
	\begin{pgfonlayer}{nodelayer}
		\node [style=none] (4) at (1.825, 0.75) {$\mathcal{O}^L$};
		\node [style=none] (5) at (-3.675, 3) {$1$};
		\node [style=none] (6) at (-3.675, -3) {$2$};
		\node [style=none] (9) at (14, 3) {$4$};
		\node [style=none] (10) at (14, -3) {$3$};
		\node [style=none] (21) at (7, 0) {};
		\node [style=none] (22) at (12.175, -1.925) {};
		\node [style=none] (29) at (12.175, 1.925) {};
		\node [style=threePoint] (30) at (10.25, 0) {$b$};
		\node [style=large green circle] (31) at (10, 0) {};
		\node [style=none] (32) at (13, 1.75) {\color{black!20!green}{$^{-1}$}};
		\node [style=s box] (34) at (8.45, 0) {$\mathbf{L}$};
		\node [style=none] (35) at (-3, 3) {};
		\node [style=threePoint] (36) at (0.075, 0) {$L_{\delta,j}$};
		\node [style=none] (37) at (-3, -3) {};
		\node [style=none] (38) at (13.25, 3) {};
		\node [style=none] (39) at (13.25, -3) {};
		\node [style=twoPoint] (40) at (3, 0) {};
		\node [style=none] (41) at (5, 0) {$\times$};
	\end{pgfonlayer}
	\begin{pgfonlayer}{edgelayer}
		\draw [style=op] (30) to (29.center);
		\draw [style=op] (30) to (22.center);
		\draw [style=op] (34) to (21.center);
		\draw [style=op] (30) to (34);
		\draw [style=op] (37.center) to (36);
		\draw [style=op] (35.center) to (36);
		\draw [style=op] (38.center) to (29.center);
		\draw [style=op] (39.center) to (22.center);
		\draw [style=op] (36) to (40);
		\draw [style=op] (41.center) to (40);
		\draw [style=op] (41.center) to (21.center);
	\end{pgfonlayer}
\end{tikzpicture}

%% file: figures/SofLKernelPairedWithStr.tikz
\begin{tikzpicture}
	\begin{pgfonlayer}{nodelayer}
		\node [style=none] (5) at (0, 3.975) {$1$};
		\node [style=none] (6) at (0, -0.025) {$2$};
		\node [style=twoPoint] (44) at (0, -2.525) {};
		\node [style=threePoint] (47) at (2.25, 1.975) {$L_{\delta,j}$};
		\node [style=threePoint] (53) at (-2.25, 1.975) {$a$};
		\node [style=none] (54) at (3.25, -0.525) {$\mathcal{O}^L$};
		\node [style=s box] (55) at (-3, -0.025) {$\mathbf{L}$};
	\end{pgfonlayer}
	\begin{pgfonlayer}{edgelayer}
		\draw [style=op, bend left=45, looseness=1.25] (47) to (44);
		\draw [style=op, bend right=45, looseness=1.25] (53) to (47);
		\draw [style=op, bend left=45, looseness=1.25] (53) to (47);
		\draw [style=op, bend right=15] (53) to (55);
		\draw [style=op, bend right] (55) to (44);
	\end{pgfonlayer}
\end{tikzpicture}

%% file: figures/LKernelPairedWithStrTwoPt.tikz
\begin{tikzpicture}
	\begin{pgfonlayer}{nodelayer}
		\node [style=none] (5) at (0, 1.975) {$1$};
		\node [style=none] (6) at (0, -2.025) {$2$};
		\node [style=threePoint] (47) at (2.25, -0.025) {$L_{\delta,j}$};
		\node [style=threePoint] (53) at (-2.25, -0.025) {$a$};
		\node [style=none] (54) at (4.25, 0.725) {$\mathcal{O}^L$};
		\node [style=s box] (55) at (-4, -0.025) {$\mathbf{L}$};
		\node [style=none] (56) at (-6, 0) {};
		\node [style=none] (57) at (5.5, 0) {};
		\node [style=none] (58) at (-5.5, 0.725) {$\mathcal{O}^{L\dagger}$};
	\end{pgfonlayer}
	\begin{pgfonlayer}{edgelayer}
		\draw [style=op, bend right=45, looseness=1.25] (53) to (47);
		\draw [style=op, bend left=45, looseness=1.25] (53) to (47);
		\draw [style=op] (47) to (57.center);
		\draw [style=op] (55) to (56.center);
		\draw [style=op] (53) to (55);
	\end{pgfonlayer}
\end{tikzpicture}

%% file: figures/twoPtFnWithDagger.tikz
\begin{tikzpicture}
	\begin{pgfonlayer}{nodelayer}
		\node [style=twoPoint] (0) at (0, 0) {};
		\node [style=none] (1) at (2, 0) {};
		\node [style=none] (2) at (-2, 0) {};
		\node [style=none] (3) at (1.5, 0.75) {$\mathcal{O}^{L}$};
		\node [style=none] (4) at (-1, 0.75) {$\mathcal{O}^{L\dagger}$};
	\end{pgfonlayer}
	\begin{pgfonlayer}{edgelayer}
		\draw [style=op] (1.center) to (0);
		\draw [style=op] (2.center) to (0);
	\end{pgfonlayer}
\end{tikzpicture}

%% file: figures/ReggeContour.tikz
\begin{tikzpicture}
	\begin{pgfonlayer}{nodelayer}
		\node [style=none] (0) at (-5, 0) {};
		\node [style=none] (1) at (5, 0) {};
		\node [style=none] (2) at (0, -5) {};
		\node [style=none] (3) at (0, 5) {};
		\node [style=pole] (4) at (0, 0) {};
		\node [style=pole] (5) at (1, 0) {};
		\node [style=pole] (6) at (2, 0) {};
		\node [style=pole] (7) at (3, 0) {};
		\node [style=pole] (8) at (4, 0) {};
		\node [style=none] (9) at (1, -0.25) {};
		\node [style=none] (10) at (1, 0.25) {};
		\node [style=none] (11) at (0, -0.25) {};
		\node [style=none] (12) at (0, 0.25) {};
		\node [style=none] (13) at (2, -0.25) {};
		\node [style=none] (14) at (2, 0.25) {};
		\node [style=none] (15) at (3, -0.25) {};
		\node [style=none] (16) at (3, 0.25) {};
		\node [style=none] (17) at (4, -0.25) {};
		\node [style=none] (18) at (4, 0.25) {};
		\node [style=none] (19) at (-0.5, -5) {};
		\node [style=none] (20) at (-0.5, 5) {};
		\node [style=none] (21) at (-0.5, 1) {};
		\node [style=none] (22) at (-1, 4) {${\color{red}\Gamma'}$};
		\node [style=none] (23) at (4.5, 0.75) {${\color{blue}\Gamma}$};
		\node [style=pole] (24) at (0.5, 2.75) {};
		\node [style=none] (25) at (0.5, 2.5) {};
		\node [style=none] (26) at (0.5, 3) {};
		\node [style=none] (27) at (1.5, 2.25) {$j(\nu)$};
		\node [style=pole] (28) at (-1, 0) {};
		\node [style=pole] (29) at (-2, 0) {};
		\node [style=pole] (30) at (-3, 0) {};
		\node [style=pole] (31) at (-4, 0) {};
		\node [style=none] (32) at (4, 4.5) {};
		\node [style=none] (33) at (4, 4) {};
		\node [style=none] (34) at (4.5, 4) {};
		\node [style=none] (35) at (4.4, 4.45) {$J$};
		\node [style=none] (36) at (0, -0.75) {$0$};
		\node [style=none] (37) at (1, -0.75) {$2$};
		\node [style=none] (38) at (2, -0.75) {$4$};
		\node [style=none] (39) at (3, -0.75) {$6$};
		\node [style=none] (40) at (4, -0.75) {$8$};
	\end{pgfonlayer}
	\begin{pgfonlayer}{edgelayer}
		\draw [style=axis] (2.center) to (3.center);
		\draw [style=axis] (0.center) to (1.center);
		\draw [style=Arrow, bend left=90, looseness=1.75] (10.center) to (9.center);
		\draw [style=anti arrow, bend right=90, looseness=1.75] (10.center) to (9.center);
		\draw [style=Arrow, bend left=90, looseness=1.75] (12.center) to (11.center);
		\draw [style=anti arrow, bend right=90, looseness=1.75] (12.center) to (11.center);
		\draw [style=Arrow, bend left=90, looseness=1.75] (14.center) to (13.center);
		\draw [style=anti arrow, bend right=90, looseness=1.75] (14.center) to (13.center);
		\draw [style=Arrow, bend left=90, looseness=1.75] (16.center) to (15.center);
		\draw [style=anti arrow, bend right=90, looseness=1.75] (16.center) to (15.center);
		\draw [style=Arrow, bend left=90, looseness=1.75] (18.center) to (17.center);
		\draw [style=anti arrow, bend right=90, looseness=1.75] (18.center) to (17.center);
		\draw [style={dashed_arrow}] (19.center) to (21.center);
		\draw [style={dashed_only}] (21.center) to (20.center);
		\draw [style={dashed_arrow}, bend right=90, looseness=1.75] (25.center) to (26.center);
		\draw [style={dashed_arrow}, bend right=90, looseness=1.75] (26.center) to (25.center);
		\draw (32.center) to (33.center);
		\draw (33.center) to (34.center);
	\end{pgfonlayer}
\end{tikzpicture}

%% file: figures/ReggeContourT.tikz
\begin{tikzpicture}
	\begin{pgfonlayer}{nodelayer}
		\node [style=none] (0) at (-5, 0) {};
		\node [style=none] (1) at (5, 0) {};
		\node [style=none] (2) at (0, -5) {};
		\node [style=none] (3) at (0, 5) {};
		\node [style=pole] (4) at (0.5, 0) {};
		\node [style=pole] (5) at (1.5, 0) {};
		\node [style=pole] (6) at (2.5, 0) {};
		\node [style=pole] (7) at (3.5, 0) {};
		\node [style=pole] (8) at (4.5, 0) {};
		\node [style=none] (9) at (1.5, -0.25) {};
		\node [style=none] (10) at (1.5, 0.25) {};
		\node [style=none] (11) at (0.5, -0.25) {};
		\node [style=none] (12) at (0.5, 0.25) {};
		\node [style=none] (13) at (2.5, -0.25) {};
		\node [style=none] (14) at (2.5, 0.25) {};
		\node [style=none] (15) at (3.5, -0.25) {};
		\node [style=none] (16) at (3.5, 0.25) {};
		\node [style=none] (17) at (4.5, -0.25) {};
		\node [style=none] (18) at (4.5, 0.25) {};
		\node [style=none] (19) at (-0.2, -5) {};
		\node [style=none] (20) at (-0.2, 5) {};
		\node [style=none] (21) at (-0.2, 1) {};
		\node [style=none] (22) at (-0.75, 4) {${\color{red}\Gamma'}$};
		\node [style=none] (23) at (4.5, 0.75) {${\color{blue}\Gamma}$};
		\node [style=pole] (24) at (0.5, 2.75) {};
		\node [style=none] (25) at (0.5, 2.5) {};
		\node [style=none] (26) at (0.5, 3) {};
		\node [style=none] (27) at (1.5, 2.25) {$j(\nu)$};
		\node [style=pole] (28) at (-0.5, 0) {};
		\node [style=pole] (29) at (-1.5, 0) {};
		\node [style=pole] (30) at (-2.5, 0) {};
		\node [style=pole] (31) at (-3.5, 0) {};
		\node [style=none] (32) at (0.5, -0.75) {};
		\node [style=none] (33) at (1.5, -0.75) {};
		\node [style=none] (34) at (2.5, -0.75) {};
		\node [style=none] (35) at (3.5, -0.75) {$7$};
		\node [style=none] (36) at (4.5, -0.75) {$9$};
		\node [style=none] (37) at (0.5, -0.75) {$1$};
		\node [style=none] (38) at (1.5, -0.75) {$3$};
		\node [style=none] (39) at (2.5, -0.75) {$5$};
		\node [style=none] (40) at (4, 4.5) {};
		\node [style=none] (41) at (4, 4) {};
		\node [style=none] (42) at (4.5, 4) {};
		\node [style=none] (43) at (4.4, 4.45) {$J$};
	\end{pgfonlayer}
	\begin{pgfonlayer}{edgelayer}
		\draw [style=axis] (2.center) to (3.center);
		\draw [style=axis] (0.center) to (1.center);
		\draw [style={dashed_anti}, bend left=90, looseness=1.75] (10.center) to (9.center);
		\draw [style={dashed_arrow}, bend right=90, looseness=1.75] (10.center) to (9.center);
		\draw [style={dashed_anti}, bend left=90, looseness=1.75] (12.center) to (11.center);
		\draw [style={dashed_arrow}, bend right=90, looseness=1.75] (12.center) to (11.center);
		\draw [style=Arrow, bend left=90, looseness=1.75] (14.center) to (13.center);
		\draw [style=anti arrow, bend right=90, looseness=1.75] (14.center) to (13.center);
		\draw [style=Arrow, bend left=90, looseness=1.75] (16.center) to (15.center);
		\draw [style=anti arrow, bend right=90, looseness=1.75] (16.center) to (15.center);
		\draw [style=Arrow, bend left=90, looseness=1.75] (18.center) to (17.center);
		\draw [style=anti arrow, bend right=90, looseness=1.75] (18.center) to (17.center);
		\draw [style={dashed_arrow}] (19.center) to (21.center);
		\draw [style={dashed_only}] (21.center) to (20.center);
		\draw [style={dashed_arrow}, bend right=90, looseness=1.75] (25.center) to (26.center);
		\draw [style={dashed_arrow}, bend right=90, looseness=1.75] (26.center) to (25.center);
		\draw (40.center) to (41.center);
		\draw (41.center) to (42.center);
	\end{pgfonlayer}
\end{tikzpicture}